\documentclass[12pt,a4paper,oneside]{book}

\usepackage[utf8]{inputenc}
\usepackage[T1]{fontenc}
\usepackage[english]{babel}   
\usepackage{setspace}         
\usepackage{geometry}         
\usepackage{anyfontsize}
\usepackage{fancyhdr}         
\usepackage[labelfont=bf]{caption}
\usepackage{enumitem}         
\usepackage{float}            
\usepackage{appendix}         
\usepackage{tcolorbox}        
\usepackage{xcolor}           
\usepackage{graphicx}         
\usepackage{longtable}
\usepackage{hyperref}         
\usepackage[nottoc]{tocbibind} 
\usepackage{csquotes}         
\usepackage{amsmath}          
\usepackage{amssymb}          
\usepackage{amsfonts}         
\usepackage{latexsym}         
\usepackage{amsthm}           
\usepackage{mathtools}        
\numberwithin{equation}{section}
\usepackage{bm}               
\usepackage{bbm}              
\usepackage{bbold}            
\usepackage{mathrsfs}
\usepackage{physics}          
\usepackage{braket}           
\usepackage{slashed}          
\usepackage{empheq}           
\usepackage{cancel}           
\usepackage{nicefrac}         
\usepackage{etoolbox}         
\usepackage{blindtext}        
\usepackage{tikz}
\usepackage{ragged2e}
\usetikzlibrary{decorations.pathmorphing}
\tikzset{snake it/.style={decorate, decoration=snake}}
\usetikzlibrary{shapes.geometric}
\usepackage[compat=1.1.0]{tikz-feynman}
\usepackage[compat=1.1.0]{tikz-feynhand}
\tikzfeynmanset{warn luatex=false}
\usepackage{comment}

\newcommand{\mathi}{{\rm i}}
\newcommand{\diff}{\mathrm{d}}
\newcommand{\must}{\mathrel{\stackrel{\makebox[0pt]{\mbox{\normalfont \small !}}}{=}}}
\def\tr{{\textrm{tr}}}
\def\Tr{{\textrm{Tr}}}

\def\la{\langle}
\def\ra{\rangle}
\def\r{\rangle}
\newcommand{\be}{\begin{equation}}
\newcommand{\ee}{\end{equation}}
\newcommand{\ba}{\begin{equation} \begin{aligned}}
\newcommand{\ea}{\end{aligned} \end{equation}}
\newcommand{\bea}{\begin{eqnarray}}
\newcommand{\eea}{\end{eqnarray}}

\newcommand{\C}{\mathcal{C}}
\newcommand{\B}{\mathcal{B}}
\newcommand{\Q}{\mathcal{Q}}
\newcommand{\J}{\mathcal{J}}

\newcommand{\balpha}{\bar{\alpha}}

\newcommand{\bL}{\bar{L}}
\newcommand{\bbeta}{\bar{\beta}}

\newcommand{\bC}{\bar{\mathcal{C}}}
\newcommand{\bB}{\bar{\mathcal{B}}}
\newcommand{\pa}{\partial}

\def\del{\Delta}
\def\ddel{{}^\bullet\! \Delta}
\def\deld{\Delta^{\hskip -.5mm \bullet}}
\def\dddel{{}^{\bullet \bullet} \! \Delta}
\def\ddeld{{}^{\bullet}\! \Delta^{\hskip -.5mm \bullet}}

\newcommand{\acr}[1]{%
  \begingroup
    \hypersetup{linkcolor=turquoise}%
    \hyperlink{abbrevlist}{\textcolor{turquoise}{#1}}%
  \endgroup
}
\def\la{\langle}
\def\ra{\rangle}
\def\cC{{\cal C}}
\def\cB{{\cal B}}

\def\cD{\mathcal{D}}
\def\cF{{\cal F}}
\def\cO{{\cal O}}
\def\cA{{\cal A}}

\def\cN{{\cal N}}

\def\cP{{\cal P}}

\def\cV{{\cal V}}

\def\del{\partial}
\def\l{\langle}
\def\r{\rangle}
\def\udot{\dot{\phantom{a}}\!}
\def\del{\Delta}
\def\ddel{{}^\bullet\! \Delta}
\def\deld{\Delta^{\hskip -.5mm \bullet}}
\def\dddel{{}^{\bullet \bullet} \! \Delta}
\def\ddeld{{}^{\bullet}\! \Delta^{\hskip -.5mm \bullet}}

\def\lpartial{ \overset{\leftarrow}{\slashed{\partial}}}
\def\rpartial{ \overset{\rightarrow}{\slashed{\partial}}}
\def \N{{\mathcal N}}
\def \J{{\mathcal J}}

\def \D{{\mathcal D}}
\def \gTr{{{\cal T}\!r}} 
\def \Y{{\mathcal Y}}
\def \gG{{\mathcal G}}
\def \g{{\mathfrak{g}}}
\newcommand{\gh}{\mathrm{gh}}

\def \q{{\mathfrak{q}}}
\def \A{{\mathcal A}}
\def \R{{\mathcal{R}}}
\def \Re{{\mathfrak{R}}}

\definecolor{jade}{RGB}{0, 168, 107}
\definecolor{turquoise}{RGB}{0, 138, 168}
\definecolor{MyBlue}{RGB}{14,41,75}
\definecolor{MyGold}{RGB}{255,200,80}
\hypersetup{
 colorlinks,
 linkcolor={jade},
 citecolor={jade},
 urlcolor={jade}
}
\makeatletter
\newcommand{\firstsectioneqnums}{%
  \renewcommand{\theequation}{\thesection.\arabic{equation}}%
  \@addtoreset{equation}{section}%
}
\newcommand{\restoreeqnums}{%
  \renewcommand{\theequation}{\thesubsection.\arabic{equation}}%
  \@addtoreset{equation}{subsection}%
}
\makeatother

\usepackage{tocloft}
\usepackage[backend=biber, citestyle=numeric-comp, bibstyle=ieee]{biblatex}
\def \diagramThree{
\begin{figure}[ht!]
\centering
\begin{tikzpicture}[scale=1.75]
    \node at (-0.5,0) {$\cA_{123}=$};
    \node at (0,0) {x};
    \node at (0,0.3) {$1$};
    \draw (0,0)--(4,0);
    \draw (2.5,1.25)--(2.5,0);
    \node at (2.3,1.25) {$2$};
    \node at (2.5,-0.1) {\small{$0$}};
    \node at (4,0) {x};
    \node at (4,0.3) {$3$};
\end{tikzpicture}
\caption{Diagrammatic representation of the color-ordered three-gluon amplitude $\cA_{123}$. The straight line denotes the insertion of an unintegrated vertex operator at $\tau=0$. The asymptotic endpoints of the worldline are depicted by a cross.}
\label{fig1}
\end{figure}
}
\def \diagramFourSmart{
\begin{figure}[ht!]
\centering
\begin{tikzpicture}[scale=1.75]
    \node at (-0.5,0) {$\cA_{1234}=$};
    \node at (0,0) {x};
    \node at (0,0.3) {$2$};
    \draw (0,0)--(4,0);
    \draw[snake it] (1.5,-1.25)--(1.5,0);
    \node at (1.3,-1.25) {$1$};
    \draw (2.5,1.25)--(2.5,0);
    \node at (2.3,1.25) {$3$};
    \node at (1.5,0.1) {\small{$\tau$}};
    \node at (2.5,-0.1) {\small{$0$}};
    \node at (4,0) {x};
    \node at (4,0.3) {$4$};
\end{tikzpicture}
\caption{Diagrammatic representation of the color-ordered four-gluon amplitude $\cA_{1234}$ with the smart choice for the worldline. The wavy line denotes the insertion of an integrated vertex operator, while, as before, the straight line denotes the one fixed by translation invariance.}
\label{fig2}
\end{figure}
}
\def \diagramFourDumb{
\begin{figure}[ht!]
\centering
\begin{tikzpicture}[scale=1]
    \node at (-1,0) {$\cA_{1234}=$};
    \node at (0,0) {x};
    \node at (0,0.3) {$1$};
    \draw (0,0)--(4,0);
    \draw[snake it] (1.5,1.25)--(1.5,0);
    \node at (1.3,1.25) {$2$};
    \draw (2.5,1.25)--(2.5,0);
    \node at (2.3,1.25) {$3$};
    \node at (1.5,-0.2) {\small{$\tau$}};
    \node at (2.5,-0.2) {\small{$0$}};
    \node at (4,0) {x};
    \node at (4,0.3) {$4$};
    \node at (4.5,0) {$+$};
    \node at (5,0) {x};
    \node at (5,0.3) {$1$};
    \draw (5,0)--(9,0);
    \draw (6.5,1.25)--(7,0)--(7.5,1.25);
    \node at (6.3,1.25) {$2$};
    \node at (7.7,1.25) {$3$};
    \node at (7,-0.2) {\small{$0$}};
    \node at (9,0) {x};
    \node at (9,0.3) {$4$};
    \node at (9.5,0) {$+$};
    \node at (10,0) {x};
    \node at (10,0.3) {$1$};
    \draw (10,0)--(14,0);
    \draw (12,0.7)--(12,0);
    \draw (11.5,1.25)--(12,0.7)--(12.5,1.25);
    \node at (11.3,1.25) {$2$};
    \node at (12.7,1.25) {$3$};
    \node at (12,-0.2) {\small{$0$}};
    \node at (14,0) {x};
    \node at (14,0.3) {$4$};
\end{tikzpicture}
\caption{Diagrammatic representation of the color-ordered four-gluon amplitude $\cA_{1234}$ with the not-so-smart choice for the worldline. As before, the wavy line denotes the insertion of an integrated vertex operator, while the straight lines denote the insertion of unintegrated vertex operators at $\tau=0$.}
\label{fig3}
\end{figure}
}
\def \ione{ 
 \begin{tikzpicture}[baseline=-\the\dimexpr\fontdimen22\textfont2\relax]
 \begin{feynhand}
 \vertex [dot] (b) at (0,0) {};
 \vertex[dot](c)at(1.4,0){};
 \draw[line width=.5mm] (b) arc [start angle=180, end angle=-180, radius=.7cm];
 \draw[line width=.5mm] (b) arc [start angle=180, end angle=-180, radius=-.7cm];
 \propag[fer] (b) to [in=90, out=90,looseness=.5] (c);
 \propag[fer] (c) to [in=-90, out=-90,looseness=.5] (b);
 \vertex [ringdot] (d) at (.2,.5) {};
 \vertex [ringdot] (d) at (1.2,.5) {};
 \end{feynhand}
 \end{tikzpicture}
 }
\def \itwo{ 
 \begin{tikzpicture}[baseline=-\the\dimexpr\fontdimen22\textfont2\relax]
 \begin{feynhand}
 \vertex [dot] (b) at (0,0) {};
 \vertex[dot](c)at(1.4,0){};
 \draw[line width=.5mm] (b) arc [start angle=180, end angle=-180, radius=.7cm];
 \draw[line width=.5mm] (b) arc [start angle=180, end angle=-180, radius=-.7cm];
 \propag[fer] (b) to [in=90, out=90,looseness=.5] (c);
 \propag[fer] (c) to [in=-90, out=-90,looseness=.5] (b);
 \vertex [ringdot] (d) at (.2,.5) {};
 \vertex [ringdot] (d) at (1.2,-.5) {};
 \end{feynhand}
 \end{tikzpicture}
 }
\def \disc{ 
 \begin{tikzpicture}[baseline=-\the\dimexpr\fontdimen22\textfont2\relax]
 \begin{feynhand}
 \vertex [dot] (b) at (0,0) {};
 \vertex [dot] (e) at (1.8,0) {};
 \vertex[dot](c)at(3.2,0){};
 \draw[line width=.5mm] (b) arc [start angle=180, end angle=-180, radius=-.7cm];
 \draw[line width=.5mm] (1.4,0) arc [start angle=-180, end angle=180, radius=-.7cm];
 \propag[fer] (e) to [in=90, out=90,looseness=.5] (c);
 \propag[fer] (c) to [in=-90, out=-90,looseness=.5] (e);
 \propag[fer] (c) to [in=90, out=90,looseness=1.5] (e);
 \propag[fer] (e) to [in=-90, out=-90,looseness=1.5] (c);
 \vertex [ringdot] (d) at (.2,-.5) {};
 \vertex [ringdot] (d) at (-.2,.5) {};
 \end{feynhand}
 \end{tikzpicture}
 }
\def \pair{
\begin{tikzpicture}[>=stealth, thick, scale=1.2]
\draw[->, dashed] (90:1) arc (90:0:1);
\draw[->, dashed] (90:1) arc (90:180:1);
\draw[dashed] (0:1) arc (0:-180:1);
\node[left] at (180:1.2) {$e^{-}$};
\node[right] at (0:1.2) {$e^{+}$};
\draw[->, line width=1pt] (2,0) -- (3.5,0);
\draw[->] (4,0.25) -- (7,0.25);
\node[above] at (5.5,0.45) {$\vec{E}$};
\fill (5.6,-0.25) circle (1pt);
\fill (5.4,-0.25) circle (1pt);
\draw[dashed, ->] (5.6,-0.25) -- (7,-0.25);   
\draw[dashed, ->] (5.4,-0.25) -- (4,-0.25);   
\node[below] at (4,-0.25) {$e^{-}$};
\node[below] at (7,-0.25) {$e^{+}$};
\end{tikzpicture}
} 
\def \EH{
\begin{tikzpicture}[thick, scale=1.4]
  \tikzset{photon/.style={decorate, decoration={snake, amplitude=2pt, segment length=7pt}}}
  \node at (-1.5,0) {$\mathcal{L}_{\mathrm{EH}} = -\frac{1}{4}F_{\mu\nu} F^{\mu\nu} \phantom{+}+$};
  \draw (0.5,0) circle (0.5);
  \node at (1.25,0) {$+$};
  \begin{scope}[xshift=2cm]
    \draw (0,0) circle (0.5);
    \draw[photon] (0,-0.5) -- (0,-1);
  \end{scope}
  \node at (2.75,0) {$+$};
    \begin{scope}[xshift=3.75cm]
    \draw (0,0) circle (0.5);
    \draw[photon] (-0.35,-0.35) -- (-0.7,-0.7);
    \draw[photon] (0.35,0.35) -- (0.7,0.7);
  \end{scope}
  \node at (4.75,0) {$+$};
    \begin{scope}[xshift=5.75cm]
    \draw (0,0) circle (0.5);
    \draw[photon] (0,0.5) -- (0,1);
    \draw[photon] (-0.43,-0.25) -- (-0.86,-0.5);
    \draw[photon] (0.43,-0.25) -- (0.86,-0.5);
  \end{scope}
  \node at (6.75,0) {$+$};
    \begin{scope}[xshift=7.75cm]
    \draw (0,0) circle (0.5);
    \foreach \angle in {45,135,225,315}{
      \draw[photon] ({0.5*cos(\angle)},{0.5*sin(\angle)}) --
                    ({0.9*cos(\angle)},{0.9*sin(\angle)});
    }
  \end{scope}
  \node at (9,0) {$+\ \dots$};
\end{tikzpicture}
}

\begin{document}

\begin{titlepage}   
\begin{tikzpicture}[remember picture,overlay]
    \fill[MyBlue] (current page.south west) rectangle (current page.north east);
\end{tikzpicture}
\color{white}
    \begin{center}
        \includegraphics[width=5cm]{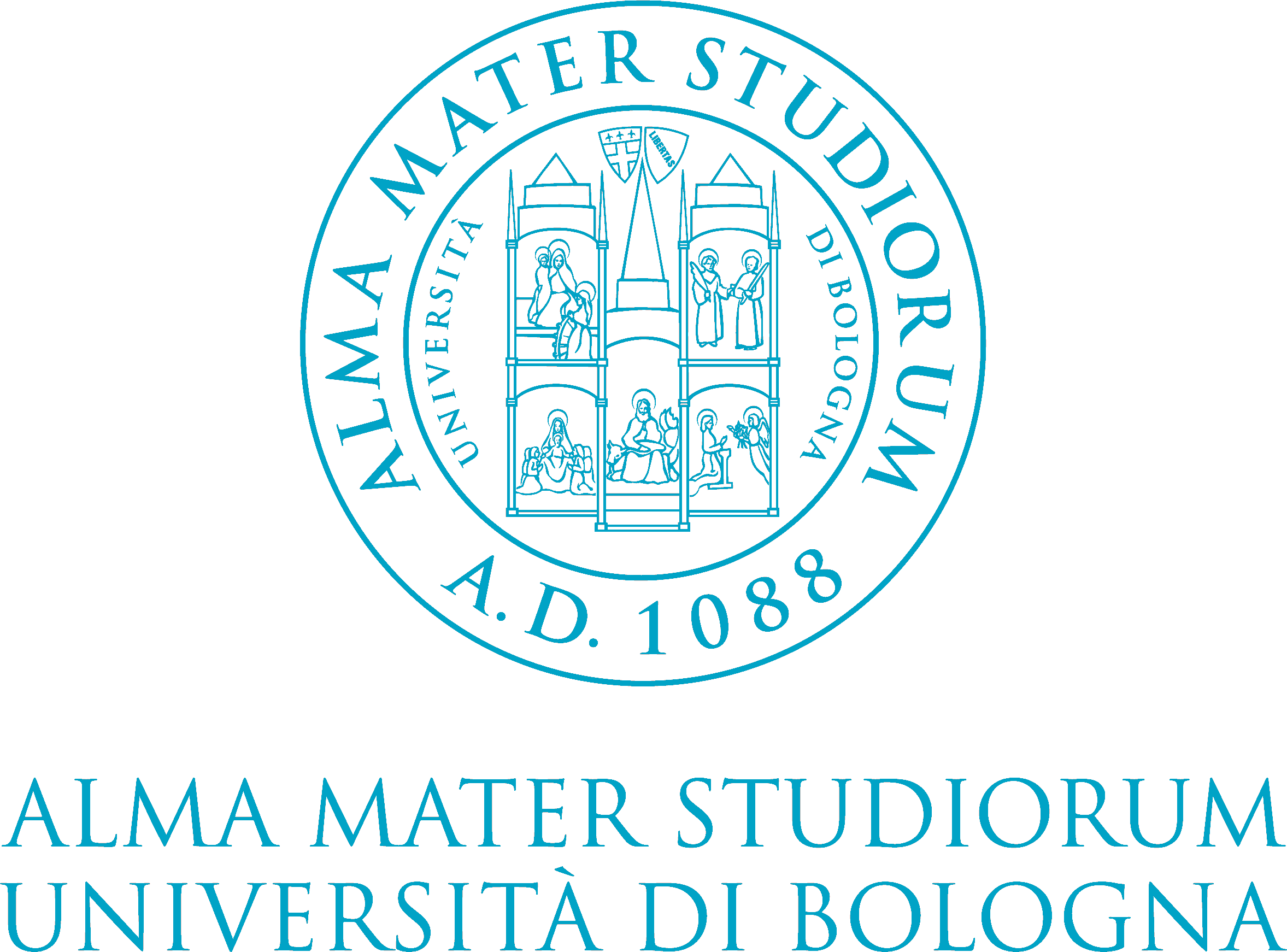}\\[1.5cm]      
{\Large \textbf{DOTTORATO DI RICERCA IN FISICA}}\\[0.2cm]
{\large ciclo XXXVIII}\\[1cm]
\end{center}  

\begin{flushleft}
    {\small 
    \textbf{Settore Concorsuale:} 02/A2 - Fisica Teorica delle Interazioni Fondamentali \\
    \textbf{Settore Scientifico Disciplinare:} FIS/02 - Fisica Teorica, Modelli e Metodi Matematici}
\end{flushleft}
\vspace{1.5cm}

\begin{center}
    {\fontsize{26}{30}\selectfont \bfseries 
    \color{MyGold}Advances in the Worldline Approach to Quantum Field Theory: Strong Fields, Amplitudes and Gravity}\\[2cm]
\end{center}
\vspace{0.8cm}

\begin{flushleft}
    {\Large Presentata da: \textbf{Filippo Fecit}}\\[2cm]
    {\Large Coordinatore Dottorato: \textbf{Prof. Alessandro Gabrielli}}\\[1.5cm]
\end{flushleft}

\begin{flushright}
    {\Large Supervisore: \textbf{Prof. Fiorenzo Bastianelli}}\\[0.2cm]
    {\Large Co-Supervisore: \textbf{Dr. Sebastián Alberto Franchino-\kern-0.25emVi\~nas}}
\end{flushright}

\vfill
\begin{center}
    {\large Esame finale anno 2026}
\end{center}

\end{titlepage}

\newpage
\thispagestyle{empty}
\mbox{}
\newpage

\newpage
\thispagestyle{empty} 
\vspace*{2cm}          
\hfill
\begin{minipage}{0.5\textwidth} 
    \raggedleft 
    \textit{To my wife,\\
    co-author of the greatest\\
    chapters of my life.}
\end{minipage}
\newpage

\newpage
\thispagestyle{empty}
\mbox{}
\newpage

\newpage
\thispagestyle{empty} 
\vspace*{2cm}          
\hfill
\begin{minipage}{0.5\textwidth} 
    \raggedleft 
    \textit{I will work on your 2$\,^{nd}$ quantization,\\when I get time.\\I really believe it is correct\\but is another example\\of the terrifying power of math\\
    to make us say things which we don’t understand\\but are true.}\\
    \vspace{12pt}
    \noindent\hrulefill\\
    Richard~Phillips~Feynman,\\
    Letter to Theodore~A.~Welton, 10 February 1947.
\end{minipage}

\clearpage
\pagenumbering{gobble}  
\null                   
\newpage

\pagenumbering{roman}  
\setcounter{page}{0}
\chapter*{\phantom{a} \hfill Abstract \hfill \phantom{a}}
\addcontentsline{toc}{chapter}{Abstract}
\begin{center}
\begin{minipage}{0.85\textwidth} 
\justifying
\noindent
This thesis is devoted to the first-quantized approach to quantum field theory, commonly known as the “Worldline Formalism”. It collects most of the works completed by the author during the PhD, illustrating the versatility and efficiency of this formalism across a broad range of physical contexts. The applications discussed fall into two broad categories: perturbative and non-perturbative analyses. In particular, the thesis investigates how quantum particles interact with strong background fields, how scattering amplitudes can be efficiently computed, and how perturbative expansions of the heat kernel can be systematically performed. These studies highlight recent advances in extending the worldline approach to increasingly complex situations. Different field theories are examined from this first-quantized perspective and are organized according to the spin of the particle under consideration. The discussion begins with the seemingly simple scalar case, progresses through spin 1, both in the abelian and the non-abelian case, and concludes with spin-2 particles, both in the massless and in the massive realizations. Through this sequence of examples, the thesis aims to demonstrate the flexibility as well as the computational power of the Worldline Formalism in addressing fundamental open problems in theoretical physics.
\end{minipage}
\end{center}

\newpage
\thispagestyle{empty}
\mbox{}
\newpage

\cleardoublepage
\tableofcontents
\cleardoublepage

\newpage
\thispagestyle{empty}
\mbox{}
\newpage

\chapter*{List of Publications}
\fancyhead[L]{} 
\fancyhead[R]{\thepage} 
\addcontentsline{toc}{chapter}{List of Publications}

This thesis is based on the following publications:\vspace{-0.4cm}

\begin{itemize}
    \item[\cite{Bastianelli:2023oca}] F.~Bastianelli, F.~Comberiati, F.~Fecit and F.~Ori,\\
    \textit{Six-dimensional one-loop divergences in quantum gravity from the $ \mathcal{N} $ = 4 spinning particle},\\
    JHEP \textbf{10} (2023), p. 152. \textnormal{DOI: }\href{https://doi.org/10.1007/JHEP10(2023)152}{10.1007/JHEP10(2023)152},\\
    $\text{[arXiv: \href{https://arxiv.org/abs/2307.09353}{2307.09353 [hep-th]}]}$ (Chapter~\ref{chap:sixth}).
    \item[\cite{Fecit:2023kah}] F.~Fecit,\\
    \textit{Massive gravity from a first-quantized perspective},\\
    Eur. Phys. J. C \textbf{84} (2024) no.4, p.445.\textnormal{DOI: }\href{https://doi.org/10.1140/epjc/s10052-024-12799-2}{10.1140/epjc/s10052-024-12799-2},\\
    $\text{[arXiv: \href{https://arxiv.org/abs/2312.15428}{2312.15428 [hep-th]}]}$ (Chapter~\ref{chap:fourth}).
    \item[\cite{Fecit:2024jcv}] F.~Fecit,\\
    \textit{Worldline path integral for the massive graviton},\\
    Eur. Phys. J. C \textbf{84} (2024) no.3, p.339.\textnormal{DOI: }\href{https://doi.org/10.1140/epjc/s10052-024-12707-8}{10.1140/epjc/s10052-024-12707-8},\\
    $\text{[arXiv: \href{https://arxiv.org/abs/2402.13766}{2402.13766 [hep-th]}]}$ (Chapter~\ref{chap:fifth}).
    \item[\cite{Fecit:2025kqb}] F.~Fecit, S.~A.~Franchino-Vi{\~n}as and F.~D.~Mazzitelli,\\
    \textit{Resummed effective actions and heat kernels: the Worldline approach and Yukawa assisted pair creation},\\
    JHEP \textbf{07} (2025), p. 041. \textnormal{DOI: }\href{https://doi.org/10.1007/JHEP07(2025)041}{10.1007/JHEP07(2025)041},\\
   $\text{[arXiv: \href{https://arxiv.org/abs/2501.17094}{2501.17094 [hep-th]}]}$ (Chapter~\ref{chap:first}).
    \item[\cite{Bastianelli:2025khx}] F.~Bastianelli, F.~Fecit and A.~Miccich{\`e},\\
    \textit{Pair production of massive charged vector bosons from the worldline},\\
    JHEP \textbf{09} (2025), p. 201. \textnormal{DOI: }\href{https://doi.org/10.1007/JHEP09(2025)201}{10.1007/JHEP09(2025)201},\\
    $\text{[arXiv: \href{https://arxiv.org/abs/2507.15943}{2507.15943 [hep-th]}]}$ (Chapter~\ref{chap:second}).
    \item[\cite{Bastianelli:2025xne}] F.~Bastianelli, R.~Bonezzi, O.~Corradini and F.~Fecit,\\
    \textit{Gluon amplitudes in first quantization},\\
    Phys. Rev. D \textbf{112} (2025) no.10, p.105015.\textnormal{DOI: }\href{https://doi.org/10.1103/krjy-csy2}{10.1103/krjy-csy2},\\
    $\text{[arXiv: \href{https://arxiv.org/abs/2508.05486}{2508.05486 [hep-th]}]}$ (Chapter~\ref{chap:third}).
\end{itemize}

Other publications by the same author:\vspace{-0.4cm}
\begin{itemize}
    \item[\cite{Bertolini:2022sao}] E.~Bertolini, F.~Fecit and N.~Maggiore,\\
    \textit{Topological BF Description of 2D Accelerated Chiral Edge Modes},\\
    Symmetry \textbf{14} (2022) no.4, p. 675. \textnormal{DOI: }\href{https://doi.org/10.3390/sym14040675}{10.3390/sym14040675},\\
    $\text{[arXiv: \href{https://arxiv.org/abs/2203.13520}{2203.13520 [hep-th]}]}$.
    \item[\cite{Farolfi:2025knq}] L.~Farolfi and F.~Fecit,\\
    \textit{The Fierz{\textendash}Pauli theory on curved spacetime at one-loop and its counterterms},\\
    Eur. Phys. J. C \textbf{85} (2025) no.3, p.356.\textnormal{DOI: }\href{https://doi.org/10.1140/epjc/s10052-025-14083-3}{10.1140/epjc/s10052-025-14083-3},\\
    $\text{[arXiv: \href{https://arxiv.org/abs/2503.11304}{2503.11304 [hep-th]}]}$.
    \item[\cite{Fecit:2025eet}] F.~Fecit and D.~Rovere,\\
    \textit{Worldline Formulations of Covariant Fracton Theories},\\
    Eur. Phys. J. C \textbf{86} (2026) no.3, p.206.\textnormal{DOI: }\href{https://doi.org/10.1140/epjc/s10052-026-15420-w}{10.1140/epjc/s10052-026-15420-w},\\
    $\text{[arXiv: \href{https://arxiv.org/abs/2508.14591}{2508.14591 [hep-th]}]}$.
\end{itemize}

\newpage
\thispagestyle{empty}
\mbox{}
\newpage

\phantomsection
\chapter*{List of Abbreviations}
\hypertarget{abbrevlist}{} 
\begin{longtable}{@{}p{3cm}p{12cm}@{}}
\textcolor{turquoise}{\textbf{QFT}} & Quantum Field Theory \\[4pt]
\textcolor{turquoise}{\textbf{EM}} & Electromagnetic \\[4pt]
\textcolor{turquoise}{\textbf{FP}} & Fierz--Pauli \\[4pt]
\textcolor{turquoise}{\textbf{QED}} & Quantum Electrodynamics \\[4pt]
\textcolor{turquoise}{\textbf{sQED}} & Scalar Quantum Electrodynamics \\[4pt]
\textcolor{turquoise}{\textbf{WQFT}} & Worldline Quantum Field Theory \\[4pt]
\textcolor{turquoise}{\textbf{SUSY}} & Supersymmetry \\[4pt]
\textcolor{turquoise}{\textbf{BRST}} & Becchi--Rouet--Stora--Tyutin \\[4pt]
\textcolor{turquoise}{\textbf{UV}} & Ultraviolet \\[4pt]
\textcolor{turquoise}{\textbf{IR}} & Infrared \\[4pt]
\textcolor{turquoise}{\textbf{BC}} & Boundary Conditions \\[4pt]
\textcolor{turquoise}{\textbf{D}} & Dirichlet \\[4pt]
\textcolor{turquoise}{\textbf{DN}} & Dirichlet--Neumann \\[4pt]
\textcolor{turquoise}{\textbf{SI}} & String-inspired \\[4pt]
\textcolor{turquoise}{\textbf{P}} & Periodic \\[4pt]
\textcolor{turquoise}{\textbf{A}} & Antiperiodic \\[4pt]
\textcolor{turquoise}{\textbf{GY}} & Gel’fand--Yaglom \\[4pt]
\textcolor{turquoise}{\textbf{LQFA}} & Locally Quadratic Field Approximation \\[4pt]
\textcolor{turquoise}{\textbf{LHS}} & Left Hand Side \\[4pt]
\textcolor{turquoise}{\textbf{RHS}} & Right Hand Side \\[4pt]
\textcolor{turquoise}{\textbf{CS}} & Chern--Simons \\[4pt]
\textcolor{turquoise}{\textbf{$\mathbf{\Phi\Pi}$}} & Faddeev--Popov \\[4pt]
\textcolor{turquoise}{\textbf{BV}} & Batalin--Vilkovisky \\[4pt]
\textcolor{turquoise}{\textbf{vDVZ}} & van Dam--Veltman--Zakharov \\[4pt]
\textcolor{turquoise}{\textbf{LMG}} & Linearized Massive Gravity \\[4pt]
\textcolor{turquoise}{\textbf{OPE}} & Operator Product Expansion \\[4pt]
\textcolor{turquoise}{\textbf{DR}} & Dimensional Regularization (on the worldline)
\end{longtable}

\newpage
\thispagestyle{empty}
\mbox{}
\newpage


\addcontentsline{toc}{part}{\hyperref[chap:intro]{Introduction and Motivations}}
\phantomsection
\chapter*{Introduction and Motivations}\label{chap:intro}
Second-quantized methods are widely employed in almost every branch of theoretical physics. Courses and textbooks on Quantum Field Theory (\acr{QFT}) are now a fundamental part of the training of any theoretical physicist. Nevertheless, exploring a subject from a different perspective often proves fruitful, offering new conceptual insights and, possibly, computational advantages. Sometimes, realizing such benefits may require taking a step back; for the sake of the present discussion, this means moving from second quantization back to first quantization. In other words, the subject of this thesis is the Worldline Formalism: in a nutshell, the framework that reformulates quantum field theory in terms of the dynamics of point particles propagating along worldlines, rather than fields defined over spacetime. Originally inspired by string-theoretic settings, the formalism has been systematically developed and applied in its modern form since the seminal works of Matthew~J.~Strassler \cite{Strassler:1992zr, Strassler:1993km}, although its conceptual roots trace back much further in time, as will be discussed in due course. Over the years, it has evolved into a versatile framework that, in certain contexts, even surpasses traditional field-theoretical techniques in efficiency.
The aim of this PhD thesis is to illustrate and substantiate these claims. It collects the author’s contributions to the worldline framework, ranging from elucidating non-trivial applications to proposing new worldline descriptions of physical phenomena.

The structure of the thesis can be viewed from two complementary perspectives.\\
On the one hand, it is naturally divided into two parts. The first part concerns the application of the Worldline Formalism to non-perturbative settings, specifically, to the study of the so-called “Schwinger effect”. The second part, by contrast, focuses on perturbative physics, covering the calculation of scattering amplitudes in Yang--Mills theory and the computation of heat kernel coefficients in (modified) gravity theories. Each part is preceded by a concise prelude that provides the necessary physical and theoretical background, together with essential references for further deepening the subjects.\\
On the other hand, the organization of the thesis may be understood following a Srednicki-like~\cite{Srednicki:2007qs} division according to the spin of the particle described by the worldline model, expressed in natural units. From this viewpoint, the work is arranged into three sections corresponding to different spin sectors.

Regardless of the chosen perspective on the structure of the thesis, a short overview of the content of each chapter is now in order. The \hyperref[chap:WF]{next chapter} offers a concise historical introduction to the Worldline Formalism, tracing its evolution and mentioning its main areas of current interest. It also discusses a few technical aspects, such as the distinction between the “bottom-up” and “top-down” approaches (as referred to by the author), and reviews the structure and quantization of a class of worldline models that will be used throughout the thesis.
\hyperref[partI]{Part~I} begins with a brief overview of the Schwinger effect and of the assisted pair production mechanism. Chapter~\ref{chap:first} focuses on the case-study of a quantum scalar particle in a classical scalar (Yukawa) background, where the pair production probability is computed, further studying how the presence of an additional rapidly varying background field affects the result. Chapter~\ref{chap:second} extends this analysis to the more realistic case of a pair of massive charged vector bosons created from the vacuum by a classical electromagnetic (\acr{EM}) background.
\hyperref[partII]{Part~II} then turns to perturbative applications. It begins by reviewing the field-theoretical setting of the heat kernel and its perturbative expansion, briefly introducing the theory of massive gravity and the issue of one-loop divergences. Chapter~\ref{chap:third} presents the computation of tree-level gluon amplitudes within the worldline approach. Chapter~\ref{chap:fourth} introduces the BRST quantization of the $\mathcal{N}=4$ spinning particle, which serves as the foundation for the computations in the subsequent two chapters. Chapter~\ref{chap:sixth} employs this model to derive the Seeley-DeWitt coefficients of perturbative quantum gravity, while Chapter~\ref{chap:fifth} extends the analysis to the Fierz--Pauli (\acr{FP}) theory on curved background.
Each chapter is accompanied at the end by a short concluding section, summarizing the main results and discussing possible directions for future developments.

\newpage
\thispagestyle{empty}
\mbox{}
\newpage
\mainmatter

\addcontentsline{toc}{part}{\hyperref[chap:WF]{The Worldline Formalism}}
\phantomsection
\chapter*{The Worldline Formalism}\label{chap:WF}
\paragraph{Concise historical context}
Dating the precise birth of the Worldline Formalism is subtle, as its development involved contributions from multiple authors over several years. Nevertheless, it is widely accepted within the worldline community that its first conceptual seeds can be traced back to around 1950. While some may argue that Julian~S.~Schwinger was the first to employ an embryonic first-quantized formulation in his seminal work \cite{Schwinger:1951nm}, there can be little doubt that the pioneering insights of Richard~P.~Feynman were of monumental importance, decisively bringing what would later be recognized as the Worldline Formalism into the theoretical landscape. Already in his 1942 PhD dissertation, Feynman developed a worldline action approach to quantum mechanics \cite{Feynman:1942us}. A few years later, his efforts culminated in a series of papers \cite{Feynman:1949hz, Feynman:1949zx, Feynman:1950ir, Feynman:1951gn} that laid the foundations of what we now know as Quantum Electrodynamics (\acr{QED}). In the meantime, however, he also introduced a path integral approach. In particular, Eq.~(11.A) of \cite{Feynman:1950ir} expresses a \acr{QFT} amplitude in terms of relativistic particle path integrals: the representation of the S-matrix of Scalar Quantum Electrodynamics (\acr{sQED}), namely the amplitude for a charged scalar particle moving, under the influence of the external potential $A_\mu(x)$, from point $x_\mu$ to $x_\mu'$ in Minkowski space, is there represented as
\begin{align}\tag{\color{jade}$\bigstar$\color{black}}\label{Feynman}
\begin{split}
        \int_{0}^{\infty} \diff T \int_{x(0)=x}^{x(T)=x'} \mathcal{D}x(\tau) \exp\bigg\{ -\frac{i}{2} \int_{0}^{T} \diff \tau\; \left( \frac{\diff x_\mu}{\diff \tau}\right)^2 -i\int_{0}^{T} \diff \tau\;\frac{\diff x_\mu}{\diff \tau} A^\mu (x(\tau))& \\ -\frac{i}{2} e^2 \int_{0}^{T} \diff \tau \int_{0}^{T} \diff \tau' \; \frac{\diff x_\mu}{\diff \tau}\frac{\diff x_\nu}{\diff \tau'} \mathcal{P}^{\mu\nu}\left[ x(\tau)-x(\tau') \right]  \bigg\}\;,
\end{split}
\end{align}
where $T$ is the so-called “Schwinger proper time” and $\mathcal{P}^{\mu\nu}$ denotes the photon propagator in configuration space. Curiously, Feynman relegated this formulation to the appendix of \cite{Feynman:1950ir}, describing it in the abstract as:
\begin{center}
“\textit{A description, in Lagrangian quantum-mechanical form, of particles satisfying the Klein-Gordon equation} [...]. \textit{It involves the use of an extra parameter analogous to proper time to describe the trajectory of the particle in four dimensions}”.
\end{center}
This suggests that Feynman himself did not view this approach as particularly promising at the time. His difficulties in developing a relativistic worldline path integral for spin-$\tfrac12$ particles, so as to include electrons and positrons,\footnote{Expressed also in a private correspondence with T.~A.~Welton \cite{letter}, later transcribed in \cite{Jacobson:2024ydt}.} may have contributed. As he later recalled in his Nobel Lecture \cite{Nobel}:
\\
\begin{center}
“\textit{Another problem on which I struggled very hard, was to represent relativistic electrons with this new quantum mechanics.} [...] \textit{And, so I dreamed that if I were clever, I would find a formula for the amplitude of a path that was beautiful and simple for three dimensions of space and one of time,} [...] \textit{I did want to mention some of the unsuccessful things on which I spent almost as much effort, as on the things that did work}”.
\end{center}
A possible candidate year for the “birth” of the modern Worldline Formalism -- or perhaps better, a “second revolution”, borrowing terminology from string theory -- can be identified around 1992. This is not to say that work on the formalism lay dormant in the intervening decades: indeed, numerous studies of relativistic particle path integrals in quantum field theory were pursued between the 1950s and 1980s, although most of these works remained conceptual rather than computational in character. 
The first hints that first-quantized techniques might offer computational advantages over the standard Feynman diagrammatic approach appeared in the early 1980s. A key milestone was the seminal work of Affleck, Alvarez, and Manton in 1982 \cite{Affleck:1981bma}, which was soon followed by remarkable applications to anomaly computations, most notably in the works of Alvarez-Gaumé and Witten \cite{Alvarez-Gaume:1983ihn} and of Friedan and Windey \cite{Friedan:1983xr}.\footnote{It may be arghued that much of the literature on heat kernel techniques in \acr{QFT} may be characterized as worldline methods as well \cite{new-book}.}

A genuine turning point, however, came from string theory, as the celebrated series of papers by Bern and Kosower \cite{Bern:1990cu, Bern:1990ux, Bern:1991an, Bern:1991aq} marked the beginning of a new phase.\footnote{Several earlier works would also merit mention: for instance, the computation of the one-loop four-gluon amplitude in $N=4$ Super Yang–Mills theory from the low-energy limit of superstring theory by Green, Schwarz, and Brink \cite{Green:1982sw}. However, in keeping with the promise of conciseness made in the introduction, we shall refrain from expanding on these here.} As is brilliantly detailed in Bern's lectures \cite{Bern:1992ad}, their goal was to systematically extract the field-theory limit ($\alpha' \to 0$, i.e. the infinite tension limit of the string) of string amplitudes, in particular to obtain the on-shell N-gluon amplitudes of Yang–Mills theory at tree and one-loop levels. The result was a remarkably compact and elegant set of parameter-integral representations that captured, in a unified manner, the contributions of entire classes of Feynman diagrams:\footnote{Here $k_i$ and $\varepsilon_i$ denote, respectively, the momentum and polarization vector of the $i$th external gluon.}
\begin{align*}
\mathcal{A}^{a_1 \dots a_N}[k_1, \varepsilon_1; \ldots; k_N, \varepsilon_N]
&= (-ig)^N \, \mathrm{tr}(T^{a_1} \ldots T^{a_N}) \,
(2\pi)^D \, \delta\,(\sum_i k_i) \\
&\phantom{=} \times
\int_0^{\infty} \diff T \, (4\pi T)^{-\nicefrac{D}{2}} \, \mathrm{e}^{-m^2 T}
\int_0^T \diff \tau_1
\cdots
\int_0^{\tau_{N-2}} \diff \tau_{N-1} \\
&\phantom{=} \times
\exp\!\left\{
\sum_{i,j=1}^{N}
\left[
\tfrac{1}{2} G^{\mathrm{B}}_{ij} \, k_i \!\cdot\! k_j
- i \dot{G}^{\mathrm{B}}_{ij} \, \varepsilon_i \!\cdot\! k_j
+ \tfrac{1}{2} \ddot{G}^{\mathrm{B}}_{ij} \, \varepsilon_i \!\cdot\! \varepsilon_j
\right]
\right\}
\Bigg|_{\text{multi-linear}}\;.
\end{align*}
This expression is an example of a “Bern-Kosower master formula”, and provides a closed representation for the color-ordered amplitude for N-gluons with a scalar in the loop. It organizes the result essentially into products of (derivatives of) bosonic worldline Green functions $G^{\mathrm{B}}(\tau_i,\tau_j)$, integrated over N-1 proper-time parameters (see \cite{Bern:1992ad, Schubert:2001he} for details). Even more remarkably, Bern and Kosower were able to dispense entirely with string theory, deriving instead a set of rules that allows one to construct these parameter integrals directly within field theory, for any number of gluons and choice of helicities.\footnote{A corresponding construction for graviton scattering from closed-string amplitudes was later developed by Bern, Dunbar, and Shimada \cite{Bern:1993wt}, as reviewed in \cite{Edwards:2022qiw}.}  These are nowadays known as “Bern-Kosower rules”, and their formulation suggested that there should have been a purely field-theoretic way of getting them. 
It was in this context that the seminal contribution of a PhD student, Matthew~J.~Strassler, appeared in  \cite{Strassler:1992zr, Strassler:1993km}. Strassler demonstrated that essentially the same parameter-integral representations obtained by Bern and Kosower could be derived directly from a non-abelian generalization of Feynman’s original worldline path integral, cf. Eq.~\eqref{Feynman}. This was achieved by manipulating the path integral into Gaussian form and using Green functions of the one-dimensional worldline theory, thus reducing the computation to straightforward Wick contractions. This established the worldline approach as a powerful computational tool, capable of dramatically simplifying multiloop calculations in gauge theories.\\
From that point onward, the Worldline Formalism entered its period of maturity. Throughout the 1990s and 2000s, different generalizations of the Bern-Kosower master formula were found \cite{Bern:1993wt, Schmidt:1994zj, Schubert:2001he, Ahmad:2016vvw, Edwards:2022qiw}. In parallel, a distinct yet complementary formulation of the worldline approach, perhaps closer in spirit to string theory, began to emerge. It is based on the quantization of mechanical systems, commonly referred to as “spinning particle models”. This path will be discussed in the second part of this chapter.
A particularly influential review of this formalism was provided by Christian~Schubert in 2001 \cite{Schubert:2001he}, which remains to this day the standard reference for practitioners in the field. More recent expositions \cite{Corradini:2015tik, Edwards:2019eby, new-book} have continued to expand and modernize the subject, reflecting the ongoing vitality of worldline methods both as a conceptual framework and as a practical computational technique in quantum field theory.

As of today, the Worldline Formalism continues to demonstrate its versatility and computational power across a broad spectrum of applications in theoretical physics. Among its most successful applications is the computation of black hole scattering amplitudes within the framework of the so-called Worldline Quantum Field Theory (\acr{WQFT}) formalism \cite{Mogull:2020sak, Jakobsen:2021smu, Jakobsen:2021zvh}. Closely related developments concern its role in the double copy program, where the worldline perspective has provided new insights e.g. in \cite{Ahmadiniaz:2019cim, Ahmadiniaz:2021ayd, Ahmadiniaz:2021fey, Bonezzi:2024fhd, Bonezzi:2025bgv}. On the non-perturbative side, advances in worldline instanton techniques have proven highly effective in describing vacuum pair production processes in electromagnetic backgrounds \cite{Dunne:2005sx, Dunne:2006st, DegliEsposti:2021its, DegliEsposti:2022yqw}, and have recently been extended to gravitational backgrounds as well \cite{Semren:2025dix}. Alongside these analytic developments, the numerical implementation of worldline methods has given rise to the “Worldline Monte Carlo” technique \cite{Nieuwenhuis:1995bfa, Gies:2001zp, Schmidt:2002mt, Franchino-Vinas:2019udt, Ahumada:2023iac}, which is an algorithm for estimating the heat kernel and remains a promising field of research.

After this brief historical overview, and before delving into the core material of the thesis, we shall now attempt to provide a concise and intuitive answer to the question:

\paragraph{What does the Worldline Formalism consist of?}
A worldline representation such as that given in Eq.~\eqref{Feynman} is, of course, familiar to any researcher who has worked with worldline-based methods, and likely just as intuitive to those experienced in worldsheet calculations in string theory. However, it may not appear as natural to readers accustomed to standard second-quantized techniques and with limited exposure to string-theoretical ideas. The historical derivation may certainly help in clarifying the formalism, yet it might still leave one wondering: “What does the Worldline Formalism truly consist of?”. The present paragraph aims to offer a concise conceptual answer to that question, serving as a guiding principle for the reading of the forthcoming chapters.

To set the stage, it will be convenient to distinguish between two complementary, though closely related, ways of performing computations within the worldline approach. We shall refer to them as “top-down” and “bottom-up” approaches, respectively. This distinction is not part of the standard nomenclature of the Worldline Formalism and is introduced here solely for expository purposes: it will help to differentiate the methodology employed in Chapter~\ref{chap:first} from those used in Chapters~\ref{chap:second}--\ref{chap:fifth}. To the author’s knowledge, no more established terminology currently exists for this purpose. 
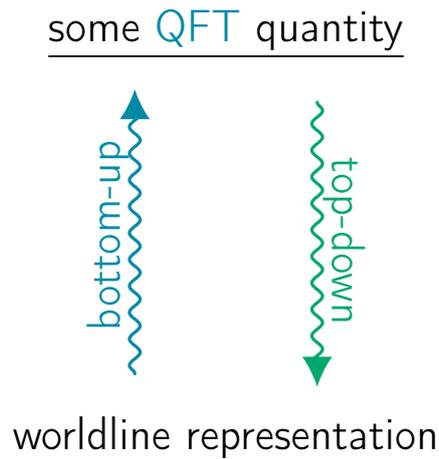
\begin{figure}[!ht]
\centering
\begin{tikzpicture}[
    font=\sffamily\Large, 
    thick, 
    scale=1.5,
    line cap=round, 
    line join=round
]
\tikzset{
    myArrowUp/.style={
        decorate, decoration={snake, amplitude=2pt, segment length=10pt},
        turquoise, very thick,
        >={Latex[length=4mm, width=4mm]},
        shorten >=-1mm
    },
    myArrowDown/.style={
        decorate, decoration={snake, amplitude=2pt, segment length=10pt},
        jade, very thick,
        >={Latex[length=4mm, width=4mm]},
        shorten >=-1mm
    }
}
\node (top) at (0,1.8) {\underline{some \acr{QFT} quantity}};
\node (bottom) at (0,-1.8) {\underline{worldline representation}};
\draw[->, myArrowUp]
    (-0.8,-1.2) -- (-0.8,1.25)
    node[midway, left=10pt, rotate=90, anchor=center, text=turquoise] {bottom-up};
\draw[->, myArrowDown]
    (0.8,1.2) -- (0.8,-1.25)
    node[midway, right=10pt, rotate=270, anchor=center, text=jade] {top-down};
\end{tikzpicture}
\caption{Schematic illustration of the distinction between the two approaches commonly employed within the worldline framework. In the “top-down” approach, one starts from a given quantum field theory quantity and reformulates it in worldline form. In the “bottom-up” approach, one instead begins with a worldline model and extracts field theoretical information from it.}\label{distinction}
\end{figure}

Fig.~\ref{distinction} graphically depicts the difference between the two approaches. Despite their different starting points, both methods ultimately pursue the same goal: to construct, and subsequently compute, a worldline representation of a \acr{QFT} quantity.\footnote{In this context, with “some \acr{QFT} quantity” we refer broadly to any object of interest such as a heat kernel, a scattering amplitude, or the effective action of a quantum field.} In a nutshell, this constitutes a succinct answer to the central question posed above. In the remainder of this section, we shall illustrate, through specific examples, how each of the two approaches realizes this idea in its own characteristic manner.

\paragraph{Top-down}
Perhaps the most straightforward way to provide a worldline representation of some \acr{QFT} quantity is to begin with the field-theoretic expression itself and manipulate it into worldline form. This procedure presupposes detailed knowledge of the underlying theory and involves non-trivial manipulations, one of which relies on the integral representation of the logarithm:
\begin{equation*}\label{log}
    \log\left(\frac{A}{B}\right)=\int_0^{\infty}\frac{\diff T}{T} \left( \mathrm{e}^{-BT} - \mathrm{e}^{-AT} \right)\;.
\end{equation*}
To make this concrete, let us anticipate the main object of interest in Chapter~\ref{chap:first}, namely the effective action of a quantum scalar field $\phi(x)$ interacting with a Yukawa classical background $V(x)$. 
According to standard field theory, the effective action $\Gamma[V]$, defined as the generating functional of the one-particle-irreducible N-point functions, is related to the operator of quantum fluctuations $\mathcal{Q}$, through the functional determinant
\begin{equation*}
\Gamma[V]=\frac12 \,\mathrm{Log} \mathrm{Det}[\Q]\ .
\end{equation*}
In the case of interest, we will take the fluctuation operator to be $\mathcal{Q}:=-\partial^2+V$. Using Schwinger proper-time parametrization~\cite{DeWitt:2003pm} (or Frullani's representation for the logarithm~\cite{Jeffreys}), the effective action can be rewritten as
\begin{equation*} \label{Gamma}
\Gamma[V]=-\frac12 \int_{0}^{\infty}\frac{{\rm d}T}{T}\int {\rm d}^Dx \, K(x,x;T)\ ,
\end{equation*}
in terms of the diagonal of the related heat kernel operator\footnote{This marks the first introduction of the heat kernel in this thesis, which will serve as a central object in many of the forthcoming applications. A full discussion of the heat kernel and its expansion will be deferred to the \hyperref[chap:introII]{prelude to Part~II}; here, we will limit ourselves to presenting only the essential details needed for \hyperref[partI]{Part~I} of the thesis.} (the integrated heat kernel or heat kernel's trace). The heat kernel operator is defined as
\begin{equation*}
K(x,x';T):=\mathrm{e}^{-T\Q}(x,x';T)\ .
\end{equation*}
As customarily, in the expression above for $\Gamma[V]$ an infinite additive constant is neglected. The heat kernel operator $K(x,x';T)$  admits a natural quantum-mechanical interpretation as the matrix elements $\braket{x|\mathrm{exp}(-T\Q)|x'}$ of the evolution operator $U(T)$ of a scalar particle, with a corresponding proper time $T$ for the evolution and a phase space Lagrangian
\begin{equation*}
L=-ip_\mu\dot{x}^\mu+p^2+ V\ ,
\end{equation*}
where the dot denotes $\dot{x}:=\frac{{\rm d}x(t)}{{\rm d}t}$. One can then represent the transition amplitude in terms of a path integral over the bosonic coordinates $x^\mu(t)$ in a first-quantized framework:
\begin{equation*} \label{path}
K(y,z;T)=\int_{x(0)=y}^{x(T)=z} \mathcal{D}x \, \mathrm{e}^{-\int_{0}^{T} {\rm d}t \left( \frac{\dot{x}^2}{4}+ V(x) \right)}\ .
\end{equation*}
This equation provides precisely the desired worldline representation of the one-loop effective action. The “top-down” construction outlined above will constitute the starting point for the analysis presented in Chapter~\ref{chap:first}.

Perhaps the most subtle aspect of this approach lies in correctly identifying the worldline Lagrangian appearing in the exponent of the path integral. While the scalar case is relatively straightforward, more complex theories might involve intricate derivations. In such situations, an alternative “bottom-up” strategy can also be employed.

\paragraph{Bottom-up}
A complementary way to approach field-theoretical computations from a first-quantized perspective consists of setting aside, at least temporarily, quantum field theory. Instead, one considers a mechanical model for a relativistic particle with gauge symmetry, which allows for a manifestly Lorentz covariant formulation. As a next step, one relates this mechanical model to its corresponding field-theoretical content. In a sense, the approach is more principled and conceptually closer in spirit to string theory.
The zoology of the existing worldline models is broad and is not straightforward to categorize every possibility precisely. Nevertheless, the majority of such models can be regarded as “spinning particle models” \cite{Corradini:2015tik}, as they describe the propagation of a relativistic particle in first quantization.\footnote{Some notable examples which fall outside this categorization include worldine models capturing topological field theories such as the Chern--Simons theory \cite{Witten:1992fb}, as well as the “superparticle models”, which has been shown to describe linearized $D=10$ super Yang--Mills \cite{Berkovits:2001rb} and $D=11$ supergravity \cite{Berkovits:2019szu} (see the \href{https://youtube.com/playlist?list=PLbcx3dKCUhgYgz7_o8JACK6Wuu3aDESVU&si=wKp7iRST8Ozj0zOh}{LACES lectures} by Nathan~Berkovits for a detailed presentation). A recent work \cite{Fecit:2025eet} has proposed a class of models that extends the worldline approach to more general gauge theories, including the covariant fracton case.} These models can be divided into two broad classes, depending on whether the spin degrees of freedom are described by means of Grassmann odd or Grassmann even variables. We now explain the main differences between these two cases and summarize their key features, anticipating a more in-depth analysis in the following chapters.
\begin{itemize}
    \item The Grassmann odd variables are spacetime vectors \emph{but worldline fermions}, usually denoted by $\psi^\mu(\tau)$, $\theta^\mu(\tau)$, or $\Xi^\mu(\tau)$ in the literature.\footnote{The variable $\tau$ represents the proper time of the particle, ranging over the interval $[0, T]$. For convenience, it is often rescaled such that $\tau \in [0, 1]$.} This choice defines models typically referred to as \textbf{$\mathbf{O(\mathcal{N})}$ spinning particles}. Although these variables are real, it is often convenient to work with their complexified form.\\
    The literature on these “fermionic models” is extensive: they have been studied and applied since the 1980s. These models exhibit $\N$ local supersymmetry (\acr{SUSY}) on the worldline, and have been shown to describe, on Minkowski, a spin $s=\frac{\N}{2}$ particle in first quantization \cite{Berezin:1976eg, Brink:1976uf, Gershun:1979fb, Howe:1988ft, Howe:1989vn, Siegel:1988ru, Bastianelli:2008nm, Corradini:2010ia, Bastianelli:2014lia, Corradini:2015tik}. They are examples of constrained Hamiltonian systems with first-class constraints \cite{VanHolten:2001nj, new-book}. These kinds of constraints, denoted as $C_\alpha$ (with $\alpha$ spanning the number of constraints), satisfy upon quantization a graded algebra of the form\footnote{We indicate with $[\cdot,\cdot\}$ the graded commutator.}
    \begin{equation*}
    [ C_\alpha, C_\beta \}=f_{\alpha\beta}{}^{\gamma}C_{\gamma}
    \end{equation*}
    with some structure functions $f_{\alpha\beta}{}^{\gamma}$. The phase-space action of such models depends on the particle spacetime coordinates $x^\mu$ together with ${\cal N}$ real fermionic superpartners $\Xi^\mu_I$.\footnote{Capital Latin letters $I,J,\dots$ are reserved for internal $SO(\N)$ indices $(I,J=1,\dots, \N)$.} In the case of a flat $d$-dimensional target spacetime, the gauged action is given by
    \begin{equation*}
    S=\int \diff \tau \left[p_\mu\dot x^{\mu}+\frac{i}{2}\Xi_{\mu}\cdot\dot\Xi^{\mu}-e\,H-i\chi^I\,q_I-a^{IJ}\,J_{IJ}\right]\ , 
    \end{equation*}
    where a dot indicates a contraction of the internal indices. This theory should be viewed as a one-dimensional field theory living on the worldline. A few remarks regarding its field content are in order.  
    The canonical coordinates $(x^\mu, p_\mu, \Xi^\mu_I)\,$ upon quantization are subject to the canonical commutation relations
    \begin{equation*} 
    [x^\mu, p_\nu]=i\,\delta^\mu_\nu\ ,\quad \{\Xi^\mu_I, \Xi^\nu_J\}=\delta_{IJ}\,\eta^{\mu\nu}\ .
    \end{equation*}
    The worldline supergravity multiplet in one dimension $(e,\chi^I, a_{IJ})$ contains the einbein $e$ which gauges worldline translations, the $\N$ gravitinos $\chi$ which gauge the worldline supersymmetry, and the gauge field $a$ for the symmetry which rotates by a phase the worldline fermions and gravitinos, the $R$-symmetry. In the action above, they act as Lagrange multipliers for the suitable first-class constraints, which are the Hamiltonian $H$, the supercharges $q_I$, and the $R$-symmetry algebra generators $J_{IJ}$, choosing a specific order for the latter to avoid ambiguities
    \begin{equation*}
    H:=\frac{1}{2} \, p^\mu p_\mu\ , \quad q_I:=\Xi_I^\mu \, p_\mu\ , \quad J_{IJ}:=i\,\Xi_{[I}^\mu \, \Xi_{J]_\mu}\ .
    \end{equation*} 
    The Hamiltonian and the supercharges together form the following one-dimensional algebra: 
    \begin{equation*}
    \{q_I, q_J\}=2\,\delta_{IJ}\,H\ ,\quad [q_I, H]=0\ .
    \end{equation*}
    These constraints have to be introduced to ensure the mass-shell condition and to eliminate negative norm states, ensuring the consistency of the model with unitarity at the quantum level \cite{Corradini:2015tik}. On the other hand, regarding the aforementioned $R$-symmetry algebra one finds:
    \begin{align*}
    [J_{IJ}, q_K] &= i \left( \delta_{JK}q_I -\delta_{IK}q_J \right)\ , \\
    [J_{IJ}, J_{KL}] &= i \left( \delta_{JK}J_{IL} -\delta_{IK}J_{JL} -\delta_{JL}J_{IK} +\delta_{IL}J_{JK} \right)\ .
    \end{align*}
    The gauging of the $R$-symmetry group is optional and can be used to constrain the model to deliver pure spin $s$ states and have the minimal amount of degrees of freedom: to count them, one can construct the path integral on the one-dimensional torus of the free spinning particles, which has been achieved for all $\N$ in Ref.~\cite{Bastianelli:2007pv}. The relations above together form the so-called $O(\N)$-\emph{extended worldline supersymmetry algebra}: it displays $\N$ supercharges $q_I$ which close on the Hamiltonian $H$ and which transform in the vector representation of $SO(\N)$.

    This model will be specialized for $\N=2$ and for $\N=4$ in the course of the thesis. Further details will be provided in due course, while we refer to the following complete review \cite{Corradini:2015tik} for an in-depth discussion.
    \item The Grassmann even variables are spacetime vectors \emph{and worldline bosons}, usually denoted by $\alpha^\mu(\tau)$ and $\bar\alpha^\mu(\tau)$ in the literature. They form a pair of complex conjugate variables. This choice defines a class of models usually referred to as \textbf{bosonic spinning particle}.\\
    The use of bosonic variables to describe particles with integer spin has a long history as well, but mainly rooted in string theory \cite{Bengtsson:1986ys, Henneaux:1987cp, Bouatta:2004kk, Hallowell:2007qk}. Indeed, it has been explored only sporadically in the context of worldline path integrals, as in \cite{Bastianelli:2009eh}. It has seen renewed interest in recent works, specifically in the context of Yang--Mills particles \cite{Bonezzi:2024emt}, and of charged massive spin-1 particles \cite{Bastianelli:2025khx}. 
    The “bosonic model” is defined by the usual set of phase-space variables with Poisson brackets
    \begin{equation*}
    \{x^\mu, p_\nu\}_{\mathrm{PB}} = \delta^\mu_\nu \;, \quad 
    \{\alpha^\mu, \bar \alpha^\nu\}_{\mathrm{PB}} = i \eta^{\mu\nu}\;.
    \end{equation*}
    The gauged worldline action is, analogously to the fermionic case, given by
    \begin{equation*}
    S=\int \diff\tau\; \left[ p_\mu \dot{x}^\mu -i \bar\alpha_\mu \dot{\alpha}^\mu-eH - \bar{u} L - u \bar{L} - a J
    \right]\;, 
    \end{equation*}
    where we introduced the worldline gauge multiplet $(e,\bar u, u, a)$ acting as a set of Lagrange multipliers that enforce the constraints
    \begin{gather*}
    H = \frac{1}{2}p^\mu p_\mu \;, \quad L = \alpha ^\mu p_\mu\;, \quad \bar{L} = \bar{\alpha}^\mu p_\mu\;, \quad J = \alpha^\mu \bar{\alpha}_\mu \;.
    \end{gather*}
    The latter satisfy a first-class Poisson-bracket algebra:
    \begin{equation*} 
    \{L,\bar L\}_{\mathrm{PB}}= 2i H \;, \quad \{J, L\}_{\mathrm{PB}}= -i L \;, \quad \{J,\bar L\}_{\mathrm{PB}}= i \bar L \;.
    \end{equation*}
    It is evident that the action and the symmetries are formally analogous to those of the fermionic spinning particle. We therefore stop here the analysis of the constraint algebra, leaving further details for the following sections. Let us emphasize that the underlying symmetry is not really a supersymmetry; however, it does play an equivalent role as will become clear in Chapter~\ref{chap:second}.      
\end{itemize}
Despite this fundamental difference in how spin degrees of freedom are represented, fermionic and bosonic models can provide equivalent worldline formulations of the same field theory. The distinction between the two is often of a technical rather than conceptual nature. For instance, the use of bosonic variables offers computational advantages when dealing with scattering amplitudes as worldline correlators of vertex operators, as detailed in Chapter~\ref{chap:third}. Conversely, fermionic variables can simplify certain calculations based on Wick contractions: a relevant example is providedby the model object of Chapter~\ref{chap:fourth} describing a spin-2 particle, dubbed $\mathcal{N}=4$ spinning particle; it exhibits a subtle double-copy-like structure that facilitates computations, as shown in Chapter~\ref{chap:sixth}.\footnote{See in particular the discussion in Appendix~\ref{appendixB2}. It is an intriguing, but not much explored, possibility to exploit this underlying double copy structure further, potentially contributing to the color-kinematics duality program.} 
 
But how do these models relate to their corresponding field theories?\\
The precise connection emerges upon quantization of the spinning particle model. We shall not delve into the details here, as the following chapters are devoted to an in-depth analysis of this subject. Let us simply anticipate that the quantization procedure involves several subtleties, since we are dealing with a constrained Hamiltonian system. In general, it can be implemented efficiently in two main ways: either following a procedure {\it à la} Dirac, or by employing the BRST formalism. In the former approach, one promotes the constraints to operators $\hat{C}_\alpha$ and defines physical states as those elements $\ket{\Psi}$ of the Hilbert space that get annihilated by all constraints:\footnote{This definition is somewhat restrictive and will be relaxed in a more thorough discussion in Chapter~\ref{chap:second}. Nevertheless, it is ultimately the defining expression that is employed in practice, cf. Eqs.~\eqref{Dirac second}-\eqref{phys cond Dirac}.}
\begin{equation*}
\hat{C}_\alpha\ket{\Psi}=0\ .
\end{equation*}
This procedure has been extensively used in Refs.~\cite{Bastianelli:2008nm, Bastianelli:2014lia} to quantize both massless and massive spinning particle models. 
The BRST method,\footnote{Which has been originally developed in the context of the path integral quantization of Lagrangian gauge theories and named after Carlo~Maria~Becchi, Alain~Rouet, Raymond~Stora and Igor~Tyutin \cite{Becchi:1975nq}.} on the other hand, will be discussed extensively in Chapter~\ref{chap:second} and in Chapter~\ref{chap:fourth}, and we shall not anticipate its details here.

These two approaches are equivalent as long as the model remains free \cite{Henneaux:1992ig}. However, the inclusion of consistent interactions or background couplings is often more efficiently achieved within the BRST framework, as pioneered in Refs.~\cite{Dai:2008bh, Bonezzi:2018box, Bonezzi:2020jjq}. For instance, regarding the fermionic models, the Dirac quantization approach historically led to the misleading conclusion that quantization is not feasible in some cases, particularly when considering generic gravitational backgrounds. This limitation was notably encountered for the $\N=4$ spinning particle, where only restricted backgrounds were initially considered consistent, until it was realized that BRST quantization provides a systematic route to more general configurations. Chapter~\ref{chap:fourth} is devoted to a thorough analysis of the matter.  

Once the worldline model is quantized and the couplings are consistently introduced, one obtains a worldline representation of a given \acr{QFT} quantity by constructing the path integral of the model. Schematically, this takes the form 
\begin{equation*} 
\Gamma[\mathfrak{B}] \sim  \int_{\mathcal{T}}
\frac{\mathcal{D}G\,\mathcal{D}X}{\mathrm{Vol(Gauge)}}\, {\rm e}^{-S[X,G; \mathfrak{B}]}\ .
\end{equation*}
The particle action depends on the worldline gauge fields $G(\tau)$ (a.k.a the set of Lagrange multipliers) and coordinates with fermionic/bosonic partners $X(\tau)$, while the overcounting from summing over gauge equivalent configurations is formally taken into account by dividing by the volume of the gauge group. We agnostically denoted by $\mathfrak{B}(x)$ the presence of background fields and did not specify the topology of the worldline $\mathcal{T}$. 
The most important topologies of the worldline are depicted in Fig.~\ref{top}. They correspond to the interval, $I$, suitable for describing the propagation of the particle, thus providing the \acr{QFT} propagator, and to the circle $S^1$, which enters in the first quantized representation of the one-loop \acr{QFT} effective action induced by relativistic particles.

\begin{figure}[!ht]
\centering
\begin{tikzpicture}[line width=0.8pt, scale=1.2]
    \draw[thick] (-1,0) -- (1,0);
    \fill (-1,0) circle (1.5pt);
    \fill (1,0) circle (1.5pt);
    \node[below] at (0,-1.2) {Propagator};
    \draw[thick] (4,0) circle (1);
    \node[below] at (4,-1.2) {Effective Action};
\end{tikzpicture}
\caption{Topologies of the worldline.}
\label{top}
\end{figure}

The “bottom-up” approach outlined here constitutes the methodology followed in Chapter~\ref{chap:second} and Chapter~\ref{chap:third}, where the bosonic spinning particle is employed, as well as in Chapters~\ref{chap:fourth}--\ref{chap:fifth}, where the $\N=2$ and $\N=4$ fermionic spinning particles are considered.

\newpage
\thispagestyle{empty}
\mbox{}
\newpage

\part{Non-Perturbative Phenomena}\label{partI}
\section*{Prelude: the Schwinger effect in a nutshell} \label{chap:introI}
\begin{figure}[h!]
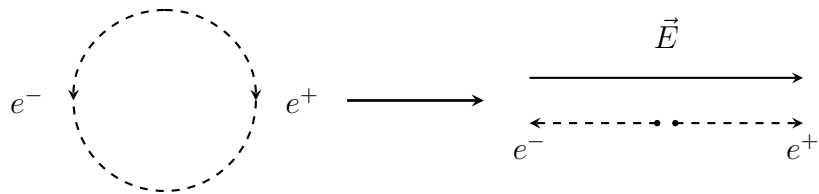

    \centering
    \pair
    \caption{Graphical representation of a virtual electron-positron pair becoming real due to the presence of strong electric field.}
    \label{ep}
\end{figure}

\setcounter{equation}{0}
\renewcommand{\theequation}{I.\arabic{equation}}
In the first part of the thesis, we explore the application of the Worldline Formalism to the realm of non-perturbative physics. Our goal is to show that, being a genuine functional approach, it naturally lends itself to efficient non-perturbative analyses. Specifically, we shall focus on the computation of the vacuum persistence probability
\begin{align}
 \vert\langle 0_{\rm out} \vert 0_{\rm in}\rangle\vert^2\;.
\end{align}
In Minkowski space, the vacuum persistence amplitude is related to the effective action by
\begin{equation}
    \langle 0_{\rm out} \vert 0_{\rm in}\rangle ={\rm e}^{i \Gamma_\mathrm{M}}\;.
\end{equation}
A remarkable feature takes place when the Minkowskian effective action $\Gamma_\mathrm{M}$ develops an imaginary part: since the vacuum persistence probability relates to 
\begin{align}
 \vert\langle 0_{\rm out} \vert 0_{\rm in}\rangle\vert^2 ={\rm e}^{- 2\operatorname{Im} \Gamma_\mathrm{M}}\;,
\end{align}
the vacuum non-persistence can be interpreted as the probability of producing particle-antiparticle pairs
\begin{equation}
    P_{\mathrm{pair}}:=1-\mathrm{e}^{- 2\operatorname{Im} \Gamma_\mathrm{M}}\;.
\end{equation}
This definition reflects the fact that the instability of the vacuum signals the emergence of states with a non-vanishing number of particles. As long as the latter are not largely populated, the pair creation probability can be approximated by
\begin{equation}
    P_{\mathrm{pair}} \approx 2 \, \mathrm{Im} \,\Gamma_{\mathrm{M}}\;.
\end{equation}
The instability of the vacuum was famously identified in the pioneering work by Euler and Heisenberg \cite{Heisenberg:1936nmg}. Even earlier, in 1931, Fritz~Sauter had demonstrated that electron-positron pairs could be produced from the vacuum in the presence of a strong electric field, interpreting the effect as a quantum tunneling \cite{Sauter:1931zz}. Dirac’s theory had already revealed that the vacuum is not empty, but filled with a sea of virtual pairs undergoing continuous creation and annihilation. In the presence of an external field, to be more precise an electric one, this vacuum becomes unstable: virtual pairs can be separated and promoted to real particles when the field is strong enough to supply the required energy (see Fig.~\ref{ep}).

\phantom{a}\\
The formal expression of this process was achieved with the tools of \acr{QED} only by Julian~S.~Schwinger in 1951 \cite{Schwinger:1951nm}. The celebrated pair production rate in a weak electric field, constant in time and spatially homogeneous, of magnitude $E$ reads  \cite{Dunne:2004nc, Schwartz:2014sze}
\begin{equation}\label{rate}
    \gamma\sim\frac{e^2 E^2}{4\pi^3} \mathrm{e}^{-\frac{m^2\pi}{eE}}\;,
\end{equation}
where $m$ and $e$ denote the electron mass and charge, respectively. From \eqref{rate}, one can extract the critical value (with the factors of $\hbar$ and $c$ temporarily restored)
\begin{equation}
    E_{\mathrm{cr}}= \frac{m^2c^3}{e\hbar}\approx 10^{18}\mathrm{\,V\,m^{-1}}\;,
\end{equation}
that an electric field must reach in order to create detectable electron-positron pairs. This value is still beyond present technological capabilities, despite all the technical advances that have taken place over the last 80 years. Consequently, an experimental proof of the Schwinger effect remains elusive. However, the latest generation of laser facilities is expected to overcome this issue in the near future, presently offering field strengths that are just a few orders of magnitude below $E_{\mathrm{cr}}$. In effect, several strong-field experiments are planned or are underway at the European XFEL~\cite{Ahmadiniaz:2024xob, LUXE:2023crk}, LASERIX~\cite{Kraych:2024wwd} and OVAL~\cite{Fan:2017fnd}, to cite a few examples.
A promising strategy to enhance the pair production rate is known as “assisted pair production” \cite{Schutzhold:2008pz}. The key idea is to introduce an additional fastly varying background, which will exert the rôle of an assisting (or catalyzing) field, in addition to the strong background one. 

To better appreciate the non-perturbative nature of the Schwinger mechanism, let us briefly recall how the rate \eqref{rate} originates from the \acr{QED} effective action $\Gamma[A_\mu]$, defined as
\begin{equation}
    \int \mathcal{D}A \;\mathrm{e}^{i\Gamma[A_\mu]} :=  \int \mathcal{D}A \mathcal{D}\bar\psi \mathcal{D}\psi \; \mathrm{e}^{i\int \diff^4x\left( -\tfrac14 F_{\mu\nu}F^{\mu\nu}+\bar\psi (i\slashed{D}-m)\psi\right)}\;.
\end{equation}
By construction, $\Gamma[A_\mu]$ describes the nonlinearities of quantum electrodynamics when the fermionic degrees of freedom of matter are integrated out. For a constant electromagnetic background, this reduces to the celebrated Euler--Heisenberg effective Lagrangian \cite{Heisenberg:1936nmg}
\begin{equation} \label{eulero}
    \mathcal{L}_{\mathrm{EH}} =-2 \int_{0}^{\infty} \frac{\diff T}{T}\frac{ \mathrm{e}^{-m^2 T}}{(4 \pi T)^2} \, \frac{eET}{\tan(eET)}\frac{eBT}{\tanh(eBT)}\;,
 \end{equation}
This Lagrangian contains a wealth of information on nonlinear quantum effects. It is non-perturbative in the coupling $e$ and resums an infinite number of one-loop N-photon diagrams (see Fig.~\ref{expamp}). In this sense, it represents one of the most famous examples of a resummed expression, encoding all invariants constructed from powers of the field strength tensor. As we shall see, the ability to obtain such closed expressions, containing infinitely many invariants, is crucial for exploring the non-perturbative regime.
\begin{figure}[!ht]
    \centering
    \EH
\caption{Diagrammatic perturbative expansion of the Euler--Heisenberg effective Lagrangian with the inclusion of the classical Maxwell Lagrangian. Note that, in QED, the diagrams with an odd number of external photon legs vanish due to Furry's theorem.}
\label{expamp}
\end{figure}
Beyond pair production, the Euler--Heisenberg Lagrangian also underlies processes such as light-by-light scattering and vacuum polarization \cite{Dunne:2012vv}. Returning to the main topic, notice that the integral in the Schwinger proper time $T$ in \eqref{eulero} is singular: the $T=0$ region signals an ultraviolet (\acr{UV}) divergence to be regularized, while infinitely many poles occur at
\begin{equation}
    T_n=\frac{n\pi}{eE}\;, \quad 0<n \in \mathbb{N}\ .
\end{equation}
These poles are responsible for the imaginary part of the effective action, leading ultimately to the non-persistence of the vacuum. Their origin can be traced to the zero modes of the operator of quantum fluctuations. While the conceptual relation between pair production and the imaginary part of the effective Lagrangian is straightforward, obtaining the Lagrangian explicitly is far from trivial. A detailed analysis of this problem, and its analogues in other field theories, will be the focus of what follows.

In the next two chapters, we will show how to obtain such Euler--Heisenberg-type Lagrangians. In particular, in Chapter~\ref{chap:first} we take the case-study of a quantum scalar field in a Yukawa scalar background by following a “top-down” approach, while Chapter~\ref{chap:second} extends this analysis to a massive charged vector boson in an electromagnetic background, adopting a “bottom-up” strategy.
\setcounter{equation}{0}
\numberwithin{equation}{section}

\newpage
\thispagestyle{empty}
\mbox{}
\newpage

\chapter*{Spin 0}
\addcontentsline{toc}{chapter}{\color{turquoise}\raisebox{0.5ex}{\rule{7.75cm}{0.4pt}} Spin 0 \raisebox{0.5ex}{\rule{7.75cm}{0.4pt}}}
\chapter{Yukawa assisted pair creation of scalar particles}\label{chap:first}
\textit{In this chapter, we illustrate the power of the Worldline Formalism in the study of non-perturbative phenomena, beginning with the seemingly simple case of a quantum scalar particle in a classical scalar Yukawa background. To obtain a worldline description of this setup, we adopt the “top-down” strategy: starting from the heat kernel operator, we derive a worldline representation of the effective action and proceed with its computation.
A central milestone of our analysis is the derivation of suitably resummed expressions: once the formalism is set up, we obtain compact forms for the effective action and the heat kernel of the scalar field, incorporating all invariants built from powers of the background potential and its first and second derivatives. Remarkably, this turns out to be rather straightforward in the first-quantized approach. Building on these results, we then analyze vacuum instability by computing the vacuum persistence amplitude and the associated Schwinger pair production probability.
We begin with the simple case of a quadratic potential and then complicate the discussion by considering non-quadratic time-dependent potentials and also spatially inhomogeneous potentials.
Finally, we refine this toy model by investigating how the inclusion of an additional, rapidly varying background can significantly enhance pair production.}

\paragraph{Conventions} We work in $D$-dimensional Euclidean space, except for Secs. \ref{chap:first:sec:non-assisted} and \ref{chap:first:sec:assisted}, where we Wick rotate to Minkowski spacetime (using the mostly plus signature); we do not distinguish the indices used in these different cases. 

\section{Worldline representation of the scalar heat kernel} \label{chap:first:sec1}
Consider a theory consisting of a single quantum scalar field $\phi(x)$, in flat spacetime and interacting with a Yukawa background (classical) field, which is described by the spacetime action
\begin{equation}
S=\frac12 \int {\rm d}^Dx \left[ (\partial \phi)^2 + V(x) \phi^2\right]\ ,
\end{equation}
where Einstein’s sum convention $\partial^2=\partial_\mu \partial^\mu$ is employed. 
Generalizations to gauge background fields or curved spacetimes are, in principle, possible. However, in this chapter, we will consider only the scalar coupling through $V(x)$, which already exhibits several key features of the Worldline Formalism. Extensions to other types of couplings, as well as to higher-spin fields, will be addressed in the concluding discussion.

We will thus focus on $V(x)$, which is an arbitrary Yukawa-type (scalar) potential and may naturally include a mass term for the scalar field. 
As customarily, the one-loop effective action $\Gamma$, which is the full effective action unless we quantize $V$ as well, is related to the operator of quantum fluctuations, $\Q:=-\partial^2 + V$:
\begin{equation}
\Gamma=\frac12 \,\mathrm{Log} \mathrm{Det}[\Q]\ .
\end{equation} 
As detailed in the \hyperref[chap:WF]{introductory section} on the Worldline Formalism, in the Schwinger proper-time parametrization, we can write the effective action as
\begin{equation} \label{chap:first:Gamma}
\Gamma=-\frac12 \int_{0}^{\infty}\frac{{\rm d}T}{T}\int {\rm d}^Dx \, K(x,x;T)\ ,
\end{equation}
in terms of the diagonal of the related heat kernel operator.
In turn, the heat kernel operator can be interpreted in quantum mechanical terms as the matrix elements $\braket{x|\mathrm{exp}(-T\Q)|x'}$ of the evolution operator $U(T)$ of an $N=0$ spinning particle. Following the paradigm of the Worldline Formalism, one can represent the transition amplitude in terms of a path integral over the bosonic coordinates $x^\mu(t)$ in a first-quantized framework.
In this chapter, we are going to follow two different approaches, depending on whether we absorb the spacetime integral into the path integral or not.
If we do not, then the effective action is computed from the coincidence point expression of
\begin{equation} \label{chap:first:path}
K(y,z;T)=\int_{x(0)=y}^{x(T)=z} \mathcal{D}x \, \mathrm{e}^{-\int_{0}^{T} {\rm d}t \left( \frac{\dot{x}^2}{4}+ V(x) \right)}\ .
\end{equation}
This approach, which involves Dirichlet (\acr{D}) boundary conditions (\acr{BC}) on the worldline, is of fundamental importance when local quantities are to be analyzed, such as the energy-momentum tensor. If instead one is interested in global quantities, as is the case for the effective action in Eq.~\eqref{chap:first:Gamma}, one can realize that the spacetime integral can be effectively incorporated into the boundary conditions of the path integration. 
A deeper discussion of this issue will be postponed to Sec.~\ref{chap:first:chap:first:sec:assisted_yukawa} and App.~\ref{chap:first:appA}, where the so-called 
string-inspired (\acr{SI}) \acr{BC} will prove helpful.

\section{Resumming first and second derivatives of the potential: the heat kernel} \label{chap:first:sec:resummation}
Having at our disposal a worldline representation for the relevant transition amplitude, cf.~\eqref{chap:first:path}, we are going to show that one can set up a perturbative expansion which already incorporates the information of all the invariants made up of the first and second derivatives of the potential. Since we are interested in working at the local level of the heat kernel, we will employ \acr{DBC} on the worldline.

As a first step, we follow Ref.~\cite{Franchino-Vinas:2023wea} and Taylor expand the scalar potential about an arbitrary point $\tilde x$,
\begin{equation} \label{Taylor}
 V(x)= V(\tilde{x})+l^\mu \partial_\mu V(\tilde{x})+\frac12 l^\mu l^\nu \partial_{\mu\nu} V(\tilde{x}) + \dots\ ,
\end{equation}
where we define the distance $l^\mu:= (x-\tilde{x})^\mu$ and employ the following shorthand notation for the higher derivatives: $\partial_{\mu_1 \dots \mu_n}V:=\partial_{\mu_1}\dots \partial_{\mu_n}V$. Identifying the base point with one of the arguments of the heat kernel, $\tilde{x} \to y$, and performing the translation $x \to x' = x-y$ in the path integral as well (we will omit the prime henceforth), the expression for the heat kernel can be recast as
\begin{equation}\label{eq:hk_resummed_initial}
K(y,z;T)=\mathrm{e}^{-T V(y)} \, \int_{x(0)=0}^{x(T)=z-y} \mathcal{D}x \, \mathrm{e}^{-\int {\rm d}t \, x^\mu \partial_\mu V(y)} \, \mathrm{e}^{-S_{\mathrm{free}}-S_{\mathrm{int}}}\ ,
\end{equation}
where the quadratic part of the worldline action will be called the free action,
\begin{equation} \label{free}
S_{\mathrm{free}}:=\frac12 \int_{0}^{T} {\rm d}t \, x^\mu \left(\frac{1}{2} \delta_{\mu\nu}\, \overleftarrow{\partial_t}\,\overrightarrow{\partial_t} +\partial_{\mu\nu} V(y)\right)x^\nu\ ,
\end{equation}
while the higher-order terms will be included in the interacting action,
\begin{align} \label{int}
\begin{split}
S_{\mathrm{int}}&:=\int_{0}^{T} {\rm d}t \, \sum_{n=3}^{\infty} \frac{1}{n!}\, x^{\mu_1} \dots x^{\mu_n} \, \partial_{\mu_1 \dots \mu_n} V(y)\ ,
\\
&=: \int_0^T {\rm d}t\, L_{\rm int} (x(t))\ .
\end{split}
\end{align}
Note that all the tensorial quantities involving the background field $V$ and its derivatives are evaluated at the initial point $y$; as a consequence, their dependence on the worldline bosonic variables $x(t)$ is factored out and the expression~\eqref{eq:hk_resummed_initial} is thus readily usable for a perturbative computation in powers of the path. 

Using $S_{\rm free}$ as the base action for the expansions is convenient as one only has to deal with a Gaussian path integral. The perturbative expansion corresponds to an expansion in the number of derivatives acting on a (single factor) $V$. 
Taking this into account, it proves useful to introduce an arbitrary external source $\eta_{\mu}(t)$ linearly coupled to the paths, as well as the corresponding generating functional of path n-point functions,
\begin{equation}
Z[\eta](y,z;T):= \int_{x(0)=0}^{x(T)=z-y} \mathcal{D}x \, \mathrm{e}^{-S_{\mathrm{free}}-\int_{0}^{T} {\rm d}t \, \eta_\mu x^\mu},
\end{equation}
which leads us to a master equation for the heat kernel:
\begin{equation} \label{HK2}
{K(y,z;T)=\left.\mathrm{e}^{-T V(y)} \, \mathrm{e}^{-\int_0^T {\rm d}t \, L_{\mathrm{int}}\left(-\frac{\delta}{\delta \eta(t)} \right)} \, Z[\eta](y,z;T)\right|_{\eta=\partial V(y)}}\ .
\end{equation}

\subsection{Generating functional} \label{chap:first:sec:sec2.1}
After an integration by parts in the worldline action, we can rewrite the generating functional as\footnote{A boundary term is omitted, as it has no bearing on the upcoming discussion of the diagonal heat kernel.}
\begin{equation} \label{gen}
Z[\eta](y,z;T)= \int_{x(0)=0}^{x(T)=z-y} \mathcal{D}x \, \mathrm{e}^{-\frac12 \int_{0}^{T} {\rm d}t \, \left(x^\mu \Delta_{\mu\nu} x^\nu+2\eta_\mu x^\mu \right)}\ ,
\end{equation}
where the action is defined in terms of a differential operator that acts on paths satisfying \acr{DBC} on the interval $[0,T]$:
\begin{equation} \label{Delta}
{\Delta_{\mu\nu} (t,t')=-\frac12 \delta_{\mu\nu}\partial^2_t\delta(t-t')+2 \, \Omega^2_{\mu\nu}(y)\delta(t-t')}\ .
\end{equation}
To simplify the notation of the upcoming results, we have introduced $\Omega_{\mu\nu}$, which is related to the second derivative of the potential in the following way:
\begin{equation} \label{Omega}
{2 \, \Omega^2_{\mu\nu}(y):=\partial_{\mu\nu} V(y)}\ .
\end{equation}
Taking all this into account, the computation of $Z[\eta]$ reduces to obtaining both the inverse and the functional determinant of the operator $\Delta_{\mu\nu}$. The explicit computation goes as follows. First, it is convenient to recast the path integral in terms of the quantum fluctuations $\hat{s}^\mu(t)$ around the classical trajectory $x_{\mathrm{cl}}^\mu(t)$,
\begin{equation} \label{2.4}
x^\mu(t):=x_{\mathrm{cl}}^\mu(t)+\hat{s}^\mu(t)\ .
\end{equation}
Secondly, the classical trajectory $x_{\mathrm{cl}}^\mu(t)$ is defined as the solution to the equations of motion imposed by the free worldline action, i.e.\footnote{Here and in the subsequent differential equations for the worldline $x(t)$ we will employ an abuse of notation, such that whenever the differential operator appears as $\Delta$, we intend it striped-off of the Dirac delta that should appear according to Eq.~\eqref{eq:BC_D}. For instance, $\Delta \, x(t)=(-\tfrac12 \partial_t^2 +2\Omega^2) \, x(t)$.}
\begin{equation}\label{eq:BC_D}
\frac{\delta S_{\mathrm{free}}}{\delta x_\mu}=0 \quad \Longrightarrow \quad \Delta^{\mu\nu}x^{\mathrm{cl}}_\nu(t)=0\ ,
\end{equation}
satisfying the following boundary conditions 
\begin{equation} \label{bc}
x_{\mathrm{cl}}^\mu(0)=0\ , \quad x_{\mathrm{cl}}^\mu(T)=(z-y)^\mu\ .
\end{equation}
This fact, together with expression~\eqref{2.4}, implies that the fluctuations satisfy vanishing \acr{DBC}, i.e.
\begin{equation} \label{DBC}
\hat{s}^\mu(0)=\hat{s}^\mu(T)=0\ .
\end{equation}

Coming back to the classical solution, it can be straightforwardly computed and is given by
\begin{equation} \label{hom}
x^\mu_{\mathrm{cl}}(t)=\left(\frac{\sinh{(2\Omega t)}}{\sinh{(2\Omega T)}}\right)^{\mu\nu}(z-y)_\nu\ ,
\end{equation}
where the tensorial character of $\Omega_{\mu\nu}$ has been explicitly shown. The resulting partition function thus reads 
\begin{equation} \label{2.9}
Z[\eta](y,z;T)= \, \mathrm{e}^{-S_{\mathrm{free}}[x_{\mathrm{cl}}]} \, \mathrm{e}^{-\int_{t} \eta_\mu x^\mu_{\mathrm{cl}}} \, \int_{\hat{s}(0)=0}^{\hat{s}(T)=0} \mathcal{D}\hat{s} \, \mathrm{e}^{-\frac12 \int_{t_1t_2} \hat{s}^\mu \Delta_{\mu\nu} \hat{s}^\nu-\int_t \eta_\mu \hat{s}^\mu}\ ,
\end{equation}
where we have introduced the following condensed notation for (multiple) integrals:\footnote{For instance, $\int_t \eta_\mu x^\mu=\int_0^T {\rm d}t \, \eta_\mu (t)x^\mu(t)$ and $\int_{t_1t_2} \hat{s} \Delta \hat{s}=\int_0^T {\rm d}t_1\int_0^T {\rm d}t_2 \, \hat{s}(t_1) \Delta(t_1,t_2) \hat{s}(t_2)$.} 
\begin{align}
 \int_{t_1t_2\cdots }:= \int_0^T {\rm d}t_1 \int_0^T {\rm d}t_2 \cdots. 
\end{align}

Performing the shift $\hat{s} \to \tilde{s}=\hat{s}+\Delta^{-1}\eta$ and a subsequent completion of squares one is lead to 
\begin{align} \label{Z}
\begin{split}
Z[\eta](y,z;T)&= \, \mathrm{e}^{-S_{\mathrm{free}}[x_{\mathrm{cl}}]} \, \mathrm{e}^{-\int_{t} \eta_\mu x^\mu_{\mathrm{cl}}} \, \mathrm{e}^{\frac12 \int_{t_1t_2} \eta^\mu \Delta^{-1}_{\mu\nu} \eta^\nu} \, \int_{\tilde{s}(0)=0}^{\tilde{s}(T)=0} \mathcal{D}\tilde{s} \, \mathrm{e}^{-\frac12 \int_{t_1t_2} \tilde{s}^\mu \Delta_{\mu\nu} \tilde{s}^\nu}
\\
&=\frac{\mathcal{C}_{\mathrm{DBC}} }{{\overline{\mathrm{Det}}^{\nicefrac{1}{2}}(\Delta)}} \, \mathrm{e}^{-S_{\mathrm{free}}[x_{\mathrm{cl}}] +\frac{1}{2} (S_1+ S_{\rm bos})}\ .
\end{split}
\end{align}
As a way to simplify the notation in Eq.~\eqref{Z}, we have introduced
actions for the linear and quadratic contributions in the external source,
\begin{align} \label{S1}
S_1[x_{\mathrm{cl}},\eta]:&=-2\int_t \, \eta_\mu(t) x^\mu_{\mathrm{cl}}(t)\ ,
\\
\label{Sbos}
S_{\mathrm{bos}}[\eta]:&=\int_{t_1 t_2} \, \eta^\mu(t_1) \Delta^{-1}_{\mu\nu}(t_1,t_2) \eta^\nu(t_2)\ ,
\end{align}
as well as the following short-hand notation 
for the quotient of functional determinants with respect to the free case: 
\begin{equation*}
 \overline{\mathrm{Det}}(\mathrm{A}) := \frac{\mathrm{Det}(\mathrm{A})}{\mathrm{Det}\left(-\tfrac12\delta_{\mu\nu}\partial^2_{\tau}\right)}\ .
\end{equation*}
Note that, in expression~\eqref{Z}, $\mathcal{C}_{\mathrm{DBC}}$ is a normalization to be determined, after the computation of the determinant and Green function of the $\Delta$ operator, from the well-known result for the Mehler kernel~\cite{Vinas:2014exa, Franchino-Vinas:2021bcl}.

In order to obtain an explicit expression for the generating functional, we begin by writing down the Green function corresponding to the operator $\Delta$:
\begin{align}\label{sGF}
\Delta^{-1}_{\mu\nu}(t,t')
&=\left[\frac{ \sinh(2\Omega t) \sinh\big(2\Omega (T-t')\big) - \Theta(t-t') \sinh(2\Omega T) \sinh\big(2\Omega (t-t')\big) }{\Omega \sinh(2\Omega T)}\right]_{\mu\nu}\ .
\end{align}
The details of its computation are given in App.~\ref{chap:first:appA} and, for future reference, we also report its following integral
\begin{align}
E_{\mu\nu}(t):&=\int_{t'} \, \Delta^{-1}_{\mu\nu}(t,t') =\left[\frac{\sinh\left(\Omega (T-t)\right)\sinh\left(\Omega t\right)}{\Omega^2\cosh\left(\Omega T\right)} \right]_{\mu\nu}\ . \label{int1} 
\end{align}

Regarding the computation of the functional determinant, this can be carried out by employing the Gel’fand--Yaglom (\acr{GY}) theorem~\cite{Gelfand:1959nq}, as generalized by Kirsten and McKane for a system of differential operators~\cite{Kirsten:2003py, Kirsten:2004qv}.\footnote{For a pedological exposition, see also the \href{https://saalburg.aei.mpg.de/wp-content/uploads/sites/25/2017/03/dunne.pdf}{Saalburg lectures} by Gerald~Dunne \cite{DunneSaalburg}.}
As a matter of completeness, the salient aspects of this procedure are given in App.~\ref{chap:first:appB}. After a direct computation, one gets the equality
\begin{equation} \label{DetS}
{\overline{\mathrm{Det}}(\Delta)=\mathrm{det}\left[ \frac{\sinh(2\Omega T)}{2\Omega T} \right]}\ .
\end{equation}
Finally, the computation of the normalization constant $\mathcal{C}_{\mathrm{DBC}}$ is straightforward; after replacing the results for $\Delta^{-1}$ and $\overline{\mathrm{Det}}(\Delta)$ into Eq.~\eqref{Z}, we obtain
\begin{equation} \label{Cs}
{\mathcal{C}_{\mathrm{DBC}} =\left( 4\pi T \right)^{-\nicefrac{D}{2}}}\ .
\end{equation}

\subsection{Resummed heat kernel} \label{chap:first:sec:sec2.2}
Recalling the formula~\eqref{HK2}, we use it in conjunction with the results of the precedent section to get the following representation for the diagonal of the heat kernel: 
\begin{equation} \label{R}
K(x,x;T)=\frac{\mathrm{e}^{-T V}}{(4\pi T)^{D/2} }\, \mathrm{det}^{-\nicefrac{1}{2}}\left(\frac{\sinh\left( 2\Omega T\right)}{2\Omega T}\right)\;\left. \mathrm{e}^{-\int_t \, L_{\mathrm{int}}\left(-\frac{\delta}{\delta \eta(t)} \right)} \, \mathrm{e}^{\frac12 S_{\mathrm{bos}}[\eta]}\right|_{\eta=\partial V}\ .
\end{equation}
The evaluation of this expression naturally proceeds by perturbatively expanding the exponent in powers of the operator-valued interacting terms contained in $L_{\mathrm{int}}$, i.e., in powers of functional derivatives $\frac{\delta}{\delta \eta}$. This can be readily used to obtain an expansion of the heat kernel up to a certain power in the proper time in an improved Schwinger--DeWitt expansion; in our case, this is measured by considering the number of derivatives acting on a single Yukawa potential. 

Explicitly, the action of the interacting exponential necessary to generate all the terms contributing up to order $T^5$ in the Schwinger--DeWitt expansion reads
\begin{align} 
\label{exp}
\begin{split} 
&\mathrm{e}^{-\int dt \, L_{\mathrm{int}}\left(-\frac{\delta}{\delta \eta(t)} \right)}\left.\mathrm{e}^{\frac12 S_{\mathrm{bos}}[\eta]}\right|_{\eta=\partial V}
\\
&\phantom{e}=\mathrm{e}^{-\int dt \, \left( W_{(3)}(t)+W_{(4)}(t)+\dots\right)} \left.\mathrm{e}^{\frac12 S_{\mathrm{bos}}[\eta]}\right|_{\eta=\partial V}
\\
&\phantom{e}=\bigg( 1-\int_t\, W_{(3)}(t)-\int_t\, W_{(4)}(t) -\int_t\, W_{(5)}(t) -\int_t\, W_{(6)}(t) -\int_t\, W_{(7)}(t)
\\
&\phantom{e}\hspace{2.5cm}-\int_t\, W_{(8)}(t) 
+\frac12 \int_{t_1t_2} \; W_{(3)}(t_1)W_{(3)}(t_2)+\cdots \bigg) \, \left. \mathrm{e}^{\frac12 S_{\mathrm{bos}}[\eta]}\right|_{\eta=\partial V}\ ,
\end{split}
\end{align}
with the vertex-generating operators given by
\begin{align}
W_{(n)}(t)&=\frac{(-1)^{n}}{n!} \partial_{\mu_1 \mu_2 \cdots \mu_n} V\, \frac{\delta^n}{\delta\eta_{\mu_1}(t)\delta\eta_{\mu_2}(t)\cdots\delta\eta_{\mu_n}(t)}\ . \label{V3}
\end{align}
The evaluation of Eq.~\eqref{exp} can be easily performed in the Worldline Formalism; once inserted in \eqref{R} we get the result
\begin{align}
K(x,x;T)=\frac{\mathrm{e}^{-T V+\partial_\mu V\left[\frac{\Omega T-\tanh\left( \Omega T \right)}{4 \Omega^3}\right]^{\mu\nu} \partial_\nu V}}{(4\pi T)^{\nicefrac{D}{2}} \mathrm{det}^{\nicefrac{1}{2}}\left(\frac{\sinh\left( 2\Omega T\right)}{2\Omega T}\right)} \, \Sigma(x,x;T)\ ,
\end{align}
where $\Sigma(x,x;T)$ contains the information on higher derivatives of the potential, which can be written in terms of worldline diagrams; more explicitly, the contributions relevant at order $T^5$ are given by manageable worldline integrals,
\begin{align} \label{upsilon}
&\Sigma(x,x;T)=1+\frac12 \partial_{\mu\nu\rho} V \, \partial_\alpha V \int_t G^{\mu\nu}(t,t) E^{\alpha\rho}(t)\nonumber \\
&-\frac{1}{8}\partial_{\mu\nu\rho\lambda} V \int_t G^{\mu\nu}(t,t) G^{\rho\lambda}(t,t)
+\frac{1}{8}\partial_{\mu\nu\rho\lambda\tau} V \, \partial_\alpha V\int_t G^{\mu\nu}(t,t)G^{\rho\lambda}(t,t) E^{\alpha\tau}(t) \nonumber \\
&+\frac{1}{24} \partial_{\mu\nu\rho} V \partial_{\alpha\beta\gamma} V \int_{t_1 t_2}\bigg[ 3 G^{\mu\nu}(1,1) G^{\rho\alpha}(1,2)G^{\beta\gamma}(2,2) +2G^{\mu\alpha}(1,2)G^{\nu\beta}(1,2)G^{\rho\gamma}(1,2)\bigg] \nonumber \\
&-\frac{1}{48}\partial_{\mu\nu\rho\lambda\tau\theta} V \int_t G^{\mu\nu}(t,t) G^{\rho\lambda}(t,t)G^{\tau\theta}(t,t) \nonumber \\
&-\frac{1}{384}\partial_{\mu\nu\rho\lambda\tau\theta\alpha\beta} V \int_t G^{\mu\nu}(t,t) G^{\rho\lambda}(t,t)G^{\tau\theta}(t,t)G^{\alpha\beta}(t,t)+\dots\ ,
\end{align}
where we denoted $G(i,j):=G(t_i,t_j)$ in the arguments of the Green function $G^{\mu\nu}(t,t'):=\Delta^{-1}_{\mu\nu}(t,t')$ and $E^{\mu\nu}(t)$ has been introduced in Eq.~\eqref{int1}. Expanding in powers of the proper time,
\begin{equation}\label{eq:generalized_coeff}
\Sigma(x,x;T)=:\sum_{j=0}^{\infty} c_j(x,x) \, T^{j}\ ,
\end{equation}
we can read the first few coefficients in the improved Schwinger--DeWitt expansion:\footnote{We postpone to the \hyperref[chap:introII]{prelude to Part~II} a proper introduction of this expansion.}
\begin{align}
c_0(x,x)&=1\ , \\
c_1(x,x)&=0\ , \\
c_2(x,x)&=0\ , \\
c_3(x,x)&=-\frac{1}{60} \partial^\mu{}_\mu{}^\nu{}_\nu V\ , \\
c_4(x,x)&=\frac{1}{30} \partial^{\mu}{}_{\mu\nu} V \, \partial^\nu V- \frac{1}{840} \partial^\mu{}_\mu{}^\nu{}_\nu{}^\rho{}_\rho V \ , \\
\begin{split}
c_5(x,x)&=\frac{17}{5040} \partial_\mu{}^\mu{}_\nu V \, \partial^\nu{}_\rho{}^\rho V +\frac{1}{840} \partial_{\mu\nu\rho} V \, \partial^{\mu\nu\rho} V\ , \\
&\phantom{=}+\frac{1}{280} \partial^{\rho\mu}{}_\mu{}^\nu{}_\nu V \, \partial_\rho V + \frac{1}{210} \partial^{\mu\nu} V \, \partial^\rho{}_{\rho\mu\nu} V -\frac{1}{15120}\partial^\mu{}_\mu{}^\nu{}_\nu{}^\rho{}_\rho{}^\lambda{}_\lambda V\ .
\end{split}
\end{align}
These coefficients are valid at the local level and, in particular, have been computed without the use of integration by parts; they are in perfect agreement with previous calculations---see Refs.~\cite{Franchino-Vinas:2023wea, Franchino-Vinas:2024wof} and references therein.

\section{Resummations for the assisted Yukawa interaction} \label{chap:first:chap:first:sec:assisted_yukawa}
As already highlighted in Ref.~\cite{Franchino-Vinas:2023wea}, the resummed expressions that we have obtained in the preceding section can be employed to analyze scenarios in which a strong field is involved.
In this realm, an exciting mechanism has been devised in Ref.~\cite{Schutzhold:2008pz}, where the rate of Schwinger pair creation was greatly enhanced by including an assisting, fastly varying field to the strong background one. 

In the following, we will consider a perturbative approach to the assisted Yukawa pair production.
Using our worldline setup, we are going to study the imaginary part of the in-out effective action, which is related to the probability of pair creation in the weak-production limit (cf. the \hyperref[chap:introI]{prelude to Part~I}). This is an alternative path to those already considered for \acr{sQED}, for which a scattering approach was considered in Ref.~\cite{Torgrimsson:2017pzs} and, as we will see, we will be able to obtain closed expressions without the need of employing saddle point approximations, which were necessary in Ref.~\cite{Torgrimsson:2018xdf}.

Let us then consider a potential of the form
\begin{equation}
 V(x)= V(x) +\epsilon \, \mathcal{V}(x) \ ,
\end{equation}
where $ V(x)$ is a strong field, while $\mathcal{V}(x)$ is a fastly varying field, whose strength is tuned by the parameter $\epsilon\ll 1$. 
Since we are interested in a global quantity (the effective action), we will employ a worldline model with string-inspired boundary conditions, which we describe in the following.

Departing from Eq.~\eqref{chap:first:Gamma} for the effective action, we recall the interpretation of the heat kernel as a transition amplitude; this suggests that the spacetime integral of the diagonal of the heat kernel can be equivalently written as a path integral over periodic (\acr{P}) trajectories,
\begin{align} \label{path3}
\begin{split}
{K}(T):&= \int {\rm d}^Dx \, K(x,x;T) 
\\
&=\oint \mathcal{D}x \, \mathrm{e}^{-\int_t \left( \frac{\dot{x}^2}{4}+ V(x) +\epsilon \, \mathcal{V}(x) \right)}\ . 
\end{split}
\end{align}
Afterwards, we introduce the loop's “center of mass”
\begin{equation} \label{CdM}
\bar{x}^\mu=\frac{1}{T} \int_t \, x^\mu(t)\ .
\end{equation}
Once it is fixed, the path integration can be reobtained by subsequently integrating over all closed loops that share the center of mass; in other words, we decompose the closed worldlines as
\begin{equation}
\oint \mathcal{D}x=\int {\rm d}^D \bar{x} \, \oint \mathcal{D}s\ ,
\end{equation}
where the paths satisfy then the string-inspired boundary conditions:
\begin{equation} \label{s}
x^\mu(t)=\bar{x}^\mu+s^\mu(t) \quad \text{with} \quad \int_t \, s^\mu(t)=0\ .
\end{equation}
For the interested reader, a more general approach to boundary conditions in the worldline is explained in App.~\ref{chap:first:appA}.

Coming back to the computation of the integrated heat kernel, for convenience we Taylor expand the strong potential $V$ about the center of mass variables:
\begin{equation}
 V(\bar{x}+s)= V(\bar{x})+s^\mu \partial_\mu V(\bar{x})+\frac12 s^\mu s^\nu \partial_{\mu\nu} V(\bar{x}) + \dots\ .
\end{equation}
This allows us to split once more the contributions of $V$ in the action~\eqref{path3} into an interacting part and a free one. Introducing the string-inspired generating functional
\begin{equation}
Z[\eta](\bar{x};T)=\oint \mathcal{D}s \, \mathrm{e}^{-\int_t \frac{\dot{s}^2}{4}+\frac12 s^\mu s^\nu \partial_{\mu\nu}V+s^\mu\eta_\mu+\epsilon \, \mathcal{V}(\bar{x}+s)}\ ,
\end{equation}
we recast the integrated heat kernel as
\begin{equation} \label{path4}
{K}(T)=\left. \int {\rm d}^D\bar{x} \, \mathrm{e}^{-TV}\, \mathrm{e}^{-\int_t L_{\mathrm{int}}\left( -\frac{\delta}{\delta \eta(t)} \right)} \, Z[\eta](\bar{x};T)\right|_{\eta=\partial V} .
\end{equation}
In the following, a perturbative expansion in the weak-field parameter $\epsilon$ is to be performed: 
\begin{equation}
Z=Z_0+\epsilon Z_1+\epsilon^2 Z_2 + \dots\ ;
\end{equation}
although each contribution $Z_i$ can be explicitly computed in the worldline, we will focus on the contributions up to order $\epsilon$, which will be enough to show the existence of the assisted effect.

\subsection{Zeroth order in $\mathcal{V}$}
At zeroth order, we need to compute
\begin{equation}
Z_0[\eta](\bar{x};T)=\oint \mathcal{D}s \, \mathrm{e}^{-\int_t \frac{\dot{s}^2}{4}+ s^\mu s^\nu \Omega_{\mu\nu}+s^\mu\eta_\mu}\ ,
\end{equation}
which is analogous to the path integral in Eq.~\eqref{2.9}, replacing the \acr{DBC} with the \acr{SI} ones. 
The relevant operator for this computation, $\Delta_{\rm SI}$, satisfies Eq.~\eqref{Delta} using $\Omega^2_{\mu\nu}(\bar{x}):=\tfrac12\partial_{\mu\nu} V(\bar{x})$ [instead of $\Omega^2_{\mu\nu}({y})$] and we should recall that its domain of definition is made of functions which obey the condition~\eqref{s}:
\begin{equation} \label{eq:Delta_SI}
{\Delta^{\mu\nu}_{\rm SI} (t,t')=-\frac12 \delta^{\mu\nu}\partial^2_t\delta(t-t')+2 \, \left(\Omega^2\right)^{\mu\nu}(\bar{x})\delta(t-t')}\ .
\end{equation}

After the replacement $s \to {s}=\tilde s+\Delta_{\rm SI}^{-1}\eta$, the generating functional can be brought into a Gaussian form, which can readily be integrated:
\begin{align} \label{3.44}
\begin{split}
Z_0[\eta](\bar{x};T)&= \mathrm{e}^{\frac12\int_{t_1t_2} \eta_\mu \left(\Delta^{-1}_{\rm SI}\right)^{\mu\nu}\eta^\nu} \, \oint \mathcal{D}\tilde{s} \, \mathrm{e}^{-\frac12\int_{t_1t_2} \tilde{s}_\mu\Delta^{\mu\nu}_{\rm SI} \tilde{s}_\nu}\\
&=\frac{\mathcal{C}_{\mathrm{SI}} \, \mathrm{e}^{\frac12 S^{\rm SI}_{\mathrm{bos}}[\eta]}}{{{\overline{\mathrm{Det}}}'}^{\nicefrac{1}{2}}(\Delta_{\mathrm{SI}})}\ .
\end{split}
\end{align}
This expression is made of three different elements. First, the exponential of the bosonic action, which is completely analogous to Eq.~\eqref{Sbos}, but for the fact that one should replace the operators with the appropriate \acr{SI} ones. 
Second, the functional determinant of the $\Delta_{\rm SI}$ operator, which can be computed by generalizing the Gel’fand--Yaglom method to operators acting on a domain of periodic functions.\footnote{The prime over the determinant means that we are excluding the constant mode: in the \acr{SI} \acr{BC} this is automatically excluded, while in the free operator this is done by hand.} As explained in App.~\ref{chap:first:appB}, the result is
\begin{equation}
\overline{\mathrm{Det}}'(\Delta_{\mathrm{SI}})=\mathrm{det}\left[ \frac{\sinh^2(\Omega T)}{\Omega^2T^2} \right]\ .
\end{equation}
Third and last, we have introduced the parameter $\mathcal{C}_{\mathrm{SI}}$; as in the Dirichlet case, it can be determined by appealing to a known result; once more, we get
\begin{align} \label{CSI}
&\mathcal{C}_{\mathrm{SI}} = \left( 4\pi T \right)^{-\nicefrac{D}{2}}\ ,
\end{align}
which also agrees with the expected result for the small proper time expansion of the integrated heat kernel~ \cite{Vassilevich:2003xt}. Note that this normalization yields a well-defined limit for a vanishing $\Omega$.
Summing all these pieces, we get
\begin{align} \label{Z0}
\begin{split}
Z_0[\eta](\bar{x};T)&= \frac{ \mathrm{e}^{\frac12 S^{\rm SI}_{\mathrm{bos}}[\eta]}}{(4\pi T)^{\nicefrac{D}{2}}\mathrm{det}^{\nicefrac{1}{2}}\left(\frac{\sinh^2(\Omega T)}{\Omega^2T^2}\right)}\ .
\end{split}
\end{align}

\subsubsection{Unassisted particle production}\label{chap:first:sec:non-assisted}
As a warm-up application of the preceding formulae, consider a strong quadratic potential, i.e. one whose higher derivatives can be neglected, $\partial^{n\geq 3}V= 0$.
This implies that the worldline interactions in Eq.~\eqref{path4} are switched off and, inserting the zeroth-order generating functional in the definition of the effective action, cf. Eq.~\eqref{chap:first:Gamma}, we obtain the following effective Lagrangian:
\begin{equation} \label{3.45}
 \mathcal{L}_{\mathrm{eff}}[V]=-\frac12 \int_0^{\infty} \frac{{\rm d}T}{T} \, \frac{\mathrm{e}^{-TV}}{\left( 4\pi T \right)^{\nicefrac{D}{2}}} \, \mathrm{det}^{\nicefrac{1}{2}}\left( \frac{\Omega^2T^2}{\sinh^2(\Omega T)} \right)\ .
\end{equation}
The first aspect worth mentioning is the fact that the proper time integral contains a singularity at $T=0$, indicating a \acr{UV} divergence that needs to be removed by renormalization. It will eventually require the inclusion of counterterms in the effective Lagrangian, whose number will depend on the spacetime dimensionality $D$. However, as we will shortly see, in our strong field case the integral has developed further poles in the $T$-plane if we work in Minkowski spacetime. 

The determinant in the equation above can be evaluated from the knowledge of the eigenvalues of the real and symmetric matrix $\Omega^2$, defined in Eq.~\eqref{Omega}. These in turn depend on the invariants $\mathrm{det}(\Omega^2)$ and $\mathrm{tr}\left(\Omega^{2j}\right)$, for $j=1,2,.., D-1$. For $D\leq 4$ it is possible to obtain analytic expressions for the eigenvalues in terms of these invariants. However, the resulting expressions become cumbersome unless $D=2$, which serves as an interesting and illuminating special case. Therefore, in what follows, we assume that the background potential depends just on two coordinates (e.g., $x_0$ and $x_1$). This results in a nontrivial $2\times 2$ block in the matrix $\Omega$, whose eigenvalues entirely determine the effective action. With a slight abuse of notation, we will continue to denote this block as $\Omega^2$.

The (real) eigenvalues of the $2\times 2$ block are given by
\begin{equation}
\lambda_\pm = -\mathfrak F\pm\sqrt{\mathfrak F^2 -\mathfrak G^2},
\end{equation}
where we have introduced the notation
\begin{equation}
\mathfrak F = -\frac{1}{2}\mathrm{tr}(\Omega^2)\, ,\quad \mathfrak G^2=\mathrm{det}(\Omega^2)\, .
\end{equation}
We have called these invariants $\mathfrak F$ and $\mathfrak G$, given that they play roles analogous to $\mathfrak{F}_{\rm QED}:=F_{\mu\nu} F^{\mu\nu}$ and $\mathfrak{G}_{\rm QED}:=\tilde F_{\mu\nu} F^{\mu\nu}$ in four-dimensional
quantum electrodynamics. 
Using these eigenvalues, the effective Lagrangian can be expressed as
\begin{equation} \label{Leff2D}
\mathcal{L}_{\mathrm{eff}}[V]=-\frac12 \int_0^{\infty} \frac{{\rm d}T}{T} \, \frac{\mathrm{e}^{-TV} }{\left( 4\pi T \right)^{\nicefrac{D}{2}}} \, \left( \frac{\sqrt{\lambda_+\lambda_-}T^2}{\sinh(\sqrt \lambda_+T)\sinh(\sqrt \lambda_-T)} \right)\ .
\end{equation}

Depending on the specific values of the eigenvalues, the effective Lagrangian can develop an imaginary part when rotated to Minkowski space. To patently see this in a particular case, let us choose $v\in\mathbb{R}$ and the strong potential in Euclidean time to be
\begin{equation}\label{eq:background_quadratic}
 V_0(x):=v^2 \, x_0^2+m^2\ ,
\end{equation}
where we have also included a mass term for convenience. As we will see, this is the analog of the constant electric field which was considered in Ref.~\cite{Franchino-Vinas:2023wea}. With this choice we have $\lambda_+=v^2$ and $\lambda_-=0$. Then, the Euclidean effective Lagrangian is given by
\begin{equation}
 \mathcal{L}_{\mathrm{eff}}[V_0]=-\frac{v}{2} \int_0^{\infty} {{ \rm d} T} \, \frac{\mathrm{e}^{-Tv^2\bar{x}_0^2-Tm^2}}{\left( 4\pi T \right)^{\nicefrac{D}{2}} {\sinh(v T)}} \ .
\end{equation}
Though the rotation to minkowski spacetime requires some care in the calculations, it can be done as follows. The Euclidean quantities $X_{\mathrm{E}}$ are related to the Minkowski ones by adding a $(-i)$ factor for every $0$th component which is involved. For example, for a vector we have $X^0_{\mathrm{E}}=-i X^0_{\mathrm{M}}$. Actually, we can keep track of the rotation by including a parameter $\phi$, so that $X_{\mathrm{E}}=\mathrm{e}^{i\phi} X_{\mathrm{M}}$; thus, expanding $\phi$ about $-\pi/2$ we can keep track of branch cuts and poles in the complex plane. The objects that we need to rotate include $v$ (which is to be counted as a derivative of the potential) and the effective action, 
which satisfies
\begin{equation}
 \Gamma_{\mathrm{M}}\Big\vert_{\text{Wick rotated}}=i \Gamma_{\mathrm{E}}\ .
\end{equation}
Taking this into account, the Minkowskian effective action at zeroth order reads
\begin{equation}\label{eq:Minkowski_nonassisted_EA}
 \Gamma^{(0)}_{\mathrm{M}}[V_0]=\frac{ \sqrt{\pi}}{2} \frac{\mathrm{Vol}_{\mathrm{(D-1)}}}{\left( 4\pi\right)^{\nicefrac{D}{2}}} \int_0^{\infty} \frac{{\rm d}T}{T^{\frac{D+1}{2}}} \, \frac{\mathrm{e}^{-m^2 T}}{\sin({v}_{\mathrm{M}} T + i 0)}\ ,
\end{equation}
where $\mathrm{Vol}_{\mathrm{(D-1)}}$ is the volume of the ($D-1$)-dimensional space and we have kept track of the rotation through the $+i 0$ contribution in the argument of the sine. 
Eq.~\eqref{eq:Minkowski_nonassisted_EA} shows nontrivial poles where the sine function vanishes, i.e. at\footnote{These are the values for which the differential operator $\Delta_{\mu\nu}$ with \acr{PBC} develops zero modes in Minkowski spacetime. Although the $n=0$ one does not correspond to a zero eigenvalue, as can be extrapolated from the spectrum (see App.~\ref{chap:first:app:det_pbc}), it is not surprising that it still represents a divergence, as it is well-known to be related to the renormalization of the theory.}
\begin{equation} \label{poles}
 T=\frac{\pi n}{{v}_{\mathrm{M}}}, \quad 0<n \in \mathbb{N} \ .
\end{equation}
On their turn, these poles generate an imaginary part in the effective action, as can be seen from the use of the residue theorem in the complex $T$ plane to perform the proper time integral,
\begin{align}\label{eq:im_ea_0}
\begin{split}
\operatorname{Im} \Gamma^{(0)}_{\mathrm{M}}
 &=\frac{\pi}{2} \frac{\mathrm{Vol}_{\mathrm{(D-1)}}}{\left( 2\pi\right)^{D}} \, v_{\mathrm{M}}^{\frac{D-1}{2}} \sum_{n=1}^\infty (-1)^{n+1} \frac{\mathrm{e}^{-\frac{m^2 \pi n }{v_{\mathrm{M}}}}}{n^{\frac{D+1}{2}}}
 \\
 &=-\frac{\pi}{2} \frac{\mathrm{Vol}_{\mathrm{(D-1)}}}{\left( 2\pi\right)^{D}} \, v_{\mathrm{M}}^{\frac{D-1}{2}} \operatorname{Li}_{\frac{D+1}{2}}\left(-\mathrm{e}^{-\frac{m^2 \pi }{v_{\mathrm{M}}}}\right)\ ,
 \end{split} 
\end{align}
where $\operatorname{Li}_s (\cdot)$ is the polylogarithm of order ${s}$.\footnote{The polylogarithm function is defined by
\begin{equation*}
	\operatorname{Li}_s (z):=\sum\limits_{n=1}^{\infty}\frac{z^n}{n^s}\;.
\end{equation*}}
As already explained, the physical interpretation of this result follows by identifying it with the vacuum persistence amplitude,
\begin{align}
 \vert\langle 0_{\rm out} \vert 0_{\rm in}\rangle\vert^2 ={\rm e}^{- 2\operatorname{Im} \Gamma_\mathrm{M}} =:1-P_{\mathrm{pair}}\ ,
\end{align}
which is reported here once again for ease of reference.

\subsubsection{Non-quadratic time-dependent potentials}
Let us consider an arbitrary time-dependent potential $V(t)$. 
Using the zeroth order result, to zeroth order also in the generalized heat kernel coefficients, we obtain what we will call the locally quadratic field approximation (\acr{LQFA}),
\begin{align}
    \begin{split}
    \label{eq:eff_time_inho}
\Gamma_{\rm M}
&= -\frac{ \operatorname{Vol_{\rm D-1}}}{2 \sqrt{2}} \int_0^{\infty} \frac{{\rm d}T}{T} \, \int_{-\infty}^{\infty} {\rm d}t\frac{\mathrm{e}^{-T \left(V(t)+m^2\right)}}{\left( 4\pi T \right)^{\nicefrac{D}{2}}} \,  \frac{\sqrt{-V''(t)}T}{ \sinh\left(\sqrt{-V''(t)} T/\sqrt{2}\right)}, \
\end{split}
\end{align}
which heuristically is expected to be valid when the potentials are slowly varying; in this sense, it is equivalent to the locally constant field approximation employed in \acr{QED}~\cite{Dunne:2006st, Fedotov:2022ely}. 
For the computation of the imaginary part of the effective action, if the potential is positive, the relevant contributions will be those for which $V''$ is negative\footnote{For a negative potential, see the discussion at the end of Sec.~\ref{chap:first:sec:spatial_pair}.}.
Importantly, one should be careful with the recipe employed to avoid the singularities that appear in the proper time integral. 
 To take care of this, we can follow as in the previous section; explained in other words, since our expressions are covariant, we can rotate the $00$th component of the metric, which we will call  $\eta_{00}$.
In order to be consistent with the $-\mathi \epsilon$ prescription, the rotation should be $\eta^{00}\to e^{-\mathi \pi}\eta^{00}$.
A further alternative is to consider the heat kernel in minkowski spacetime and employ its imaginary proper time expansion~\cite{DeWitt:2003pm}. After a proper time Wick rotation, this implies a $T+\mathi \epsilon$ prescription that will be used in what follows (it will be especially useful for the space-dependent potentials).

Coming back to Eq.~\eqref{eq:eff_time_inho}, its imaginary part can be readily computed to be
\begin{align}
 \begin{split} &\operatorname{Im} \frac{\Gamma_{\rm M}}{\operatorname{Vol_{\rm D-1}}}   = \frac{\pi}{2} \, \sum_{r=1}^{\infty} (-1)^{r+1}  \int^{\infty}_{-\infty} {\rm d}t\,  \Theta(V''(t))
  \\
 &\hspace{4cm}\times \left(\frac{\sqrt{V''(t)} }{ 4 \sqrt{2} \pi^2  r }\right)^{\nicefrac{D}{2}}  \mathrm{exp}\left(\frac{-\sqrt{2}\pi r\left(V(t)+m^2\right)}{\sqrt{V''(t)}}\right)\, .
\end{split}
\end{align}
 We can go beyond \acr{LQFA}  by including higher derivative terms, i.e. by multiplying the integrand in the equation above by using the improved Schwinger--DeWitt coefficients obtained in the \acr{SI} approach (see App.~\ref{chap:first:app:coeff_SI}). To lowest order,  we obtain
\begin{align}
\begin{split}\label{eq:im_time_inho}
  \operatorname{Im} \frac{\Gamma_{\rm M}}{\operatorname{Vol_{\rm D-1}}}=& \frac{\pi}{2} \,
\sum_{r=1}^{\infty} (-1)^{r+1} \int^{\infty}_{-\infty} {\rm d}t\,  \Theta\big(V''(t)\big) \left(\frac{\sqrt{V''(t)} }{ 4\sqrt{2}\pi^2 r}\right)^{\nicefrac{D}{2}}   \\
& \times \mathrm{exp}\left(\frac{-\sqrt{2} \pi r\left(V(t)+m^2\right)}{\sqrt{V''(t)}}\right) \left(1-\frac{2^{\nicefrac{3}{4}}\pi^3}{288} \frac{V^{(4)}(t)}{(V'')^{\nicefrac{3}{2}}}+\cdots\right).
\end{split}
\end{align}

As a particular case, consider an oscillatory potential; having in mind the analogy with the electromagnetic case, which in essence is $(F^2)_{\mu\nu}\to \Omega_{\mu\nu}$, a plausible form is
\begin{align}
    V(t)=\frac{V_0}{\omega^2} \sin^2(\omega t), \quad V_0>0,
\end{align}
which additionally reduces to the quadratic potential when $\omega\to 0$.
The Hessian matrix is defined in terms of the single element
\begin{align}
    V''(t)= 2{V_0} \cos(2\omega t);
\end{align}
inserting it  in Eq.~\eqref{eq:im_time_inho} and noting that the periodicity of the potential can be absorbed into the length $L_0=\frac{2\pi}{\omega}$,
we are led to the result
\begin{align}\label{eq:im_time_sine}
    \begin{split}
&\operatorname{Im} \frac{\Gamma_{\rm M}}{L_0\operatorname{Vol_{\rm D-1}}} 
= \, \sum_{r=1}^{\infty}  \frac{(-1)^{r+1}\pi}{4} \int^{\pi/4}_{-\pi/4} {\rm d}t \left(\frac{\sqrt{V_0 \cos(2 t)} }{ 4\pi^2 r }\right)^{\nicefrac{D}{2}}
\\
&\hspace{1cm} \times \mathrm{exp}\left(- \pi r \frac{ \sqrt{V_0}}{\omega^2}  \frac{\left( \sin^2 t+\gamma^2\right)}{ \sqrt{ \cos(2 t)}}\right)  \left(1+\frac{\pi^3}{36} \frac{\omega^2}{ \sqrt{V_0} } \frac{1}{\sqrt{\cos (2 t)}}+\cdots\right) \,,
\end{split}
\end{align}
where we have introduced Keldysh's adiabaticity parameter~\cite{Keldysh:1965ojf} :
\begin{align}
    \gamma:=\frac{m\omega}{\sqrt{V_0}}.
\end{align}
Note that the first correction in Eq.~\eqref{eq:im_time_sine} is positive and also reflects the large-potential character of our expansion, which is not in powers of $\gamma$, but in terms of the dimensionless, small parameter $\omega^2/\sqrt{V_0}$. For extreme fields,  one could work out the time integral using Laplace's theory, which effectively leads to a parabolic approximation. Indeed, physically speaking, the expression~\eqref{eq:im_time_sine} can be seen as an improvement over simply approximating the harmonic function with a periodic array of quadratic potentials centered around its minima.
Additionally, in the small $\omega$ limit the quadratic result is  reobtained; this can be seen either by using Laplace's theory in the previous equation or by rescaling in $\omega$,
\begin{align}
    \begin{split}\label{eq:im_time_oscillating}
\operatorname{Im} \frac{\Gamma_{\rm M}}{L_0 \operatorname{Vol_{\rm D-1}}}  
&\overset{\phantom{\omega\to 0}}{=}\frac{ \pi\omega }{4} \, \sum_{r=1}^{\infty} (-1)^{r+1}
\\
&\hspace{1cm}\times\int^{\pi/4\omega}_{-\pi/4\omega} {\rm d}t \left(\frac{\sqrt{V_0 \cos(2\omega t)} }{ 4\pi^2 r }\right)^{\nicefrac{D}{2}} \mathrm{exp}\left(\frac{-\pi r \left(\frac{V_0}{\omega^2} \sin^2(\omega t)+m^2\right)}{\sqrt{V_0 \cos(2\omega t)}}\right) 
\\
&\overset{\omega\to 0}{=}\frac{ \pi\omega}{4} \, \sum_{r=1}^{\infty} (-1)^{r+1} \left(\frac{v }{ 4\pi^2 r }\right)^{\nicefrac{D}{2}} \frac{1}{\sqrt{r v}}  \mathrm{exp}\left(\frac{-\pi r m^2}{v}\right),
\end{split}
\end{align}
where we have identified $\sqrt{V_0}\to v$.
Although we have not managed to obtain a closed analytic formula for Eq.~\eqref{eq:im_time_sine}, one can truncate the series at any desired order and readily numerically perform the integrations, at least for massive fields, for which a decreasing exponential behaviour is guaranteed.

As a last comment, we would like to emphasize once more that we expect Eq.~\eqref{eq:im_time_sine} to be a good approximation for small frequency $\omega$. If an all-scalar version of the worldline instantons technique could be developed, it could be used to check all these (and the following section's) results and complement them with expressions encompassing all the derivative contributions (but just the first large mass one). 

\subsubsection{Spatial particle creation}\label{chap:first:sec:spatial_pair}
The framework is slightly more involved when considering the spatial case. For simplicity, we will analyze the case depending on just one variable, say $x_3$, so that
\begin{align}
    \begin{split}
    \label{eq:eff_sapce_inho}
\Gamma_{\rm M}
&= -\frac{\mathi \operatorname{Vol_{\rm D-1}}}{2\sqrt{2}} \int_0^{+\infty} \frac{{\rm d}T}{T} \, \int_{-\infty}^{\infty} {\rm d}x_3\frac{\mathrm{e}^{-T \left(V(x_3)+m^2\right)}}{\left( 4\pi T \right)^{\nicefrac{D}{2}}} \,  \frac{\sqrt{V''(x_3)}T}{ \sinh\left(\sqrt{V''(x_3)} T/\sqrt{2}\right)}. \
\end{split}
\end{align}
In this setup, for weak fields instabilities arise when $\partial^2_3 V$ becomes negative; the difficulty resides in the fact that the instabilities are already present at the level of the Euclidean effective action and, thus, the imaginary proper time prescription described in the previous section is required in order to obtain a sensible answer. 

To simplify the discussion further, let us consider the quadratic case 
\begin{align}
    V(x_3)=v^2 \, x_3^2+m^2,
\end{align}
leaving the sign of $v^2$ for the moment undetermined; note that the situation with the inverted quadratic potential mimics what happens in the electromagnetic homogeneous field in the spatial gauge. 
Then, following the lines of the previous section we arrive at
\begin{align}\label{eq:eff_spatial_har}
    \frac{\Gamma_\mathrm{M}[V]}{\operatorname{Vol}_{\rm D-1}}=-\frac{\mathi}{2} \int_0^{\infty} \frac{{\rm d} T}{T} \, \int^{\infty}_{-\infty}{\rm d}x_3\frac{\mathrm{e}^{-T(v^2{x}_3^2+m^2)}}{\left( 4\pi T \right)^{\nicefrac{D}{2}}} \,  \frac{v T}{\sinh(v T)}\ .
\end{align}
At this point, we have to consider the two possible alternatives separately. If $v^2>0$, all the integrals in Eq.~\eqref{eq:eff_spatial_har} are well defined and are real, i.e. there is no creation of pairs. This is of course logical: since the potential is static, no energetic source is available to create the pairs. If instead $v^2<0$, the sinh in the denominator becomes a sine and singularities appear. 
However, the integral in $x_3$ becomes ill-defined; this can be seen as a consequence of the fact that the effective mass $V+m^2$ is negative in an infinite region. 
A way to circumvent this problem in this simple case is to perform first the integral in $x_3$ (with $v^2>0$) and afterwards appeal to an analytical continuation in\footnote{One should employ the rule of thumb that the rotation should not interfere with the $\mathi \epsilon$ Feynman rule.} $v$. The result obtained with this prescription agrees with the one using the imaginary proper time, and gives the same pair creation probability as for the quadratic time dependence. This agreement resembles the situation in the \acr{EM} homogeneous case. 

As a last example, consider another spatial, now oscillating background, 
\begin{align}\label{eq:spatial_harmonic}
    V(x_3)=\frac{V_0}{\omega^2} \sin^2(\omega x_3). 
\end{align}
If $V_0>0$, then the only change with respect to the time-dependent field is in the integration region over $x_3$, which is effectively shifted to the complementary interval in the period of $\cos (2\omega x_3)$;\footnote{In this case, $\rm Vol_{D}=\mathit{L}_3 Vol_{D-1}$, where $L_3$ denotes a spatial cutoff.} after shifting the integral one gets:
\begin{align}\label{eq:im_space_sine}
    \begin{split}
\operatorname{Im} \frac{\Gamma_{\rm M}}{\operatorname{Vol_{\rm D}}} 
=& \, \sum_{r=1}^{\infty}  \frac{(-1)^{r+1} \pi}{4}
\\
&
\times \int^{\pi/4}_{-\pi/4} {\rm d}x_3 \left(\frac{\sqrt{V_0 \cos(2x_3)} }{ 4\pi^2 r }\right)^{\nicefrac{D}{2}}  \mathrm{exp}\left(\frac{-\pi r \left(\frac{\sqrt{V_0}}{\omega^2} \cos^2(x_3)+\frac{m^2}{\sqrt{V_0}}\right)}{\sqrt{ \cos(2x_3)}}\right)
\,. 
\end{split}
\end{align}
Since our expansion is valid for strong potentials, the integral in the coordinate $x_3$ can be done using Laplace's theory; contrary to the time-dependent scenario,  one notices that $\cos^2 x_3\geq 1/2$ in the integration region, and thus the integral is exponentially suppressed in the strong field expansion. This is consistent with the fact that for a positive potential one does not expect particle creation to take place. This can actually be discussed in more general terms: whenever the potential is strictly positive and its second derivative is negative, an exponential suppression is present in the imaginary part of the effective action in the \acr{LQFA}. 

Taking into account our discussion for the quadratic case, one would expect that the harmonic potential~\eqref{eq:spatial_harmonic} with $V_0<0$ should resemble the most an electromagnetic space-dependent field. However, in this scenario there is a caveat: for strong (negative) fields the integral in Eq.~\eqref{eq:eff_sapce_inho} develops infrared (\acr{IR}) singularities. These are related to the divergent behaviour for large $T$ and could imply a further source of imaginary contributions in the effective action; they are of course not present for \acr{QED}, since gauge invariance precludes the presence of the gauge potential in the effective action. One could still appeal for example to analytical continuations to analyze the \acr{IR} singularities, as done for the quadratic field; however, for the oscillatory potential the integrals can not all be done explicitly and the same can be said for the most frequently analyzed pulses in the literature. 

For the sake of completeness, let us briefly discuss an analogue expansion in \acr{QED}; a thorough analysis will be  left for the future. In such a case, the first derivative contribution, i.e. the equivalent of the $\omega^2/\sqrt{V_0}$ term in Eq. \eqref{eq:im_time_sine}, is given by the $c_3^{\rm QED}$ coefficient. For a background field which is exclusively electric ($F_{0i}=E_i(x)$ and $F_{ij}=0$), it can be shown that 
\begin{align}
    c_3^{\rm QED}=\frac{2}{5}\left[ \partial_j E_i  \partial^j E^i - \partial_0 E_i \partial^0 E^i \right],
\end{align}
cf. the $\bar{o}_3$ coefficient in Ref.~\cite{Navarro-Salas:2020oew}.
This expression clearly displays a sign flip when shuffling from purely spatial to time-dependent potentials, meaning that the pair production probability is enhanced in the latter case, while it decreases for the former. This seems to be in agreement with the threshold which has been observed using worldline instantons for potentials depending on spatial coordinates~\cite{Dunne:2006st}.

\subsection{First and higher orders in $\mathcal{V}$} \label{chap:first:sec3.4.2}
Before analyzing the first-order contribution in $\epsilon$, we will sketch how the computation to any order can be done. 
At $n$th order in the assisting potential, the contribution to the generating functional is
\begin{align}\label{eq:Zn}
\begin{split}
&Z_n[\eta](\bar{x};T)=(-1)^n\oint \mathcal{D}s \, \int_{t_1 t_2\cdots}\, \mathrm{e}^{-\int_t \frac{\dot{s}^2}{4}+ s^\mu s^\nu \Omega_{\mu\nu}+s^\mu\eta_\mu} \, \mathcal{V}(\bar{x}+s(t_1))\mathcal{V}(\bar{x}+s(t_2))\cdots
\\
&=(-1)^n\int \left[\prod_{j=1}^n \hat{\rm d}q_j \tilde{\mathcal{V}}(q_1) \right] 
\int_{t_1\cdots}\, \oint \mathcal{D}s \, \mathrm{e}^{-i \sum_{l=1}^n q_l \cdot \bar{x}-\int_t \frac{\dot{s}^2}{4}+s^\mu s^\nu \Omega_{\mu\nu}+s^\mu\eta_\mu+is_\mu\sum_{i=1}^n q_i^\mu \delta(\tau-t_i)} \, \ ,
\end{split}
\end{align}
where we have Fourier transformed $\mathcal{V}$ as
\begin{equation} \label{FT}
\mathcal{V}(x)=\int\hat{\rm d}q \, \mathrm{e}^{-i q \cdot x } \, \tilde{\mathcal{V}}(q)\ 
\end{equation}
and used the short-hand notation $\hat{\rm d}q:=\frac{{\rm d}^Dq}{(2\pi)^D}$. 
The expression \eqref{eq:Zn} is still manageable, since introducing explicitly the classical solution to the equations of motion of the worldline,
\begin{equation} \label{eq:}
s^\mu \to s^\mu_{\mathrm{cl}} + \hat{s}^\mu\ ,
\end{equation}
the exponent becomes quadratic in the quantum fluctuations.
A direct calculation shows that we need to find the solution of
\begin{equation}
(\Delta_{\rm SI})_{\mu\nu} \, s_{\mathrm{cl}}^\nu(\tau)=-\eta_\mu(\tau)-\mathfrak{d}_\mu(\tau)\ ,
\end{equation}
where $\mathfrak{d}$ contains all the momentum inhomogeneities:
\begin{align}
 \mathfrak{d}^\mu(\tau):= i \sum_{i=1}^n q^\mu_i\delta(\tau-t_i)\ .
\end{align}
This can be simply solved by considering the Green function of the operator, $\mathcal{G}:=\Delta_{\rm SI}^{-1}$, which is computed in App.~\ref{chap:first:appA}:
\begin{align}
\mathcal{G}^{\mu\nu} (t,t')
&= \left[-\frac{1}{2T\Omega^2}+ \frac{\cosh{(\Omega(T+2t-2t'))}}{2\Omega\sinh{(\Omega T)}}- \Theta(t-t') \, \frac{\sinh{(2\Omega(t-t'))}}{\Omega} \right]^{\mu\nu}\!\! ,
\end{align}
so that
\begin{align}
s^\mu_{\mathrm{cl}}(\tau)&=-i \sum_{i=1}^n \mathcal{G}^{\mu}{}_{\nu}(\tau,t_i) q^\nu_{i} -\int_{t'} \, \mathcal{G}^{\mu\nu}(\tau,t')\eta_\nu(t')\label{eq:classical_solution_SI}\ .
\end{align}
After replacing Eq.~\eqref{eq:classical_solution_SI} in the expression for $Z_n$, we obtain 
\begin{align}
&Z_n[\eta](\bar{x};T) \nonumber
\\
&=(-1)^n\int \left[\prod_{j=1}^n \hat{\rm d}q_j \tilde{\mathcal{V}}(q_1) \right]\, \mathrm{e}^{-i \sum_{l=1}^n q_l \cdot \bar{x}} 
\int_{t_1t_2\cdots} \oint \mathcal{D}s \, \mathrm{e}^{-\frac12 \int_{t t'} \hat{s}^\mu \Delta_{\mu\nu} \hat{s}^\nu
+\frac12 \int_{t t'} \left( \eta +\mathfrak{d}\right)_\mu \mathcal{G}^{\mu\nu} \left( \eta +\mathfrak{d}\right)_\nu} \, \nonumber
\\
&=(-1)^n\frac{\mathrm{e}^{-TV}}{\left( 4\pi T \right)^{\nicefrac{D}{2}}} \, \mathrm{det}^{\nicefrac{1}{2}}\left( \frac{\Omega^2T^2}{\sinh^2(\Omega T)} \right) \nonumber
\\
&\hspace{2cm}\times
 \int \left[\prod_{j=1}^n \hat{\rm d}q_j \tilde{\mathcal{V}}(q_1) \right]\, \mathrm{e}^{-i \sum_{l=1}^n q_l \cdot \bar{x}} 
\int_{t_1t_2\cdots}\, \mathrm{e}^{\frac12 \int_{t t'} \left( \eta +\mathfrak{d}\right)_\mu \mathcal{G}^{\mu\nu} \left( \eta +\mathfrak{d}\right)_\nu} \ .
\end{align}
If we are just interested in the contribution without worldline interactions, this formula can be further simplified to
\begin{align}\label{eq:Zn_final}
\begin{split}
Z_n[\eta](\bar{x};T)
&=(-1)^n\frac{\mathrm{e}^{-TV}}{\left( 4\pi T \right)^{\nicefrac{D}{2}}} \, \mathrm{det}^{\nicefrac{1}{2}}\left( \frac{\Omega^2T^2}{\sinh^2(\Omega T)} \right)
\\
&\hspace{2cm}\times\int \left[\prod_{j=1}^n \hat{\rm d}q_j \tilde{\mathcal{V}}(q_1) \right]\, \mathrm{e}^{-i \sum_{l=1}^n q_l \cdot \bar{x}} 
\int_{t_1t_2\cdots}\, \mathrm{e}^{\frac12 \int_{t t'} \mathfrak{d}_\mu \mathcal{G}^{\mu\nu} \mathfrak{d}_\nu} \ ,
\end{split}
\end{align}
given that the \acr{SI} Green function satisfies
\begin{align}
 \int_{t'} \mathcal{G}^{\mu\nu}(t,t') \partial_\mu V(\bar{x})=0\,\ .
\end{align}
It is important to note that, in Eq.~\eqref{eq:Zn_final}, an integral in $\bar{x}$ is not simply going to give a Dirac delta, which would imply momentum conservation, because of our nonperturbative approach. Indeed, there are also implicit dependences on $\bar{x}$ through $V$ and $\Omega$; the former will be a crucial difference with respect to the Abelian gauge field case.

Let us now focus on the first order contribution in the assisting potential. 
The relevant computation reads
\begin{align} \label{3.57}
\begin{split}
Z_1[\eta](\bar{x};T)=&- \frac{\mathrm{e}^{-TV}}{\left( 4\pi T \right)^{\nicefrac{D}{2}}} \, \mathrm{det}^{\nicefrac{1}{2}}\left( \frac{\Omega^2T^2}{\sinh^2(\Omega T)} \right)
\\
&\hspace{1.5cm}\times\int \hat{\rm d}q\,\mathrm{e}^{-i q \cdot \bar{x}-\frac12 q_\mu \mathcal{E}^{\mu\nu}q_\nu} \, \tilde{\mathcal{V}}(q) \int_{t_1} \, \mathrm{e}^{\frac12 S_{\mathrm{bos}}[\eta]+i q \int_t \mathcal{G}(t,t_1) \, \eta(t)}\ ,
\end{split}
\end{align}
with
\begin{equation} \label{EE}
\mathcal{E}^{\mu\nu}:=\mathcal{G}^{\mu\nu}(t,t)=\left[ \frac{\Omega T \coth{(\Omega T)}-1}{2\Omega^2 T}\right]^{\mu\nu}\ .
\end{equation}
Once these results are replaced in Eq.~\eqref{path4}, we can perturbatively recast the trace of the heat kernel as 
\begin{align}\label{eq:trace_HK_assisted_1order}
\begin{split} 
\tilde{K}(T)=-\frac{1}{\left( 4\pi T \right)^{\nicefrac{D}{2}}} &\int {\rm d}^D\bar{x} \; \mathrm{e}^{-TV} \, \mathrm{det}^{\nicefrac{1}{2}}\left( \frac{\Omega^2T^2}{\sinh^2(\Omega T)} \right)
\\
\times&\int \hat{\rm d}q\,\mathrm{e}^{-i q \cdot \bar{x}-\frac12 q_\mu \mathcal{E}^{\mu\nu}q_\nu} \, \tilde{\mathcal{V}}(q) \int_{t_1}
\, \big(1+ \Sigma^{(1)}(\bar{x};T,t_1)\big)\ ,
\end{split}
\end{align}
where higher-derivatives contributions in the Yukawa potential are encoded in $\Sigma^{(1)}(\bar{x};T,t_1)$.
If we split them according to the number of momenta they contain, 
\begin{equation} 
\Sigma^{(1)}(\bar{x};T,t_1)=\Sigma^{(1)}_0(\bar{x};T)+\Sigma^{(1)}_1(\bar{x};T,t_1)+\cdots \ ,
\end{equation}
we get the first terms
\begin{align}
\Sigma^{(1)}_0(\bar{x};T)&=-\frac{T}{8}\partial_{\mu\nu\rho\lambda} V \, \mathcal{E}^{\mu\nu} \mathcal{E}^{\rho\lambda} -\frac{T}{48}\partial_{\mu\nu\rho\lambda\tau\theta} V \, \mathcal{E}^{\mu\nu} \mathcal{E}^{\rho\lambda} \mathcal{E}^{\tau\theta}
 \nonumber \\
&\hspace{1cm}-\frac{T}{384}\partial_{\mu\nu\rho\lambda\tau\theta\alpha\beta} V \, \mathcal{E}^{\mu\nu} \mathcal{E}^{\rho\lambda} \mathcal{E}^{\tau\theta} \mathcal{E}^{\alpha\beta}\nonumber
\\
&\hspace{1cm}+\frac{1}{12} \partial_{\mu\nu\rho} V \partial_{\alpha\beta\gamma} V \, \int_{a b} \mathcal{G}^{\mu\alpha}(a,b)\mathcal{G}^{\nu\beta}(a,b)\mathcal{G}^{\rho\gamma}(a,b) 
+\dots\ ,
\\
\begin{split}
\Sigma^{(1)}_1(\bar{x};T,t_1)&=\frac{i}{2} \,\partial_{\mu\nu\rho} V \, q_{\alpha} \int_t \mathcal{G}^{\mu\nu}(t,t) \mathcal{G}^{\rho\alpha}(t,t_1) \\
&\hspace{1cm}+\frac{i}{6} \, \partial_{\mu\nu\rho\lambda\tau} V \, q_{\alpha}\int_t \mathcal{G}^{\mu\nu}(t,t)G^{\rho\lambda}(t,t) \mathcal{G}^{\tau\alpha}(t,t_1)+\dots\ ,
\end{split}
\end{align}
where we reported only the terms contributing to order $T^5$ once expanded in powers of $T$.
These results can be compared with the known expansion for integrated heat kernels, the difference with the local results in Sec.~\ref{chap:first:sec:resummation} consisting just in boundary contributions.

\subsubsection{Gauss assisted pair creation at first order}\label{chap:first:sec:assisted}
The results from the previous section can be readily used to analyze the effect of assisted pair creation. 
For simplicity, let us consider a minimal model in which the background, assisted field is quadratic in the coordinates, so that the integrated heat kernel simplifies to
\begin{align}\label{eq:HK_gaussian_assisted}
\begin{split} 
\tilde{K}(T)&=-\frac{T}{\left( 4\pi T \right)^{\nicefrac{D}{2}}} \int {\rm d}^D\bar{x} \; \mathrm{e}^{-TV} \, \mathrm{det}^{\nicefrac{1}{2}}\left( \frac{\Omega^2T^2}{\sinh^2(\Omega T)} \right)
\int \hat{\rm d}q\,\mathrm{e}^{-i q \cdot \bar{x}-\frac12 q_\mu \mathcal{E}^{\mu\nu}q_\nu} \, \tilde{\mathcal{V}}(q) \ .
\end{split}
\end{align}
Without losing generality, we can further restrict the potential to be only time-dependent, as we have done in Eq.~\eqref{eq:background_quadratic}:
\begin{equation}
 V_0(x)=v^2 \, x_0^2+m^2\ , \quad v=\text{const}\ .
\end{equation}
As assisting field we propose instead a Gaussian (infinitely wide) pulse, which is frequently employed in the description of experimental setups: 
\begin{equation}
 \mathcal{V}(x)=\frac{\omega^2}{\pi} \, \mathrm{e}^{-\omega^2 x_0^2}\ .
\end{equation}
Replacing these profiles for the fields into Eq.~\eqref{eq:HK_gaussian_assisted} and performing the integral in $\bar{x}_0$ and in the energy $q_0$, we get a compact expression for the trace of the heat kernel:
\begin{align}
\begin{split}
 \tilde{K}(T)
 &=-\frac{\mathrm{Vol}_{\mathrm{(D-1)}} }{(4\pi T)^{\nicefrac{D}{2}}} \frac{ \omega T^{\nicefrac{3}{2}} }{\sinh{\left( vT \right)}} \, \mathrm{e}^{-m^2T}\,\left( \frac{1}{\omega^2}+\frac{\coth{(vT)}}{v} \right)^{-\nicefrac{1}{2}}\ .
\end{split}
\end{align}
Using this expression, the Euclidean effective action is immediately obtained. As we have already performed at zeroth order in the assisting field, in the present case we can analyze the Minkowskian scenario by performing a Wick rotation, the only difference being that the frequency $\omega$ shall be also included among the quantities that have to be Wick-rotated. 
Keeping track once more of the necessary regularizations resulting from the rotation, we get
\begin{align}
\begin{split}
 \Gamma_\mathrm{M}^{(1)}
 &=-\frac{\mathrm{Vol}_{\mathrm{(D-1)}} \epsilon \omega_\mathrm{M}}{2(4\pi)^{\nicefrac{D}{2}}} \int_0^\infty \frac{{\rm d}T}{T^{\nicefrac{(D-1)}{2}}}
 \frac{\mathrm{e}^{-m^2T}}{\sin{\left( v_\mathrm{M} T + i 0\right)}} \, \left( \frac{1}{\omega_\mathrm{M}^2}+\frac{\cot{( v_\mathrm{M}T)}}{v_\mathrm{M}} -i0 \right)^{-\nicefrac{1}{2}}.
\end{split}
\end{align}
The analytic structure of the integrand in this expression is severely influenced by the square root factor. On the one hand, notice that the zeros of the sine are not poles of the integrand, given that the cotangent inside the square root partially compensates the divergence, rendering it integrable. On the other hand, its argument vanishes periodically in $T$, generating an infinite number of branch cuts in the complex $T$ plane, the $-i0$ resulting from the Wick rotation telling us on which side of the cut the integral shall be performed. 
To simplify the discussion, let us use $v_\mathrm{M}$ as scale and define the following complete set of dimensionless parameters:
\begin{align}
\bar{\omega}:&= \frac{\omega_{\rm M}}{v_\mathrm{M}^{\nicefrac{1}{2}}}\ ,
 \\
\bar{m}:&=\frac{m}{v_\mathrm{M}} \ ,
\\
\bar{T}^*:&= v_\mathrm{M} T^*\ ,
\end{align}
where $T^*$ is the first positive root of $\bar{\omega}^{-2}+\cot{( \bar{T}^*)}=0$. Then, the cuts from the cotangent give rise to an imaginary part in the effective action, whose analytic expression can be straightforwardly found to be
\begin{align}
 \begin{split}\label{eq:im_ea_assisted}
\operatorname{Im} \Gamma_\mathrm{M}^{(1)}
 &=\frac{\pi\mathrm{Vol}_{\mathrm{(D-1)}} }{2(2\pi)^{{D}{}} } {v_\mathrm{M}}^{\nicefrac{(D-1)}{2}} \epsilon \mathrm{e}^{-\bar{m}^2 \bar{T}^* } \bar{\omega}
 \int_{0}^{\pi- \bar{T}^*} {\rm d}T \frac{\mathrm{e}^{-\bar{m}^2T}}{\sin{\left( T + \bar{T}^* \right)}}
\\
&\hspace{1cm}\times \, \left| \bar{\omega}^{-2}+\cot(T+\bar{T}^*)\right|^{-\nicefrac{1}{2}} \Phi\left(-\mathrm{e}^{-\pi \bar{m}^2}, \frac{D-1}{2}, \frac{T+\bar{T}^*}{\pi}\right),
 \end{split}
 \end{align}
where $\Phi(\cdot,\cdot, \cdot)$ is the Lerch transcendent function,
\begin{align}
 \Phi(z,s,a):=\sum_{n=0}^{\infty} \frac{z^n}{(a+n)^s}.
\end{align}

\begin{figure}[ht]
\begin{center}
 \begin{minipage}{0.48\textwidth}
 \includegraphics[width=1.0\textwidth,height=0.8\textwidth]{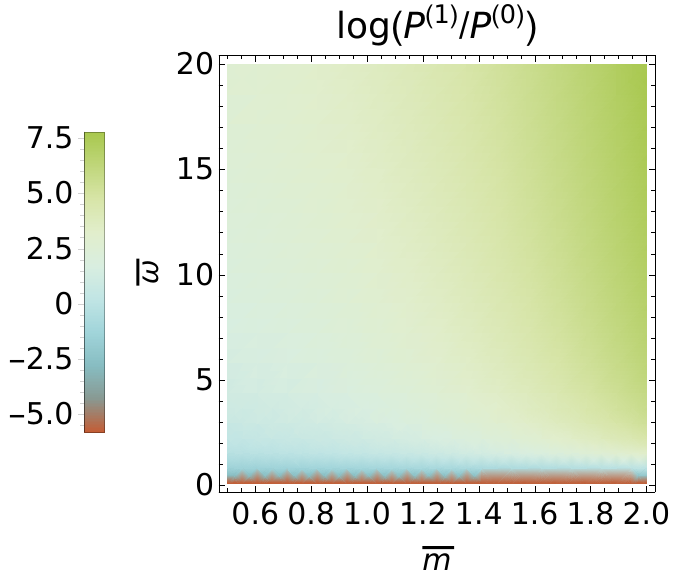}
 \end{minipage}
 \hspace{0.02\textwidth}
 \begin{minipage}{0.48\textwidth}
 \vspace{0.4cm}\includegraphics[width=0.95\textwidth]{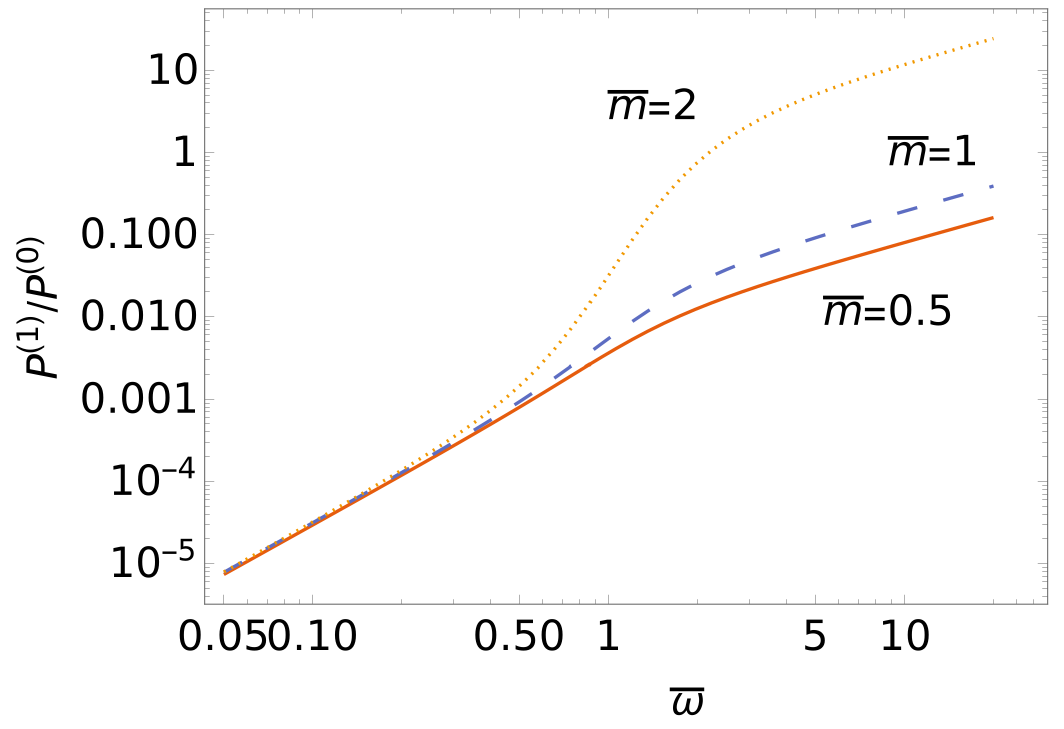}
 \end{minipage}
 \caption{The left panel is a density plot of $\log\left(\frac{P^{(1)}}{P^{(0)}}\right)$ as a function of $\bar m$ and $\bar \omega$, taking $\epsilon=10^{-3}$ and $D=4$. The right panel shows $\frac{P^{(1)}}{P^{(0)}}$ as a function of $\bar\omega$, for three different values of the dimensionless mass: $\bar m=2$ (yellow, dotted line), $\bar m=1$ (blue, dashed line) and $\bar m=0.5$ (red, continuous line). }
 \label{fig:pair_creation}
 \end{center}
\end{figure}

Importantly, $T^*$ always belongs to the interval $T^*\in [\pi/2, \pi)$, since we consider $v_\mathrm{M},\omega_{\mathrm M}>0$; this implies that the exponential decay of the pair creation effect can be greatly softened by the assisted field.
Such a softening can be readily observed by inspecting the ratio of the pair creation probability at first assisted order, $P^{(1)}$, with respect to the non-assisted case, $P^{(0)}$. In the left panel of Fig.~\ref{fig:pair_creation}, one can observe a density plot of $\log\left(\frac{P^{(1)}}{P^{(0)}}\right)$ as a function of $\bar m$ and $\bar\omega$; the integral in the proper time $T$ has been performed numerically, while the value $\epsilon=10^{-3}$ has been chosen so that in the depicted region we roughly satisfy the small potential criterium $\epsilon\bar\omega^2\ll 1$.

As expected, the ratio of probabilities rapidly increases as a function of the rescaled frequency $\bar\omega$. 
Indeed, the higher the frequency $\omega$ the larger the available energy to catalyze the production of pair creation. One can also see that, for the depicted values of the parameters, the ratio also increases with the mass; the reason is that the transseries structure in Eq.~\eqref{eq:im_ea_assisted} is shifted with respect to that in Eq.~\eqref{eq:im_ea_0}, thanks to the exponential prefactor ${\mathrm{e}}^{-\bar m ^2 \bar T^*}$.

We have also depicted the double logarithmic plot of the ratio of pair creation assisted and non-assisted probabilities as a function of $\bar\omega$ in the right panel of Fig.~\ref{fig:pair_creation}, for a few values of the rescaled mass. For large values of $\bar\omega$, the behaviour is linear, as dictated by the prefactor in Eq.~\eqref{eq:im_ea_assisted}. Going to smaller values, there is a transition to a $\bar\omega^2$ behaviour. For small $\bar\omega$ the hierarchy of the curves with respect to the mass is kept, thanks to the following effect: in this region the softening ${\mathrm{e}}^{-\bar m ^2 \bar T^*}$ plays no role, the Lerch function becomes essentially a polylogarithm and the relevant ratio is governed by the dimensionality of the two polylogarithms.

\section{Final remarks}\label{chap:first:sec:conclusion}
In this chapter, we have shown how the Worldline Formalism can be used to perform nonperturbative computations for a scalar quantum field coupled to a Yukawa background. In particular, using Dirichlet boundary conditions on the worldline, we have seen that a local, resummed expression for the heat kernel is obtainable, cf. Sec.~\ref{chap:first:sec:resummation}. In this way one can straightforwardly compute the first generalized heat kernel coefficients, which agree with those recently found in the literature. Instead, in Sec.~\ref{chap:first:chap:first:sec:assisted_yukawa} we have developed what would be combined Parker--Toms and Barvinsky--Vilkovisky resummations, the former being used for the strong field background and the latter for the assisting field. Since the ultimate goal was to obtain the pair production probability, the so-called string-inspired boundary conditions proved to be more efficient. From the master formula \eqref{path4}, an expansion of the integrated heat kernel to include higher derivative terms for arbitrary fields can be obtained by direct replacement. 
This enables one to compute the imaginary part of the Minkowskian effective action and, ergo, the instabilities of the vacuum in terms of the probability of pair creation. The advantage of our method over other available techniques (such as the worldline instantons~\cite{Dunne:2006st}) is that our method provides a formula that can be computed for arbitrary, inhomogeneous fields to any desired level of precision. Using these results, we have seen that the assistance occurs already at the first order in the assisting field. To illustrate this effect, we have worked out in detail the calculations for a Gaussian assisting field; compared to the non-assisted effect, the transseries structure of the pair creation probability is translated, leaving room for a large exponential enhancement. On the contrary, in \acr{sQED} one has to resort to at least second order in order to observe the assisted effect~\cite{Torgrimsson:2018xdf}, because gauge invariance precludes a term like $\mathrm{e}^{-TV}$ in Eq.~\eqref{eq:HK_gaussian_assisted}.

Our analysis can be generalized in several directions. First, contributing to the discussion on a possible new effect of pair creation in the presence of gravitational fields~
\cite{Ferreiro:2023jfs,Hertzberg:2023xve,Akhmedov:2024axn, Boasso:2024ryt}. Using the curved spacetime version of the Wordline Formalism, such a situation is tractable.
In this scenario, we would also expect the possibility to have a first order assisted pair creation, if a mixed of strong/fastly-varying potentials is used. In this sense, our computations are of great help, since at least in a perturbative approach, the computation in a curved spacetime is totally analogous to a scalar computation~\cite{vandeVen:1997pf}.
Second, the interaction of fields with higher spins is also accessible in the worldline; the corresponding resummed expressions could already be inferred from our present results.
Finally, understanding whether further resummations of derivatives are possible would be a further step in the analysis of nonperturbative phenomena, in particular related to the transseries structure of the corresponding physical effects.

\addcontentsline{toc}{chapter}{Appendices}
\section*{Appendices}
\begin{subappendices}

\section{Boundary conditions and Green functions in the worldline} \label{chap:first:appA}
In this appendix, we will discuss some of the technicalities mentioned throughout the chapter regarding boundary conditions; first we will consider the case of the free operator, i.e. the pure kinetic one, and will compute the corresponding Green functions. Afterwards, we will extend the results to the Yukawa case.

\subsection{Kinetic operator and zero mode}
The kinetic term of the worldline theory that we study corresponds to the free worldline action
\begin{equation}
S_{\mathrm{kin}}=\frac12 \int_t \, \delta_{\mu\nu} \, \dot{x}^\mu(t)\dot{x}^\nu(t)=-\frac12 \int_{t_1t_2} \, x^\mu(t_1)\left[ 
\delta_{\mu\nu} \partial^2_{t_1} \delta(t_1-t_2) \right] x^\nu(t_2)\ ,
\end{equation}
where the boundary terms arising from the integration by parts vanish both for Dirichlet and periodic \acr{BC}.
In particular, we are interested in the case where the trajectories are periodic, since we will see that the vanishing \acr{DBC} can be reobtained as a special case.

The associated differential operator,
\begin{equation}
K_{\mu\nu}(t,t')=-\frac12 \delta_{\mu\nu} \partial^2_{t} \delta_\mathrm{P}(t-t'),
\end{equation}
is defined on the interval $t,t'\in[0,T]$ and has a zero mode on the circle: periodic boundary conditions allow for the presence of constant paths, preventing thus $K_{\mu\nu}$ to be invertible. Note that $\delta_\mathrm{P}(t-t')$ is the Dirac delta function on the space of periodic functions on $[0,T]$ \cite{Bastianelli:2002qw}; the subscript has been dropped throughout the work for notational convenience.

One way to solve the invertibility issue consists of factoring out the zero mode, i.e. to decompose the generic periodic path into the zero mode $\bar{x}^\mu$ and a quantum fluctuation $y^\mu(t)$, so that
\begin{equation} \label{fact}
x^\mu(t)=\bar{x}^\mu+y^\mu(t)\ .
\end{equation}
This amounts to introducing a redundant (pure gauge) field $\bar{x}^\mu$ in the worldline theory; indeed, there is now a shift symmetry, which is actually a gauge symmetry:
\begin{align}
\begin{split} \label{gauge}
\delta \bar{x}^\mu &=\xi^\mu\ , \\
\delta y^\mu(t)&=-\xi^\mu\ .
\end{split}
\end{align}
Differently to what happens in the background field method~\cite{Abbott:1980hw}, here both fields are considered dynamical.\footnote{That is, they are integrated over in the path integral and the overcounting of physically distinct configurations must be avoided by a gauge-fixing procedure.} This explains why the shift symmetry \eqref{gauge} is promoted to a \emph{gauge symmetry}, and therefore must be gauge-fixed. The gauge-fixing procedure can be performed by means of BRST methods, as analyzed in Refs.~\cite{Bastianelli:2003bg, Bastianelli:2009eh, Corradini:2018lov, Fecit:2023kah} within the Worldline Formalism. The outcome of the discussion is that one can introduce a gauge condition of the form
\begin{equation} \label{gf}
\int_t \, \rho(t)\, y(t)=0\ ,
\end{equation}
where $\rho(t)$ is usually called the \emph{background charge} (in analogy with electrostatics) and is normalized to
\begin{equation}
\int_t \, \rho(t)=1\ .
\end{equation}
The gauge-fixing \eqref{gf} allows to invert the free kinetic operator of the fluctuations $y^\mu$. It is clear that different choices of background charge will result, in general, in different boundary conditions for the quantum fluctuations and propagators. Nevertheless, the $\rho$-independence of the path integral is still guaranteed by the BRST symmetry.\footnote{To be more precise, the worldline partition function can be expressed as an integral over the zero mode
\begin{equation*}
 Z=\int {\rm d}^D\bar{x} \, z^{(\rho)}(\bar{x},\rho)\ ,
\end{equation*}
where the partition function density $z^{(\rho)}$ may depend on the gauge-fixing choice $\rho(t)$, but this can only happen through total derivatives, which must then integrate to zero \cite{Bastianelli:2003bg}.}

More precisely, the Green function $\mathfrak{G}(t,t')$ of the operator $-\tfrac12\partial^2_t$ acting on fields constrained by the equation \eqref{gf} depends on $\rho(t)$ and satisfies
\begin{equation} \label{MS}
{-\frac12\partial_t^2 \, \mathfrak{G}(t,t')=\delta(t-t')-\rho(t)}\ .
\end{equation}
Let us discuss the most prominent choices for the background charge for the {free} theory.

\begin{itemize}
 \item \textbf{\acr{DBC}} - The choice 
 \begin{equation}
 \rho(t)=\delta(t)
 \end{equation}
 is tantamount to factorizing out the zero mode as in \eqref{fact} where $\bar{x}^\mu$ is identified with a base-point in the target space along the loop. Indeed, such a background charge produces, from \eqref{gf}, vanishing Dirichlet boundary conditions for the fluctuations: 
 \begin{equation}
 y(0)=y(T)=0\ .
 \end{equation}
 In this scenario, i.e. when working in the space of functions with vanishing \acr{DBC} on the interval $[0,T]$, the Dirac delta function is actually $\delta(t-t')$ and vanishes at the boundaries $t,t'=\lbrace 0, T\rbrace $ \cite{Bastianelli:2002qw}. The delta function will still be denoted as usual, since no confusion should arise. The free propagator is given by
 \begin{align}
 \begin{split} \label{freeop}
 G^{(f)}(t,t')&=-|t-t'|+\left(t+t'\right)-\frac{2t t'}{T}\ ,
 \end{split}
 \end{align}
 which can be obtained by solving
 \begin{equation}
 -\frac12\partial_\tau^2G^{(f)}(t,t')=\delta(t-t')\ ,
 \end{equation}
 with boundary conditions
 \begin{equation}
 G^{(f)}(0,t')= G^{(f)}(T,t')=0\ .
 \end{equation}
 
 \item \textbf{\acr{SI}} - The choice 
 \begin{equation}
 \rho(\tau)=\frac{1}{T}
 \end{equation}
 gives rise to the so-called “string-inspired” boundary conditions, as the factorization \eqref{fact} of the zero mode is akin to the customary practice in string theory, namely to identify $\bar{x}^\mu$ with the “center of mass” of the worldline \eqref{CdM} and separate it from the quantum fluctuations. The resulting boundary conditions for the latter are
 \begin{equation}
 \int_t \, y(t)=0\ ,
 \end{equation}
 while the free propagator is given by
 \begin{equation} \label{freeop2}
 \mathcal{G}^{(f)}(t,t')=-|t-t'|+\frac{(t-t')^2}{T}+\frac{T}{6}\ .
 \end{equation}
 It satisfies the equation 
 \begin{equation}
 -\frac12\partial_t^2\mathcal{G}^{(f)}(t,t')=\delta(t-t')-\frac{1}{T}\ ,
 \end{equation}
 with the conditions
 \begin{align}
 \int_t \, \mathcal{G}^{(f)}(t,t')=0\ , \quad \mathcal{G}^{(f)}(0,t')= \mathcal{G}^{(f)}(T,t')\ .
 \end{align}
 Note that $\mathcal{G}^{(f)}(t,t')$ actually depends only on $t-t'$, while $G^{(f)}(t,t')$ is a true function of two variables.
\end{itemize}

\subsection{Green function with a Yukawa coupling}
Let us now consider quadratic interactions, namely the linear, one-dimensional differential operator
\begin{equation} \label{Delta'}
\Delta_{\mu\nu} (\tau,\tau')=-\frac12 \delta_{\mu\nu}\partial^2_{t_1}\delta(t_1-t_2)+2 \, \Omega^2_{\mu\nu}(y)\delta(t_1-t_2)\ .
\end{equation}
The corresponding Green functions are described in the following (we omit Lorentz indices throughout the calculations).

\begin{itemize}
 \item \textbf{\acr{DBC}} - We start with the computation of Sec.~\ref{chap:first:sec:resummation}, namely with the case of vanishing \acr{DBC} for the fluctuations $\hat{s}$, cf. Eq.~\eqref{DBC}. The Green function $G^{\mu\nu}(t,t'):=\Delta^{-1}_{\mu\nu}$ is defined by the following differential equation,
\begin{equation} \label{greq}
\Delta(t,t') \, G(t',t'')=\delta(t-t'')\ ,
\end{equation}
together with the boundary conditions
\begin{equation}
G (0,t')=G (T,t')=0\ .
\end{equation}
We can build an Ansatz starting from the solution of the homogeneous equation associated with \eqref{greq}, namely
\begin{align}
G (t,t')=\begin{cases}
 A(t') \, \mathrm{e}^{2\Omega t}+B(t') \, \mathrm{e}^{-2\Omega t} \quad \text{if} \quad t < t' \\
C(t') \, \mathrm{e}^{2\Omega t}+D(t') \, \mathrm{e}^{-2\Omega t} \quad \text{if} \quad t' \leq t
\end{cases}\ .
\end{align}
Imposing the usual conditions on the (dis)continuity of the (first derivative of) the Green function at $t=t'$, we can relate the four unknown functions $A,B,C,D$ to get
\begin{equation}
G(t,t')=\frac{ \sinh(2\Omega t) \sinh\big(2\Omega (T-t')\big) - \Theta(t-t') \sinh(2\Omega T) \sinh\big(2\Omega (t-t')\big) }{\Omega \sinh(2\Omega T)}\ ,
\end{equation}
where one should not forget that the Green function carries Lorentz indices $G^{\mu\nu}(t,t')$. 
Note that, in the case of vanishing strong field ($\Omega \to 0$), the Green function reduces to the free propagator \eqref{freeop}.

 \item \textbf{\acr{SI}} - Let us move on to the computation of Sec.~\ref{chap:first:chap:first:sec:assisted_yukawa}, where string-inspired boundary conditions are enforced for the fluctuations $\hat{s}$. The Green function $\mathcal{G}^{\mu\nu}(t,t')$ associated with the linear, one-dimensional differential operator $\Delta^{\mu\nu}_{\rm SI}(t)$ is defined by the differential equation
\begin{equation}
\Delta_{\rm SI}(t,t') \, \mathcal{G}(t',t'')=\delta(t'-t'')-\frac{1}{T}\ ,
\end{equation}
and must satisfy the following conditions:
\begin{align} \label{A.33}
 \int_t \, \mathcal{G}(t,t')=0\ , \quad \mathcal{G}(0,t')= \mathcal{G}(T,t')\ .
 \end{align}
The final result for the Green function reads
\begin{align}
\begin{split}
\mathcal{G} (t,t')=\frac{1}{2\Omega} \bigg[&\frac{1}{\sinh({\Omega T})}\bigg(- \frac{\sinh(\Omega T)}{\Omega T}+\cosh\Big( \Omega(T+2t-2t' ) \Big) \bigg) 
\\
&-\Theta(t-t') \,2\,\sinh \Big({2\Omega (t-t')} \Big) \bigg]\ .
\end{split}
\end{align}
As already mentioned, the \acr{SI} propagator depends only on the difference of its two arguments: for instance, note that the coincidence limit reads
\begin{equation}
\mathcal{G} (t,t)=\frac{\Omega T \coth{(\Omega T)}-1}{2\Omega^2 T}\ .
\end{equation}
In the case of vanishing assisted field ($\Omega \to 0$) the Green function reduces to the free propagator \eqref{freeop2}, as expected. 
\end{itemize}

\section{Computation of the functional determinants and generalized Gel'fand--Yaglom theorem} \label{chap:first:appB}
The Gel’fand--Yaglom theorem \cite{Gelfand:1959nq} and its extensions \cite{Kirsten:2003py, Kirsten:2004qv, Kirsten:2005di} have been extensively applied in the context of the Worldline Formalism \cite{Dunne:2006st, DegliEsposti:2022yqw, DegliEsposti:2024upq}. Let us recall the main statement of the theorem: given a one-dimensional operator defined on an interval $z \in [0,T]$ with vanishing Dirichlet boundary conditions
\begin{equation} \label{chap:first:Laplace}
\left[ -\frac{{\rm d}^2}{{\rm d} z^2}+V(z) \right] \psi(z)=\lambda \, \psi(z)\ , \quad \text{with} \quad \psi(0)=\psi(T)=0\ ,
\end{equation}
there is no need to explicitly know its eigenvalues (not even one of them) if one desires to compute its functional determinant~\cite{Dunne:2007rt}. The only required information is the boundary value of the unique solution to the initial value problem
\begin{equation}
\left[ -\frac{{\rm d}^2}{{\rm d}z^2}+V(z) \right] \Phi(z)=0\ , \quad \text{with} \quad \Phi(0)=0\ , \quad \dot{\Phi}(0)=1\ ,
\end{equation}
which satisfies
\begin{equation} \label{chap:first:GY1D}
\mathrm{Det} \left[ -\frac{{\rm d}^2}{{\rm d}z^2}+V(z) \right] \propto \Phi(T)\ .
\end{equation}

The result~\eqref{chap:first:GY1D} can be extended to more general boundary conditions and for higher-dimensional differential operators. Indeed, consider a family of Sturm--Liouville type operators of the form
\begin{equation} \label{L1}
L_i=-\frac{{\rm d}}{{\rm d}z}\left(P_i(z)\frac{{\rm d}}{{\rm d}z}\right) \, \mathbb{1}_{r \times r} +V_i(z)
\end{equation}
where $\mathbb{1}_{r \times r}$ is the $r \times r$ identity matrix and $V_i$ can also be matrix-valued. Then, the ratio of the functional determinants of the operator $L_1$ relative to that of another operator $L_2$ of the same type can be expressed in terms of four $2r \times 2r$ matrices $M$, $N$ and $Y_i$, which are built as follows~\cite{Kirsten:2004qv}.
The matrices $Y_i$ contain the $2r$ solutions $\mathbf{u}^{(j)}(z)$ of the eigenvalue problem\footnote{We omit a subscript $i$ in the eigenfunctions for readability reasons.} $L_i \mathbf{u}^{(j)}=\lambda \mathbf{u}^{(j)}$ with $\lambda\to 0$ and their weighted first derivatives 
\begin{align}\label{eq:weighted_derivatives}
 \mathbf{v}^{(j)}(z):=P_i(z)\frac{{\rm d}}{{\rm d}z}\mathbf{u}^{(j)}(z)\ ;
\end{align}
using these vectors as (semi)-column entries, we have 
\begin{equation}
Y_i:=\left(
\begin{array}{ccc} \mathbf{u}^{(1)} & \dots & \mathbf{u}^{(2r)} \\ \mathbf{v}^{(1)} & \dots & \mathbf{v}^{(2r)} \\
\end{array}\right)\ .
\end{equation}
For simplicity, one generally chooses the initial conditions as\footnote{For instance, in the case $r=1$ we have two one-dimensional solutions $u^{1,2}(z)$ with first derivatives $v^{1,2}(z)$, whose initial conditions read $u^{(1)}(0)=1 \, , v^{(1)}(0)=0, u^{(2)}(0)=0$ and $v^{(2)}(0)=1$.}
\begin{equation} \label{BCY}
Y(0)=\mathbb{1}_{2r \times 2r}\ . 
\end{equation}
On the other hand, the matrices $M$ and $N$ are fixed by the choice of the boundary conditions; in fact, the latter can be recast in full generality as
\begin{equation}
M \, Y(0)+ N \, Y(T)=0\ .
\end{equation}
As an example, in the case of vanishing \acr{DBC} these matrices reduce to
\begin{equation}
M_{\rm D}=\left(
\begin{array}{cc} \mathbb{1}_{r \times r} & \mathbb{0}_{r \times r} \\ \mathbb{0}_{r \times r} & \mathbb{0}_{r \times r} \\
\end{array}\right)\ , \quad 
N_{\rm D}=\left(
\begin{array}{cc} \mathbb{0}_{r \times r} & \mathbb{0}_{r \times r} \\ \mathbb{1}_{r \times r} & \mathbb{0}_{r \times r} \\
\end{array}\right)\ ,
\end{equation}
while in the case of periodic boundary conditions they read
\begin{equation} \label{MN}
M_{\rm P}=\left(
\begin{array}{cc} \mathbb{1}_{r \times r} & \mathbb{0}_{r \times r} \\ \mathbb{0}_{r \times r} & \mathbb{1}_{r \times r} \\
\end{array}\right)\ , \quad 
N_{\rm P}=\left(
\begin{array}{cc} -\mathbb{1}_{r \times r} & \mathbb{0}_{r \times r} \\ \mathbb{0}_{r \times r} & -\mathbb{1}_{r \times r} \\
\end{array}\right)\ .
\end{equation}

Employing this notation, the generalized \acr{GY} theorem states that the ratio of the functional determinants satisfies
\begin{equation} \label{GY}
\frac{\mathrm{Det}(L_1)}{\mathrm{Det}(L_2)}=\frac{\mathrm{det}(M+N \, Y_1(T))}{\mathrm{det}(M+N \, Y_2(T))}\ .
\end{equation}
Let us now show how to use this formalism for the interacting differential operator $\Delta_{\mu\nu}$ defined in Eq.~\eqref{Delta}. 

\subsection{Dirichlet boundary conditions} \label{chap:first:appB.2}
For \acr{DBC}, relevant to Sec.~\ref{chap:first:sec:resummation}, the formula \eqref{GY} reduces to the evaluation of the determinant of an $r \times r$ solutions-submatrix:
\begin{equation}
\mathrm{Det}(\Delta)=\mathrm{det}(\mathbf{u}^{(r+1)} \, \dots \, \mathbf{u}^{(2r)})\ .
\end{equation}
Thus, we don't need to know the whole set of $2r$ solutions of the homogeneous equation; instead, we only need the $r=D$ solutions $\varphi_\mu^{(\rho)}$ which have initial conditions
\begin{equation}
\varphi_\mu^{(\rho)}(0)=\mathbb{0}\ , \quad \dot{\varphi}_\mu^{(\rho)}(0)=\delta^\rho_\mu\ .
\end{equation}
Building from the one-dimensional solution, an Ansatz of the form
\begin{equation}
\varphi_\mu^{(\rho)}(z)=\left(A \, \mathrm{e}^{2\Omega z} \right)^\rho{}_\mu+\left(B \, \mathrm{e}^{-2\Omega z} \right)^\rho{}_\mu\ ,
\end{equation}
one obtains
\begin{equation}
\varphi_\mu^{(\rho)}(z)=\left[ \left.\frac{1}{2\Omega}\sinh(2\Omega z) \right]^\rho{}\right._\mu\ .
\end{equation}
Using as a reference operator the free kinetic operator $L_2=-\tfrac12\delta_{\mu\nu}\partial^2_z$, whose $Y$-matrix can be computed effortlessly, the final result for the functional determinant reads
\begin{equation}
{\overline{\mathrm{Det}}(\Delta)=\mathrm{det}\left[ \frac{\sinh(2\Omega T)}{2\Omega T} \right]}\ .
\end{equation}
As a double check of this result, we can calculate the determinant as a $\zeta$-regularized infinite product of the eigenvalues of the operator $\Delta$ divided by those of the free operator. Using a simple Fourier expansion to solve the eigenvalue equation, we get
\begin{equation}
 \overline{\mathrm{Det}}(\Delta)= \prod_{j=1}^{r} \prod\limits_{n =1}^{\infty} \left[ 1+\left(\frac{2\Omega^{(j)} T}{\pi n}\right)^2\right] =\mathrm{det}\left[ 
\frac{\sinh{(2\Omega T)}}{2\Omega T} \right]\ ,
\end{equation}
where $\Omega^{(j)}$ are the eigenvalues of the matrix $\Omega^{\mu}{}_{\nu}$ and we have used the well-known result
\begin{equation}
 \prod\limits_{n=1}^{\infty}\left( 1+\frac{x^2}{\pi^2n^2} \right)=\frac{\sinh{(x)}}{x}\ .
\end{equation}
Note in particular that the free limit is regular:
\begin{equation}
 \lim_{\Omega\to0} \, \overline{\mathrm{Det}}(\Delta)=1\ .
\end{equation}

\subsection{Periodic boundary conditions} \label{chap:first:app:det_pbc}
The \acr{PBC}, relevant to Sec.~\ref{chap:first:chap:first:sec:assisted_yukawa}, require more carefulness; let us first show the exemplificative non-matricial case, i.e. $r=1$. 

\newtheorem{theorem}{Theorem}
\newtheorem{proposition}{Proposition} 

\begin{proposition}

Assuming Eq.~\eqref{MN} for the matrices $M$ and $N$, and considering an arbitrary operator of the form given in Eq.~\eqref{L1}, the relevant determinant for the \acr{GY} formula in the one-dimensional ($r=1$) case reduces to
\begin{align}
\begin{split} \label{r=1}
\mathrm{det}\Big(M+N \, Y_1(T)\Big)
&=2-\mathrm{tr}\left(Y_{r=1}(T)\right)
= 2-2\cosh(2\Omega T) \ .
\end{split}
\end{align}
\end{proposition}

\begin{proof}
In general, a direct calculation actually gives
\begin{align}
\begin{split}
\mathrm{det}\Big(M+N \, Y_1(T)\Big)&=1-\left.\left( u^{(1)}+v^{(2)} \right)\right|_{z=T}+\left.(u^{(1)}v^{(2)}-v^{(1)}u^{(2)})\right|_{z=T} \label{B20} \\
&=1- \mathrm{tr}\left(Y_{r=1}(T) \right)+\mathrm{det}\left( Y_{r=1} (T)\right)\ ,
\end{split}
\end{align}
with the matrix of solutions given by
\begin{equation}
Y_{r=1}(z)=\left(
\begin{array}{cc} u^{(1)}(z) & u^{(2)}(z) \\ v^{(1)}(z) & v^{(2)}(z) \\
\end{array}\right)\ .
\end{equation}
One can recognize in the \acr{RHS} of \eqref{B20} the Wronskian
\begin{equation}
\mathcal{W}(u^{(1)},u^{(2)})(z)=u^{(1)}(z)v^{(2)}(z)-v^{(1)}(z)u^{(2)}(z)\ ,
\end{equation}
evaluated at $z=T$, which is associated with the two independent solutions $u^{1,2}(z)$ of the eigenvalue equation for a Sturm--Liouville operator.
It is a stablished result that this type of Wronskian is constant;
therefore, extracting from the initial conditions its value at $z=0$,
\begin{equation}
\mathcal{W}(u^{(1)},u^{(2)})(0)=1\ ,
\end{equation}
we also have that $\mathcal{W}(u^{(1)},u^{(2)})(T)=1$; hence, using the explicit solutions $u^{(1)}$ and $v^{(2)}$, Eq.~\eqref{r=1} follows.
\end{proof}

It is not hard to convince oneself that the previous one-dimensional result can be directly generalized to higher dimensions.

\begin{theorem}\label{th:GY_periodic}
Consider an operator of the form given in Eq.~\eqref{eq:Delta_SI}, defined on the domain of periodic functions. Be $u_a^{(b)}$ the $a$th component of the $b$th solution of the homogeneous equation $L_1 \mathbf{u}^{(b)}=0$ and $\mathbf{v}^{(b)}$ its corresponding weighted derivative, as in Eq.~\eqref{eq:weighted_derivatives}. 
Then, the relevant determinant for the \acr{GY} theorem satisfies
\begin{align}\label{eq:theorem}
\mathrm{det}\Big(M+N \, Y_{\Delta, \mathrm{PBC}}(T)\Big)
=\det \big(2-2\cosh(2\Omega T)\big)\ .
\end{align}
\end{theorem}

\begin{proof}
Note first that, in contrast to the \acr{DBC}, to compute the functional determinant using the generalized \acr{GY} theorem we do need the whole set of $2r$ solutions of the homogeneous equation
\begin{equation}
\Delta_{\rm SI}^{\mu\nu} \, \Phi_\nu^{(\Pi)}=0\ , \quad \Phi^{(\Pi)}_\mu=\begin{cases}
 \phi_\mu^{(\rho)} \quad \text{if} \quad \Pi=1,\dots,r \\
 \psi_\mu^{(\rho)} \quad \text{if} \quad \Pi=r+1,\dots,2r
\end{cases}\ .
\end{equation}
 Considering Eq.~\eqref{BCY} as initial conditions, namely 
\begin{align}
\phi_\mu^{(\rho)}(0)&=\delta^\rho_\mu\ , \quad \frac{{\rm d}\phi}{{\rm d}z}_\mu^{(\rho)}(0)=\mathbb{0}\ ,
\\
\psi_\mu^{(\rho)}(0)&=\mathbb{0}\ , \quad \frac{{\rm d}\psi}{{\rm d}z}_\mu^{(\rho)}(0)=\delta^\rho_\mu\ ,
\end{align}
an explicit computation gives
\begin{equation}
\phi_\mu^{(\rho)}(z)=\dot{\psi}_\mu^{(\rho)}(z)=\Big[\cosh(2\Omega z)\Big]^\rho{}\Big._\mu\ .
\end{equation}
Note then that in arbitrary dimensions we can straightforwardly prove that the $n$th power of the matrix $Y_\Delta$ is
\begin{align}
 Y^n_{\Delta, \mathrm{PBC}} (T)= \begin{pmatrix}
 \cosh(2n\Omega T) & \sinh(2n\Omega T)
 \\
 \sinh(2n\Omega T) & \cosh(2n\Omega T)
 \end{pmatrix},
\end{align}
and consequently, from Eq.~\eqref{GY}, we have the equalities
\begin{align}
 \begin{split}
 \det(\mathbb{1}_{2r} - Y_{\Delta, \mathrm{PBC}}) 
 &= \mathrm{e}^{\tr \log [2-2 \cosh(2\Omega T)]}
 = \det \big(2-2\cosh(2\Omega T)\big) \ ,
 \end{split}
\end{align}
which indeed agree with the \acr{RHS} of Eq.~\eqref{eq:theorem}. 
\end{proof}
As a corollary of Th.~\ref{th:GY_periodic}, recall that the difference between \acr{P} and \acr{SI} \acr{BC} resides just in the omission of the constant mode in the latter; in the free case, this mode corresponds to a zero mode. The quotient of $\Delta_{\rm SI}$ with the the free operator, after extracting the zero mode of the latter (denoted with a prime in the Det), is thus given by
\begin{equation} \label{Rr}
{\overline{\mathrm{Det}}'(\Delta_{\rm SI})=\mathrm{det}\left[ \frac{\sinh^2(\Omega T)}{\Omega^2T^2} \right]}\ .
\end{equation}
This is consistent with the fact that we have omitted the constant mode, so that the free limit $\Omega\to 0$ is non-vanishing. As an alternative check of the \acr{GY} result, we can directly evaluate the determinant of $\Delta_{\rm SI}$ as a product of its eigenvalues; using a Fourier expansion and dividing by the free eigenvalues (we exclude the vanishing one), we get the following $\zeta$-regularized infinite product
\begin{equation} \label{B.48}
 \overline{\mathrm{Det}}'
 (\Delta_{\rm SI})= \mathrm{det}\left[ \prod\limits_{n =1}^{\infty} \left( 1+\frac{\Omega^2T^2}{\pi^2n^2} \right)^2 \, \right]=\mathrm{det}\left[ \frac{\sinh^2(\Omega T)}{\Omega^2 T^2} \right]\ ,
\end{equation}
which confirms Eq.~\eqref{Rr}.

\section{Generalized Schwinger--DeWitt coefficients in the string-inspired approach}\label{chap:first:app:coeff_SI}
The first generalized Schwinger--DeWitt coefficients computed using the \acr{SI} \acr{BC}, defined in the expansion
\begin{align}
    \Sigma^{(1)}_0({x};T) =:\sum_{j=0}^{\infty} c^{\mathrm{SI}}_j(x) \, T^{j}\ ,
\end{align}
are given by
\begin{align}
c^{\mathrm{SI}}_0(x)&=0\ , \\
c^{\mathrm{SI}}_1(x)&=0\ , \\
c^{\mathrm{SI}}_2(x)&=0\ , \\
c^{\mathrm{SI}}_3(x)&=-\frac{1}{288}\partial_{\mu}{}^{\mu}{}_{\nu}{}^{\nu} V  \ ,
\\
c^{\mathrm{SI}}_4(x)&=-\frac{1}{10368}\partial_{\mu}{}^\mu{}_\nu{}^\nu{}_\rho{}^\rho V \ ,
\\
c^{\mathrm{SI}}_5(x)&= 
\frac{1}{45360} \partial_{\mu\nu\rho} V \partial^{\mu\nu\rho} V  
+ \frac{1}{4320} \partial_{\mu}{}^{\mu}{}_{\rho\lambda} V \partial^{\rho\lambda} V 
-\frac{1}{497664}\partial_{\mu}{}^\mu{}_\nu{}^\nu{}_\rho{}^\rho{}_\sigma{}^\sigma V  \ .
\end{align}
Recall that these coefficients are only valid at the integrated level, i.e. they are related via integration by parts to the $c_i$ coefficients in Eq.~\eqref{eq:generalized_coeff}.
\end{subappendices}

\newpage
\thispagestyle{empty}
\mbox{}
\newpage

\chapter*{Spin 1}
\addcontentsline{toc}{chapter}{\color{turquoise}\raisebox{0.5ex}{\rule{7.75cm}{0.4pt}} Spin 1 \raisebox{0.5ex}{\rule{7.75cm}{0.4pt}}}
\chapter{Pair production of massive charged vector bosons}\label{chap:second}
\textit{In this chapter, we continue our investigation of non-perturbative physics within the Worldline Formalism by considering a massive spin-1 particle interacting with an electromagnetic background. Unlike the previous chapter, here we adopt a “bottom-up” strategy and introduce a worldline model augmented by bosonic oscillators to encode the spin degrees of freedom, a framework that can accommodate particles of arbitrary integer spin. We first review how quantization in the spin-1 sector -- carried out both through Dirac’s method and via BRST quantization -- reproduces the free Proca field theory. We then couple the model to an external \acr{EM} field and show that Maxwell’s equations for the background arise as the consistency condition ensuring the nilpotency of the BRST charge in the spin-1 sector.
Encouraged by this result, which establishes the viability of the particle model, we proceed to construct the path integral quantization of the worldline action for a charged spin-1 particle on the circle. This leads to the one-loop effective Lagrangian for a constant electromagnetic background induced by a massive charged vector boson. As expected, the result exhibits a vacuum instability, which we quantify by deriving the pair-production rate for the vector bosons, thereby recovering known results from quantum field theory.}

\paragraph{Conventions} Whether we work in Minkowski or Euclidean signature, and in which spacetime dimension, will be specified as needed throughout the text for convenience.

\section{The free bosonic spinning particle}\label{sec1}
The worldline model known as the bosonic spinning particle has already been introduced in the \hyperref[chap:WF]{introductory section on the Worldline Formalism}. In this section, we shall provide a more detailed presentation and establish our notation. It is defined by the usual set of phase-space variables $(x^\mu, p_\mu)$, augmented by an additional pair of complex bosonic variables $(\alpha^\mu, \balpha^\mu)$. The former represent Cartesian coordinates and conjugate momenta of a relativistic particle, while the latter are needed to account for the spin degrees of freedom. Both pairs are functions of the proper time, which we take to range as $\tau\in[0,1]$. The kinetic term
\begin{equation}
 S_{\mathrm{kin}}=\int \diff\tau \left( p_\mu \dot{x}^\mu -i \bar\alpha_\mu \dot{\alpha}^\mu\right)
\end{equation}
defines the phase space symplectic structure and fixes the Poisson brackets to 
\begin{equation}
\{x^\mu, p_\nu\}_{\mathrm{PB}} = \delta^\mu_\nu 
\;, \quad 
\{\alpha^\mu, \bar \alpha^\nu\}_{\mathrm{PB}} = i \eta^{\mu\nu}
\;.
\end{equation}
As it stands, the model is not unitary, as upon quantization negative norm states will be generated by the $(x^0, p^0, \alpha^0, \balpha^0)$ variables. Moreover, the model, as we shall discuss, contains particle excitations of any integer spin, and one needs to eliminate some further degrees of freedom to describe a single particle with fixed spin. Both problems can be addressed by gauging suitable constraints: the gauged worldline action we are interested in is given by
\begin{equation} \label{massless bosonic action}
S=\int \diff\tau\; \left[ p_\mu \dot{x}^\mu -i \bar\alpha_\mu \dot{\alpha}^\mu-eH - \bar{u} L - u \bar{L} - a J
\right]\;, 
\end{equation}
where we introduced the worldline gauge multiplet $(e,\bar u, u, a)$ acting as a set of Lagrange multipliers 
that enforce the constraints that, in the massless case, read
\begin{gather}\label{constraints}
 H = \frac{1}{2}p^\mu p_\mu \;, \quad L = \alpha ^\mu p_\mu\;, \quad \bar{L} = \bar{\alpha}^\mu p_\mu\;, \quad J = \alpha^\mu \bar{\alpha}_\mu \;.
\end{gather}
The latter satisfy a first-class Poisson-bracket algebra:
\begin{equation} \label{algebra}
\{L,\bar L\}_{\mathrm{PB}}= 2i H \;, \quad \{J, L\}_{\mathrm{PB}}= -i L \;, \quad \{J,\bar L\}_{\mathrm{PB}}= i \bar L \;.
\end{equation}
The phase-space functions \eqref{constraints} play rather different roles.
\begin{itemize}
 \item The role of $(H, L, \bL)$ constraints is to remove the negative-norm states and \emph{must} be gauged to make the model consistent with unitarity. The Hamiltonian constraint $H$ corresponds to the mass-shell condition for massless particles, and generates $\tau$-reparametrization in phase space, while the remaining pair, $L$ and  $\bL$, generates “bosonic” supersymmetries.\footnote{We deliberately use this terminology since \eqref{algebra} is formally similar to a \acr{SUSY} algebra, except for the fact that $L$ and $\bar L$ are bosonic rather than fermionic.}
\item The $J$ constraint is a $U(1)$ generator which rotates the bosonic oscillators by a phase; its gauging is optional as far as unitarity is concerned. However, upon quantization, it projects the  Hilbert space onto the physical 
subspace with a specific occupation number, describing the degrees of freedom of a particle with maximal spin $s$. For this to happen, one must add a Chern--Simons term on the worldline
with the Chern--Simons (\acr{CS}) coupling fine-tuned according to the value of the spin $s$ one wants to achieve.
This approach has been discussed extensively 
 in Ref.~\cite{Bastianelli:2013pta, Bastianelli:2015iba}. 
\end{itemize}
To make explicit the gauge symmetries enjoyed by \eqref{massless bosonic action}, one has to compute the action of the constraints on generic phase-space functions $F$ via the Poisson brackets: $\delta F = \{ F, V\}_{\mathrm{PB}}$. Considering the linear combination of the constraints $V=\epsilon H + \bar \xi L + \xi \bar L +\phi J$ with gauge parameters $(\epsilon, \bar \xi, \xi, \phi)$, the corresponding transformations are
\begin{subequations}\label{gauge trans0}
\begin{align}
\delta x^\mu &= \epsilon p^\mu + \xi \bar{\alpha}^\mu + \bar{\xi} \alpha^\mu\;, \label{time-transl}\\
\delta p_\mu &= 0\;, \\
\delta \alpha^\mu &= i\xi p^\mu + i \phi \alpha^\mu\;, \label{gauge1} \\
\delta \bar{\alpha}^\mu &= -i\bar{\xi} p^\mu - i \phi \bar{\alpha}^\mu\;. \label{gauge2}
\end{align}
\end{subequations}
For the action \eqref{massless bosonic action} to be invariant, the gauge fields must transform as follows
\begin{subequations}
\begin{align}
\delta e &= \dot{\epsilon} + 2i u \bar{\xi} - 2i \bar{u} \xi\;, \\
\delta u &= \dot{\xi} - i a \xi + i \phi u\;,  \label{u}
\\
\delta \bar{u} &= \dot{\bar{\xi}} + i a \bar{\xi} - i \phi \bar{u}\;, 
\label{bar u}
\\
\delta a &= \dot{\phi}\;.
\end{align}
\end{subequations}
The need for the worldline constraints to enforce unitarity remains somewhat obscure in this setup. 
To review and clarify this point, it may be beneficial to perform a brief lightcone analysis.

\subsection{Lightcone analysis}\label{chap:second:lightcone}
Despite the loss of manifest covariance, a lightcone analysis allows for a direct calculation of the number of propagating physical degrees of freedom. It is a well-known method, employed in many worldline models, see e.g. \cite{Siegel:1988yz, Bastianelli:2014lia, Bastianelli:2015tha}. We define lightcone coordinates $x^{\pm}$ in $D$ spacetime dimensions by
\begin{equation}
 x^\mu=(x^+,x^-,x^a)\;, \quad \text{with} \quad x^{\pm}=\frac{1}{\sqrt{2}}(x^0 \pm x^{D-1})\;,
\end{equation}
where $x^{a=1,\dots,D-2}$ are the transverse directions. The line element reads $\diff s^2=-2 \diff x^+ \diff x^-+\diff x^a \diff x^a$; for any vector $V^\mu$, $V^+=-V_-$ and $V^-=-V_+$.

The guiding idea behind the lightcone analysis is to remove negative-norm states by implementing a gauge-fixing that isolates the physical degrees of freedom, which in turn leads to a well-defined Hilbert space upon quantization. To do that, let us first assume motion with $p^+\neq0$ and consider the Hamiltonian constraint 
\be
H= \frac{1}{2}p^\mu p_\mu= -p^+p^- + \frac12 p^ap^a=0 \;.
\ee
It generates time translations, see \eqref{time-transl}. These symmetries   
are gauge-fixed by imposing the lightcone gauge 
\be
x^+=\tau \;.
\ee
Correspondingly, the Hamiltonian constraint is solved for the momentum $p^-$, conjugate to $x^+$,
\be
p^-  = \frac{1}{2p^+} p^ap^a \;.
\ee
At this point, the remaining independent phase-space variables are $(x^-, p^+)$ and $(x^a,p^a)$. The lightcone gauge has the flavor of nonrelativistic mechanics, but the model is fully relativistic. A Hilbert space can be constructed by quantizing these independent variables. 
On top of these variables, there are also the relativistic oscillators, which may also lead to negative norms.
That this does not happen (the so-called no-ghost theorem) is again made explicit by completing the lightcone gauge
fixing.
The gauge symmetries generated by $L$ and $\bar L$, see Eqs. 
\eqref{gauge1} and \eqref{gauge2}, are fixed by setting 
\begin{equation}
 \alpha^+=0\;, \quad \bar\alpha^+=0\;,
 \label{lcgf}
\end{equation}
while the constraints $L=\bar L=0$ are solved explicitly by expressing the 
variables conjugate to \eqref{lcgf} in terms of the remaining independent variables
\begin{align}
  \bar{\alpha}^-=\frac{1}{p^+}\bar\alpha^a p_a
  \;, \quad 
 \alpha^-=\frac{1}{p^+}\alpha^a p_a
 \;.
\end{align}
The conjugated pairs $(\bar\alpha^-,\alpha^+)$ and $(\bar\alpha^+,\alpha^-)$ are thus eliminated as independent phase-space coordinates, highlighting the fact that the only independent physical oscillators are the transverse ones $(\bar\alpha^a,\alpha^a)$. They produce states with positive norm upon quantization, as can be inferred by promoting their Poisson brackets to commutation relations
\begin{equation} \label{ab}
 [\bar\alpha^a,\alpha^b
 ]=\delta^{ab}\;,
\end{equation}
which are realized on a Fock space, where $\alpha^a$ act as creation operators while  $\bar\alpha^a$ as destruction operators, 
thus yielding a unitary spectrum of massless particles that decompose into irreps of the little group $SO(D-2)$. 

To conclude this section, we report the (partially) gauge-fixed worldline Lagrangian 
\begin{equation}
 \mathcal{L}= p_- \dot{x}^- + p_a \dot{x}^a - \frac{1}{2p^+} p^ap^a 
 -i \bar\alpha_a \dot{\alpha}^a 
 - a \, \bar{\alpha}_a \alpha^a \;.
\end{equation}
At this stage, it only remains to address the further constraint related to the worldline gauge field~$a(\tau)$, but this has no relevance to the no-ghost theorem.

\subsection{Mass from dimensional reduction}
In this chapter, our main interest is to describe \emph{massive} spinning particles. One way to introduce the mass consists 
of the dimensional reduction of a higher-dimensional massless theory. We take the theory \eqref{massless bosonic action} to live in $(D+1)$-dimensions and gauge the direction $x^D$ by imposing the first-class constraint
\begin{equation}
 p_D=m\;,
\end{equation}
with $m$ the mass of the particle. We further define $(\beta, \bbeta) \coloneqq (\alpha^D, \balpha^D)$, which inherit the following Poisson brackets
\begin{equation}
\{\beta, \bar \beta\}_{\mathrm{PB}} = i \;.
\end{equation}
The constraints \eqref{constraints} get modified by the presence of the mass:\footnote{From now on, we take spacetime indices to run as $\mu=0,\dots,D-1$ where $D$ denotes the number of spacetime dimensions.}
\begin{gather}  \label{new constraints}
 H = \frac{1}{2}(p^\mu p_\mu + m^2) \;, \quad L = \alpha ^\mu p_\mu +\beta m\;, \quad \bar{L} = \bar{\alpha}^\mu p_\mu 
 +\bar \beta m\;, \quad J_c=\alpha^\mu \bar{\alpha}_\mu + \beta \bbeta - c\;,
\end{gather}
where we redefined the $U(1)$ constraint as $J_c = J - c $ for future convenience. The constant $c$
is sometimes called Chern--Simons coupling. Note that, importantly, they still satisfy the first-class algebra \eqref{algebra} despite the mass improvement. The gauge transformations \eqref{gauge trans0} are enriched by
\begin{subequations} \label{gauge trans}
\begin{align}
\delta \beta &= i \xi m + i \phi \beta\;, \label{gauge trans1} \\
\delta \bar{\beta} &= -i \bar{\xi} m - i \phi \bar{\beta}\;.
\end{align}
\end{subequations}
\paragraph{Lightcone analysis}
The lightcone gauge is implemented just as in the massless case: 
in particular, the bosonic supersymmetries are used to fix $\alpha^+= \bar\alpha^+=0$ once again, and the constraints $L=\bar L=0$ are solved by 
\begin{align}
 \alpha^-=\frac{1}{p^+}\left(\alpha^a p_a+m\beta\right)\;, \quad \bar{\alpha}^-=\frac{1}{p^+}\left(\bar\alpha^a p_a+m\bar\beta\right)\;,
\end{align}
thus eliminating the longitudinal oscillators. 
Differently from the massless case, the presence of the extra pair of $\beta$-oscillators produces a sum of irreps\footnote{This can be explicitly seen by implementing the $J_c$ constraint {\it à la} Dirac \cite{Bastianelli:2014lia}.} of the $SO(D-2)$ group that fill irreps of the $SO(D-1)$ rotation group, corresponding to the polarizations of massive spin particles in $D$ spacetime dimensions.

\subsection{Dirac quantization}
Upon covariant quantization, the worldline coordinates obey the following commutation relations fixed by their classical Poisson brackets
\begin{gather}\label{canon comm}
 [x^\mu, p_\nu]= i \delta^{\mu}_{\nu}\;, \quad [\bar{\alpha}^\mu, \alpha^\nu]= \eta ^{\mu \nu}\;, \quad [\bar{\beta}, \beta]= 1\;,
\end{gather}
and the first-class algebra becomes
\begin{equation}\label{op algebra}
 [\bar{L}, L] = 2H \;, \quad [J_c, L] = L\;, \quad [J_c, \bar{L}] = -\bar{L}\;. 
\end{equation}
Note that ordering ambiguities emerge only for the constraint $J_c$.
We have defined the quantum $J_c$ operator by a symmetric quantization prescription, so that
\begin{align}
J_c  &= \frac12 (\alpha_\mu \bar{\alpha}^\mu + \bar{\alpha}^\mu  \alpha_\mu + \beta \bbeta + \bbeta \beta) -c\;,
\cr
&=  \alpha_\mu \bar{\alpha}^\mu + \beta \bbeta  +  \frac{D+1}{2} -c\;,
\cr
&= N_{\alpha} + N_{\beta} - s\;,
\label{quantum Jc}
\end{align}
where we have used the commutation relations and introduced the usual number operators 
\be
N_{\alpha} =  \alpha_\mu \bar{\alpha}^\mu \;,
\quad  N_{\beta}=\beta \bbeta\;,
\ee
and related the \acr{CS} coupling $c$ to the real number $s$ by setting
\be 
c= \frac{D+1}{2} + s  \;.
\label{2.25}
\ee
The algebra \eqref{op algebra} is now easily obtained. 
At this point, it is worth mentioning that the relation between the \acr{CS} coupling $c$ and the physical value of the spin $s$ generally depends on the quantization scheme adopted.

The Hilbert space $\mathcal{H}_{\mathrm{matter}}$ of the “matter” sector, i.e. the one  associated with the $(x,p)$ coordinates and $(\alpha,\bar\alpha,\beta,\bar\beta)$ oscillators, is realized as a tensor product of the representations of the algebras \eqref{canon comm}
\begin{equation} \label{Hm}
\mathcal{H}_{\mathrm{matter}}=\mathcal{H}_{\mathcal{M}}\otimes \mathcal{H}_{(\alpha,\beta)}\;.
\end{equation}
Specifically, we represent it by identifying the states in $\mathcal{H}_{\mathcal{M}}$ as the smooth functions of $x^\mu$, while we construct $\mathcal{H}_{(\alpha,\beta)}$ as the Fock space with vacuum defined by 
\begin{equation}
 (\bar\alpha^\mu,\bar\beta)\,\ket{0}=0\;.
\end{equation}
The decomposition of a generic state $\ket{\varphi}$ is thus written in terms of coefficients corresponding to rank-$s$ symmetric tensors:
\begin{equation}
 \ket{\varphi} = \sum_{r,p = 0}^{\infty} \ket{\varphi^{(r,p)}}= \sum_{r,p = 0}^{\infty} \frac{1}{r! p!} \, \varphi^{(r,p)}_{\mu_1 ... \mu_r}(x) \otimes \alpha ^{\mu_1} \dots \alpha ^{\mu_r} \beta^p \ket{0}\;. 
\end{equation}
The quantization may proceed either following a procedure {\it à la} Dirac or by using BRST techniques; either way, we can deal with the gauge symmetries without abandoning manifest covariance, obtaining at last a positive-definite physical Hilbert space. We start with the former method, leaving the BRST analysis for the dedicated section.

The physical Hilbert space in the Dirac (also known as Dirac--Gupta--Bleuler) scheme is determined by asking the constraints to have null matrix elements for arbitrary physical states  
$|\varphi\rangle$ and $|\chi\rangle$  
\begin{equation}\label{Dirac second}
 \braket{\chi |(H, L, \bL, J_c)|\varphi} = 0\;.
\end{equation}
This can be satisfied by requiring 
\begin{equation}\label{phys cond Dirac}
 H\ket{\varphi} = \bL\ket{\varphi} = J_c \ket{\varphi} = 0
\end{equation}
for any physical state  $|\varphi\rangle$,
since then also $\langle \varphi | \bar L = 0 $, 
as $\bL$ is the hermitian conjugate of $L$. Recalling now that at the quantum level 
\be
J_c= N_{\alpha} + N_{\beta} - s
\ee
where the number operators $N_{\alpha}$ and $N_{\beta}$
count the occupation number of the $\alpha$ and $\beta$ oscillators
in the Fock space, we see 
that the quantum $J_c$ constraint selects precisely states with occupation number $s$, which must be 
a nonnegative integer.
Incidentally, we notice that the \acr{CS} coupling must be quantized to get a nontrivial solution of the constraint, and therefore a nontrivial quantum theory. 
The upshot is that the quantum $J_c$ constraint
reduces the Hilbert space $\mathcal{H}_{\mathrm{matter}}$ to the subspace with occupation number $s$ for the oscillators. 
The condition \eqref{phys cond Dirac} defines physical states which are seen to form equivalence classes
\begin{equation}
 \ket{\varphi} \sim \ket{\varphi} + \ket{\varphi_{\rm null}}\;,
\end{equation}
where  $\ket{\varphi_{\rm null}}$ is  a null state of the form 
\begin{gather}
\ket{\varphi_{\rm null}} = L \ket{\xi}  \;, \qquad \text{with} \quad 
H \ket{\xi} = \bL \ket{\xi} =  (J_c+1) \ket{\xi} = 0 \;.
\end{gather}
These null states are physical, but have zero norm and vanishing overlap with any other physical state.
They give rise to redundancies or residual “gauge symmetries” of the state $\ket{\varphi}$. 

Let us make the conditions for the case $s=1$, which is of interest to us, explicit. 
A generic state at occupation number $s=1$ is given by
\begin{equation}
 \ket{\psi} = W_\mu(x) \alpha ^\mu \ket{0} -i \varphi(x) \beta \ket{0}
\end{equation}
and the physicality conditions \eqref{phys cond Dirac} translate into the following set of equations, denoting $\Box := \partial^\mu \partial_\mu$,
\begin{align}
 (\Box - m^2) W_\mu &= 0\;,  \label{box proca dirac}\\ 
 (\Box - m^2) \varphi &= 0\;,  \\
 \partial^\mu W_\mu + m \varphi &= 0 \;, \label{trans proca Dirac}
\end{align}
with gauge symmetries related to  null states  given by
\begin{gather}
 \delta W_\mu = \partial_\mu \xi\;, \quad \delta \varphi = - m\xi\;.
\end{gather}
Using the gauge symmetry to set $\varphi(x)=0$, we recover the standard  Fierz--Pauli equations for a massive spin-1 field $W_\mu(x)$. 

\subsection{Counting degrees of freedom \label{degrees of freedom}}
In this section, we aim to use the path integral to count the number of degrees of freedom propagated by the massive model for different values of the \acr{CS} coupling. 
As a byproduct, this will provide the overall normalization of the effective action we intend to study in later sections.

To count the number of degrees of freedom, we consider the one-loop effective action obtained by 
path integrating the free action on worldlines with the topology of a circle $S^1$ (the loop). After fixing 
the overall normalization to match the scalar case, we will get the number of degrees of freedom in the other
sectors of the worldline theory.

Thus, we consider the following path integral
\begin{align}
\Gamma= \int_{S^1} \frac{DG DX}{\rm Vol (Gauge)}\, \mathrm{e}^{iS[X,G]} \;,
\label{2.39}
\end{align}
where $G=(e,\bar u, u, a)$ denotes the gauge fields, whereas $X= (x^\mu ,p_\mu, \alpha^\mu, \bar \alpha_\mu, \beta, \bar \beta)$ collectively denotes all dynamical variables parametrizing the phase space. 
The action is similar to the one in \eqref{massless bosonic action} with 
the additional $(\beta,\bar\beta) $ oscillator and with
the constraints in \eqref{new constraints}, namely
\be
\begin{aligned}
S=\int  \diff\tau \biggl [ & p_\mu \dot{x}^\mu -i \bar\alpha_\mu \dot{\alpha}^\mu -i \bar\beta \dot{\beta}
-\frac{e}{2}(p^\mu p_\mu + m^2) 
- \bar{u} ( \alpha ^\mu p_\mu +\beta m)
 - u  (\bar{\alpha}^\mu p_\mu  +\bar \beta m)
 \cr &
 - a (\alpha^\mu \bar{\alpha}_\mu + \beta \bbeta - c) \biggr ] \;.
 \end{aligned}
\ee
Periodic boundary conditions are understood to implement the path integral on the circle.

The overcounting from summing over gauge equivalent configurations is formally taken into account by dividing by the volume of the gauge group. We use the Faddeev--Popov ($\acr{\Phi\Pi}$) method to extract the latter, and gauge-fix the worldline gauge fields to constant moduli 
\begin{equation}\label{gauge fixed G}
 G=(e,\bar u, u, a) \ \ \to\ \ \hat{G}=(2T, 0, 0, \theta)\;.
\end{equation}
Here $T$ is once again the Schwinger proper time, playing the role of the modulus related to the einbein $e(\tau)$, and corresponding to the gauge-invariant worldline length $\int_0^1 \diff\tau \, e$. 
The modulus $\theta\in[0,2\pi]$ is associated with the worldline $U(1)$ gauge field $a(\tau)$ and parametrizes the gauge invariant Wilson loop $\mathrm{e}^{-i \int_0^1\diff\tau \, a}$. 
It is responsible for the reduction of the Hilbert space to a given spin sector. On the other hand, the gauge fields $(u,\bar u)$ can be gauge-fixed to zero:
 this value can always be reached by inverting the differential operator that relates these fields to their respective gauge parameters  $(\xi,\bar \xi)$, as shown in Eqs. \eqref{u} and \eqref{bar u}.
 This inversion fails only at the point $\theta =0$, which, however, can be handled through a limiting procedure, as in the standard 
  $\mathcal{N}=2$ particle case \cite{Bastianelli:2005vk, Bastianelli:2005uy}.  Therefore, $(u,\bar u)$  carry no moduli.
As a reminder, moduli generically parametrize gauge-invariant field configurations that must be integrated over in the path integral.
  
We prefer to work in the Euclidean version of the theory, so that we first pass to configuration space by eliminating the momenta $p_\mu$,
Wick rotate the action with $\tau \to -i\tau$, taking into account also the rotation of the gauge field $a\to ia$, and we get the Euclidean worldline action
\begin{equation} \label{Ecugfaction}
S_{\mathrm{E}}[X,\hat{G}]=\int  \diff\tau \left [ \frac{1}{4T} \dot x^2 + \alpha_\mu (\partial_\tau + i\theta) \balpha^\mu 
+ \beta (\partial_\tau +i\theta) \bbeta 
+m^2T -i c \theta  \right ]\;.
\end{equation}
The final expression of the path integral on worldlines with the topology of a circle -- dubbed “worldloop” -- can thus be recast as
\begin{align} \label{2.43}
\Gamma =  - \int_0^{\infty} \frac{\diff T}{T} \mathrm{e}^{-m^2 T}
\int
\frac{\diff^D\bar{x}}{(4\pi T)^{\nicefrac{D}{2}}}\ {\rm DoF}(c,D)\;,
\end{align}
where we extracted the dependence on the zero modes $\bar{x}^\mu$ of the coordinates by setting 
\begin{equation}
    x^\mu(\tau)=\bar{x}^\mu+t^\mu(\tau)\;, \quad \text{with} \quad t^\mu(0)=t^\mu(1)=0\;,
    \label{split}
\end{equation}
with the quantum fluctuations $t^\mu(\tau)$ satisfying Dirichlet boundary conditions, evaluated the free path integral that produces functional determinants, and denoted by 
DoF($c,D$)  the number of (complex) degrees of freedom, that acquires the expression
\begin{equation} \label{2.45}
{\rm DoF}(c,D) = k \int_0^{2\pi} \frac{\diff\theta}{2\pi }\, \mathrm{e}^{i c \theta}\;
{\rm Det}\left(\partial_\tau -i\theta\right) 
{\rm Det}\left(\partial_\tau +i\theta\right) 
\left[{\rm Det}(\partial_\tau +i\theta)\right]^{-D-1}\;,
\end{equation}
with $k$ an overall normalization to be fixed later on. 
The value DoF $=1$ corresponds to a complex scalar, as seen by comparing with \acr{QFT} expressions.
In this formula, the first two functional determinants are the $\acr{\Phi\Pi}$ ones,
whereas the third one is due to the path integration over the bosonic oscillators. All determinants are evaluated with periodic boundary conditions:
using 
\begin{equation}\label{det}
{\rm Det}\left(\partial_\tau + i\theta\right) = 2i \sin \left(\frac{\theta}{2}\right)\;,
\end{equation}
see for example \cite{Bastianelli:2007pv, Dai:2008bh}, 
setting the \acr{CS} coupling to\footnote{The shift from the value given in \eqref{2.25} is due to the contribution of the ghost fields. For convenience, we now indicate the degrees of freedom by 
${\rm DoF}(s,D)$, which highlights the dependence on the value of the spin $s$, rather than on the \acr{CS} coupling $c$.
This should not cause any confusion.}
$c = \frac{D - 1}{2} + s$, and fixing $k=- 1$ as overall normalization, we find the following expression for the number of degrees of freedom
\begin{align}
{\rm DoF}(s,D) &=  \int_0^{2\pi} \frac{\diff\theta}{2\pi }\; \mathrm{e}^{i \left(\frac{D-1}{2}+s\right) \theta} \;\left(2i \sin \frac{\theta}{2}\right )^{1-D}\;.
\end{align}
To evaluate it, we find it more convenient to recast it in terms of the Wilson loop variable
$w:=\mathrm{e}^{-i\theta}$, so that
\begin{equation}
{\rm DoF}(s,D) 
= 
\oint \frac{\diff w}{2\pi i}\, \frac{1}{w^{s + 1}}\, \frac{1}{(1-w)^{D-1}}\;.
\end{equation}
Deforming the contour to exclude the singular point $w=1$, while taking care of the pole in $w=0$, we get
\begin{align}
\begin{split}
&{\rm DoF}(0,D) = 1\;, \\ 
&{\rm DoF}(1,D) = ( D-1)\;, \\
&{\rm DoF}(2,D) =  \frac{D(D-1) }{2}\;, \\
& \cdots \\
&{\rm DoF}(s,D) =\frac {(D-1) D\cdots (D+s-2)}{s!}\;,
\end{split}
\end{align}
which indeed describes the degrees of freedom of a reducible (for $s \geq 1$) representation of the little group $SO(D-1)$
as carried by a symmetric tensor with $s$ indices. It corresponds  to the propagation of a 
  multiplet of massive particles of decreasing 
 spin $s, s-2, s-4, \cdots, 0$ for even $s$, and $s, s-2,\cdots, 1$ for odd $s$. 
 This matches the results seen in the lightcone gauge.

\section{Coupling to electromagnetism} \label{sec2} 
The BRST formalism is especially well-suited for analyzing the constraints required for consistent background interactions. For this reason, we briefly review the free particle in this framework and then examine its interaction with an electromagnetic background.

\subsection{Free BRST analysis} \label{sec2.1}
We proceed with the BRST quantization focusing only on the subalgebra of \eqref{op algebra} generated by $(H, L, \bar{L})$. The constraint $J_c$ is treated on different footings: it is imposed as a constraint on the
BRST Hilbert space, defining a restricted Hilbert space where the cohomology of the BRST operator will be analyzed.

The Hilbert space is enlarged to realize the fermionic ghost-antighost pairs of operators 
\begin{gather}
 \{b, c\} = 1 \;, \quad \{\mathcal{B},\bar{\mathcal{C}}\} = 1 \;, \quad \{\bar{\mathcal{B}},\mathcal{C}\} = 1\;,
\end{gather}
associated with the $(H, L, \bar{L})$ constraints, respectively. We assign them the following ghost numbers: $\mathrm{gh}(c, \bar{\mathcal{C}},\mathcal{C} ) = +1\,, \;\mathrm{gh}(b, \bar{\mathcal{B}},\mathcal{B} ) = -1$. The BRST charge associated with a first-class system is readily constructed. In the present case, it takes the form
\begin{equation} \label{Q}
 \mathcal{Q} = cH + \bar{\mathcal{C}} L + \mathcal{C} \bar{L} -2\mathcal{C} \bar{\mathcal{C}} b \;.
\end{equation}
It is an anticommuting, ghost number $+1$, nilpotent operator by construction. It is hermitian provided that
\begin{gather}
 c^\dagger\ = c \;, \quad b^\dagger = b \;, \quad \C^\dagger = \bC \;, \quad \B^\dagger = \bB \;. 
\end{gather}
The matter sector Hilbert space \eqref{Hm} is extended to the BRST Hilbert space $\mathcal{H}_{\mathrm{BRST}}$ by a tensor product with the ghost sector, associated with the $(c,b,\mathcal{B},\bar{\mathcal{C}}, \mathcal{C},\bar{\mathcal{B}})$ operators. The latter is constructed 
as a Fock space on the ghost vacuum defined by 
\begin{equation}
(b,\bar{\mathcal{C}},\bar{\mathcal{B}})\,\ket{0}_{\mathrm{gh}}=0\;.
 \end{equation}
Since all ghosts are Grassmann odd, $\mathcal{H}_{\mathrm{gh}}$ is finite dimensional.\footnote{In particular, the Fock vacuum $\ket{0}_{\mathrm{gh}}$ can be mapped into the “physical vacuum” $\ket{1}_{\mathrm{gh}}$, with the concept of “physicality” to be defined shortly,  
by
\begin{equation*}
 \ket{1}_{\mathrm{gh}}:=\mathcal{B}\ket{0}_{\mathrm{gh}}\ ,
\end{equation*} 
see e.g. the discussion in Ref.~\cite{Bengtsson:2004cd, Bonezzi:2024emt}. The vacuum $\ket{1}_{\mathrm{gh}}$ turns out to be the correct one to consider in order to create external states by inserting vertex operators in the worldline path integral, as illustrated in the upcoming Chapter~\ref{chap:third}.} A generic state $\ket{\Phi}$ in the BRST-extended Hilbert $\mathcal{H}_{\mathrm{BRST}}$ space reads
\begin{gather}\label{gen state}
 \ket{\Phi} = \sum_{s,p = 0}^{\infty} \sum_{q,r,t = 0}^{1} c^q \mathcal{C}^r \mathcal{B}^t\ket{\Phi^{(s,p)(q,r,t)}} 
\end{gather}
where
\begin{gather}
 \ket{\Phi^{(s,p)(q,r,t)}} = \frac{1}{s! p!} \, \Phi^{(s,p)(q,r,t)}_{\mu_1 ... \mu_s}(x)\, \alpha ^{\mu_1} \dots \alpha ^{\mu_s} \beta^p \ket{0}\;,
\end{gather}
with $\ket{0}$ now denoting the full BRST vacuum. With this choice, the conjugate momenta act as derivatives:
\begin{equation}
 p_\mu=-i\partial_\mu\;, \quad \bar\alpha^\mu=\partial_{\alpha^\mu}\;, \quad \bar\beta=\partial_{\beta}\;, \quad b=\partial_c\;, \quad \bar{\mathcal{C}}=\partial_\mathcal{B}\;, \quad \bar{\mathcal{B}}=\partial_{\mathcal{C}}\;.
\end{equation}
We now introduce a couple of operators, $G$ and $\J_s$, to further restrict the full BRST Hilbert space. These are the ghost number operator
\begin{align} 
 G &= cb + \mathcal{C} \bar{\mathcal{B}} - \mathcal{B} \bar{\mathcal{C}}\;, \qquad [G,\Q] = \Q \;,
\end{align}
and the (shifted) occupation number operator $\J_s$
\begin{align}
 \J_s &= \alpha_\mu \bar{\alpha}^\mu + \beta \bar{\beta} + \mathcal{C} \bar{\mathcal{B}} + \mathcal{B} \bar{\mathcal{C}} - s\;, \quad [\Q,\J_s] = 0\;.
\end{align}
They commute between themselves, $[G,\J_s]=0$.  
The ghost number operator grades the BRST Hilbert space according to the ghost number, and the commutator $ [G,\Q] = \Q$ manifests that the BRST charge has ghost number 1.
The occupation number operator  $\J_s$ also grades the Hilbert space according to its eigenvalues and
can be used as a constraint to project the Hilbert space onto the subspace with fixed occupation number $s$.\footnote{We choose an antisymmetric quantization's prescriptions for fermionic operators. In combination with the Weyl ordering for the bosonic operators previously discussed, the effect is to shift the \acr{CS} coupling as $c = \frac{D - 1}{2} + s$.
This relation has already been used in the path integral construction (see footnote 4), which evidently involves a regularization consistent with this ordering prescription.} 

We can exploit the operators above simultaneously -- since $[G, \J_s] = 0$ -- to select states in $\mathcal{H}_{\mathrm{BRST}}$ with a precise ghost and occupation number. The physical states are identified as elements of the BRST cohomology
\begin{gather} \label{cohomology}
 \Q \ket{\Phi} = 0\;,  \quad \ket{\Phi} \sim \ket{\Phi} + \Q \ket{\Lambda} 
\end{gather}
restricted to the subspace with vanishing eigenvalues of the ghost number and shifted occupation number operators, i.e.
\begin{gather}
 G\ket{\Phi}=\J_s \ket{\Phi} = 0\;.
\end{gather}

Our interest lies in the first-quantized description of a massive spin-1 particle: this is achieved by choosing $s=1$. 
An arbitrary wavefunction  at zero ghost number and with $s=1$ is then given by
\begin{equation}\label{eq: s=1 state}
 \ket{\psi} = W_\mu(x) \alpha ^\mu \ket{0} -i\varphi (x) \beta \ket{0} +f(x) c \mathcal{B} \ket{0}\;,
\end{equation}
where the complex fields $W_\mu(x)$, $ \varphi (x) $, and $f(x)$ must be further constrained by Eq.~\eqref{cohomology} to represent the physical states of the theory.
From the closure equation, i.e., the first one in \eqref{cohomology}, we obtain
\begin{subequations}\label{Proca0}
\begin{align}
\left( \Box-m^2 \right)W_\mu-2i\partial_\mu f&=0\;, \\
\left( \Box-m^2 \right)\varphi+2imf&=0\;, \\
\partial_\mu W^\mu+m\varphi-2if&=0\label{algebraic}\;,
\end{align}
\end{subequations}
which, upon eliminating the auxiliary field $f(x)$, represent the field equations of the Proca field in the St\"uckelberg formulation
\begin{subequations}\label{Proca}
\begin{align}
\left( \Box-m^2 \right)W_\mu-\partial_\mu\partial \cdot W
-m\partial_\mu\varphi&=0\;, \\
\Box \varphi+m\partial_\mu W^\mu&=0\;,
\end{align}
\end{subequations}
where the dot “$\cdot$” indicates contraction over spacetime indices. The latter equations enjoy a gauge symmetry, which, from \eqref{cohomology}, reads
\begin{equation}
 \delta\ket{ \psi} = Q \ket{\Lambda}\;, \quad \text{with} \quad \ket{\Lambda} = i \xi(x) \mathcal{B} \ket{0}\;,
\end{equation}
i.e.
\begin{equation}\label{chap:second:gauge}
 \delta W_\mu=\partial_\mu\xi\;, \quad \delta\varphi=-m \xi\;,
\end{equation}
which is the well-known St\"uckelberg gauge symmetry.

A few comments are in order: (\textit{i}) the wavefunction \eqref{eq: s=1 state} can be interpreted as a spacetime Batalin--Vilkovisky (\acr{BV}) “string field” displaying only
the classical fields out of the minimal \acr{BV} spectrum of the Proca theory, along with an auxiliary field.\footnote{The complete minimal \acr{BV} spectrum is obtained by relaxing the condition $G\ket{\psi}=0$, see for instance \cite{Carosi:2021wbi, Fecit:2023kah}.} The Grassmann parities and ghost numbers of the field components are all equal to zero.
Note the presence of the St\"uckelberg scalar $\varphi$, which restores the $U(1)$ gauge symmetry \cite{Stueckelberg:1957zz}, originally broken due to the introduction of the mass.
(\textit{ii}) In the so-called \emph{unitary gauge}, namely setting the St\"uckelberg field to zero, one reduces the field equations to the standard Fierz--Pauli system for the massive spin-1 field $W_\mu(x)$. (\textit{iii}) Taking the massless limit produces from \eqref{Proca} a pair of decoupled equations: one for a free-propagating massless vector field $W_\mu(x)$ and one for a massless scalar field $\varphi(x)$. This is tantamount to the fact that the theory of massive spin-1 does not suffer from the so-called “van Dam--Veltman--Zakharov (\acr{vDVZ}) discontinuity”,\footnote{Roughly speaking, the \acr{vDVZ} discontinuity implies that that the massive theory, however small the mass of the particles may be, leads to different physical predictions that do not smoothly recover the results of the massless theory.} differently from the massive spin 2 case \cite{vanDam:1970vg, Zakharov:1970cc}, which is covered in Chapter~\ref{chap:fourth}.

\subsection{Consistent electromagnetic coupling}\label{coupling}
The coupling of the worldline to an abelian background field $A_\mu(x)$ in spacetime (with coupling constant $q$) is achieved by covariantizing the $(L,\bar L)$ constraints as follows
\begin{gather}
\label{cov constr}
 L \rightarrow \alpha^\mu \pi_\mu + \beta m\;, \quad \bL \rightarrow \balpha^\mu \pi_\mu + \bbeta m\;,
\end{gather}
where the covariantized momentum $\pi_\mu$ 
with coupling constant $q$ is defined by
\begin{equation}
\pi_\mu = p_\mu -q A_\mu\;.
\end{equation}
It becomes the covariant derivative in the coordinate representation, $\pi_\mu= -i(\partial_\mu -iq A_\mu)=-i D_\mu$.
The new constraints do not form a first-class algebra anymore: while the bosonic supersymmetry charges do commute into a possibly deformed Hamiltonian
\begin{align}
 [\bL, L] &= \pi^2 + m^2 +\balpha^\mu \alpha^\nu \tilde{F}_{\mu\nu}=:2 H_{- 1/2}\;,
\end{align}
where we have denoted $[\pi_\mu, \pi_\nu]=-[D_\mu, D_\nu]=  iq F_{\mu\nu}=:\tilde{F}_{\mu\nu}$, we find that the remaining commutators read
\begin{align}
 [L, H_{-1/2}] &= i \alpha^\mu \pa^\nu \tilde{F}_{\nu \mu} + \frac{3}{2}\alpha^\mu \tilde{F}_{\mu \nu} \pi^\nu +\frac{i}{2} \balpha^\nu \alpha^\rho \alpha^\mu \pa_\mu \tilde{F}_{\rho \nu} \;, \\
 [\bL, H_{-1/2}] &=i \balpha^\mu \pa^\nu \tilde{F}_{\nu \mu} + \frac{3}{2}\balpha^\mu \tilde{F}_{\mu \nu} \pi^\nu +\frac{i}{2} \balpha^\nu \alpha^\rho \balpha^\mu \pa_\mu \tilde{F}_{\rho \nu}\;,
\end{align}
and do not allow for a suitable redefinition of the constraints to form a first-class algebra. Thus, we expect the associated BRST charge to fail to be nilpotent, which indicates an inconsistency of the interacting worldline theory at the quantum level. We then proceed tentatively and try an ansatz
\begin{equation}\label{Q defor}
 \mathcal{Q}_A = c H_\kappa + S^\mu \pi_\mu + \bar{\mathcal{C}}\beta m + \mathcal{C} m \bar{\beta} - Mb\;,
\end{equation}
with a deformed Hamiltonian
\begin{equation}\label{Hdef}
 H_\kappa =\frac{1}{2}\left(\pi ^2 + m^2 + 2 \kappa \alpha ^\mu \balpha^\nu \tilde F_{\mu\nu} \right) 
\end{equation}
that contains a non-minimal coupling with constant $\kappa$ to be conveniently fixed, and then compute
\begin{equation}\label{Q2}
 \mathcal{Q}_A^2 =-\frac{2 \kappa + 1}{4} M S^{\mu \nu} \tilde{F}_{\mu \nu} +c [H_\kappa, S^\mu \pi_\mu] \;,
\end{equation}
where we used the shorthand notations
\begin{gather}
 S^\mu = \alpha^\mu \bar{\mathcal{C}} + \bar{\alpha}^\mu \mathcal{C}\;, \quad S^{\mu \nu} = \alpha^\mu \bar{\alpha}^\nu - \alpha^\nu \bar{\alpha}^\mu\;, \quad M = 2 \mathcal{C} \bar{\mathcal{C}}\;.
\end{gather}
In general, \eqref{Q2} is not zero, except for the trivial case of vanishing field strength, which manifests the inconsistency of coupling massive spin $s$ particles, with generic $s$, to an \acr{EM} background. This is also the case for massless particles, as already discussed in Ref.~\cite{Bonezzi:2024emt}. However, restricting the occupation number to be $s\leq1$, the nilpotency condition simplifies:\footnote{This can be inferred by counting the number of annihilation operators in $ \mathcal{Q}_A^2$: if there are two or more, they annihilate the physical wavefunction for $s=1$.}
\begin{align}
\begin{split} \label{Q^2}
 \mathcal{Q}_A^2\big\vert_{s=0,1} &= c [H_\kappa, S^\mu \pi_\mu]\big\vert_{s=0,1}\\
 &=-\frac{i c}{2}\left(\partial_\mu \tilde{F}^{\mu \nu} S_\nu + 2 i(1 -\kappa) \tilde{F}^{\mu \nu} \pi_\mu S_\nu -\kappa \, S^\nu S^{\alpha \beta} \partial_\nu \tilde{F}_{\alpha \beta}\right)\big\vert_{s=0,1}\;.
\end{split}
\end{align}
For the $s=0$ sector, this expression is automatically zero regardless of any condition on the background electromagnetic field, as this operator contains destruction operators sitting on the right that annihilate the $s=0$ wave function (recall the expressions for the operators $S^\mu$ and $S^{\mu\nu}$).
Physically, this expresses the fact that spinless particles can be consistently coupled to off-shell abelian background fields. 
As for the massive spin-1 sector, cf. \eqref{eq: s=1 state}, the previous equation further simplifies to
\begin{equation}
 \mathcal{Q}_A^2\big\vert_{s=1}=\frac{q}{2} c(\alpha^\nu \bar{\mathcal{C}} - \bar{\alpha}^\nu \mathcal{C}) \partial^\mu (\partial_\mu A_\nu - \partial _\nu A_\mu)\;,
\end{equation}
having set $\kappa=1$ to achieve this result: then, nilpotency of the deformed BRST charge requires the background $A_\mu(x)$ to be on-shell, i.e.
\begin{equation}
 \partial^\mu F_{\mu\nu}=\partial^\mu(\partial_\mu A_\nu - \partial_\nu A_\mu) \must 0\;.
\end{equation}
This is enough to prove the consistency of the coupling.

Let us notice that the mass does not obstruct the nilpotency, namely, it does not seem to carry substantial differences with respect to the massless case. To be more precise, $m$ does not \emph{explicitly} enter in the BRST algebra for any spin $s$, but it may obstruct the nilpotency for higher-spin particles, starting from the spin 2 case as discussed in Ref.~\cite{Fecit:2023kah} for the gravitational coupling. 
How the no-go theorem about massless charged particles \cite{Weinberg:1980kq} appears from a worldline perspective is at the moment unclear to us.

\section{Effective action in electromagnetic background}\label{sec3}
In this section, we employ the worldline model to compute the one-loop effective action induced by a charged spin-1 particle in a constant \acr{EM} background.

The worldline representation of the effective action is derived by following the same approach as in the free case (cf. Section \ref{degrees of freedom}), which in particular determines the overall normalization of the path integral. 
Given the BRST analysis presented in Section \ref{coupling}, we are naturally led to treat the $s=0$ and $s=1$ cases simultaneously. 
This approach allows for a direct comparison, with the spinless case serving as a check on the novel spin-1 contribution within the first-quantized framework.

As the interacting worldline action, we take the covariantized version of the gauge-fixed free action in Euclidean configuration space \eqref{Ecugfaction}
with covariantized constraints \eqref{cov constr} and deformed Hamiltonian $H_{1}$ \eqref{Hdef}, i.e. (factoring out the $m^2T -i c \theta$ constant term)
\begin{equation}
\begin{split}
 S_{\mathrm{E}}[X, \hat{G}; A] = \int \diff \tau \left[\frac{\dot{x}^2}{4T} - i q A^\mu \dot{x}_\mu + \alpha^\mu \left( \delta_{\mu \nu} \left(\frac{\diff}{\diff\tau}+ i \theta \right) + 2 i q T F_{\mu \nu} \right) \balpha^\nu + \beta \left(\frac{\diff}{\diff\tau} + i \theta \right) \bbeta 
\right]\;.
\end{split}
\end{equation}
We restrict our analysis to four spacetime dimensions and consider a constant electromagnetic field as the on-shell background. Under these conditions, we derive the one-loop effective action of the Euler–Heisenberg type induced by a massive spin-1 particle. This effective action is given 
by the path integral on the circle of the gauge-fixed action and takes the form
\begin{equation}\label{path integral}
 \Gamma[A] =  \int_{0}^{\infty} \frac{\diff T}{T} \mathrm{e}^{-m^2 T} 
 \int_{0}^{2\pi} \frac{\diff\theta}{2\pi} 
 \mathrm{e}^{i c \theta} 
 \,{\rm Det}\left(\partial_\tau -i\theta\right) {\rm Det}\left(\partial_\tau +i\theta\right) \int_{\mathrm{PBC}} DX \,\mathrm{e}^{-S_\mathrm{E}[X,\hat{G}; A]} \;,
\end{equation}
with measure in moduli space and determinants already fixed by the free case, see Eqs. \eqref{2.43} and \eqref{2.45}, and
with the \acr{CS} coupling fixed to  $c=\frac32 +s$.
Recalling the coordinate split in Eq.~\eqref{split},
 we use the Fock--Schwinger gauge around $\bar{x}$ for the background field, i.e.
\begin{equation}\label{FS}
 (x-\bar{x})^\mu A_\mu (x)=0\;,
\end{equation}
to express derivatives of the gauge potential at the point $\bar{x}$ in terms of derivatives of the field strength tensor
\begin{equation}\label{FS2}
 A_\mu(\bar{x}+t)=\frac12 t^\nu F_{\nu\mu}(\bar{x})+\dots\;,
\end{equation}
where the higher-derivative terms hidden inside the dots vanish since we focus on the constant electromagnetic background case. 
Then, the path integral becomes Gaussian, and it simplifies to
\begin{equation}
\begin{split}\label{path integral 2}
 \Gamma[A]& = \int\diff^4 \bar{x} \int_{0}^{\infty} \frac{\diff T}{T} \frac{\mathrm{e}^{-m^2 T}}{(4\pi T)^2} \int_{0}^{2\pi} \frac{\diff \theta}{2\pi} \, \mathrm{e}^{i\left(\frac{3}{2}+ s\right) \theta} \;4 \sin^2\left(\frac{\theta}{2}\right)\int_{\mathrm{DBC}} Dt \, \mathrm{e}^{-S_{t}[X,\hat{G};A]} \\ 
 &\phantom{=\;} \int_{\mathrm{PBC}} D\alpha D\balpha \, \mathrm{e}^{-S_{\alpha}[X,\hat{G};A]} \int_{\mathrm{PBC}} D\beta D\bbeta \, \mathrm{e}^{-S_\beta[X,\hat{G}]}\;,
\end{split}
\end{equation} 
where we factored out the normalization of the free particle path integral, and where we defined
\begin{align}
 S_t[X,\hat{G};A] &= \int \diff\tau \, \frac{1}{2} t^\mu \Delta_{\mu\nu}^{(t)} t^\nu\;, \quad \text{with} \quad \Delta_{\mu\nu}^{(t)}=- \frac{1}{2T} \delta_{\mu \nu}\frac{\diff^2}{\diff\tau^2} - i q F_{\mu \nu}\frac{\diff}{\diff\tau}\;,\\
 S_\alpha[X,\hat{G};A] &= \int \diff\tau \, \alpha^\mu \Delta_{\mu\nu}^{(\alpha)}\balpha^\nu\;, \quad \hspace{-0.05cm}\text{with} \quad \Delta_{\mu\nu}^{(\alpha)}=\delta_{\mu\nu}\left(\frac{\diff}{\diff\tau} + i \theta \right) + 2i q T F_{\mu \nu}\;,\\
 S_\beta[X,\hat{G}] &= \int \diff\tau \beta \Delta^{(\beta)}\bbeta\;, \quad \hspace{0.45cm}\text{with} \quad \Delta^{(\beta)} =\frac{\diff}{\diff\tau} + i \theta\;,
\end{align}
in order to highlight the three differential operators whose functional determinants have to be computed as a result of the path integration over the variables $X(\tau)$. We apply the Gel’fand--Yaglom theorem \cite{Gelfand:1959nq} to compute the first one, leaving details and conventions in Appendix \ref{appA}, while the remaining two can be directly inferred from the previous result \eqref{det}. Using the required  boundary conditions, 
as  indicated in Eq.~\eqref{path integral 2},  we get
\begin{align}
 \mathrm{Det}\left(\Delta_{\mu\nu}^{(t)} \right)&= {\rm det} \left(\frac{\sin(q T F_{\mu \nu})}{q T F_{\mu \nu}}\right) \;, \label{Det1}\\
 \mathrm{Det}\left( \Delta_{\mu\nu}^{(\alpha)} \right)&= {\rm det}\left[2i \sin\left(\frac{\theta}{2}\delta_{\mu \nu} + q T F_{\mu \nu} \right)\right]\;, \label{Det3} \\
 \mathrm{Det}\left(\Delta_{\mu\nu}^{(\beta)} \right)&=2i \sin\left(\frac{\theta}{2}\right)\label{Det2}\;,
\end{align}
having already extracted the zero modes for the $x$-coordinates. Our final expression is 
\begin{equation}
 \Gamma[A] = \int \diff^4 \bar{x} \int_{0}^{\infty} \frac{\diff T}{T}\frac{ \mathrm{e}^{-m^2 T}}{(4 \pi T)^2} \,{\rm det}^{-\nicefrac{1}{2}} \left(\frac{\sin(q T F_{\mu \nu})}{q T F_{\mu \nu}}\right)\, I_s(T, A)\;,
\end{equation}
where all that is left to do is to perform the modular integration in $\theta$ for a given value of spin $s$:
\begin{equation}\label{I_s}
I_s(T, A) = \int_{0}^{2\pi} \frac{\diff\theta}{ 2\pi i} \, \mathrm{e}^{i\left(\frac{3}{2}+ s\right)\theta} \, 2\sin\left(\frac{\theta}{2}\right) \, {\rm det}^{-1} \left[2i \sin\left(\frac{\theta}{2}\delta_{\mu \nu} + q T F_{\mu \nu} \right)\right] \;.
\end{equation}
It is convenient to recast the determinants above by diagonalizing the (Euclidean) field strength tensor,\footnote{Explicitly: $F_{4 i} = -i E_i, \ F_{ij} = \epsilon_{ijk} B_k, \ (i,j = 1, 2, 3)$.} given that its eigenvalues are
\begin{gather}\label{diag1}
 \lambda_1 = K_- \;, \quad \lambda_2 = i K_+ \;, \quad \lambda_3 = - K_- \;, \quad \lambda_4 = -i K_+ \;,
\end{gather}
having defined $K_{\pm} = \sqrt{\sqrt{\mathcal{F} ^2 + \mathcal{G}^2} \pm \mathcal{F}}$ in terms of the Maxwell invariants
\begin{gather}\label{diag2}
 \mathcal{F} = \frac{1}{4} F_{\mu \nu} F^{\mu \nu} = \frac{\vec{B}^2 - \vec{E}^2 }{2}\;,  
  \quad \mathcal{G} = -\frac{i}{8} \varepsilon_{\mu\nu\rho\sigma}F^{\rho\sigma} F^{\mu \nu} = \vec{E} \cdot \vec{B} \;.
\end{gather}
The modular integration in the Wilson variable $w=\mathrm{e}^{-i\phi}$ is then
\begin{equation}
 I_s(T, A) = \oint \frac{\diff w}{2 \pi i} 
 \frac{1}{w^{s +1} }
 \frac{w-1 }{\left(1 + w^2-2 w \mathcal{K}_+\right) \left(1 + w^2-2 w \mathcal{K}_-\right)}\;,
\end{equation} 
where $\mathcal{K_{+}}= \cosh( 2 q T K_{+})$ and $\mathcal{K_{-}}= \cos( 2 q T K_{-})$.

We now have all the ingredients to investigate the effective action $\Gamma[A] = \int \diff^4 \bar{x}\, \mathcal{L}[A]$
for spin $s=0,1$.
 In particular:
\begin{enumerate}[label=(\roman*)]
 \item The scalar case $s=0$, which corresponds to \acr{sQED}, comes from the simple pole at $w=0$
 \begin{align}
I_0(T, A) &= \mathrm{Res} \left[ \frac{w - 1}{w \left(1 + w^2-2 w \mathcal{K}_+\right) \left(1 + w^2-2 w \mathcal{K}_-\right)}\right]_{w=0} =  -1\;,
 \end{align}
 hence it correctly reproduces the celebrated Weisskopf Lagrangian \cite{Weisskopf:1936hya}
 \begin{equation}
 \mathcal{L}_{s=0}[A] =- \int_{0}^{\infty} \frac{\diff T}{T}\frac{ \mathrm{e}^{-m^2 T}}{(4 \pi T)^2} \, \frac{q^2 T^2 K_- K_+ }{\sinh(qTK_+)\sin(qTK_-) }\;.
 \end{equation}
 
 \item The massive spin 1 case instead arises form the double pole at $w=0$
 \begin{align}
 I_1(T, A) &= \mathrm{Res} \left[ \frac{w - 1}{w^{2} \left(1 + w^2-2 w \mathcal{K}_+\right) \left(1 + w^2-2 w \mathcal{K}_-\right)}\right]_{w=0} = 1 - 2(\mathcal{K}_+ + \mathcal{K}_-)\;,
 \end{align}
 leading to
 \begin{equation}\label{effect action}
 \mathcal{L}_{s=1}[A] =\int_{0}^{\infty} \frac{\diff T}{T}\frac{ \mathrm{e}^{-m^2 T}}{(4 \pi T)^2} \, \frac{q^2 T^2 K_- K_+ }{\sinh(qTK_+)\sin(qTK_-) } \, \left[1- 2\cosh( 2 q T K_+) -2 \cos( 2 q T K_-) \right]\;.
 \end{equation}
  \end{enumerate}
This last expression corresponds to the Euler--Heisenberg effective Lagrangian for a massive charged vector boson in a constant electromagnetic background. It was originally derived in 1965 by Vanyashin and Terent’ev, starting from a quantum field theory of vector electrodynamics \cite{Vanyashin:1965ple}. 
 In contrast, our derivation employs a self-consistent first-quantized approach, which offers a more direct and transparent computation than the conventional second-quantized formalism. This constitutes the main result we set out to obtain using the worldline method.
  
 Our approach offers a natural framework for exploring possible extensions. For instance, one could interpret our final expression as the result of a locally constant field approximation \cite{Fedotov:2022ely} and investigate corrections by systematically including higher-order terms in \eqref{FS2}. This would likely involve following a procedure similar to that of Ref.~\cite{Fecit:2025kqb} for performing perturbative corrections from the worldline, ultimately leading to the determination of the generalized heat kernel coefficients computed in Refs.~\cite{Franchino-Vinas:2023wea, Franchino-Vinas:2024wof}. We leave this analysis to future work.
 
 As a final note, let us report the perturbative expression given by an expansion in the particle's electric charge $q$ 
 \begin{equation}
 \mathcal{L}_{s=1}[A] =\int_{0}^{\infty} \frac{\diff T}{T}\frac{ \mathrm{e}^{-m^2 T}}{(4 \pi T)^2} \, 
 \left(-3 + \frac{7}{4} q^2 T^2 \, \mathrm{tr}[F_{\mu \nu}^2] + \frac{5}{32} q^4 T^4 \,\mathrm{tr}^2[F_{\mu \nu}^2] - \frac{27}{40} q^4 T^4 \, \mathrm{tr}[F_{\mu \nu}^4] + \mathcal{O}(q^6)\right)\;.
 \end{equation}
 The first two terms give divergent contributions, the first one being an infinite vacuum energy, while the second one corresponds to the one-loop divergence in the photon self-energy, and they should be renormalized away. On the other hand, the last two terms are finite and give rise to the quartic interaction’s contributions once integrated in the proper time. Thus, the leading terms of the renormalized effective (Euclidean) Lagrangian, with the tree-level Maxwell term included,
  are expressed as 
\begin{equation}
 \mathcal{L}^{\mathrm{ren}}_{s=1}[A] = \frac{1}{4} F_{\mu \nu} F^{\mu \nu} + \frac{q^4}{16 \pi^2 m^4} \left(\frac{5}{32} (F_{\mu \nu} F^{\nu \mu})^2 -\frac{27}{40} F^{\mu \nu} F_{\nu \rho} F^{\rho \sigma} F_{\sigma \mu}\right) +\cdots
\end{equation}
which shows the leading vertices for the scattering of light by light.

An overall minus sign arises upon continuation back to Minkowski spacetime.
Inserting this sign, the Lagrangian in Minkowski spacetime can be written in the more explicit form
\begin{equation}\begin{aligned}
 \mathcal{L}^{\mathrm{ren}}_{s=1}[A] &= - \frac{1}{4} F_{\mu \nu} F^{\mu \nu} + \frac{q^4}{16 \pi^2 m^4} \left(-\frac{5}{32} (F_{\mu \nu} F^{\nu \mu})^2 +\frac{27}{40} F^{\mu \nu} F_{\nu \rho} F^{\rho \sigma} F_{\sigma \mu}\right)+ \cdots
 \cr 
 &= \frac12 (\vec{E}^2 - \vec{B}^2)
 + \frac{\alpha^2}{40 m^4} \left ( 29 (\vec{E}^2 - \vec{B}^2)^2 + 108 (\vec{E}\cdot \vec{B})^2 \right ) 
 + \cdots
 \end{aligned}\end{equation}
where, for ease of comparison with the literature, we have introduced the fine-structure constant $\alpha =\frac{q^2}{4\pi}$ in natural units, and used the relations
\begin{equation}
F_{\mu \nu} F^{\mu \nu}= 2 (\vec{B}^2 - \vec{E}^2)\;, \quad F^{\mu \nu} F_{\nu \rho} F^{\rho \sigma} F_{\sigma \mu}= 2 (\vec{E}^2 - \vec{B}^2)^2 + 4 (\vec{E}\cdot \vec{B})^2\;,
\end{equation}
to obtain the second line. 

It may be interesting to compare this result with the more widely known results for the
spin-0 and spin-$\frac{1}{2}$ cases, which we include here for convenience:
\begin{equation}\begin{aligned}
 \mathcal{L}^{\mathrm{ren}}_{s=0}[A] 
&= -\frac{1}{4} F_{\mu \nu} F^{\mu \nu} + \frac{q^4}{16 \pi^2 m^4} \left( \frac{1}{288} (F_{\mu \nu} F^{\nu \mu})^2 +\frac{1}{360} F^{\mu \nu} F_{\nu \rho} F^{\rho \sigma} F_{\sigma \mu}\right) 
 + \cdots
\cr
&= \frac12 (\vec{E}^2 - \vec{B}^2)
+ \frac{\alpha^2}{360 m^4} \left ( 7 (\vec{E}^2 - \vec{B}^2)^2 + 4 (\vec{E}\cdot \vec{B})^2 \right ) + \cdots
\end{aligned}\end{equation}
and
\begin{equation}\begin{aligned}
  \mathcal{L}^{\mathrm{ren}}_{s=\frac12}[A] 
 &= - \frac{1}{4} F_{\mu \nu} F^{\mu \nu} + \frac{q^4}{16 \pi^2 m^4} \left(-\frac{1}{32} (F_{\mu \nu} F^{\nu \mu})^2 +\frac{7}{90} F^{\mu \nu} F_{\nu \rho} F^{\rho \sigma} F_{\sigma \mu}\right) + \cdots
\cr
&= \frac12 (\vec{E}^2 - \vec{B}^2)+ \frac{2 \alpha^2}{45 m^4} \left ( (\vec{E}^2 - \vec{B}^2)^2 + 7 (\vec{E}\cdot \vec{B})^2 \right ) + \cdots \;.
\end{aligned}\end{equation}
They arise from the Weisskopf and Euler–Heisenberg effective Lagrangians, respectively.
 
\subsection{Production of massive spin-1 particle pairs}\label{sec3.2}
In this section, we compute the rate for the Schwinger pair production of massive charged spin-1 particles in a constant external electric field $\vec{E}$. For easier reference, we shall recall that if the effective action in the presence of a classical background field assumes a non-vanishing imaginary contribution, this has the physical interpretation of an instability of the quantum field theory vacuum: quantitatively, the Minkowskian effective action is related to the vacuum persistence probability by 
\begin{align}
 \vert\langle 0_{\rm out} \vert 0_{\rm in}\rangle\vert^2 =\mathrm{e}^{- 2\operatorname{Im} \Gamma_\mathrm{M}}\;.
\end{align}
The effective action \eqref{effect action} with   $K_+ =0$ and $K_- =E$, with  $E$ being the modulus of the electric field,
reduces to
\begin{equation}\label{effect action E B}
 \Gamma[A] =\int \diff^4 \bar{x}\int_{0}^{\infty} \frac{\diff T}{T}\frac{ \mathrm{e}^{-m^2 T}}{(4 \pi T)^2} \, \frac{q T E }{\sin(qTE) }\, [-1 - 2\cos( 2 q T E) ]\;.
\end{equation}
Apparently, it is a real quantity, but the presence of poles in the $T$-integral signals that this is not the case.
To extract its imaginary part, we go back to Minkowski spacetime  via a Wick rotation, using $T \rightarrow iT$, 
$\mathcal{L} \rightarrow -\mathcal{L}$, 
to obtain the Minkowskian effective Lagrangian 
\begin{equation}
 \mathcal{L}[A]= \int_{0}^{\infty} \frac{\diff T}{T}\frac{ \mathrm{e}^{- i m^2 T}}{(4 \pi T)^2} \left(- 3 \frac{ i q T E }{\sin( i q T E) } + 4  (i q T E) \sin( i q T E) \right)\;. 
\end{equation}
For certain values of proper time, the integral develops poles in the $T$-plane, which in turn produce an imaginary part of the Minkowskian effective action. In fact, from
\begin{equation}
  \operatorname{Im}\mathcal{L}[A] = \frac{\mathcal{L}[A] - \mathcal{L}^*[A]}{2i} = \frac{1}{2i} \int_{- \infty}^{+\infty} \frac{\diff T}{T}\frac{ \mathrm{e}^{- i m^2 T}}{(4 \pi T)^2} \left(- 3 \frac{ i q T E }{\sin( i q T E) } + 4  (i q T E) \sin( i q T E) \right)\; , 
\end{equation}
one finds that the contour must be closed in the lower half-plane, and the imaginary part is determined by the residues 
at the poles of the first integrand function, located at\footnote{They correspond to the zero modes of the differential operator $\Delta_{\mu\nu}^{(t)}$ in Minkowski spacetime except for the value $n=0$, which indicates a \acr{UV} divergence as discussed at the end of the previous section.}
\begin{equation}
 T = -i \frac{\pi n}{q E}\;, \quad 0<n \in \mathbb{N}\; .
\end{equation} 
The final result is
\begin{equation}
 \operatorname{Im}\mathcal{L}[A] = \frac{3}{16 \pi^3}(q E)^2 \, \sum\limits_{n=1}^{\infty} (-1)^{n+1}\frac{\mathrm{e}^{-\frac{m^2\pi n}{q E}}}{n^2} \,\;. 
\end{equation}

In conclusion, the probability for massive spin-1 particle-antiparticle pair production in the presence of a constant electric field per unit of volume and time $\mathcal{P}:=P/\Delta V \Delta \mathcal{T}$ can be written as
\begin{equation}
 \mathcal{P}_{\mathrm{pair}} \approx -\frac{3}{8 \pi^3}(q E)^2\, \operatorname{Li}_{2}\left(-\mathrm{e}^{-\frac{m^2 \pi }{q E}}\right)\;, 
 \end{equation}
where $\operatorname{Li}_2 (\cdot)$ is the polylogarithm of order ${2}$.
As already noted in Ref.~\cite{Vanyashin:1965ple}, this probability corresponds to three times the probability of the production of pairs of scalar particles with mass $m$.

\section{Final remarks} \label{sec6}
In this chapter, we have shown how to construct worldline actions for a massive, charged, spin-1 particle. These actions carry the necessary gauge symmetries to ensure a unitary description at the quantum level. A crucial feature is the gauging of the oscillator number operator, appropriately shifted by a Chern--Simons coupling, which enforces the projection onto the spin-1 sector.
A BRST analysis reveals that the spin-1 sector can couple to an external electromagnetic field in both formulations, provided the field satisfies the vacuum Maxwell equations. This condition suffices for applications such as computing scattering amplitudes involving external photons, whose asymptotic states obey these equations. We employed this coupling to analyze the system's behavior in a constant \acr{EM} field.
The spin-0 sector admits coupling to the electromagnetic field without requiring any conditions on the field. In contrast, other sectors corresponding to higher-spin particles do not admit \acr{EM} couplings within our worldline framework. By studying the path integral of the spin-1 particle action on the circle, we obtained the one-loop effective action. For a constant electromagnetic background, we explicitly computed the corresponding effective Lagrangian of the Euler-Heisenberg type. From this, we derived the pair production rate for spin-1 particle-antiparticle pairs in a constant electric field. Our results fully agree with the quantum field theory result originally presented in Ref.~\cite{Vanyashin:1965ple}, which we recover here using a purely worldline approach, independent of any second-quantized formalism.
Additional, more recent works discussing the effective Lagrangian induced by spin-1 particles and related matters, include 
\cite{Batalin:1976uv, Skalozub:1976tr, Dittrich:1983ej, Reuter:1984zk, Blau:1988iz,         
Jikia:1993tc, Preucil:2017wen, Henriksson:2021ymi, Alviani:2024sxx}.
Alternative worldline formulations have been explored in the literature, notably in Ref.~\cite{Reuter:1996zm}, where worldline techniques are employed to express the heat kernel representation of the spin-1 \acr{QFT} effective action as worldline path integrals, allowing for a more efficient computation, including the evaluation of worldline functional determinants.
Our results add to the effort of constructing wordline methods without using key input from \acr{QFT}, 
following the path employed in the study of Yang--Mills theory \cite{Dai:2008bh, Bastianelli:2025xx}, gravity \cite{Bonezzi:2018box, Bonezzi:2020jjq, Bastianelli:2019xhi, Bastianelli:2022pqq, Bastianelli:2023oca, Fecit:2023kah, Fecit:2024jcv}, and scalar theories \cite{Bonezzi:2025iza}.

\addcontentsline{toc}{chapter}{Appendix}
\section*{Appendix}
\begin{subappendices}

\section{Functional determinant from the Gel'fand--Yaglom theorem}\label{appA}
In this appendix, we detail the computation of the functional determinant \eqref{Det1} associated with the differential operator
\begin{align}\label{op}
 \Delta_{\mu\nu}^{(t)}(\tau,\tau')&=\left [
 - \frac{1}{2T} \delta_{\mu \nu}\frac{\diff^2}{\diff\tau^2}  - i q F_{\mu \nu}\frac{\diff}{\diff\tau}\right ]\delta(\tau-\tau')\;,
\end{align}
using (generalizations of) the Gel'fand--Yaglom theorem \cite{Gelfand:1959nq}. For convenience, we restate here the main theorem, previously given in Appendix~\ref{chap:first:appB}, in order to adapt the notation to the present chapter. 

In its simplest formulation, the \acr{GY} theorem states that given a one-dimensional second-order differential operator defined on the interval $z \in [0,1]$ and the associated eigenvalue problem with vanishing Dirichlet boundary conditions
\begin{equation} \label{Laplace}
\left[ -\frac{{\rm d}^2}{{\rm d} z^2}+V(z) \right] \psi(z)=\lambda \, \psi(z)\ , \quad \text{with} \quad \psi(0)=\psi(1)=0\;,
\end{equation}
if we can solve the initial value problem
\begin{equation}
\left[ -\frac{{\rm d}^2}{{\rm d}z^2}+V(z) \right] \Phi(z)=0\ , \quad \text{with} \quad \Phi(0)=0\ , \quad \dot{\Phi}(0)=1\;,
\end{equation}
then the boundary value of the solution determines the functional determinant of the differential operator
\begin{equation}
\mathrm{Det} \left[ -\frac{{\rm d}^2}{{\rm d}z^2}+V(z) \right] \propto \Phi(1)\;.
\end{equation}
The precise definition of the functional determinant actually involves the ratio of two determinants, and should be understood in this sense \cite{Dunne:2007rt}; in this chapter, we compute the determinants relative to the corresponding determinant for the free operators upon extracting the zero modes of the latter. For our purposes, we need the generalization of the theorem for higher-dimensional operators: we report the main formulae, while referring the reader to the references \cite{Kirsten:2003py, Kirsten:2004qv, Fecit:2025kqb} for more details. 

To compute the functional determinant of $\Delta_{\mu\nu}^{(t)}$ acting on the quantum fluctuations $t^\mu(\tau)$ with Dirichlet boundary conditions, one needs to solve the associated homogeneous differential equation for the solution with initial conditions
\begin{equation}\label{initDBC}
\varphi_\mu^{(\rho)}(0)=0\;, \quad \dot{\varphi}_\mu^{(\rho)}(0)=\delta^\rho_\mu\;,
\end{equation}
and then 
\begin{equation}
 \mathrm{Det}\left(\Delta_{\mu\nu}^{(t)}\right)=\mathrm{det}\left[\varphi_\mu^{(\rho)}(1) \right]\;.
\end{equation} 
Using a simple trick, we can recast the operator \eqref{op} into a form suitable for a direct application of the theorem. In particular, it is not hard to show that\footnote{We thank Markus~B.~Fr\"ob for discussions on this point.}
\begin{equation}
 \Delta_{\mu\nu}^{(t)}(\tau,\tau') \propto \left[ \mathrm{e}^{-iq F T \tau} \left(-\frac12\frac{\diff^2}{\diff\tau^2}\delta(\tau-\tau') -\frac12 q^2 F^2\delta(\tau-\tau') \right) \mathrm{e}^{i q FT \tau} \right]_{\mu\nu}\;,
\end{equation}
from which one sees that its functional determinant can be equivalently computed by calculating the determinant of the operator inside round brackets.\footnote{The prefactors hidden in “$\propto$” are taken care of by normalizing with the corresponding free operator.} Solving for the associated homogeneous equation with initial conditions \eqref{initDBC} we get
\begin{equation}
 \varphi_\mu^{(\rho)}(z)=\left[ \frac{\sin(q FTz)}{q FT} \right]^\rho_\mu\;,
\end{equation}
which confirms Eq.~\eqref{Det1}.
\end{subappendices}

\newpage
\thispagestyle{empty}
\mbox{}
\newpage

\part{Perturbative Physics}\label{partII}
\section*{Prelude: the heat kernel technique and the Fierz--Pauli theory} \label{chap:introII}
\setcounter{equation}{0}
\renewcommand{\theequation}{II.\arabic{equation}}
In the second part of this thesis, we explore the application of the Worldline Formalism to perturbative physics, illustrating its use through several examples. The first example concerns the computation of color-ordered tree-level amplitudes in Yang--Mills theory. Before delving into this, we shall provide a brief overview of the worldline computational framework. The subsequent examples focus instead on the perturbative expansion of the heat kernel. To this end, we introduce some basic concepts and outline the main ideas underlying this technique, thereby setting the stage for its applications in the following sections. These applications concern two gravitational theories. The first one, discussed in Chapter~\ref{chap:sixth}, is the well-known Einstein--Hilbert theory, which we shall not anticipate further here. The second one, treated in Chapter~\ref{chap:fifth}, is a modified theory of gravity, known as “massive gravity”. In this thesis, however, we shall restrict our attention to its linearized formulation, which may be viewed simply as a theory describing a massive spin-2 “meson”, rather than an ambitious attempt to give mass to the graviton. We will review a few of its key features before turning to the original material.

\paragraph{Yang--Mills from the worldline}
Chapter~\ref{chap:third} is devoted to the computation of color-ordered tree-level amplitudes in Yang--Mills theory. We adopt a “bottom-up” approach, starting from the bosonic spinning particle and constructing its path integral, this time formulated on the infinite line rather than on the circle. The choice of an infinite line geometry automatically implements the LSZ reduction \cite{Laenen:2008gt,Bonocore:2020xuj,Mogull:2020sak,Bonezzi:2025iza}, thereby allowing one to extract on-shell amplitudes directly from the worldline path integral. The infinite line constitutes the main worldline that connects two external points and can be chosen arbitrarily. It describes asymptotically free vector particles. Vertices linking the worldline to the remaining external (asymptotic on-shell) particle states emerge in part from the perturbative expansion of the particle action, leading to insertions of vertex operators. Additionally, insertion of additional composite vertex operators is needed to obtain the complete BRST-invariant color-ordered amplitudes. Although this construction may appear elementary, the approach developed in Chapter~\ref{chap:third} provides a robust foundation for deriving scattering amplitudes purely within the worldline framework.\footnote{The overall framework is conceptually similar to the one proposed in \cite{Dai:2008bh}, where the fermionic particle was used.} Furthermore, we further explore the role played by the BRST symmetry. First, it is employed to analyze the general structure and properties of both integrated and unintegrated vertex operators contributing to the construction of the color-ordered amplitudes. BRST symmetry also serves as the guiding principle for identifying whether additional vertex operators, which we refer to as “pinch operators”, must be inserted to restore BRST invariance when it is broken in perturbatively constructed tree-level amplitudes. Secondly, we show how the decoupling of BRST-exact operators in correlation functions corresponds to the target-space Ward identities, similarly to what occurs in string theory. 

\paragraph{The heat kernel method}
The heat kernel has already been briefly introduced in \hyperref[partI]{Part~I} of the thesis. However, since it plays a central role in the forthcoming chapters, we take the opportunity here to review its main features and its use in the computation of one-loop divergences, which is a procedure commonly referred to as the “heat kernel technique”. This well-established method in mathematical physics provides a perturbative framework for studying second-order differential operators on Riemannian manifolds \cite{Avramidi2015}, and it has been applied extensively in quantum field theory, with a particular interest in quantum gravity starting from the work of DeWitt, \cite{DeWitt:1964mxt, DeWitt:1984sjp, DeWitt:2003pm}, as for example in Refs.~\cite{Barvinsky:1985an, Fradkin:1985am, Avramidi:2000bm, Vassilevich:2003xt}. The starting point is the same as that of the “top-down” approach to the Worldline Formalism: the one-loop effective action expressed in terms of a functional determinant as
\begin{equation}\label{II-Gamma1}
\Gamma[\Phi]=\frac{1}{2}\log{\text{sDet}{\mathcal{Q}}}
= \frac{1}{2}{\text{sTr} \log {\mathcal{Q}}}\;,
\end{equation}
where “$\text{sDet}$” is the Berezin functional superdeterminant and “sTr” the functional supertrace (see for example Ref.~\cite{Avramidi:2000bm}). We write this expression in its most general form, without specifying whether the underlying field theory is bosonic or fermionic. In what follows, we assume that it is defined on a $D$-dimensional Euclidean manifold. The quantum fluctuations operator $\mathcal{Q}$ is an elliptic second-order differential operator, which can take the general form
\begin{equation}\label{II-DiffOp}
\mathcal{Q}= -\nabla_{\scriptscriptstyle \! \! ({\cal A})}^2 -V\;.
\end{equation}
The operator $\mathcal{Q}$ is here taken to act on a scalar field $\phi$ which carries a representation of the gauge field ${\cal A}_\mu$, contained in the connection that defines the full covariant derivative $\nabla^{\scriptscriptstyle ({\cal A})}_\mu = \nabla_\mu + {\cal A}_\mu$. It has an associated gauge field strength $\Omega_{\mu\nu}$, defined by the commutator of the covariant derivatives on the scalar field $\phi$, i.e.
\begin{equation}
    [\nabla^{\scriptscriptstyle ({\cal A})}_\mu, \nabla^{\scriptscriptstyle ({\cal A})}_\nu]\phi = \Omega_{\mu\nu}\,\phi\;.    
\end{equation}
The Laplacian is defined as usual by $\nabla_{\scriptscriptstyle \! \! ({\cal A})}^2 =g^{\mu\nu} \nabla^{\scriptscriptstyle ({\cal A})}_\mu\nabla^{\scriptscriptstyle ({\cal A})}_\nu$ and we consider a potential $V$ which is matrix valued, just like the gauge field ${\cal A}_\mu$. Thus, the elliptic second-order differential operator $\mathcal{Q}$ generally depends on the metric $g_{\mu\nu}$ and on the matrix-valued potentials ${\cal A}_\mu$ and $V$. 

In the usual Schwinger proper-time parametrization, \eqref{II-Gamma1} becomes
\begin{equation}\label{II-Gamma1_2}
\Gamma = \frac{1}{2}{\text{sTr} \log {\mathcal{Q}}} =-\frac{1}{2}\int_0^\infty \frac{\diff{T}}{T}\ \text{sTr} \exp{(-T\mathcal{Q})}\ .
\end{equation}
The operator $\exp{(-T \mathcal{Q})}$ is known as the heat semigroup of the operator $\mathcal{Q}$, and its integral kernel is the heat kernel $K(x,x';T)$. It is usually needed at coinciding points $x'\to x$ for most of the physical applications, as the quantity of interest is typically the trace of the operator. Consequently, we will focus on this case in the following. Plugging the explicit form of $\mathcal{Q}$ given by \eqref{II-DiffOp} inside \eqref{II-Gamma1_2},\footnote{It is customary to introduce an \acr{IR} regulator by shifting $\mathcal{Q} \to \mathcal{Q} + m^2$, corresponding to the addition of a mass term for otherwise massless fields.} we have
\begin{equation}\label{II-Gamma1bis}
\Gamma=-\frac{1}{2}\int_0^\infty \frac{\diff{T}}{T}\ \exp{(-T m^2)}\int \diff{^Dx}\ \sqrt{g}\ \text{str} \, {K(x,x;T)}\ .
\end{equation}
Note that the leftover supertrace “str” is now to be performed only over the remaining discrete indices carried by the representation of the field $\phi$ on which $\mathcal{Q}$ acts upon.

For small Euclidean proper time $T\to 0^+$, the heat kernel, at coinciding points, admits the asymptotic expansion \cite{DeWitt:1964mxt, DeWitt:1984sjp, DeWitt:2003pm}
\begin{equation}\label{II-HDMS}
K(x,x;T)\sim (4\pi T)^{-\nicefrac{D}{2}}\sum_{j=0}^{\infty} T^j a_j(x)\ ,
\end{equation}
where the heat kernel coefficients $a_j(x)$, also known as Seeley-DeWitt coefficients,\footnote{These coefficients appear in the literature under various names. In this thesis, we use the terms “heat kernel coefficients”, “Seeley-DeWitt coefficients”, and “Schwinger-DeWitt coefficients” interchangeably.} can be expressed in terms of the metric and gauge invariants of the manifold. 
For an intuitive derivation of the expansion, see, for instance, \href{https://www-th.bo.infn.it/people/bastianelli/4-Heat-Kernel-FT1-17-18.pdf}{these} lecture notes by Fiorenzo~Bastianelli \cite{BastianelliHeatkernel}, while for pedagogical treatments refer to \cite{Vassilevich:2003xt, Avramidi2015, Ori:2023, Ferrero:2023xsf}.
With the aid of \eqref{II-HDMS}, we are able to identify the divergences of the one-loop effective action \eqref{II-Gamma1bis} precisely with a subset of the Seeley-DeWitt coefficients, namely the ones that produce a divergence in the proper time integration at $T\to 0$ in
\begin{equation}\label{II-Gamma1ter}
\Gamma=-\frac{1}{2}\int_0^\infty \frac{\diff{T}}{T}\ \mathrm{e}^{-m^2T}\int \frac{\diff{^Dx}\,\sqrt{g}}{(4\pi T)^{\frac{D}{2}}}\ \operatorname{str}{\sum_{j=0}^{\infty} T^j a_j(x)}\ .
\end{equation}
For example, at $D=4$ the diverging terms are associated with the coefficients $a_0(x), a_1(x), a_2(x)$, while at $D=6$ also $a_3(x)$ leads to an additional divergence (the logarithmic divergence in that dimension). 

The computation of these coefficients using worldline techniques will be the subject of Chapter~\ref{chap:sixth} and Chapter~\ref{chap:fifth}, except for Section~\ref{chap:sixth:sec3'} where a genuine heat kernel calculation is performed as a cross-check.

Let us conclude with an important remark concerning the expansion \eqref{II-HDMS}. In principle, such an expansion does not exist in the same simple form when $\mathcal{Q}$ is a \emph{non-minimal} operator. In the heat kernel context, a second-order operator is termed non-minimal when the part of the operator with the highest derivatives -- the principal symbol \cite{Vassilevich:2003xt} -- has a non-trivial matrix structure. For instance, an operator of Laplace type
\begin{equation}
 \Delta=-g^{\mu\nu}\nabla_\mu\nabla_\nu+V\ ,
\end{equation}
is a \emph{minimal} operator, as the principal symbol is the metric. A \emph{non-minimal} operator presents a more general principal symbol: for instance, the kinetic operator of the Proca field on curved spacetime has the form \cite{Buchbinder:2017zaa, Ruf:2018vzq}
\begin{equation}
\Delta^{\mu\nu}=(-\nabla^2+m^2) \,g^{\mu\nu}+\nabla^\mu \nabla^\nu +R^{\mu\nu}\ ,
\end{equation}
and the offending term $\nabla^\mu \nabla^\nu$ is responsible for the complicated structure of the leading derivative terms. This situation will arise in our analysis of Linearized Massive Gravity (\acr{LMG}). Several approaches exist for reducing such operators to minimal form, the most prominent being the generalization of the Schwinger-DeWitt algorithm developed by Barvinsky and Vilkovisky \cite{Barvinsky:1985an}. Alternatively, one may adopt a path integral approach, as in \cite{Dilkes:2001av, Duff:2001zz, Duff:2002sm, Farolfi:2025knq}. A recent notable advance in this direction is given in \cite{Sauro:2025sbt}, which provides model-independent expressions for the Seeley-DeWitt coefficients of general non-minimal second-order operators. In this thesis, we aim to bypass these technical complications by employing a suitable worldline model, offering further evidence for the efficiency of first-quantized techniques in contrast to traditional second-quantized methods.

\paragraph{Massive gravity and the Fierz--Pauli theory}
Massive gravity has long attracted considerable interest in theoretical physics as a compelling modification of general relativity, ever since its first formulation by Fierz and Pauli in 1939 \cite{Fierz:1939zz, Fierz:1939ix}. The \acr{FP} theory describes the propagation of a massive spin $2$ field carrying\footnote{Although the FP theory historically arose as the first attempt to modify gravity by introducing a massive mediator of the gravitational interaction, one may disregard its gravitational interpretation and instead view $h_{\mu\nu}(x)$ as describing a massive spin $2$ meson propagating on flat or curved backgrounds, possibly interacting with a background Einstein metric. See \cite{Mazuet:2018ysa} for a related discussion.}
\begin{equation}
 \# {\rm DoF}=\frac{(D+1)(D-2)}{2}
\end{equation}
degrees of freedom in $D$ spacetime dimensions. Its simplest, second-order, formulation can be expressed in terms of an action principle involving a symmetric and traceful rank-2 tensor field $h_{\mu\nu}(x)$ with a flat spacetime background:
\begin{align} \label{FP}
 S_{\mathrm{FP}}[h_{\mu\nu}]&= \frac12 \int \diff^Dx \, \Bigg[ \underbrace{\partial_\rho h_{\mu\nu} \partial^\rho h^{\mu\nu}-\partial_\mu h \partial^\mu h -2 \partial_\mu h_{\nu\rho}\partial^\nu h^{\mu\rho}+2\partial_\mu h^{\mu\nu} \partial_\nu h}_{=: \, \mathcal{L}_{m=0}[h_{\mu\nu}]}+\underbrace{ m^2\left( h_{\mu\nu} h^{\mu\nu}-h^2\right)}_{=: \, \mathcal{L}_{m}[h_{\mu\nu}]} \Bigg]\ ,
\end{align}
with $h \equiv h^\mu_\mu$. Let us review some technical aspects of the theory, which will later be revisited through the lens of the Worldline Formalism, starting from Eq.~\eqref{FP}. The massless action $S_{m=0}[h_{\mu\nu}]:= \tfrac12 \int \diff^Dx \, \mathcal{L}_{m=0}[h_{\mu\nu}] $ is invariant under the gauge symmetry\footnote{In our notation, parentheses and brackets denote symmetrization and antisymmetrization with a conventional factor, e.g.
$(ab) := \tfrac{1}{2}(ab + ba)$ and $[ab] := \tfrac{1}{2}(ab - ba)$.}
\begin{equation}
 \delta h_{\mu\nu}=\partial_\mu \xi_\nu+\partial_\nu \xi_\mu=2\partial_{(\mu}\xi_{\nu)}\ ,
\end{equation}
where $\xi^\mu(x)$ is a spacetime dependent gauge parameter. The mass term $\mathcal{L}_{m}[h_{\mu\nu}]$ breaks the gauge symmetry, as usual in any gauge field theory. It is actually not the most general mass term that can be added to the invariant action $S_{m=0}[h_{\mu\nu}]$ respecting Lorentz invariance and power counting \cite{Gambuti:2021meo}. However, the choice of the relative $-1$ coefficient between the $h_{\mu\nu} h^{\mu\nu}$ and $h^2$ contractions, known as \emph{Fierz--Pauli tuning}, is the only combination ensuring the propagation of the correct number of degrees of freedom; any other choice would inevitably lead to the presence of a ghost mode with negative energy. The equations of motion can be expressed in terms of the so-called \emph{Fierz--Pauli system}:
\begin{align}
 \left(\partial^2-m^2 \right)h_{\mu\nu}(x)=0\ , \label{KG} \\
 \partial^\mu h_{\mu\nu}(x)=0\ , \label{trans} \\
 h(x)=0\ , \label{trace}
\end{align}
which shows that the tensor field corresponds to a unitary irreducible representation of the Poincar\'e group with mass $m$ and spin $2$. Violating the \acr{FP} tuning would lead to the loss of \eqref{trace}, and a ghost-like scalar mode inside $h_{\mu\nu}$ becomes propagating \cite{Hinterbichler:2011tt}. The situation becomes even more intricate on a curved background. Referring to Ref.~\cite{Farolfi:2025knq} for a detailed analysis, let us mention that the bottom line is that \acr{FP} action can be consistently extended only on Einstein spacetimes \cite{Hinterbichler:2011tt, deRham:2014zqa, Aragone:1971kh, Deser:2006sq} as\footnote{We denote the metric tensor in bold in order to hide the spacetime indices, $\mathbf{g}:=g_{\mu\nu}(x)$. Moreover, $g \equiv \det(g_{\mu\nu})$ and $R(\mathbf{g})$ is the Ricci scalar.}
\begin{equation} \label{FPcurved}
S_{\mathrm{FP}}[h_{\mu\nu};\mathbf{g}]= \frac12 \int \diff^Dx \sqrt{g} \Big[ \mathcal{L}_{m=0}[h_{\mu\nu};\mathbf{g}]+\mathcal{L}_{m}[h_{\mu\nu};\mathbf{g}] \Big]\;,
\end{equation}
where the mass term remains unchanged, while the massless Lagrangian now reads
\begin{equation} 
 \mathcal{L}_{m=0}[h_{\mu\nu};\mathbf{g}] = \nabla_\rho h_{\mu\nu} \nabla^\rho h^{\mu\nu}-\nabla_\mu h \nabla^\mu h -2 \nabla_\mu h_{\nu\rho}\nabla^\nu h^{\mu\rho}+2\nabla_\mu h^{\mu\nu} \nabla_\nu h-\frac{2R}{D}\left( h_{\mu\nu} h^{\mu\nu}-\frac12 h^2\right)\ 
\end{equation}
and is invariant under the gauge symmetry
\begin{equation} \label{gaugecurvo}
 \delta h_{\mu\nu}=2\nabla_{(\mu}\xi_{\nu)}\ .
\end{equation}
The action can be recast in the following convenient form, assuming that the fields decay quickly enough at infinity when integrating by parts,
\begin{equation}
 S_{\mathrm{FP}}[h_{\mu\nu};\mathbf{g}]= \frac12 \int \diff^Dx \sqrt{g} \, h_{\mu\nu} \, \Delta^{\mu\nu\rho\sigma}_{(\mathrm{FP})} \, h_{\rho\sigma} 
\end{equation}
in terms of the differential operator
\begin{align} \label{Delta2}
 \Delta^{\mu\nu\rho\sigma}_{(\mathrm{FP})}=\left(-\nabla^2 +m^2\right) \left( g^{\mu(\rho}g^{\sigma)\nu} -g^{\mu\nu} g^{\rho\sigma}\right)+2R^{\mu(\rho\sigma)\nu}+\frac{R}{D}g^{\mu\nu}g^{\rho\sigma}-2g^{\rho\sigma}\nabla^{(\mu}\nabla^{\nu)}+2g^{(\mu(\rho}\nabla^{\nu)}\nabla^{\sigma)}\ .
\end{align}
This operator is clearly non-minimal, which complicates its use in heat kernel computations. It will therefore be advantageous to construct a worldline model that bypasses these difficulties, as will be developed in Chapter~\ref{chap:fourth}.

For the sake of brevity, we shall only mention a few additional features of the Fierz--Pauli theory without entering into further details. Among them is the well-known \acr{vDVZ} discontinuity \cite{vanDam:1970vg, Zakharov:1970cc}, namely the different predictions between linear general relativity and the \acr{FP} theory in the massless limit, and the presence of a Boulware--Deser ghost, a problematic extra degree of freedom which inevitably appears at the non-linear level \cite{Boulware:1972yco}. To this date, only a promising non-linear theory of massive gravity free from the Boulware--Deser instability has been formulated, which is known as dRGT theory, named after the work of Claudia~de~Rham, Gregory~Gabadadze, and Andrew~Tolley \cite{deRham:2010ik, deRham:2010kj, deRham:2011rn}. The interested reader may refer to excellent reviews for a comprehensive exploration of this subject \cite{Hinterbichler:2011tt, deRham:2014zqa, Schmidt-May:2015vnx}. 

Finally, we anticipate that massive gravity, like general relativity, does not escape the issue of one-loop divergences. This is unsurprising, since both are non-renormalizable theories: their Lagrangians are built upon the Einstein--Hilbert action, which is known to produce diverging terms at one-loop that cannot be absorbed into the parameters of the action \cite{tHooft:1974toh, VanNieuwenhuizen:1977ca, Critchley:1978kb, Christensen:1979iy}. The quantum aspects of massive gravity, and in particular its one-loop divergences, have been extensively studied through field-theoretical techniques \cite{Dilkes:2001av, Buchbinder:2012wb, deRham:2013qqa, Ferrero:2023xsf}, and will be revisited within the worldline approach in Chapter~\ref{chap:fifth}.
\setcounter{equation}{0}
\numberwithin{equation}{section}

\newpage
\thispagestyle{empty}
\mbox{}
\newpage

\chapter{Tree-level gluon amplitudes}\label{chap:third}

\textit{In this chapter, we begin to show how the Worldline Formalism can be adapted to perform perturbative quantum field theory calculations. We start with the case of tree-level scattering amplitudes, where the method shares many parallels with string theory. Specifically, we compute gluon amplitudes at tree level as worldline correlators of vertex operators in a suitable worldline model. In doing so, we adopt a purely “bottom-up” worldline approach, which may be viewed as complementary to the original framework developed by Bern and Kosower. The model under consideration corresponds to the massless version of the bosonic spinning particle model introduced in the previous chapter, which, as already demonstrated, admits a consistent truncation to describe a pure spin-1 particle. The particle is coupled to a non-abelian background field: interactions are once again incorporated through BRST methods, which here also provide a systematic way to extract the vertex operators as deformations of the BRST charge. BRST invariance thus plays a central role in ensuring the consistency of the tree-level amplitudes. Finally, we discuss connections with similar worldline constructions and comment on the potential relevance of this framework for uncovering the structures underlying the double-copy program in gauge and gravitational theories.}

\paragraph{Conventions} The target space of our worldline model is a $D$-dimensional Minkowski spacetime, and Greek letters will be used to denote Lorentz indices.

\section{The bosonic spinning particle in a Yang--Mills background} \label{chap:third:sec2}
We begin by reviewing some features of the free bosonic spinning particle model. Although the first subsection partly overlaps with material from the previous chapter, we include it here once again in view of the forthcoming construction of the path integral on the open line. In this context, certain subtleties arise that were not addressed in the earlier discussion, and which will play an important role in the present analysis.

\subsection{The free spinning particle} \label{chap:third:sec2.1}
The symplectic term of the free worldline action is
\begin{equation}
S_{\rm symp}=\int_0^1 d\tau\,\Big[p_\mu\dot x^\mu-i\,\bar\alpha_\mu\dot\alpha^\mu\Big]\;. 
\end{equation}
It defines the canonical Poisson brackets for the particle's coordinates and momenta $(x^\mu,p_\mu)$ joined by the bosonic oscillator pair $(\alpha^\mu,\bar\alpha_\mu)$. The elementary Poisson brackets are given by
 \be
 \{x^\mu, p_\nu\} = \delta^\mu_\nu 
 \;, \qquad 
 \{\alpha^\mu, \bar \alpha_\nu\} = i \delta^\mu_\nu \;.
  \ee
The following conserved phase-space functions
\begin{equation}
H = \frac{1}{2}\,p^2 \;, \quad
L = \alpha^\mu p_\mu \;, \quad
\bar L = \bar\alpha^\mu p_\mu\;,
\end{equation}
form a closed algebra under Poisson brackets
\begin{equation} \label{constraint algebra}
\{L,\bar L\}= 2i H\;,
\end{equation}
while other independent brackets vanish. As previously noted, this algebra is first-class and bears a close resemblance to a supersymmetry algebra, except that $L$ and $\bar L$ are bosonic rather than fermionic. 
It corresponds to a contraction of the $sl(2, \mathbb{R})$ algebra \cite{Bengtsson:1986ys,Bouatta:2004kk}.
The charges ($H, L, \bar L)$ generate worldline symmetries: $H$ is the generator of $\tau$-translations, while 
$L$ and $\bar L$  mix the spacetime coordinates 
$x^\mu$ with the oscillator variables $\alpha^\mu$ and $\bar \alpha^\mu$, respectively.  
Their structure is reminiscent of the Virasoro generators $L_0$ and $L_{\pm1}$ 
 of the bosonic open string.

Let us recall that, upon quantization, the states of the theory are interpreted as those corresponding to massless particles, with the spin degrees of freedom carried by the oscillator variables $\alpha^\mu$. 
 To this end, one must impose the mass-shell constraint $p^2=0$ by gauging the Hamiltonian $H$, 
 and eliminate remaining unphysical states by gauging the “Virasoro-like” generators  $L$ and $\bar L$.
This gauging plays a role analogous to the gauging of the Virasoro algebra in string theory, where lightcone oscillator modes must be removed to avoid negative-norm states.

The full worldline action, incorporating these gauge symmetries, is given by
\begin{equation}
S=\int_0^1 d\tau\,\Big[p_\mu\dot x^\mu-i\,\bar\alpha^\mu\dot\alpha_\mu-e\,H-\bar u\,L-u\,\bar L\Big] \;,
\label{act-1}
\end{equation}
where $e(\tau)$ is the einbein enforcing the mass-shell condition $H=0$,
and  $u(\tau)$ and $\bar u(\tau)$
are complex Lagrange multipliers enforcing $\bar L= L=0$.
The action is invariant under local worldline reparametrizations and the bosonic analog 
of supersymmetry, with infinitesimal transformations
\ba
&\delta x^\mu=\epsilon\,p^\mu+\xi\,\bar\alpha^\mu+\bar\xi\,\alpha^\mu\;,  
\hskip .8cm 
\delta p_\mu=0\;,
\\
&\delta\alpha^\mu=i\,\xi\,p^\mu\;,
\hskip 2.9cm 
\delta\bar\alpha^\mu=-i\,\bar\xi\,p^\mu\;,
\\
&\delta e=\dot\epsilon+2i\,u\,\bar\xi-2i\,\bar u\,\xi\;, 
\hskip 1.2cm 
\delta u=\dot\xi\;, 
\hskip 2cm 
\delta\bar u=\dot{\bar\xi}\;,
\label{2.6}
\ea
where $\epsilon(\tau)$ is the local parameter for worldline reparametrizations, while $\xi(\tau)$
and $\bar\xi(\tau)$ are complex gauge parameters for the bosonic symmetry.

The quantum theory corresponding to this action has been analyzed 
in Ref.~\cite{Bonezzi:2024emt} to covariantly identify the physical sector 
of the Hilbert space, using both the Dirac quantization method 
and the more general BRST approach.\footnote{Recall that Chapter~\ref{chap:second} carried out the same analysis but in the case of the massive model.} 
The resulting spectrum was found to include an infinite tower of bosonic spinning particles. 
This fact emerges by recalling that the quantized phase-space coordinates obey the commutation relations
\be
 [x^\mu, p_\nu] = i \delta^\mu_\nu 
 \;, \qquad 
 [\bar \alpha_\nu, \alpha^\mu ] =  \delta_\nu^\mu\;,
 \label{comm-rel}
  \ee
with the $\alpha^\mu$ oscillators that act as creation operators on the Fock vacuum $|0\ra$.
The covariant analysis carried out in Ref.~\cite{Bonezzi:2024emt} shows that, 
at occupation number $s$, defined 
according to the number operator $N= \alpha^\mu  \bar \alpha_\mu$, the Hilbert space contains particle's excitations of spin 
  $s, s-2,\cdots, 0$ for even $s$, and $s, s-2,\cdots, 1$ for odd $s$. 
  A quick way to see this is by adopting the light-cone gauge, as discussed in Chapter~\ref{chap:second}, in which the bosonic creation operators $\alpha^i$
 carry only transverse indices $i=1,\cdots, D-2$. Let us indeed recall the result: at level  $s$,  the action of $s$ such creation operators on the Fock vacuum
 yields states corresponding to a totally symmetric wavefunction $\phi_{{i_1}\cdots{i_s}}(x)$, which includes all possible traces. Iteratively factorizing the traces shows that  the wavefunction naturally accommodates spin states $s,s-2,\cdots$ 
down to either 0 or 1, depending on whether  $s$ is even or odd. We refer to Section~\ref{chap:second:lightcone} for further details.
 
In the rest of this section, we construct a path integral for this model.
One can anticipate some difficulties in the construction. 
Considering a finite line, the gauge transformation laws in \eqref{2.6} suggest that all the gauge fields
$(e, u, \bar u)$ should each acquire a modulus,
with the set of moduli collectively denoted by  $ (2T, u_0, \bar u_0)$, where $T$ is the usual Schwinger proper time. 
After gauge-fixing the gauge fields to 
the corresponding moduli, i.e. setting
$(e, u, \bar u) =(2T, u_0, \bar u_0)$, one finds that 
the action \eqref{act-1} reduces to
\begin{equation} 
S_{\mathrm{gf}}=\int_0^1 d\tau\,\Big[p\cdot\dot x-i\,\bar\alpha\cdot\dot\alpha-Tp^2-\bar u_0\, p\cdot\alpha -u_0\,p\cdot\bar \alpha\Big] \;.
\label{act-2}
\end{equation}
An integration over the moduli, with  $T\in \mathbb{R}^+$ 
and $u_0, \bar u_0 \in \mathbb{R}$,
produces for constant $p, \alpha, \bar \alpha$ 
\begin{equation}
\frac{1}{p^2}\ \delta(p\cdot\alpha)\ \delta(p\cdot\bar \alpha)\;,
\end{equation}
which, on top of delivering the massless propagator, seems to enforce the transversality of the oscillators. However, we are going to argue that, for our purposes, we do not need to consider the moduli for $u$ and $\bar u$. 
The reason is that we intend to restrict the model to obtain the propagation of particles of fixed spin $s$ only, and in particular $s=1$, to describe massless gluons.
To achieve such a projection, it is useful to gauge the global $U(1)$ symmetry, generated by the number charge $N=\alpha^\mu \bar\alpha_\mu$,
and add a Chern--Simons coupling,  to obtain a constraint that projects physical states to have
occupation number $s$ in the quantum setting.

Therefore, we gauge the global $U(1)$ symmetry 
\ba
\alpha' &= e^{i \phi} \alpha \;,    
\qquad
 \bar \alpha'= e^{-i\phi} \bar \alpha\;,  \\
u' &= e^{i\phi}u \;,
\qquad  
  \bar u'= e^{-i\phi} \bar u\;, \\
\ea
and add a Chern--Simons coupling with charge $q$, left unspecified for the moment, to be related to the spin $s$. We arrive at the modified action
\begin{equation}
S=\int_0^1 d\tau\,\Big[p_\mu\dot x^\mu-i\,\bar\alpha_\mu\dot\alpha^\mu-e\,\frac 12 p^2 -\bar u \, \alpha^\mu p_\mu 
-u\ \bar\alpha^\mu p_\mu- a \, 
( \alpha^\mu \bar\alpha_\mu - q)
\Big]\;.
\label{act-3}
\end{equation}
Here $(e,\bar u, u, a)$ are the full set of worldline gauge fields that gauge the charges $(H, L, \bar L, J)$ defined by
\begin{equation}
H=\frac12\,p^2\;,
\quad 
L=\alpha^\mu p_\mu \;,
\quad 
\bar L=\bar\alpha^\mu p_\mu\;, 
\quad 
 J=\alpha^\mu \bar\alpha_\mu -q \;. 
 \label{charges}
\end{equation} 
They satisfy the following Poisson bracket algebra 
\begin{equation}
\{L,\bar L\}= 2i H \;, \quad   \{J, L\}= -i L \;, \quad  \{J,\bar L\}= i \bar L\;,
\end{equation}
while other brackets vanish. It is a first-class algebra and thus its gauging is consistent.
Denoting by  $(\epsilon, \bar \xi, \xi, \phi)$ the gauge parameters 
for the gauge fields above, we get the following infinitesimal transformation rules for the gauge symmetries of the action \eqref{act-3}

\ba
&\delta x^\mu=\epsilon\,p^\mu+\xi\,\bar\alpha^\mu+\bar\xi\,\alpha^\mu\;,  
\hskip .8cm 
\delta p_\mu=0\;,
\\
&\delta\alpha^\mu=i\,\xi\,p^\mu+i \,\phi\, \alpha^\mu    
\;,
\hskip 1.61cm 
\delta\bar\alpha^\mu=-i\,\bar\xi\,p^\mu -i \,\phi \,\bar \alpha^\mu\;,
\\
&\delta e=\dot\epsilon+2i\,u\,\bar\xi-2i\,\bar u\,\xi\;, 
\hskip 1.2cm 
\delta a=\dot{\phi}\;,
\\
&\delta u=\dot\xi -i\, a\, \xi +i\, \phi \,u
\;,  
\hskip 1.7cm 
\delta\bar u=\dot{\bar\xi} +i\, a\, \bar \xi -i\, \phi \, \bar u \;.
\ea

Notice that the $U(1)$ gauge field $a$ admits a modulus, $a(\tau)=\theta$,
which in turn prevents the emergence of moduli for $u$ and $\bar u$. 
A quantum massless spin-$s$ particle is obtained by quantizing precisely this action. Before proceeding with the path integral quantization, let us clarify the relation between the Chern--Simons coupling $q$ and the spin $s$. This relation generally depends on the quantization scheme that is adopted, which we make explicit in the course of our discussion.

Using canonical quantization, we recall that the quantum operators satisfy the commutation relations in \eqref{comm-rel} with $\alpha$ acting as creation operator. Then, 
of the charges in \eqref{charges}, 
only $J$ exhibits potential ordering ambiguities. We resolve them by adopting a symmetric ordering prescription for the oscillators. This is our chosen quantization scheme and yields the quantum operator
\begin{align}
J &= \frac{1}{2} (\alpha^\mu \bar{\alpha}_\mu + \bar{\alpha}_\mu \alpha^\mu) - q
\cr
&= \alpha^\mu \bar{\alpha}_\mu + \frac{D}{2} - q
\cr
&= N - s\;,
\label{quantumJ}
\end{align}
where we have used the commutation relations of the oscillators, recognized the number operator
\be
N = \alpha^\mu \bar{\alpha}_\mu 
\label{2.16}
\ee
which counts the occupation number in the Fock space of states, and relates the Chern--Simons coupling $q$ to the real parameter $s$ by
\be
q = \frac{D}{2} + s \;.
\ee
Imposing the constraint {\it à la} Dirac to select a generic physical state $|\phi\rangle$,
\be
J |\phi\rangle = 0 \qquad \Rightarrow \qquad N |\phi\rangle = s |\phi\rangle\;,
\ee
shows that physical states must have occupation number $s$. This corresponds to having maximal spin $s$.
Later, we specialize to $s = 1$, as we intend to focus on spin-1 particles. However, for the time being, we keep $s$ arbitrary and observe that it must be a non-negative integer in order to find nontrivial solutions to the constraint equation, and allow for a nontrivial quantum theory. This illustrates the well-known fact that Chern--Simons couplings are quantized.

Now, let us proceed with the worldline path integration and derive the propagator for a fixed spin-$s$ particle (together with its lower spin tail, as discussed above).
We consider the path integral on the finite line, with fixed initial and final states chosen to be
 momentum eigenstates for the $(x,p)$-variables and coherent states for the bosonic oscillators. This leads to 
the path integral representation for the particle propagator in momentum space
 \begin{equation}
_{out}\la p', \alpha | p, \bar \alpha \ra_{in}= \int \frac{DX D G}{\text{Vol(Gauge)}}\, 
e^{iS}\;,
 \end{equation}
which uses the action in \eqref{act-3}. 
On the left-hand side, we have indicated by $p$ and $p'$ the momentum eigenstates and $\alpha $ and $\bar\alpha$ the coherent states for the oscillators
(no confusion should arise with this notation).
Also, we have collectively denoted the dynamical fields by $X=(x^\mu, p_\mu, \alpha^\mu, \bar \alpha_\mu)$
and the gauge fields by $G=(e, \bar u,u, a)$. 
Note that, to impose boundary conditions on $p$, 
one must add boundary terms to the action. Their final effect is to change the symplectic term $p_\mu\dot x^\mu$
  into $- \dot p_\mu x^\mu$, see Appendix \ref{chap:third:Appendix:A}.

At this stage, we gauge-fix the gauge fields to their respective moduli, setting  $(e, \bar u, u, a) = (2T, 0,0, \theta)$, where $T$ and $\theta$ denote the moduli.
The latter must be integrated over a suitable range of values and with a measure with 
a normalization that can be fixed later.
This way, we arrive at an expression for the path integral on the finite line of the form
 \begin{equation}
_{out}\la p', \alpha | p, \bar \alpha \ra_{in}\sim 
\int dT \int d \theta\
e^{i s\theta}\
\text{Det}(\partial_\tau -i\theta)\, \text{Det}(\partial_\tau +i \theta)
  \int {DX}\,  e^{iS_{\text{gf}}}\;,
 \end{equation}
where the gauge-fixed action in the exponent reads
 \begin{equation}
S_{\text{gf}}= 
\int_0^1 d\tau\,\Big[-\dot p_\mu x^\mu-i\,\bar\alpha_\mu(\partial_\tau - i \theta) \alpha^\mu- T p^2  \Big] 
-i \bar \alpha(0) \alpha(0)\;.
\label{action-4}
\end{equation}
Here, we have extracted from the action the constant Chern--Simons term, whose coupling $q$ gets renormalized to the physical value $s$. Also, we have
included the Faddeev--Popov determinants, 
and inserted the boundary terms in the action, needed
to match the quantum numbers of the external states.
Performing the path integral, and normalizing appropriately the  integration over the moduli, while setting their correct range, we obtain 
  \ba \label{1.17}
_{out}\la p', \alpha | p, \bar \alpha \ra_{in} &\sim 
\int_0^\infty dT \int_0^{2\pi} \frac{d \theta}{2\pi} \ 
(2\pi)^D \delta^{D}(p-p') \,
e^{is\theta}\, e^{e^{-i\theta } \bar\alpha \alpha}\,  e^{-iTp^2}
\\
&= (2\pi)^D \delta^{D}(p-p') \frac{(-i)}{p^2} \int_0^{2\pi} \frac{d \theta}{2\pi} \, 
 e^{e^{-i\theta } \bar\alpha \alpha + is\theta}\;.
 \ea
Note that the ghost determinant, which has vanishing Dirichlet boundary conditions, becomes $\theta$ independent and can be normalized to 1, see Appendix \ref{chap:third:Appendix:A} for details.\footnote{The result is similar to that in eq. \eqref{A.15}, but with $\alpha=\bar \alpha= 0$, which selects the vacuum as the external states.}
Finally, performing the $\theta$-integration, we find
  \ba
  _{out}\la p', \alpha | p, \bar \alpha \ra_{in} =
 (2\pi)^D \delta^{D}(p-p') \frac{(-i )}{p^2} \frac{(\bar\alpha \alpha)^s}{s!} 
\;.
 \label{spins}
 \ea
In particular, for spin 1, obtained by setting $s=1$, we get
  \ba
_{out}\la p', \alpha | p, \bar \alpha \ra_{in} =
 (2\pi)^D \delta^{D}(p-p')\, \frac{-i\eta_{\mu\nu}}{p^2}\, \bar\alpha^\mu \alpha^\nu \;.
 \ea
We recognize the propagators of massless fields of spin $s$ in the Feynman gauge. The independent variables $\alpha$ and $\bar \alpha$ parametrize the coherent states and can be readily interpreted as the external polarizations. 

This concludes our discussion of the path integral on the finite line for the free spin-$s$ particle. A natural extension involves introducing a set of ghost fields (henceforth denoted $\cal B$ $\bar {\cal C}$  and $\bar {\cal B}$ $\cal C$) to exponentiate the $\acr{\Phi\Pi}$ determinants. Then, considering external states also for them  allows to explore the full BRST Hilbert space.

A final comment concerns the issue of gauge-fixing $u$ and $\bar u$. We observe that these gauge fields can be defined so as not to possess associated moduli,
so that the path integral for the original action \eqref{act-1} can be performed without the need to gauge
the $U(1)$ symmetry. This leads to the propagation of the full set of higher spin fields in the Feynman gauge.
Indeed, the path integral produces
  \ba
 _{out}\la p', \alpha | p, \bar \alpha \ra_{in} =
(2\pi)^D \delta^{D}(p-p') \frac{(-i)}{p^2} \,
 e^{\bar\alpha \alpha}\;,
 \label{allspins}
 \ea
which corresponds precisely to the sum over all spins of Eq. \eqref{spins}.

\subsection{Non-abelian background} \label{chap:third:sec2.2}
To compute tree-level gluon amplitudes, we couple the bosonic spinning particle to a Yang--Mills background field 
\begin{equation}
\cA_\mu:=A_\mu^a\,T_a\;,
\end{equation}
where $T_a$ are anti-hermitian generators; we will consider color-ordered amplitudes, thus the Lie algebra generators will not play much of a role in the following.
We extend the gauge constraints by minimal coupling $p_\mu\rightarrow p_\mu-i\cA_\mu$, and by adding a non-minimal term with 
coupling constant $\kappa$ to the Hamiltonian,
\ba \label{Covariant constraints}
L_A= \alpha^\mu  (p_\mu-i \cA_\mu) 
\;, \quad \bar L_A=  \bar \alpha^\mu  (p_\mu-i \cA_\mu)\;, \quad H_A= \frac12 (p-i \cA)^2 -\kappa\, \cF_{\mu\nu}\,\alpha^\mu \bar\alpha^\nu\;,
\ea
with $\cF_{\mu\nu}=\del_\mu\cA_\nu-\del_\nu\cA_\mu+[\cA_\mu,\cA_\nu]$.
The constraint algebra \eqref{constraint algebra} is broken by the field strength $\cF_{\mu\nu}$ for any value of $\kappa$, both at the quantum and classical level. In fact, from the perspective of the symplectic structure of the algebra, the presence inside $H_A$ of the full field strength tensor (with the commutator term) is not obvious at all. This issue and its resolution were discussed in Refs~\cite{Bastianelli:2013pta, Edwards:2016acz,Corradini:2016czo}. Here, however, the approach is different and based on the construction of a nilpotent BRST operator built from a precise Hamiltonian constraint. This ensures, in particular, that suitable spin sectors of the model can be quantized consistently, using BRST techniques; we shall briefly review how such a construction works.

In canonical BRST quantization, one extends the theory by adding fermionic antighost/ghost pairs $(\cB, \bar \cC)$, $(\bar \cB, \cC)$, $(b, c)$
associated with the constraints $(L_A , \bar L_A,  H_A)$, respectively, with anticommutation relations defined by
\be
 \{\mathcal{B},\bar{\mathcal{C}}\} = 1 \;, \quad 
\{\bar{\mathcal{B}},\mathcal{C}\} = 1\;,  \quad 
   \{b, c\} = 1 \;.
\ee 
The quantum BRST operator corresponding to the constraints \eqref{Covariant constraints} is then given by
\begin{equation}\label{QA}
\begin{split}
Q_A&=-2\,c\, H_A+i\,\bar\cC L_A+i\,\cC\bar L_A-\cC\bar\cC\,b\\
&=c\,\big(\cD^\mu\cD_\mu+2\kappa\, \cF_{\mu\nu}\,\alpha^\mu \bar\alpha^\nu\big)+(\bar\cC\alpha^\mu+\cC\bar\alpha^\mu)\cD_\mu-\cC\bar\cC\,b\;,
\end{split}    
\end{equation}
where $\cD_\mu=\del_\mu+\cA_\mu$ and we identified $p_\mu := -i\del_\mu$, keeping for simplicity the same notation for classical and quantum variables. The BRST operator $Q_A$ is hermitian if one assigns suitable hermiticity properties to the ghosts
\be
 c^\dagger\ = c \;, \quad b^\dagger = b \;, \quad \cC^\dagger = -\bar \cC \;, \quad \cB^\dagger = -\bar \cB \;. 
\ee
The extended BRST Hilbert space, on which the BRST charge $Q_A$ operates, has a double grading. 
One corresponds to the usual ghost number. 
The other one is associated with the integer $U(1)$ charge given by the number charge ${\cal N}$, which counts the number of $\alpha^\mu$ oscillators as in \eqref{2.16}, appropriately extended to account for the charge of the ghost oscillators:
\begin{equation}\label{BRST U1}
\cN:=\alpha^\mu\bar\alpha_\mu+\cC\bar\cB+\cB\bar\cC\;.    
\end{equation}
In the previous section, when constructing the free path integral, we gauged the $N$ charge in \eqref{2.16} 
to project on the spin-1 sector. Here, its extended version will still be used as a projector, but we prefer to keep it outside the BRST operator.
It is defined to obey $[Q_A,\cN]=0$, so that it becomes consistent in the following to restrict  the study of the cohomology of the BRST operator 
to a pre-fixed $U(1)$ charge sector.
Quantization is consistent if $Q_A^2=0$. According to the analysis of Ref.~\cite{Bonezzi:2024emt} (see also Chapter~\ref{chap:second} for a similar analysis but in the massive case), we have the following situation, depending on the spin sector of the theory: 
\begin{itemize}
\item If the background field $\cA_\mu$ is off-shell, only the spin zero sector of the model (i.e., the subspace of the Hilbert space with $\cN=0$) can be quantized consistently. In this case, the coupling $\kappa$ is irrelevant, as the corresponding non-minimal term acts trivially. This sector simply describes a colored scalar field in the adjoint representation of the gauge group.
\item If the background field $\cA_\mu$ is on-shell, i.e. it obeys the nonlinear Yang--Mills equations $\cD^\mu\cF_{\mu\nu}=0$, and $\kappa=1$, the model can be quantized also in the $\cN=1$ sector. This is the case we are interested in, where the worldline describes a gluon in a Yang--Mills background.
\end{itemize}

In order to extract the gluon vertex operators and study the path integral, we set $\kappa=1$ and proceed with a Hamiltonian BRST gauge-fixing of the classical action.
Upon denoting graded coordinates and momenta by $Z^I:=(x^\mu,\alpha^\mu;c,\bar\cC,\cC)$ and $\cP_I:=(p_\mu,-i\bar\alpha_\mu;-ib,-i\cB,-i\bar\cB)$, respectively, we choose $\Psi=iTb$ as the gauge-fixing fermion related to the gauge choice
$(e,u,\bar u)=(2T,0,0)$. Thus, the gauge-fixed action reads
\begin{equation}\label{coupled-action}
\begin{split}
S&=\int_0^1 d\tau\,\Big[\dot Z^I\cP_I+\big\{Q_A,\Psi\big\}_{\rm PB} 
\Big]\\
&=\int_0^1 d\tau\,\Big[p_\mu\dot x^\mu-i\,\bar\alpha_\mu\dot\alpha^\mu+ib\dot c+i\cB\dot{\bar\cC}+i\bar\cB\dot\cC-2TH_A 
\Big] \;.    
\end{split}    
\end{equation}
Integrating out the momenta, we finally obtain the sigma model action
\begin{equation}\label{Sigma model}
S=S_{\text{free}} +S_{\text{int}}\;,
\end{equation}
where the free Lagrangian action, including the ghosts, reads\footnote{The parameter $\tau$ is understood to be shifted and rescaled.}
\begin{equation} \label{Free action}
S_{\text{free}}=\int_{-T/2}^{T/2}\!\!\!d\tau\Big[\tfrac14\,\dot x^2-i\,\bar\alpha_\mu\dot\alpha^\mu+ib\dot c+i\cB\dot{\bar\cC}+i\bar\cB\dot\cC
\Big] \;,
\end{equation}
while the interaction terms are collected in
\begin{equation} \label{Interacting action}
   S_{\text{int}} =\int_{-T/2}^{T/2}\!\!\!d\tau\Big[i\cA_\mu(x)\dot x^\mu+2\cF_{\mu\nu}(x)\alpha^\mu\bar\alpha^\nu
\Big] \;.
\end{equation}
One can derive vertex operators from the above action in the usual way, upon specializing the background gauge field to a sum of plane waves, with definite momentum and polarization. These will be the so-called
integrated vertex operators, which describe the emission/absorption of a gluon from the worldline interior. One can already see from \eqref{Interacting action} the need for a two-gluon vertex, arising from the coupling to the non-abelian field strength.

In the previous subsection, we have attributed the spin-1 degrees of freedom to the worldline endpoints by employing coherent states for the $\alpha^\mu$ oscillators. In the following, we will find it more convenient to take simpler boundary conditions, corresponding to a suitable vacuum state, and create \emph{all} external gluon states by means of vertex operators. To do so, the vertex operators extracted from the sigma model action \eqref{Interacting action} are not sufficient; therefore, in order to discuss vertex operators more generally, together with the issue of ghost zero modes, we will briefly revert to canonical BRST quantization.

\subsection{Vertex operators from canonical BRST quantization} \label{chap:third:sec2.3}
Here we will briefly revisit the vertex operators as consistent deformations of the free BRST charge. In particular, we will review how the consistent coupling to the background $\cA_\mu$ translates into the BRST invariance of vertex operators. We will then motivate the definition of integrated vertex operators in the canonical language.

Let us collectively denote bosonic oscillators and ghost variables as
\begin{equation}
\alpha^M:=(\cC,\alpha^\mu,\cB)\;,\quad\bar\alpha_M:=(\bar\cB,\bar\alpha_\mu,\bar\cC)\;,    
\end{equation}
according to their $U(1)$ charge: $[\cN,\alpha^M]=\alpha^M$ and $[\cN,\bar\alpha_M]=-\bar\alpha_M$.
In canonical quantization, we define the BRST “oscillator vacuum” $\ket{0}$ as the Fock vacuum annihilated by the operators $\bar\alpha_M$ and by the antighost $b$. In the path integral formulation, the boundary conditions corresponding to this vacuum at both endpoints of the worldline are $\bar\alpha_M(-T/2)=0$, $\alpha^M(T/2)=0$ and $b(\pm T/2)=0$.
Taking also $\dot x^\mu(\pm T/2)=0$, it is natural to view the full BRST vacuum as the tensor product of an eigenstate of zero momentum with $\ket{0}$, i.e.
$\ket{0,0}=\ket{k=0} \otimes\ket{0}$. We thus have
\begin{equation}
\bar\alpha_M\ket{0,0}=0\;,\quad b\ket{0,0}=0\;,\quad p_\mu\ket{0,0}=0\;.   
\end{equation}
Although $\ket{0,0}$ is not normalizable, we will always create states with nonzero momentum by suitable vertex operators at the boundary. These states have the standard plane wave normalization 
\begin{equation}
\bra{k',0}c\ket{k,0}=(2\pi)^D\delta^D(k-k')\;,    
\end{equation}
where we recall that the basic overlap for the $bc$ system is $\bra{0}c\ket{0}=1$.

The subspace of the Hilbert space we are interested in is the $\cN=1$ sector, which describes free spin-1 particles. A generic state in this sector can be written as
\begin{equation}\label{spin one state}
\l x|\psi\r=\psi_M(x)\,\alpha^M\ket{0}+\chi_M(x)\,\alpha^M\,c\ket{0}\;, 
\end{equation}
in the position representation. 
Any such state is annihilated by normal ordered operators with more than one $\bar\alpha_M$, a feature that will be crucial for a consistent coupling to Yang--Mills backgrounds or, equivalently, for the BRST invariance of the vertex operators.

We define the \emph{unintegrated} canonical vertex operator for the gauge field as the linear deformation of the covariant BRST operator \eqref{QA}:
\begin{equation}
Q_A=Q+\cV_A+\cO(A^2)\;,
\end{equation}
where the free (${\cal A}$-independent) BRST differential reads
\begin{equation}\label{Q free}
Q=c\,\square+(\bar\cC\alpha^\mu+\cC\bar\alpha^\mu)\del_\mu-\cC\bar\cC\,b\;,    
\end{equation}
while the linear vertex operator is given by
\begin{equation}
\begin{split}
\cV_A&=(\bar\cC\alpha^\mu+\cC\bar\alpha^\mu)\cA_\mu+c\,\big(2\cA^\mu\del_\mu+\del\cdot\cA+2\,f_{\mu\nu}\,\alpha^\mu \bar\alpha^\nu\big)\;,
\end{split}    
\end{equation}
upon denoting $f_{\mu\nu}=\del_\mu\cA_\nu-\del_\nu\cA_\mu$ the abelian part of the field strength.
Expanding $Q_A^2$ at first order in $\cA_\mu$ produces
\begin{equation}\label{QV}
\{Q,\cV_A\}=c\,(\alpha^\nu\bar\cC-\cC\bar\alpha^\nu)\,\del^\mu f_{\mu\nu}-3\,\cC\alpha^\mu\bar\cC\bar\alpha^\nu\,f_{\mu\nu}-2\,c\,(\cC\alpha^\nu\bar\alpha^\mu\bar\alpha^\rho+\alpha^\mu\alpha^\nu\bar\cC\bar\alpha^\rho)\,\del_\mu f_{\nu\rho}  \;, 
\end{equation}
written in normal ordering. We see that the second and third terms in \eqref{QV} are of the form $\cO^{MN}\bar\alpha_M\bar\alpha_N$, with two annihilation operators on the right. For this reason, they kill any state in the $\cN=1$ sector. Since both $Q$ and $\cV_A$ commute with $\cN$, the restriction to any $\cN=s$ subspace can be explicitly enforced by the projector 
\begin{equation}
\cP_s:=\int_0^{2\pi}\frac{d\theta}{2\pi}\,e^{i\theta(\cN-s)} := \delta_{\cN,s}\;,\quad\cP_s^2=\cP_s\;.    
\end{equation}
Projecting \eqref{QV} to the spin one subspace, we thus obtain
\begin{equation}\label{QV is VQ}
\{Q,\cV_A\}\,\cP_1=c\,(\bar\cC\alpha^\nu-\cC\bar\alpha^\nu)\,\del^\mu f_{\mu\nu}\,\cP_1\;.    
\end{equation}
If the gauge field $\cA_\mu$ obeys the linearized Yang--Mills equations, the unintegrated vertex operator is BRST invariant under projection. It can thus be used to create asymptotic gluon states that satisfy such free equations.

For the path integral on the circle, it is necessary to enforce the constraint by inserting the projector $\cP_1$, which is equivalent to gauging the $U(1)$ symmetry. However, when computing tree-level processes on an open worldline, it is sufficient to have initial and final states with $\cN=1$ to ensure that no state with $\cN\neq1$ is created at any stage. This is so because $\cN$ is conserved in the process, as all vertex operators have charge zero and $\cN$ commutes with the gauge-fixed Hamiltonian. 

In view of computing the path integral on the open line, the above discussion brings a small subtlety. In order to create all external states in the process by inserting vertex operators, the Fock vacuum $\ket{0}$ is not the correct boundary state, since it has charge $\cN=0$. In order to generate physical states with vertex operators, the correct vacuum state is
\begin{equation}
\ket{1}:=\cB\ket{0}\;,\quad\bra{1}:=\bra{0}\bar\cB \;.
\end{equation}
This state has $\cN=1$, obeys $Q\ket{1}=0$ and can be viewed as a constant Yang--Mills ghost. We will later present the boundary conditions corresponding to the physical vacuum $\ket{1}$.

We now come to discuss integrated vertex operators. To this end, we start by splitting the unintegrated vertex $\cV_A$ according to its dependence on the reparametrization ghost $c$:
\begin{equation}
\cV_A=W_A+c\,V_A\;,
\end{equation}
where
\begin{equation}
\begin{split}
W_A&=(\bar\cC\alpha^\mu+\cC\bar\alpha^\mu)\cA_\mu\;,\\  V_A&=2\cA^\mu\del_\mu+\del\cdot\cA+2\,f_{\mu\nu}\,\alpha^\mu \bar\alpha^\nu\;.
\end{split}    
\end{equation}
The (integrand of) the integrated vertex operator is defined via $\{b,\cV_A\} :=  V_A$. The rationale for this definition is that, if $\{Q,\cV_A\}=0$, the BRST transformation of $V_A$ is given by
\begin{equation}
[Q,V_A]=[\square,\cV_A]\;,    
\end{equation}
upon using $\{Q,b\}=\square$. As we have previously discussed, if $\cA_\mu$ obeys the linearized field equations $\{Q,\cV_A\}$ is not identically zero, but does vanish on the projected Hilbert space.

To integrate the vertex operator $V_A$, we have to introduce time dependence by switching to the Heisenberg picture. Upon rescaling the gauge-fixed Hamiltonian to $p^2=-\square$, we define time-dependent vertex operators via
\begin{equation}
\cV_A(\tau):=e^{iH\tau}\cV_A\,e^{-iH\tau}=e^{ip^2\tau}\cV_A\,e^{-ip^2\tau}\;,    
\end{equation}
with analogous formulas for any other operator. Since $[Q,H]=0$, we can see that the BRST transformation of $V_A(\tau)$ is a total derivative:
\begin{equation}
[Q,V_A(\tau)]=-[p^2,\cV_A(\tau)]=i\,\frac{d}{d\tau}\cV_A(\tau)    \;.
\end{equation}
The BRST variation of the integrated vertex operator is then given by
\begin{equation}
\left[Q,\int_{\tau_1}^{\tau_2} d \tau\,V_A(\tau)\right]=i\,\cV_A(\tau_2)-i\cV_A(\tau_1)\;.    
\end{equation}

Typically, this fails to vanish when the boundaries of the integral reach other insertions. Perhaps, this is the most direct way to see the need for nonlinear vertex operators: they must be added to cure the failure of BRST invariance at these insertion points.

The basic, unintegrated, bilinear vertex can be extracted from the part of $Q_A$ \eqref{QA} quadratic in the gauge field, namely
\begin{equation}\label{V2 BRST}
\cV_{A,A}=c\,\big(\cA^2+2\,[\cA_\mu,\cA_\nu]\alpha^\mu\bar\alpha^\nu\big)  \;.  
\end{equation}
We will discuss the role of this vertex in the computation of amplitudes as we go along.

\subsection{Lagrangian vertex operators and color stripping} \label{chap:third:sec2.3.1}
In order to compute color-ordered amplitudes, we have to strip off color from the vertex operators and take the gauge field to represent a single gluon state. For both $\cV_A$ and $V_A$, which are linear in $\cA_\mu$, this is simply done by removing the Lie algebra generator and taking the field to be a plane wave with definite momentum and polarization:
\begin{equation}
\cA_\mu(x)=A_\mu^a(x)\,T_a\quad\longrightarrow\quad \epsilon_\mu\,e^{ik\cdot x}\;.    
\end{equation}
The derivative operator $-i\del_\mu$, when appearing in Weyl-ordered expressions, reduces in the path integral approach
to the classical canonical momentum $p_\mu$ without further terms.
For the free action \eqref{Free action} this coincides with $\frac12\,\dot x^\mu$, but in the sigma model \eqref{Sigma model} it is rather given by $\frac12\,\dot x^\mu+i\cA^\mu$. This is ultimately the reason why the nonlinear vertices differ between the Hamiltonian and Lagrangian path integrals. This correspondence allows us to identify
\begin{equation}
2\cA^\mu\del_\mu+\del\cdot\cA\quad\longrightarrow \quad e^{ik\cdot x}\big(2\epsilon^\mu\del_\mu+ik\cdot\epsilon\big)\quad\longrightarrow \quad i\epsilon_\mu\dot x^\mu\,e^{ik\cdot x}\;, 
\end{equation}
where we used the fact that the operator expression $2\,\cA^\mu\del_\mu+\del\cdot\cA$ is already Weyl ordered.\footnote{Weyl ordering is tied to the construction of a regulated, background gauge-invariant, path integral, see e.g. \cite{Bastianelli:2006rx}.}
Proceeding in the same way for all terms in $\cV_A$ and $V_A$, we find the one-gluon color-stripped vertex operator
\begin{equation}\label{Vertex operators}
\cV_{\epsilon,k}(\tau)=W_{\epsilon,k}(\tau)+c(\tau)\,V_{\epsilon,k}(\tau)\;,
\end{equation}
where we recall that $V_{\epsilon,k}(\tau)$ is to be used for integrated vertices:
\begin{equation}
\begin{split} \label{Vertex operator V}
W_{\epsilon,k}(\tau)&=\epsilon_\mu\Big(\cC(\tau)\,\bar\alpha^\mu(\tau)+\bar\cC(\tau)\,\alpha^\mu(\tau)\Big)\,e^{ik\cdot x(\tau)}\;,\\
V_{\epsilon,k}(\tau)&=\Big(i\,\epsilon_\mu\,\dot x^\mu(\tau)+2i\,f_{\mu\nu}(k)\,\alpha^\mu(\tau)\,\bar\alpha^\nu(\tau)\Big)\,e^{ik\cdot x(\tau)}\;,
\end{split}
\end{equation}
and we defined $f_{\mu\nu}(k)=k_\mu\epsilon_\nu-k_\nu\epsilon_\mu$.

The above vertex operators are not enough to compute arbitrary amplitudes. 
As already mentioned, further nonlinear vertex operators are needed. The simplest one is bilinear and can be seen directly from the action or the BRST charge. In its Hamiltonian form, this bilinear vertex is extracted from the BRST charge \eqref{QA} and is given by $\cV_{A,A}$ in  \eqref{V2 BRST}.
To properly strip off color, we assign the two gluon states respecting the order in which the generators are removed:
\begin{equation}
\begin{split}
\cA^2+2\,[\cA_\mu,\cA_\nu]\alpha^\mu\bar\alpha^\nu&=\big(A^a\cdot A^b+4\,A^a_\mu A^b_\nu\,\alpha^{[\mu}\bar\alpha^{\nu]}\big)T_aT_b\\
&\longrightarrow\quad\big(\epsilon_1\cdot\epsilon_2+4\,\epsilon_{1\mu} \epsilon_{2\nu}\,\alpha^{[\mu}\bar\alpha^{\nu]}\big)\,e^{i(k_1+k_2)\cdot x}\;.
\end{split}    
\end{equation}
The bilinear operator corresponding to the second term
\begin{equation} \label{bl}
V_{12}(\tau)=4\,\epsilon_{1\mu} \epsilon_{2\nu}\,\alpha^{[\mu}(\tau)\bar\alpha^{\nu]}(\tau)\,e^{i(k_1+k_2)\cdot x(\tau)}  \;,  
\end{equation}
where we have used the shorthand notation $V_{ij}=V_{\epsilon_i,k_i;\epsilon_j,k_j}$,
can be clearly derived from the action \eqref{Interacting action}. The first term corresponding to $\cA^2$, however, is absent from the Lagrangian action. This is because the canonical momentum is given by $\frac12\,\dot x^\mu+i\cA^\mu$. The effects of the bilinear vertex $\epsilon_1\cdot\epsilon_2\,e^{i(k_1+k_2)\cdot x(\tau)}$ are correctly taken into account by contractions of the $\dot x^\mu$ factors with each other in products $V_{\epsilon_1,k_1}(\tau)\,V_{\epsilon_2,k_2}(\sigma)$. The vertex operator \eqref{bl} is thus the only two-gluon vertex to be used in the Lagrangian path integral, as indeed implied by the interacting part of the sigma model action \eqref{Interacting action}.

Let us comment on the color structure of the bilinear vertex. In computing full amplitudes, including color, the bilinear vertex operator for particles $i$ and $j$ contributes a color factor of the form ${\rm tr}(\cdots T_iT_j\cdots)$. This implies that the two gluons created by $V_{ij}$ are necessarily \emph{color-adjacent}, which means that, in computing color-ordered amplitudes, $V_{ij}$ is needed only if gluons $i$ and $j$ are adjacent in the cyclic ordering $12\ldots n$.

Finally, as we shall discuss, additional nonlinear vertex operators are required for the construction of amplitudes. While the linear and quadratic vertex operators are already encoded in the action -- or equivalently, in the BRST charge -- further composite vertex operators, which attach subtrees to a main worldline, arise from the requirement of BRST-invariant amplitudes, thereby ensuring the consistency of the theory. Although a derivation of these operators from purely worldline considerations remains perhaps incomplete, the analysis presented in Ref.~\cite{Bonezzi:2025iza} for the scalar theory contains the essential ingredients. We will return to this issue in the next sections.

\section{Tree-level gluon amplitudes} \label{chap:third:sec3}
We now apply the formalism developed in the previous sections to compute gluon scattering amplitudes at tree level. To this end, following the analysis of Ref.~\cite{Bonezzi:2025iza}, we take the worldline to have infinite length, so that the proper time $T$ serves purely as an infrared regulator and drops out of all computations. The amplitudes are obtained by evaluating suitable correlation functions with insertions of the vertex operators, previously derived by coupling the worldline to a Yang--Mills background field. 
Before delving into the explicit calculations, we briefly outline how correlators are defined and evaluated on open worldlines in our formalism. In what follows, all external gluons are taken to be on-shell, though not necessarily transverse.

\subsection{Correlation functions on open lines} \label{chap:third:sec3.1}
To project on states of zero momentum, the coordinate trajectories are split as
\begin{equation}\label{BC Neumann z}
x^\mu(\tau)=x^\mu_0+z^\mu(\tau)\;,\quad \dot z^\mu(\pm T/2)=0\;, 
\end{equation}
where $x_0^\mu$ is the zero mode, which enforces momentum conservation. Coming to the oscillator vacuum, recall that the boundary conditions corresponding to $\ket{0}$ at both $\tau=\pm T/2$ are
\begin{equation}
\bra{0}\ldots\ket{0}\quad\longleftrightarrow \quad\bar\alpha_M(-T/2)=0\;,\quad\alpha^M(T/2)=0\;,\quad b(\pm T/2)=0\;.   
\end{equation}
On the other hand, the physical vacuum $\ket{1}=\cB\ket{0}$ is characterized by $(\cB,b,\bar\cB)\ket{1}=0$. Upon denoting all ghosts and antighosts as $C^i:=(\cC,c,\bar\cC)$ and $B_i:=(\bar\cB,b,\cB)$, the boundary conditions corresponding to the vacuum $\ket{1}$ are then
\begin{equation}\label{BC for amplitudes}
\bra{1}\ldots\ket{1}\quad\longleftrightarrow \quad\bar\alpha_\mu(-T/2)=0\;,\quad\alpha^\mu(T/2)=0\;,\quad B_i(\pm T/2)=0\;.   
\end{equation}
Since all ghosts and antighosts have the same boundary conditions, we can rewrite the free Lagrangian action \eqref{Free action} as
\begin{equation}\label{Free action for amplitudes}
S_{\text{free}}=\int_{-T/2}^{T/2}\!\!\!d\tau\Big[\tfrac14\,\dot x^2-i\,\bar\alpha_\mu\dot\alpha^\mu+iB_i\dot C^i
\Big] \;.        
\end{equation}
Note that the basic overlap for the physical vacuum needs three ghost insertions:
\begin{equation}
\bra{1}\cC\,c\,\bar\cC\ket{1}=1\;.
\end{equation}
In the path integral, this is reflected by the fact that all $C^i(\tau)$ admit a zero mode, which needs to be saturated. The basic expectation value for the ghosts is thus given by
\begin{equation}\label{Ghost saturation}
\l\cC(\tau_1)\,c(\tau_2)\,\bar\cC(\tau_3)\r=1\;.   
\end{equation}
From the quadratic action \eqref{Free action for amplitudes} with boundary conditions \eqref{BC Neumann z} and \eqref{BC for amplitudes}, we derive the following two-point functions (the ghost zero mode saturation is left implicit)
\begin{equation}\label{z-z}
\begin{split}
\l\bar\alpha^\mu(\tau)\alpha^\nu(\sigma)\r&=\eta^{\mu\nu}\theta(\tau-\sigma)\;,\\
\l z^\mu(\tau)z^\nu(\sigma)\r&=-i\eta^{\mu\nu}G(\tau,\sigma)\;,\\
\l \dot z^\mu(\tau)z^\nu(\sigma)\r&=-i\eta^{\mu\nu}\udot G(\tau,\sigma)\;,\\
\l \dot z^\mu(\tau)\dot z^\nu(\sigma)\r&=-i\eta^{\mu\nu}\udot G\udot (\tau,\sigma)\;,
\end{split}    
\end{equation}
where we denote derivatives with respect to the left and right variables by left and right dots, respectively. The $z$-propagator and its derivatives read
\begin{equation}\label{z-z-1}
\begin{split}
G(\tau,\sigma)&=|\tau-\sigma|-\frac{1}{T}\,(\tau-\sigma)^2-\frac{2}{T}\,\tau\sigma-\frac{T}{6}\;,\\
\udot G(\tau,\sigma)&=\text{sgn}(\tau-\sigma)-\frac{2}{T}\,\tau\;,\\
\udot G\udot(\tau,\sigma)&=-2\,\delta(\tau-\sigma)\;.
\end{split}    
\end{equation}
In particular, notice that the Neumann boundary condition $\udot G(\pm T/2,\sigma)=0$ holds only at separate points. At coincident points one rather has $\udot G(\pm T/2,\pm T/2)=\mp1$. This reflects the Weyl ordering of operators, that is implicit in the path integral.\footnote{ For instance, the quantum operator corresponding to $x^\mu p^\nu$ is $\frac12\,(\hat x^\mu\hat p^\nu+\hat p^\nu\hat x^\mu)$.} This fact will be relevant when computing the three-gluon amplitude.

\subsection{Color-ordered amplitudes} \label{chap:third:sec3.2}
We now illustrate the construction above with explicit computations, starting from the simplest test of the model: the three-gluon amplitude. Then, we move on to the calculation of the color-ordered four-point amplitude, which we perform in two distinct ways. First, we select the worldline connecting two of the least color-adjacent particles, which we regard as a “smart” choice for computational reasons, as should be clear shortly. Then, we reproduce the same calculation by choosing the “not-so-smart” worldline, namely the one that connects two color-adjacent particles. Throughout this section, we implicitly factor out the momentum-conserving delta function, which arises from the integral over the zero-mode $x_0^\mu$. Every $x^\mu(\tau)$ in the vertex operators can thus be substituted with the fluctuation $z^\mu(\tau)$.

\subsubsection{The three-gluon amplitude} \label{chap:third:sec3.2.1}
Although the three-gluon amplitude vanishes for real momenta due to kinematical reasons, by taking the external gluons to be on-shell but not necessarily transverse ($k_i^2=0$, $k_i\cdot\epsilon_i\neq0$), we obtain a nontrivial expression that can be directly compared with the standard field theory result. 

To saturate the ghost zero modes, we need all three vertex operators to be unintegrated. This has a clear geometric interpretation: two vertex operators are fixed at asymptotic times $\tau=\pm\infty$ to create the incoming and outgoing worldline states, while the third one is fixed at an intermediate time using the rigid translation invariance on the infinite line. To compute the color-ordered amplitude, we need to insert the three vertex operators with a fixed (cyclic) order in time (see figure \ref{fig1}). The remaining expectation value is taken with respect to $z^\mu$ and the other variables and gives the color-ordered amplitude as
\begin{equation}
\cA_{123}=\big\l\cV_1(+\infty)\cV_2(0)\cV_3(-\infty)\big\r\;,  
\end{equation}
where we used the shorthand $i=\{\epsilon_i,k_i\}$ for particle labels. 

    \diagramThree

Using the split $\cV_i(\tau_i)=W_i(\tau_i)+c(\tau_i)\,V_i(\tau_i)$, the above expression gives three contributions
(as one needs only one $c$-ghost insertion for obtaining a non-vanishing result) 
\begin{equation}\label{A3 split}
\begin{split}
\cA_{123}&=\big\l W_1(+\infty)\,c(0)V_2(0)W_3(-\infty)\big\r+\big\l W_1(+\infty)W_2(0)\,c(-\infty)V_3(-\infty)\big\r\\    
&\phantom{=}+\big\l c(+\infty)V_1(+\infty)W_2(0)W_3(-\infty)\big\r \;.    
\end{split}
\end{equation}
The expression \eqref{Vertex operator V} for $V_i(\tau_i)$ suggests that $V_i(\pm\infty)=0$, thanks to the boundary conditions. As we have already mentioned, however, this is not true due to the self-contraction
\begin{equation}
\epsilon\cdot\dot z(\tau)\,e^{ik\cdot z(\tau)}\sim k\cdot\epsilon\udot G(\tau,\tau)\,e^{ik\cdot z(\tau)}+\cdots\;,  
\end{equation}
and $\udot G(\tau,\tau)$ does not vanish when $\tau\rightarrow\pm\infty$. This term does vanish for transverse gluons, which motivates us to consider non-transverse polarizations in order to study its effect.

Let us compute separately the three contributions in \eqref{A3 split}. All correlators have a common “Koba-Nielsen”
factor, given by the contraction of the plane waves $e^{ik_i\cdot z(\tau_i)}$, which is trivial, due to momentum conservation and the on-shell condition:
\begin{equation}
\begin{split}
\exp\Big\{\tfrac{i}{2}\sum_{ij}k_i\cdot k_j\,G(\tau_i,\tau_j)\Big\}&=\exp\Big\{i\sum_{i<j}k_i\cdot k_j\,|\tau_i-\tau_j|\Big\}\\
&=\exp\left\{-\frac{iT(k_1^2+k_3^2)}{2}\right\}\to1\;,
\end{split}
\end{equation}
where we used $T/2\rightarrow\infty$ as an infrared regulator. We further use \eqref{Ghost saturation} for saturating the ghost zero modes. This only gives relative signs between the three contributions. The first term in \eqref{A3 split} yields
\begin{equation}
\begin{split}
&\big\l W_1(+\infty)\,c(0)V_2(0)W_3(-\infty)\big\r\\
&=\Big\l\epsilon_1\cdot\bar\alpha(+\infty)\,\big[i\dot z(0)\cdot\epsilon_2+4i\,k_2^{[\mu}\epsilon_2^{\nu]}\alpha_\mu(0)\bar\alpha_\nu(0)\big]\epsilon_3\cdot\alpha(-\infty)e^{i\sum_ik_i\cdot z(\tau_i)}\Big\r\\
&=\epsilon_1\cdot\epsilon_3\,i\epsilon_2\cdot\sum_ik_i\udot G(0,\tau_i)+4i\,k_2^{[\mu}\epsilon_2^{\nu]}\epsilon_{1\mu}\epsilon_{3\nu}\\
&=i\Big(\epsilon_1\cdot\epsilon_3\,\epsilon_2\cdot(k_3-k_1)+2\,\epsilon_1\cdot k_2\,\epsilon_3\cdot\epsilon_2-2\,\epsilon_1\cdot \epsilon_2\,\epsilon_3\cdot k_2\Big)\;,
\end{split}    
\end{equation}
where in the second line we removed the ghost factor. The above result is already the correct amplitude for transverse gluons. The remaining two terms add the contributions from the longitudinal modes:
\begin{equation}
\begin{split}
&\big\l W_1(+\infty)W_2(0)\,c(-\infty)V_3(-\infty)\big\r\\
&=-\Big\l\epsilon_1\cdot\bar\alpha(+\infty)\,\epsilon_2\cdot\alpha(0)\,i\epsilon_3\cdot\dot z(-\infty)\,e^{i\sum_ik_i\cdot z(\tau_i)}\Big\r\\
&=-\epsilon_1\cdot\epsilon_2\,ik_3\cdot\epsilon_3\udot G(-\infty,-\infty)=-i\,\epsilon_1\cdot\epsilon_2\,k_3\cdot\epsilon_3\;,
\end{split}    
\end{equation}
where we used the fact that only the term with $\dot z^\mu$ in $V_i(\pm\infty)$ contributes. The last term in \eqref{A3 split} similarly gives
\begin{equation}
\begin{split}
&\big\l c(+\infty)V_1(+\infty)W_2(0)W_3(-\infty)\big\r\\
&=-\Big\l i\epsilon_1\cdot\dot z(+\infty)\,\epsilon_2\cdot\bar\alpha(0)\,\epsilon_3\cdot\alpha(-\infty)\,\,e^{i\sum_ik_i\cdot z(\tau_i)}\Big\r\\
&=-\epsilon_2\cdot\epsilon_3\,ik_1\cdot\epsilon_1\udot G(+\infty,+\infty)=i\epsilon_2\cdot\epsilon_3\,k_1\cdot\epsilon_1\;.
\end{split}    
\end{equation}
By using momentum conservation, we can write the full result in the manifest cyclic form
\begin{equation}\label{A123}
\begin{split}
\cA_{123}&=
i\,\Big[\epsilon_1\cdot\epsilon_2\,\epsilon_3\cdot(k_1-k_2)+\epsilon_2\cdot\epsilon_3\,\epsilon_1\cdot(k_2-k_3)+\epsilon_3\cdot\epsilon_1\,\epsilon_2\cdot(k_3-k_1)\Big]\;.
\end{split}    
\end{equation}
This is the correct three-point amplitude even for non-transverse gluons, as it coincides with the color-ordered cubic vertex seen in the second quantized picture.

\subsubsection{The four-gluon amplitude}  \label{chap:third:sec3.2.2}
For the four-point color-ordered amplitude, we start by selecting the “smart” worldline, i.e. the one connecting particles 4 in the past with particle 2 in the future (or, equivalently, the one connecting particles 1 and 3, see figure \ref{fig2}).\footnote{It seems that the most convenient choice is a worldline that connects two of the least color-adjacent particles.} The advantage in this choice of worldline comes from the fact that we are allowed to integrate $V_1(\tau)$ over the entire line, without altering the cyclic order of the particles. We are thus led to evaluate
\begin{equation}
\cA_{1234}=
\Big\l\cV_2(+\infty)\, \cV_3(0) \, \int_{-\infty}^{+\infty} {\hskip -6mm d\tau}\, V_1(\tau)\, \cV_4(-\infty)
\Big\r\;,  
\end{equation}
which, for transverse gluons ($V_{\epsilon,k}(\pm\infty)=0$), reduces to 
\begin{equation}
\cA_{1234}=
\Big\l W_2(+\infty)\, c(0)\,V_3(0) \, \int_{-\infty}^{+\infty} {\hskip -6mm d\tau}\, V_1(\tau)\, W_4(-\infty)
\Big\r\;.
\end{equation}
Considering the overlap $\l \cC\, c\ \bar\cC\r =1 $, it simplifies to 
\begin{equation}
\cA_{1234}=
\Big\l 
\epsilon_2\cdot\bar\alpha(+\infty) e^{i k_2\cdot z(+\infty)}\, 
 V_3(0) \, \int_{-\infty}^{+\infty} {\hskip -6mm d\tau}\, V_1(\tau)\, 
 \epsilon_4\cdot \alpha(-\infty) e^{i k_4\cdot z(-\infty)}
 \Big\r\;.
\end{equation} 
The overall exponentials of the vertex operators will lead by Wick contractions to 
\begin{equation}
\begin{split}
\big\l 
e^{i \sum_i  k_i\cdot z(\tau_i)}  
 \big\r\
&=
e^{-\frac12 \sum_{ij}  k_i\cdot \l z(\tau_i)  z(\tau_j) \r \cdot k_j} 
 =
  e^{\frac{i}{2}  \sum_{ij}  k_i \cdot k_j G(\tau_i,\tau_j)}
\\
&=
 e^{i  \sum_{i<j}  k_i \cdot k_j |\tau_i -\tau_j|}
 =
 e^{\frac{i}{2} (- s_{12} \tau + s_{23} \tau + s_{31} |\tau|) }\;,
\end{split}
\end{equation} 
 where we have used the basic two-point functions in \eqref{z-z} and \eqref{z-z-1},
 noted that only the absolute value in $G(\tau_i,\tau_j)$ contributes, 
 and used the Mandelstam variables 
\begin{equation}
s_{ij}= (k_i+k_j)^2 = 2 k_i \cdot k_j \;.
\end{equation}
To display the separation in two channels, the final answer is better written using the on-shell value of $s_{31}=-s_{12}-s_{23}$, leading to 
\begin{equation}\label{Bosonic core}
\big\l 
e^{i \sum_i  k_i\cdot z(\tau_i)}  
 \big\r\
 = e^{\theta(\tau)(-i \tau s_{12}) +\theta(-\tau)(i \tau s_{23})} =
 \theta(\tau)\, e^{ -i \tau s_{12}} +
 \theta(-\tau)\, e^{  i \tau s_{23}} \;.
\end{equation} 
The integral in $\tau$ will split into the two regions $\tau>0$ and $\tau<0$, corresponding to the $s$- and $t$-channels, respectively (recall that there is no $u$-channel in the color-ordered amplitude).

    \diagramFourSmart

The remainder of the correlation function yields the numerator contributions. To compute these, we can further split the vertex operators $V_i(\tau)$ into their “scalar” and “spin” parts as
\begin{equation} \label{chap:third:split}
V_{i}(\tau)=V^{\rm scal}_{i}(\tau)+V^{\rm spin}_{i}(\tau)\;
\end{equation}
with
\begin{equation}
\begin{split}
V^{\rm scal}_{i}(\tau)&=i\,\epsilon_i\cdot\dot z(\tau)\,e^{ik_i\cdot z(\tau)}\;,\\
V^{\rm spin}_{i}(\tau)&=4i\,k_i^{[\mu}\epsilon_i^{\nu]}\alpha_\mu(\tau)\,\bar\alpha_\nu(\tau)\,e^{ik_i\cdot z(\tau)}\;.  
\end{split}    
\end{equation}
Omitting the contraction \eqref{Bosonic core} of the plane waves, these terms contribute to the correlator as follows:
\begin{equation}
\begin{split} \label{4.50}
&\big\l 
\cdots\,V^{\rm scal}_3(0)\, V^{\rm scal}_1(\tau)\cdots\big\r=4\,\theta(\tau)\,\epsilon_2\cdot\epsilon_4\,\epsilon_1\cdot k_2\,\epsilon_3\cdot k_4+4\,\theta(-\tau)\,(1\leftrightarrow3)-2i\,\delta(\tau)\,\epsilon_2\cdot\epsilon_4\,\epsilon_1\cdot\epsilon_3\;,\\[3mm] 
&\big\l 
\cdots\,V^{\rm scal}_3(0)\, V^{\rm spin}_1(\tau)\cdots
 \big\r+\big\l 
\cdots\,V^{\rm spin}_3(0)\, V^{\rm scal}_1(\tau)\cdots\big\r=\\
&4\,\theta(\tau)\,\Big[\big(\epsilon_2\cdot k_3\,\epsilon_3\cdot\epsilon_4-\epsilon_2\cdot \epsilon_3\,k_3\cdot\epsilon_4\big)\,\epsilon_1\cdot k_2-\big(\epsilon_2\cdot k_1\,\epsilon_1\cdot\epsilon_4-\epsilon_2\cdot \epsilon_1\,k_1\cdot\epsilon_4\big)\,\epsilon_3\cdot k_4\Big]+4\,\theta(-\tau)\,(1\leftrightarrow3)\;,\\[3mm]
&\big\l 
\cdots\,V^{\rm spin}_3(0)\, V^{\rm spin}_1(\tau)\cdots\big\r=\\
&4\,\theta(\tau)\,\Big[\epsilon_4\cdot k_3\big(\epsilon_3\cdot\epsilon_1\,k_1\cdot\epsilon_2-\epsilon_3\cdot k_1\,\epsilon_1\cdot\epsilon_2\big)-\epsilon_4\cdot \epsilon_3\big(k_3\cdot\epsilon_1\,k_1\cdot\epsilon_2-k_3\cdot k_1\,\epsilon_1\cdot\epsilon_2\big)\Big]+4\,\theta(-\tau)\,(1\leftrightarrow3)\;,
\end{split}    
\end{equation}
where $+\,(1\leftrightarrow3)$ means to exchange simultaneously $(\epsilon_1,k_1)$ with $(\epsilon_3,k_3)$. Summing the three contributions above, multiplied by the common scalar factor \eqref{Bosonic core}, and performing the $\tau$ integral gives the final result. This can be organized upon introducing the following products:
\begin{equation}
\begin{split} \label{C-inf prod}
m_2^\mu(\epsilon_i,\epsilon_j)&:=i\,\Big(2\,\epsilon_i\cdot k_j\,\epsilon_j^\mu-2\,\epsilon_j\cdot k_i\,\epsilon_i^\mu+(k^\mu_i-k^\mu_j)\,\epsilon_i\cdot\epsilon_j\Big)\;,\\
m_3^\mu(\epsilon_i,\epsilon_j,\epsilon_k)&:=\epsilon_i\cdot\epsilon_j\,\epsilon_k^\mu+\epsilon_k\cdot\epsilon_j\,\epsilon_i^\mu-2\,\epsilon_i\cdot\epsilon_k\,\epsilon_j^\mu\;.
\end{split}    
\end{equation}
These products encode (for transverse gluons) the color-ordered Feynman rules for the cubic and quartic vertices, respectively \cite{Zeitlin:2008cc, Bonezzi:2022yuh, Bonezzi:2023xhn}, and play a central role in the Berends-Giele recursion relations \cite{Berends:1987me, Selivanov:1997aq, Mizera:2018jbh, Lopez-Arcos:2019hvg}. In terms of the products above, one can rewrite the result coming from the path integral as
\begin{equation}
\cA_{1234}=\frac{i}{s_{12}}\,m_2(\epsilon_1,\epsilon_2)\cdot m_2(\epsilon_3,\epsilon_4)+\frac{i}{s_{23}}\,m_2(\epsilon_2,\epsilon_3)\cdot m_2(\epsilon_4,\epsilon_1)+i\,\epsilon_2\cdot m_3(\epsilon_3,\epsilon_4,\epsilon_1)\;,    
\end{equation}
which highlights the two exchange diagrams and the contact quartic term. This is the correct color-ordered four-gluon amplitude. It is customary to blow up the quartic vertex and define kinematic numerators as
\begin{equation}
\begin{split}
n_{12}&:=m_2(\epsilon_1,\epsilon_2)\cdot m_2(\epsilon_3,\epsilon_4)+\frac{s_{12}}{3}\,\big[m_3(\epsilon_1,\epsilon_2,\epsilon_3)-m_3(\epsilon_2,\epsilon_1,\epsilon_3)\big]\cdot\epsilon_4\;,\\
n_{23}&:=m_2(\epsilon_2,\epsilon_3)\cdot m_2(\epsilon_1,\epsilon_4)+\frac{s_{23}}{3}\,\big[m_3(\epsilon_2,\epsilon_3,\epsilon_1)-m_3(\epsilon_3,\epsilon_2,\epsilon_1)\big]\cdot\epsilon_4\;,\\
n_{31}&:=m_2(\epsilon_3,\epsilon_1)\cdot m_2(\epsilon_2,\epsilon_4)+\frac{s_{31}}{3}\,\big[m_3(\epsilon_3,\epsilon_1,\epsilon_2)-m_3(\epsilon_1,\epsilon_3,\epsilon_2)\big]\cdot\epsilon_4\;,
\end{split}    
\end{equation}
where the various channels are sometimes identified with the standard names for the Mandelstam variables for four points: $12\rightarrow s$, $23\rightarrow t$, $31\rightarrow u$. Upon using the following symmetry properties of the products:
\begin{equation}
\begin{split}
m_2(\epsilon_i,\epsilon_j)+m_2(\epsilon_j,\epsilon_i)&=0\;,\\
m_3(\epsilon_i,\epsilon_j,\epsilon_k)-m_3(\epsilon_k,\epsilon_j,\epsilon_i)&=0\;,\\
m_3(\epsilon_i,\epsilon_j,\epsilon_k)+m_3(\epsilon_j,\epsilon_k,\epsilon_i)+m_3(\epsilon_k,\epsilon_i,\epsilon_j)&=0\;,
\end{split}    
\end{equation}
the color-ordered amplitude can be recast in the compact form
\begin{equation}
\cA_{1234}=i\,\left[\frac{n_s}{s}-\frac{n_t}{t}\right] \;.
\end{equation}

We shall now illustrate how a different choice for the worldline would lead to the same result, albeit with comparatively greater effort.
\paragraph{Not-so-smart worldline}
Choosing instead the worldline connecting the color-adjacent particles 1 and 4 in figure \ref{fig2}, we have to compute three different contributions
\begin{equation}
\cA_{1234}=\cA_{1234}^{\mathrm{I}}+\cA_{1234}^{\mathrm{II}}+\cA_{1234}^{\mathrm{III}}\;,
\end{equation}
corresponding to three different diagrams (see figure \ref{fig3}). Explicitly
\begin{align}
    \cA_{1234}^{\mathrm{I}}&=-\Big\l\cV_1(+\infty) \, \int_{0}^{+\infty} {\hskip -6mm d\tau}\, V_2(\tau) \, \cV_3(0) \, \cV_4(-\infty)\Big\r\ , \label{AI}\\
    \cA_{1234}^{\mathrm{II}}&= i\,\Big\l\cV_1(+\infty) \, c(0) V_{23}(0)\, \cV_4(-\infty)
\Big\r\ , \label{AII} \\
    \cA_{1234}^{\mathrm{III}}&= i\,\Big\l\cV_1(+\infty) \, c(0)\,V_{23}^{\rm pinch}(0)\, \cV_4(-\infty) \Big\r\; , \label{AIII}
\end{align}
where the phases are given by a factor of $i$ for every insertion corresponding to a real interaction vertex. The first contribution $\cA_{1234}^{\mathrm{I}}$ arises from integrating the linear vertex for particle 2, with the range restricted to preserve the cyclic ordering of particles. Since gluons 2 and 3 are color-adjacent, the bilinear vertex $V_{23}(\tau)$ is also needed, as discussed in sec.~\ref{chap:third:sec2.3.1}. Using translation invariance, we place it at $\tau=0$. 
These two contributions, however, are not sufficient to produce a gauge invariant amplitude, as we will demonstrate in sec.~\ref{chap:third:sec4}. The missing part, which we denoted $\cA_{1234}^{\mathrm{III}}$ above, involves a composite operator, $V_{23}^{\rm pinch}(\tau)$, which attaches to the worldline the cubic subtree formed by particles 2 and 3. This pinch operator is nothing but a vertex operator $V_{\epsilon_{23},k_{23}}$ of the standard form \eqref{Vertex operator V}, with momentum $k_2+k_3$ and multiparticle polarization given by $\epsilon^\mu_{23}=\tfrac{1}{s_{23}} m_2^\mu(\epsilon_2,\epsilon_3)$, to be further discussed in the next section. 
    \diagramFourDumb

We start the computation with the first diagram $\cA_{1234}^{\mathrm{I}}$ \eqref{AI}, whose calculation is akin to the one for the smart worldline. Taking transverse gluons and recalling the overlap $\l \cC\, c\ \bar\cC\r =1$, it simplifies to
\begin{equation} \label{4.60}
\cA_{1234}^{\mathrm{I}}=-\big\l 
\epsilon_1\cdot\bar\alpha(+\infty) e^{i k_1\cdot z(+\infty)} \, \int_{0}^{+\infty} {\hskip -6mm d\tau}\, V_2(\tau)\, V_3(0) \,
 \epsilon_4\cdot \alpha(-\infty) e^{i k_4\cdot z(-\infty)}
 \big\r\;,
\end{equation} 
with the Koba-Nielsen scalar factor reducing to 
\begin{equation} \label{chap:third:fact}
\big\l 
e^{i \sum_i  k_i\cdot z(\tau_i)}  
 \big\r\
= e^{-i \, s_{12} \tau}\;.
\end{equation} 
Splitting the vertex operators into scalar and spin parts as in \eqref{chap:third:split}, and omitting the contraction of the plane waves, the numerator contributions are given by the following correlators
\begin{equation}
\begin{split}
&\big\l 
\cdots\,V^{\rm scal}_2(\tau)\,V^{\rm scal}_3(0)\,\cdots\big\r=-2i \, \delta(\tau) \, \epsilon_1 \cdot \epsilon_4 \, \epsilon_2 \cdot \epsilon_3 +4 \,  \epsilon_1 \cdot \epsilon_4 \, \epsilon_2 \cdot k_1 \, \epsilon_3 \cdot k_4\;,\\[3mm] 
&\big\l 
\cdots\,V^{\rm spin}_2(\tau)\,V^{\rm scal}_3(0)\,\cdots
 \big\r+\big\l 
\cdots\,V^{\rm scal}_2(\tau)\,V^{\rm spin}_3(0)\,\cdots\big\r=\\
&-4\left[ \epsilon_3 \cdot k_4 \left( \epsilon_1 \cdot k_2 \, \epsilon_2 \cdot \epsilon_4 - \epsilon_1 \cdot \epsilon_2 \, k_2 \cdot \epsilon_4 \right) - \epsilon_2 \cdot k_1 \left( \epsilon_1 \cdot k_3 \, \epsilon_3 \cdot \epsilon_4 -\epsilon_1 \cdot \epsilon_3 \, k_3 \cdot \epsilon_4 \right)\right]\;,\\[3mm]
&\big\l 
\cdots\,V^{\rm spin}_2(\tau)\,V^{\rm spin}_3(0)\,\cdots\big\r=\\
&-4\left[ k_3 \cdot \epsilon_4 \left( \epsilon_1 \cdot \epsilon_2 \, k_2 \cdot\epsilon_3 - \epsilon_2 \cdot \epsilon_3 \, \epsilon_1 \cdot k_2  \right) - \epsilon_3 \cdot \epsilon_4 \left( \epsilon_1 \cdot \epsilon_2 \, k_2 \cdot k_3 -\epsilon_1 \cdot k_2 \, \epsilon_2 \cdot k_3 \right) \right]\;,
\end{split}    
\end{equation}
which is the same as in \eqref{4.50} but for $\tau>0$ and with $(1\leftrightarrow2)$. Multiplying by the common scalar factor \eqref{chap:third:fact} and integrating over $\tau$ we get
\begin{align}\nonumber
&\big\l 
\cdots\, V^{\rm scal}_2(\tau)\,V^{\rm scal}_3(0)\,\cdots\big\r=i \epsilon_1 \cdot \epsilon_4 \, \left(\epsilon_2 \cdot \epsilon_3+\frac{4 \epsilon_2 \cdot k_1 \, \epsilon_3 \cdot k_4}{s_{12}}\right)\;,\\[3mm]\nonumber 
&\big\l 
\cdots\,V^{\rm spin}_2(\tau)\,V^{\rm scal}_3(0)\,\cdots
 \big\r+\big\l 
\cdots\,V^{\rm scal}_2(\tau)\,V^{\rm spin}_3(0)\,\cdots\big\r=\\\nonumber
&-\frac{4i}{s_{12}}\left[ \epsilon_3 \cdot k_4 \left( \epsilon_1 \cdot k_2 \, \epsilon_2 \cdot \epsilon_4 - \epsilon_1 \cdot \epsilon_2 \, k_2 \cdot \epsilon_4 \right) - \epsilon_2 \cdot k_1 \left( \epsilon_1 \cdot k_3 \, \epsilon_3 \cdot \epsilon_4 -\epsilon_1 \cdot \epsilon_3 \, k_3 \cdot \epsilon_4 \right) \right]\;,\\[3mm]\nonumber
&\big\l 
\cdots\,V^{\rm spin}_2(\tau)\,V^{\rm spin}_3(0)\,\cdots\big\r=\\
&-\frac{4 i}{s_{12}}\left[  k_3 \cdot \epsilon_4 \left( \epsilon_1 \cdot \epsilon_2 \, k_2 \cdot\epsilon_3 - \epsilon_2 \cdot \epsilon_3 \, \epsilon_1 \cdot k_2  \right) - \epsilon_3 \cdot \epsilon_4 \left( \epsilon_1 \cdot \epsilon_2 \, k_2 \cdot k_3 -\epsilon_1 \cdot k_2 \, \epsilon_2 \cdot k_3 \right) \right]\;.
\end{align}
The second diagram \eqref{AII}, with the bilinear vertex 
\begin{equation}
V_{23}(0)=\Big(2\,\epsilon_{2}\cdot\alpha(0)\,\epsilon_3\cdot\bar\alpha(0)-2\,\epsilon_{3}\cdot\alpha(0)\,\epsilon_2\cdot\bar\alpha(0)\Big)\,e^{i(k_2+k_3)\cdot x(0)}  \ ,  
\end{equation}
yields the local contribution
\begin{equation}
\cA_{1234}^{\mathrm{II}}=2i \left( \epsilon_1 \cdot \epsilon_2 \, \epsilon_3 \cdot \epsilon_4-\epsilon_1 \cdot \epsilon_3 \, \epsilon_2 \cdot \epsilon_4 \right)\ .
\end{equation}
We can recast the following partial contribution in a clearer form through the products \eqref{C-inf prod}: 
\begin{equation}
    \cA_{1234}^{\mathrm{I}}+\cA_{1234}^{\mathrm{II}}=\frac{i}{s_{12}} \, m_{2}(\epsilon_1 ,\epsilon_2) \cdot m_{2}(\epsilon_3 ,\epsilon_4) +i\,\epsilon_2\cdot m_3(\epsilon_3,\epsilon_4,\epsilon_1)\ . 
\end{equation}
Finally, the third contribution \eqref{AIII} can be computed with the following pinch operator
\begin{equation}
  V_{23}^{\rm pinch}(0)=\frac{i}{s_{23}} m_{2 \,\mu}(\epsilon_2 ,\epsilon_3) \left(\dot{x}^\mu(0) +4 (k_2+k_3)_\nu \, \alpha^{[\nu}(0)\bar{\alpha}^{\mu]} (0) \, e^{i(k_2+k_3)\cdot x(0)} \right)\;.
\end{equation}
Since the scalar factor coming from the plane waves is $1$, this contribution reduces to
\begin{equation}
\begin{split}
&\left.\big\l 
\cdots\,V_{23}^{\rm pinch}(0)\,\cdots\big\r\right|_{\rm scal}=\frac{i}{s_{23}} \, \epsilon_1 \cdot \epsilon_4 \,  m_{2}(\epsilon_2 ,\epsilon_3) \cdot (k_4-k_1)\;,\\[3mm] 
&\left.\big\l 
\cdots\,V_{23}^{\rm pinch}(0)\,\cdots\big\r\right|_{\rm spin}=\frac{2i}{s_{23}} m_{2 \, \mu}(\epsilon_2 ,\epsilon_3) \left[ \epsilon_1 \cdot (k_2+k_3) \, \epsilon_4^\mu -\epsilon_4 \cdot (k_2+k_3) \, \epsilon_1^\mu  \right] \;, 
\end{split}    
\end{equation}
which yields the $t$-channel part of the amplitude
\begin{equation}
\begin{split}
\cA_{1234}^{\mathrm{III}}&=-\frac{1}{s_{23}} \, m_{2 \, \mu}(\epsilon_2 ,\epsilon_3) \left[\epsilon_1 \cdot \epsilon_4 \, (k_4-k_1)^\mu +2\epsilon_1 \cdot (k_2+k_3) \, \epsilon_4^\mu -2\epsilon_4 \cdot (k_2+k_3) \, \epsilon_1^\mu  \right]\\
&=\frac{i}{s_{23}} \, m_{2}(\epsilon_2 ,\epsilon_3) \cdot m_{2}(\epsilon_4 ,\epsilon_1)\;.
\end{split}    
\end{equation}
The final result is thus given by the sum of the three individual contributions, reproducing the expected expression
\begin{equation}
\cA_{1234}=\frac{i}{s_{12}}\,m_2(\epsilon_1,\epsilon_2)\cdot m_2(\epsilon_3,\epsilon_4)+\frac{i}{s_{23}}\,m_2(\epsilon_2,\epsilon_3)\cdot m_2(\epsilon_4,\epsilon_1)+i\,\epsilon_2\cdot m_3(\epsilon_3,\epsilon_4,\epsilon_1)\;.  
\end{equation}
This shows that the worldline connecting two color-adjacent gluons requires as many correlators as field theory Feynman diagrams. In addition, we have verified that the outcome does not depend on the choice of the main worldline.

\subsection{Path integral representation for Berends-Giele currents}  \label{chap:third:sec3.3}
As just seen in the latter case, the pinching contribution, where gluons 2 and 3 are fused into a cubic subtree, requires the introduction of the composite field
\begin{equation}\label{A23}
A_{23}^\mu(x)=\frac{1}{k_{23}^2}\,m_2^\mu(\epsilon_2,\epsilon_3)\,e^{ik_{23}\cdot x}  \;.  
\end{equation}
The form of $A_{23}^\mu$ is usually determined by the Bern-Kosower pinching rules, or from the field theory color-ordered Feynman rules. In this last part, we will show that the composite field can be computed directly from a worldline path integral, with different boundary conditions.

In order to describe nonlinear fields via worldline correlators, we follow the scalar example studied in Ref.~\cite{Bonezzi:2025iza}, and consider a semi-infinite worldline. We take the affine parameter $\tau\in(-\infty,0]$, so that the worldline has one real boundary and an asymptotic one. Since the proper length of the line is still infinite, there is no modulus for the einbein, which we gauge fix to $e(\tau)=2$. At the asymptotic past we want, as before, to project onto the zero-momentum vacuum $\ket{1}=\cB\ket{0}$, so that we can create an asymptotic gluon state by inserting $\cV_{\epsilon,k}(-\infty)$. The vacuum $\ket{1}$ corresponds to the asymptotic boundary conditions
\begin{equation}
\dot x^\mu(-\infty)=0\;,\quad\bar\alpha^\mu(-\infty)=0\;,\quad\bar\cB(-\infty)=\cB(-\infty)=b(-\infty)=0 \;.   
\end{equation}
At $\tau=0$ instead, we want to project onto a position eigenstate which, in order to support a vector field, is at the same time a coherent state for $\alpha^\mu$: $\bra{x,\alpha}c$. This bra state corresponds to the boundary conditions
\begin{equation}
x^\mu(0)=x^\mu\;,\quad\alpha^\mu(0)=\alpha^\mu\;,\quad \cB(0)=\cC(0)=c(0)=0\;.    
\end{equation}
Note that the appearance of the $c$ ghost in the bra state $\bra{x,\alpha}c$, corresponding to the boundary condition $c(0)=0$, is due to the absence of $c$ zero-modes on the semi-infinite line.  
The BRST invariant action with these boundary conditions is given by
\begin{equation}
S_{\rm DN}=\int_{-\infty}^0\!\!\!d\tau\,\Big[\tfrac14\,\dot x^2-i   \,\bar\alpha\cdot\dot\alpha+i\,b\dot c+i\,\bar\cB\dot\cC+i\,\bar\cC\dot\cB\Big]\;,    
\end{equation}
where the subscript \acr{DN} stands for Dirichlet--Neumann. 

Coherently with the boundary conditions, the only ghost that admits a zero mode is $\bar\cC$, so that the saturation requires $\big\l\bar\cC(\tau)\big\r=1$. To compute correlation functions, we expand the fields as backgrounds plus fluctuations
\begin{equation}
x^\mu(\tau)=x^\mu+z^\mu(\tau)\;,\quad\alpha^\mu(\tau)=\alpha^\mu+\kappa^\mu(\tau)\;,\quad\bar\alpha^\mu(\tau)=\bar\kappa^\mu(\tau)\;,    
\end{equation}
with vanishing (asymptotic) boundary conditions
\begin{equation}
z^\mu(0)=0\;,\quad\dot z^\mu(-\infty)=0\;,\quad\kappa^\mu(0)=0\;,\quad\bar\kappa^\mu(-\infty)=0\;.    
\end{equation}
These lead to the two-point functions
\begin{equation}
\begin{split}
\l z^\mu(\tau)\,z^\nu(\sigma)\r&=-i\,\eta^{\mu\nu}\,G_{\rm DN}(\tau,\sigma)\;,\quad \l \bar\kappa^\mu(\tau)\,\kappa^\nu(\sigma)\r=\eta^{\mu\nu}\,\theta(\tau-\sigma)\;,\\
G_{\rm DN}(\tau,\sigma)&=|\tau-\sigma|+(\tau+\sigma)\;,\quad \udot G_{\rm DN}(\tau,\sigma)=\text{sgn}(\tau-\sigma)+1\;,
\end{split}
\end{equation}
with the Dirichlet--Neumann propagator obeying $G_{\rm DN}(0,\sigma)=0$ and $\udot G_{\rm DN}(-\infty,\sigma)=0$.

As a first example, we compute the one-point function of an unintegrated vertex operator placed at $\tau=-T$, to be sent to infinity. The path integral over the fluctuations is normalized to one, and we denote correlation functions by $\l\cdots\r_{\rm DN}$ to remind us of the nontrivial boundary conditions. The correlator yields
\begin{equation}
\begin{split}
\Big\l\cV_{\epsilon,k}(-T)\Big\r_{\rm DN}&=e^{ik\cdot x}\Big\l\epsilon\cdot\big(\alpha+\kappa(-T)\big)\,e^{ik\cdot z(-T)}\bar\cC(-T)\Big\r_{\rm DN}\\
&=e^{ik\cdot x}\epsilon\cdot \alpha\Big\l e^{ik\cdot z(-T)}\bar\cC(-T)\Big\r_{\rm DN}=\alpha^\mu\epsilon_\mu\,e^{ik\cdot x}e^{-iTk^2}\;.
\end{split}    
\end{equation}
We see that, as for the infinite line, if the momentum of the asymptotic state is off-shell, the correlator vanishes in the limit $T\rightarrow\infty$, while for $k^2=0$ it gives $\alpha^\mu\epsilon_\mu\,e^{ik\cdot x}$, thus reproducing a linear on-shell state.

As a more interesting application, we try to reproduce the nonlinear field $A_{23}^\mu$ in \eqref{A23} by inserting $\cV_3(-\infty)$ and integrating $V_2(\tau)$ over the line. Since both describe asymptotic particles, the momenta are on-shell and their polarizations are transverse. Taking care of the ghost zero mode saturation, the correlator reads
\begin{equation}
\begin{split}
i\int_{-\infty}^0\!\!\!d\tau\,\Big\l V_2(\tau)\cV_{3}(-\infty)\Big\r_{\rm DN}&=i\,e^{ik_{23}\cdot x}\Big\l \Big(i\epsilon_2\cdot\dot z(\tau)+4ik_2^{[\mu}\epsilon_2^{\nu]}\big(\alpha_\mu+\kappa_\mu(\tau)\big)\bar\kappa_\nu(\tau)\Big)e^{ik_2\cdot z(\tau)}\\
&\phantom{=}\times\epsilon_3\cdot\big(\alpha+\kappa(-\infty)\big)e^{ik_3\cdot z(-\infty)}\bar\cC(-\infty)\Big\r_{\rm DN}   \;.
\end{split}    
\end{equation}
The contraction of the two plane waves yields the common factor
\begin{equation}
\big\l e^{ik_2\cdot z(\tau)}e^{ik_3\cdot z(-\infty)}\big\r_{z}=e^{ik_{23}^2\tau}\;,
\end{equation}
where the subscript refers to the path integral over $z$. The term proportional to $\alpha\cdot\epsilon_3$ contributes as
\begin{equation}
\begin{split}
i\alpha\cdot\epsilon_3\Big\l e^{ik_2\cdot z(\tau)}e^{ik_3\cdot z(-\infty)}i\epsilon_2\cdot\dot z(\tau)\Big\r_{z}&=-\alpha\cdot\epsilon_3\,\epsilon_2\cdot k_3\,e^{ik_{23}^2\tau}\udot G_{\rm DN}(\tau,-\infty)\\
&=-2\,\alpha\cdot\epsilon_3\,\epsilon_2\cdot k_3\,e^{ik_{23}^2\tau}\;,
\end{split}    
\end{equation}
while the term with $\epsilon_3\cdot\kappa(-\infty)$ yields
\begin{equation}
\begin{split}
-4  \,k_2^{[\mu}\epsilon_2^{\nu]}\Big\l e^{ik_2\cdot z(\tau)}e^{ik_3\cdot z(-\infty)}\big(\alpha_\mu+\kappa_\mu(\tau)\big)\bar\kappa_\nu(\tau)\,\epsilon_3\cdot\kappa(-\infty)\Big\r_{z,\kappa}=-4\,k_2^{[\mu}\epsilon_2^{\nu]}\,\alpha_\mu\epsilon_{3\nu}\,e^{ik_{23}^2\tau}\;.
\end{split}    
\end{equation}
Finally, integrating over $\tau$ we obtain
\begin{equation}
\begin{split} \label{BGtilde}
i\int_{-\infty}^0\!\!\!d\tau\,\Big\l V_2(\tau)\cV_{3}(-\infty)\Big\r_{\rm DN}&=\frac{2i}{k_{23}^2}\,\alpha_\mu\Big(\epsilon_2\cdot k_3\,\epsilon_3^\mu-\epsilon_3\cdot k_2\,\epsilon_2^\mu+k_2^\mu\epsilon_2\cdot\epsilon_3\Big)\,e^{ik_{23}\cdot x}\\&=\frac{1}{k_{23}^2}\,\alpha_\mu\Big(m_2^\mu(\epsilon_2,\epsilon_3)+i\,k_{23}^\mu\,\epsilon_2\cdot\epsilon_3\Big)\,e^{ik_{23}\cdot x}\\
&=\alpha_\mu\,\Big(A_{23}^\mu(x)+\del^\mu\lambda_{23}(x)\Big)=\alpha_\mu\,\tilde A_{23}^\mu(x)\;,
\end{split}    
\end{equation}
with the composite gauge parameter
\begin{equation}\label{composite lambda}
\lambda_{23}(x)=\frac{1}{k_{23}^2}\,\epsilon_2\cdot\epsilon_3\,e^{ik_{23}\cdot x}\;.   
\end{equation}
We have thus shown that the path integral with Dirichlet--Neumann boundary conditions reproduces the desired field $A_{23}^\mu$, modulo a linear gauge transformation.
We conclude by noting that, in the computation of the four-gluon amplitude, the difference between $\tilde A_{23}^\mu$ and $A_{23}^\mu$ drops from the correlation function, thus resulting in the same amplitude. 

\section{Ward identities}  \label{chap:third:sec4}
To establish that the amplitudes computed with this method are gauge invariant, we have to discuss the BRST symmetry of the path integral. Let us recall that to compute amplitudes we use the free action
\begin{equation}
S_{\text{free}}=\int_{-\infty}^{+\infty}\!\!\!d\tau\Big[\tfrac14\,\dot x^2-i\,\bar\alpha_\mu\dot\alpha^\mu+iB_i\dot C^i
\Big] \;,        
\end{equation}
with boundary conditions
\begin{equation}
\dot x^\mu(\pm\infty)=0\;,\quad \alpha^\mu(+\infty)=0\;,\quad\bar\alpha^\mu(-\infty)=0\;,\quad B_i(\pm\infty)=0\;.   
\end{equation}
The above action with these boundary conditions is invariant under the BRST transformations
\begin{equation}\label{BRST transf lagrangian}
\begin{split}
sx^\mu&=c\,\dot x^\mu-i(\cC\bar\alpha^\mu+\bar\cC\alpha^\mu)\;,\\
s\alpha^\mu&=\tfrac12\,\cC\dot x^\mu\;,\quad s\bar\alpha^\mu=-\tfrac12\,\bar\cC\dot x^\mu\;,\\
sc&=i\,\cC\bar\cC\;,\quad s\cC=0\;,\quad s\bar\cC=0\;,\\
sb&=\tfrac{i}{4}\,\dot x^2\;,\quad s\cB=\tfrac12\,\alpha\cdot\dot x-i\,\cC b\;,\quad s\bar\cB=\tfrac12\,\bar\alpha\cdot\dot x+i\,\bar\cC b\;.
\end{split}    
\end{equation} These transformations can be modified by adding trivial terms that vanish on-shell as, for instance, $c\dot c$ which can be added to $sc$. In \eqref{BRST transf lagrangian} we chose the simplest form, with no such terms. The BRST differential so defined is nilpotent only on-shell: $s^2\approx0$, where by $\approx$ we denote equality up to equations of motion. This is expected from the gauge-fixed theory in Lagrangian form. If one reintroduces momenta $p_\mu$, the differential $s$ can be made nilpotent off-shell. The Noether charge corresponding to the BRST symmetry is given by
\begin{equation}
Q=-\tfrac14\,c\,\dot x^2+\tfrac{i}{2}\,(\cC\bar\alpha+\bar\cC\alpha)\cdot\dot x-\cC\bar\cC\, b\;,    
\end{equation}
which yields the free BRST operator \eqref{Q free} upon canonical quantization. In the path integral formulation, one can use the charge $Q$ inside correlation functions to generate BRST transformations. More specifically, given a functional $F$, its BRST transformation is given by the equal time commutator
\begin{equation}
[Q,F](\tau)=i\big(sF\big)(\tau)\;.    
\end{equation}

\subsection{Operator product expansion and equal-time commutators} \label{chap:third:sec4.1}
To discuss equal time commutators and other operator relations in the path integral formulation, we start with a brief detour into the operator product expansion (\acr{OPE}). Similar to string theory, we will call operator product the product of functions, viewed inside correlators in the path integral:
\begin{equation}
F(\tau)\,G(\sigma):=\big\l\cdots F(\tau)\,G(\sigma)\cdots\big\r\;. 
\end{equation}
These are then computed using Wick contractions and the basic two-point functions. The simplest example is the \acr{OPE} of two $z^\mu$ fluctuations, yielding
\begin{equation}
z^\mu(\tau)\,z^\nu(\sigma)=-i\,\eta^{\mu\nu}\,G(\tau,\sigma)+:z^\mu(\tau)\,z^\nu(\sigma):\;.    
\end{equation}
The normal ordering symbol amounts to the prescription of not performing Wick contractions within it. This immediately leads to $\l\,:\!F(z)\!:\,\r=F(0)$, where $F(z)$ is an analytic function of $z^\mu$. Another simple example is the \acr{OPE} of $\dot z^\mu$ (related on shell to the momentum operator via $p^\mu=\frac12\,\dot z^\mu$) with a normal-ordered function of $z$:
\begin{equation}\label{zdot F OPE}
\dot z^\mu(\tau)\,:F\big(z(\sigma)\big):\;=-i\,\left(\text{sgn}(\tau-\sigma)-\frac{2\tau}{T}\right):\del^\mu F\big(z(\sigma)\big):+ :\dot z^\mu(\tau)F\big(z(\sigma)\big):\;,    
\end{equation}
where we used the two-point function $\udot G(\tau,\sigma)=\text{sgn}(\tau-\sigma)-2\tau/T$.
The \acr{OPE} can then be used to define equal time commutators via the regularized prescription
\begin{equation}\label{Equal time commutator}
\begin{split}
[A,B](\tau)&=\lim_{\epsilon\rightarrow0}\Big(A(\tau+\epsilon)-A(\tau-\epsilon)\Big)\,B(\tau)=\lim_{\epsilon\rightarrow0}\int_{\tau-\epsilon}^{\tau+\epsilon}\!\!\!\!\!\!d\sigma\;\dot A(\sigma)\,B(\tau)\\
&=\lim_{\epsilon\rightarrow0}A(\tau)\,\Big(B(\tau-\epsilon)-B(\tau+\epsilon)\Big)=-\lim_{\epsilon\rightarrow0}\int_{\tau-\epsilon}^{\tau+\epsilon}\!\!\!\!\!\!d\sigma\; A(\tau)\,\dot B(\sigma)\;,
\end{split}    
\end{equation}
where the equivalence of the two definitions complies with the (graded) antisymmetry of the commutator.
An important point of the above definition is that one takes the limit $\epsilon\rightarrow0$ \emph{after} computing the \acr{OPE} at separate points. In particular, this assumes that the \emph{only} operators in the interval $[\tau-\epsilon,\tau+\epsilon]$ are $A$ and $B$, for otherwise the result would not yield just the commutator $[A,B](\tau)$. This is an important subtlety that will play a role in the following. 

As an example, we can compute the commutator $[\dot z^\mu,:\!F(z)\!:]$. Using \eqref{zdot F OPE} and the definition for the commutator we obtain
\begin{equation}
[\dot z^\mu,:\!F(z)\!:](\tau)=-2i\,:\del^\mu F\big(z(\tau)\big):+\lim_{\epsilon\rightarrow0}\Big(:\dot z^\mu(\tau+\epsilon)F\big(z(\tau)\big):-:\dot z^\mu(\tau-\epsilon)F\big(z(\tau)\big):\Big)\;.    
\end{equation}
Naively, $\dot z^\mu(\tau+\epsilon)=\dot z^\mu(\tau)$, since $\ddot z^\mu(\tau)\approx0$ on-shell. Inside correlation functions, $\ddot z^\mu(\tau)$ contributes only to contact terms, since $\ddot z(\tau)\,z(\sigma)\sim\delta(\tau-\sigma)$. Now, it is important that in the regularized definition \eqref{Equal time commutator} one assumes that no other operator sits in the interval $[\tau-\epsilon,\tau+\epsilon]$. This ensures that $\lim_{\epsilon
\rightarrow0}:\dot z^\mu(\tau+\epsilon)F\big(z(\tau)\big):$ is regular, finally yielding
\begin{equation}
[\dot z^\mu,:\!F(z)\!:](\tau)=-2i\,:\del^\mu F\big(z(\tau)\big): \;,   
\end{equation}
which is the expected result. One can similarly define the operator product at coincident points via Weyl ordering:
\begin{equation}
(AB)(\tau):=\lim_{\epsilon\rightarrow0}\,\frac12\,\Big(A(\tau+\epsilon)+A(\tau-\epsilon)\Big)\,B(\tau)\;,    
\end{equation}
which reproduces the \acr{OPE} computed with the two-point functions at coincident points. For instance, taking the coincidence limit of \eqref{zdot F OPE}, one has
\begin{equation}\label{Weyl ordered dot z F}
\dot z^\mu(\tau)\,:F\big(z(\tau)\big):\;=\frac{2i\tau}{T}\,:\del^\mu F\big(z(\tau)\big):+ :\dot z^\mu(\tau)F\big(z(\tau)\big):\;.   
\end{equation}

A second useful example is the commutator of $\dot z^2$ with functions of $z$, since $\frac14\,\dot z^2$ corresponds to the gauge-fixed Hamiltonian. Let us mention that one has to use the normal ordered expression $:\dot z^2(\tau):$ to avoid divergences from self-contractions. Keeping this in mind, we omit explicit normal ordering symbols in the initial functions. To compute the commutator, we evaluate the \acr{OPE} at separate points:
\begin{equation}
\frac14\,\dot z^2(\tau\pm\epsilon)\,F\big(z(\tau)\big)=-\frac14\,\left(\pm1-\frac{2\tau}{T}\right)^2:\square F(z):-\frac{i}{2}\,\left(\pm1-\frac{2\tau}{T}\right)\,:\dot z^\mu\del_\mu F(z):+\frac14\,:\dot z^2F(z):    
\end{equation}
where on the right-hand side all $z^\mu$ are evaluated at $\tau$. Above, we have already taken the limit $\epsilon\rightarrow0$, taking into account that it is regular inside normal ordered expressions. The commutator is readily evaluated as
\begin{equation}
\tfrac14\,[\dot z^2,F(z)]=\frac{2\tau}{T}\,:\square F(z):-i:\dot z^\mu\del_\mu F(z):\;,    
\end{equation}
where, again, all $z^\mu$ are $z^\mu(\tau)$. The above expression may look somewhat unfamiliar because of the $\square F$ contribution linear in time. However, looking at the result \eqref{Weyl ordered dot z F} one can see that this is exactly the Weyl ordered expression at coincident points, so that
\begin{equation}
\tfrac14\,[\dot z^2,F(z)]=-i\,\dot z^\mu\del_\mu F(z)=-i\,\frac{dF}{d\tau}\;,    
\end{equation}
which is the Heisenberg equation of motion.

\subsection{BRST cohomology of vertex operators} \label{chap:third:sec4.2}
We can now show that on-shell vertex operators with transverse gluons are elements of the BRST cohomology. We start by showing that vertex operators with polarization $\epsilon_\mu=ik_\mu$ are trivial, manifesting the on-shell residual gauge symmetry. Recall the unintegrated vertex operator $\cV_{\epsilon,k}(\tau)$:
\begin{equation}
\begin{split}
\cV_{\epsilon,k}(\tau)&=W_{\epsilon,k}(\tau)+c(\tau)\,V_{\epsilon,k}(\tau)\;,\\
W_{\epsilon,k}(\tau)&=\epsilon_\mu\Big(\cC(\tau)\,\bar\alpha^\mu(\tau)+\bar\cC(\tau)\,\alpha^\mu(\tau)\Big)\,e^{ik\cdot x(\tau)}\;,\\
V_{\epsilon,k}(\tau)&=i\,\epsilon_\mu\Big(\dot x^\mu(\tau)+2\,k\cdot\alpha(\tau)\,\bar\alpha^\mu(\tau)-2\,k\cdot\bar\alpha(\tau)\,\alpha^\mu(\tau)\Big)\,e^{ik\cdot x(\tau)}\;.
\end{split}
\end{equation}
Taking the polarization to be proportional to momentum, we have
\begin{equation}
\cV_{ik,k}(\tau)=i\big(\cC\bar\alpha+\bar\cC\alpha\big)\cdot k\,e^{ik\cdot z}-c\,k\cdot\dot z\,e^{ik\cdot z}\;,    
\end{equation}
where we have removed the zero mode $x_0$, which in amplitudes only gives momentum conservation. Using the \acr{OPE} as described in the previous section, we see that $\cV_{ik,k}(\tau)$ is BRST-exact: 
\begin{equation}
\cV_{ik,k}(\tau)=[Q,e^{ik\cdot z}](\tau) \;.   
\end{equation}
In a similar fashion, the integrated vertex operator $V_{ik,k}(\tau)$ is a total derivative:
\begin{equation}
V_{ik,k}(\tau)=i\,\frac{d}{d\tau}\left(e^{ik\cdot z(\tau)}\right)\;.    
\end{equation}

Coming now to BRST closure, we compute the $Q$-commutator of $\cV_{\epsilon,k}$. Using the on-shell condition and transversality, we obtain
\begin{equation}\label{QV OPE}
\{Q,\cV_{\epsilon,k}\}=-3\,:f_{\mu\nu}(z)\,\alpha^\mu\bar\alpha^\nu\cC\bar\cC:-2i\,:c\,(\cC\bar\alpha+\bar\cC\alpha)\cdot k\,f_{\mu\nu}(z)\,\alpha^\mu\bar\alpha^\nu:\;,    
\end{equation}
where $f_{\mu\nu}(z)=2i\,k_{[\mu}\epsilon_{\nu]}e^{ik\cdot z}$. This is nothing but the functional version of \eqref{QV}. We will now show that the right-hand side above vanishes inside all correlation functions. In particular, we want to prove that
\begin{equation}\label{Normal order vanishing}
\big\l\cdots:F(\tau)\bar\alpha^\mu(\tau)\bar\cC(\tau):\cdots\big\r=0\;,\quad\l\cdots:F(\tau)\bar\alpha^\mu(\tau)\bar\alpha^\nu(\tau):\cdots\r=0\;,  
\end{equation}
for any functional $F(\tau)$ and with arbitrary insertions of vertex operators outside the normal ordering. This is the path integral version of the statement, discussed in sec. \ref{chap:third:sec2.3}, that normal ordered operators with more than one barred oscillator annihilate every state in the $\cN=1$ sector of the Hilbert space.

To prove the first relation, we first notice that the equation of motion $\del_\tau{\bar\cC}=0$ holds everywhere on the line, since there are no insertions of the conjugate antighost $\cB$. We can then push the ghost to the infinite past $\bar\cC(\tau)=\bar\cC(-\infty)$, where it saturates its zero mode. Now we write $\bar\alpha^\mu(\tau)$ as
\begin{equation}
\bar\alpha^\mu(\tau)=\int_{-\infty}^\tau d\sigma\,\del_\sigma\bar\alpha^\mu(\sigma)\;,    
\end{equation}
using the boundary condition $\bar\alpha^\mu(-\infty)=0$. The integral above is zero almost everywhere, since the equation of motion $\del_\tau\bar\alpha^\mu=0$ holds except at insertions of $\alpha^\nu$, where it gives a contraction with a delta function. At this stage, we need to invoke the $U(1)$ invariance of the vertex operators: every insertion of $\alpha^\nu$ necessarily comes with either a ghost $\bar\cC$, or another $\bar\alpha^\rho$. In the first case, the insertion does not contribute, since there is already a factor of $\bar\cC$. In the second case, the above integral produces a factor of $\bar\alpha^\rho$ at the insertion point. One then starts the same procedure with $\bar\alpha^\rho$ until it reaches the asymptotic past, which concludes the proof. An analogous reasoning can be used to prove the second equation in \eqref{Normal order vanishing}.

Since \eqref{Normal order vanishing} holds in all correlation functions, we use it as an operator equation in the sense of \acr{OPE}.  
We have thus shown that, inside correlation functions, 
 the unintegrated vertex operators are $Q$-closed:
\begin{equation}
\{Q,\cV_{\epsilon,k}\}=0\;.    
\end{equation}
Using the same arguments, in particular that equations of motion hold when regularizing commutators by point splitting, one shows that the $Q$-commutator of the integrated vertex operator is a total derivative:
\begin{equation}\label{Q Vintegrated}
[Q,V_{\epsilon,k}]=i\,\frac{d}{d\tau}\,\cV_{\epsilon,k}\;.    
\end{equation}
In the following, we will show how to use these relations to prove Ward identities.

\subsection{Ward identities from BRST invariance} \label{chap:third:sec4.3}
To see how to prove Ward identities, we start from the somewhat trivial example of the three-point amplitude $\cA_{123}$.
For transverse gluons, the amplitude \eqref{A123}
\begin{equation}
\cA_{123}=i\,\Big[\epsilon_1\cdot\epsilon_2\,\epsilon_3\cdot(k_1-k_2)+\epsilon_2\cdot\epsilon_3\,\epsilon_1\cdot(k_2-k_3)+\epsilon_3\cdot\epsilon_1\,\epsilon_2\cdot(k_3-k_1)\Big]\;,
\end{equation}
vanishes when taking any polarization $\epsilon^\mu_i$ to be proportional to the corresponding momentum $k^\mu_i$, which is the Ward identity. In the worldline formulation, the amplitude is given by the correlator
\begin{equation}
\cA_{123}=\big\l\cV_1(+\infty)\,\cV_2(0)\,\cV_3(-\infty)\big\r\;.    
\end{equation}
If we take the particle 1 to have $\epsilon^\mu_1=ik^\mu_1$, the corresponding vertex operator is exact: $\cV_1(\tau)=[Q,e^{ik_1\cdot z}](\tau)$. The asymptotic vacuum $\ket{1}$ is BRST invariant, which translates to the boundary condition $Q(\pm\infty)=0$. Since the vertex operator $\cV_1$ is inserted at $T/2\rightarrow+\infty$, the equal time commutator reduces to
\begin{equation}
\cV_1(T/2)=-\lim_{\epsilon\rightarrow0}e^{ik_1\cdot z(T/2)}Q(T/2-\epsilon)\;.    
\end{equation}

The idea is to “move” the BRST charge to the asymptotic past, where it annihilates the vacuum. To do so, we use the fact that $Q$ is conserved, so that $\dot Q=0$ except at insertions of other operators. Since there is no operator insertion until $\tau=0$, we have $Q(T/2-\epsilon)=Q(\tau_*)$ for any $\tau_*\in(0,T/2-\epsilon]$. To move $Q$ across $\cV_2(0)$ we pick a commutator contribution at $\tau=0$. After that, $Q(\tau)$ remains constant for negative times until it approaches $\cV_3(-T/2)$: 
\begin{equation}
Q(T/2-\epsilon)\,\cV_2(0)=-\cV_2(0)\,Q(-T/2+\epsilon)+\{Q,\cV_2\}(0)=-\cV_2(0)\,Q(-T/2+\epsilon)\;,    
\end{equation}
upon using the closure of $\cV_2(0)$. Since $\cV_3$ is at the boundary, and $Q(-\infty)=0$, approaching it from the left coincides with the commutator:
\begin{equation}
\lim_{\epsilon\rightarrow0}Q(-T/2+\epsilon)\,\cV_3(-T/2)=\{Q,\cV_3\}(-T/2)=0\;,
\end{equation} 
which proves the Ward identity $\cA_{k_123}=0$.

For the next example, we consider the four-point amplitude, computed with the smart worldline:
\begin{equation}
\cA_{1234}=\int_{-\infty}^{+\infty}\!\!\!\!\!\!d\tau \,\Big\l\cV_2(+\infty)\,\cV_3(0)\,V_1(\tau)\,\cV_4(-\infty)\Big\r\;.   
\end{equation}
Upon taking $\epsilon_2^\mu=ik_2^\mu$ we have again that the corresponding vertex operator is $Q$-exact:
\begin{equation}
\cV_2(T/2)=-\lim_{\epsilon\rightarrow0}e^{ik_2\cdot z(T/2)}Q(T/2-\epsilon)\;.    
\end{equation}
We can use the same arguments to push $Q$ to the asymptotic past, except when it has to go through the integrated vertex $V_1(\tau)$, picking a contribution from the commutator. Upon regularizing the asymptotic time with $T/2\rightarrow+\infty$ we obtain
\begin{equation}
\begin{split}
-\lim_{\epsilon\rightarrow0}\,&\Big\l e^{ik_2\cdot z(T/2)}Q(T/2-\epsilon) \,\cV_3(0)\,V_1(\tau)\,\cV_4(-T/2)\Big\r\\
&=\Big\l e^{ik_2\cdot z(T/2)}\cV_3(0)\,[Q,V_1](\tau)\,\cV_4(-T/2)\Big\r=i\,\Big\l e^{ik_2\cdot z(T/2)}\cV_3(0)\,\dot\cV_1(\tau)\,\cV_4(-T/2)\Big\r \;,  
\end{split}    
\end{equation}
where we used \eqref{Q Vintegrated}. Upon integrating over $\tau$, we end up with
\begin{equation}
\cA_{1k_234}=\lim_{T\rightarrow\infty}i\,\Big\l e^{ik_2\cdot z(T/2)}\cV_3(0)\,\big(\cV_1(T/2)-\cV_1(-T/2)\big)\,\cV_4(-T/2)\Big\r   
\end{equation}
where the notation $\cA_{1k_234}$ highlights the use of a longitudinal polarization for particle 2.
Regardless of the detailed structure, the above expression vanishes in the limit $T\rightarrow\infty$, because it has off-shell plane waves at asymptotic times. To see this, let us focus on the first term above with $\cV_1(T/2)$. The Wick contraction of the plane waves gives
\begin{equation}
\big\l e^{i(k_1+k_2)\cdot z(T/2)}e^{ik_3\cdot z(0)}e^{ik_4\cdot z(-T/2)} \big\r=\exp\left\{-\tfrac{i}{2}\,(k_1+k_2)^2T\right\}\xrightarrow{T\rightarrow\infty}0\;,    
\end{equation}
since $(k_1+k_2)^2=s\neq0$. We thus see that the Ward identity $\cA_{1k_234}=0$ holds, thanks to the unrestricted integration range of $V_1(\tau)$. From this discussion, one can also see that off-shell correlation functions (for which $T$ is a modulus) do not obey linear Ward identities.

\subsubsection{Ward identity for the color-adjacent worldline}
Using the not-so-smart worldline connecting particles 1 and 4, the four-gluon color-ordered amplitude requires three separate correlation functions. Recalling that every vertex (except the ones at infinity) requires an additional factor of $i$, these are given by \eqref{AI}, \eqref{AII} and \eqref{AIII}
\begin{align}
    \cA_{1234}^{\mathrm{I}}&=(i)^2\Big\l\cV_1(+\infty) \, \int_{0}^{+\infty} {\hskip -6mm d\tau}\, V_2(\tau) \, \cV_3(0) \, \cV_4(-\infty)\Big\r\;,\\
    \cA_{1234}^{\mathrm{II}}&= i\,\Big\l\cV_1(+\infty) \, \cV_{23}(0)\, \cV_4(-\infty)
\Big\r\;, \\
    \cA_{1234}^{\mathrm{III}}&= i\,\Big\l\cV_1(+\infty) \, \cV_{23}^{\rm pinch}(0)\, \cV_4(-\infty) \Big\r\;.
\end{align}
As we have discussed in the previous section, the second contribution $\cA_{1234}^{\mathrm{II}}$ comes from the insertion of the quadratic vertex operator
\begin{equation}\label{V23}
\cV_{23}(\tau)=4\,c(\tau)\,\epsilon^{[\mu}_{2}\epsilon_{3}^{\nu]}\alpha_\mu(\tau)\,\bar\alpha_\nu(\tau)\,e^{i(k_2+k_3)\cdot x(\tau)}   \;,
\end{equation}
corresponding to the quartic vertex. The third contribution, instead, comes from “pinching” the gluons 2 and 3 into a cubic subtree attached to the worldline. This is done by inserting an unintegrated vertex operator \eqref{Vertex operators}, corresponding to a nonlinear field with momentum $k_{23}^\mu=k^\mu_2+k^\mu_3$ and composite polarization
\begin{equation}
\epsilon_{23}^\mu=\frac{1}{k_{23}^2}\,m_2^\mu(\epsilon_2,\epsilon_3)\;,   
\end{equation}
written in terms of the product \eqref{C-inf prod}.

The necessity for the second and third contributions to the amplitude can be understood from studying its Ward identities. To this end, we take the first gluon to have the pure gauge polarization $\epsilon_1^\mu=ik_1^\mu$, so that $\cV_1=[Q,e^{ik_1\cdot x}]$. Following the procedure described before, the three correlation functions contribute as
\begin{equation}
\begin{split}
\cA_{k_1234}^{\mathrm{I}}&=i\int_{0}^{\infty}\!\!\!\!d\tau\,\Big\l e^{ik_1\cdot z(+\infty)}\,\frac{d}{d\tau}\cV_2(\tau) \, \cV_3(0) \, \cV_4(-\infty)\Big\r=-i\,\Big\l e^{ik_1\cdot z(+\infty)}\,\cV_2(0) \, \cV_3(0) \, \cV_4(-\infty)\Big\r\;,\\
\cA_{k_1234}^{\mathrm{II}}&= -i\,\Big\l e^{ik_1\cdot z(+\infty)} \,\{Q,\cV_{23}(0)\}\, \cV_4(-\infty)
\Big\r\;,\\
\cA_{k_1234}^{\mathrm{III}}&= -i\,\Big\l e^{ik_1\cdot z(+\infty)} \,\{Q,\cV_{23}^{\rm pinch}(0)\}\, \cV_4(-\infty)
\Big\r\;.
\end{split}    
\end{equation}
Compared to the smart worldline, the first correlator does not vanish, since the $\tau$ integral receives a boundary contribution from $\cV_2(\tau)$ colliding with $\cV_3(0)$. Instead of computing the full correlation functions, we will determine the \acr{OPE} $\cV_2(0)\cV_3(0)$ at coinciding points, and show that it cancels $\{Q,\cV_{23}(0)+\cV_{23}^{\rm pinch}(0)\}$. 

For later convenience, let us rewrite the unintegrated vertex operator in the compact form
\begin{equation}
\cV_i=\epsilon^\mu_{i}\big(S_\mu+i\,c\,\dot x_\mu\big)e^{ik_i\cdot x}+2i\,c\,k^\mu_i\epsilon^\nu_i\,S_{\mu\nu}\,e^{ik_i\cdot x}\;,    
\end{equation}
where the time dependence is left implicit and we have introduced the ghost-number one vector $S^\mu$ and the Lorentz spin generator $S^{\mu\nu}$, defined by
\begin{equation}
S^\mu=\bar\cC\alpha^\mu+\cC\bar\alpha^\mu\;,\quad S^{\mu\nu}=\alpha^\mu\bar\alpha^\nu-\alpha^\nu\bar\alpha^\mu \;.   
\end{equation}
Notice that, for $k_i^2=0$ and $k_i\cdot\epsilon_i=0$, the vertex operator is normal ordered.
In computing the \acr{OPE}, we will leave equal-time products of $\dot x^\mu$ and functions of $x^\mu$ as they are, without taking contractions. This corresponds to the Weyl ordering of the related operators. For instance, the product $e^{ik_i\cdot x(\tau)}e^{ik_j\cdot x(\tau)}=e^{i(k_i+k_j)\cdot x(\tau)}$ is Weyl ordered, not normal ordered, unless the propagator $G(\tau,\tau)$ vanishes. For equal-time products of $\alpha^\mu$ and $\bar\alpha^\nu$ we will use instead the \acr{OPE} at coincident times
\begin{equation}
\bar\alpha^\mu(\tau)\,\alpha^\nu(\tau)=\frac12\,\eta^{\mu\nu}+:\bar\alpha^\mu(\tau)\,\alpha^\nu(\tau):    \;,
\end{equation}
to take advantage of the fact that normal ordered expressions with more than one barred field vanish in all correlation functions due to charge conservation, c.f. \eqref{Normal order vanishing}. Some useful examples are
\begin{equation}
\begin{split}
S^\mu S^\nu&=-:\cC\bar\cC(\alpha^\mu\bar\alpha^\nu-\alpha^\nu\bar\alpha^\mu):\;=0 \;,\\
S^{\mu\nu}S^\rho&=\tilde S^{[\mu}\eta^{\nu]\rho}+:S^{\mu\nu} S^\rho:\;= \tilde S^{[\mu}\eta^{\nu]\rho}\;,
\end{split}    
\end{equation}
where in the final results we have discarded all terms vanishing inside correlators, and we have defined a second ghost-number one vector
\begin{equation}
\tilde S^\mu= \bar\cC\alpha^\mu-\cC\bar\alpha^\mu\;.   
\end{equation}

Armed with these relations, we can determine the equal-time \acr{OPE} of the two vertex operators $\cV_2(0)$ and $\cV_3(0)$. Omitting the time dependence, we have
\begin{equation}
\begin{split}
\cV_2\cV_3&=\Big(\epsilon^\mu_{2}\big(S_\mu+i\,c\,\dot x_\mu\big)+2i\,c\,k^\mu_2\epsilon^\nu_2\,S_{\mu\nu}\Big)\Big(\epsilon^\rho_{3}\big(S_\rho+i\,c\,\dot x_\rho\big)+2i\,c\,k^\rho_3\epsilon^\sigma_3\,S_{\rho\sigma}\Big)e^{ik_{23}\cdot x} \\
&=2i\,c\Big(\epsilon^{[\mu}_{2}\epsilon^{\nu]}_{3}\dot x_\mu S_\nu+k_2^{[\mu}\epsilon_2^{\nu]}\tilde S_\mu\epsilon_{3\nu}-k_3^{[\mu}\epsilon_3^{\nu]}\tilde S_\mu\epsilon_{2\nu} \Big)e^{ik_{23}\cdot x}\\
&=c\,\tilde S_\mu\,m_2^\mu(\epsilon_2,\epsilon_3)\,e^{ik_{23}\cdot x}+2i\,c\,\epsilon^{[\mu}_{2}\epsilon^{\nu]}_{3}\Big(\dot x_\mu S_\nu+\tilde S_\mu k_{23\nu}\Big)e^{ik_{23}\cdot x}\;,
\end{split}    
\end{equation}
The quadratic vertex operator \eqref{V23} can be written as
\begin{equation}
\cV_{23}=2\,c\,\epsilon^\mu_2\epsilon^\nu_3\,S_{\mu\nu}\,e^{ik_{23}\cdot x}\;,
\end{equation}
and its equal-time commutator with $Q$ can be computed as $\{Q,\cV_{23}\}=is(\cV_{23})$, where the BRST differential $s$ on the worldline variables is given by \eqref{BRST transf lagrangian}. Its action on functions is defined by the Leibniz rule, with the resulting expressions corresponding to Weyl ordering. Discarding again all terms with two barred fields in normal ordering, we obtain
\begin{equation}
is(\cV_{23})=-2i\,c\,\epsilon^{[\mu}_{2}\epsilon^{\nu]}_{3}\Big(\dot x_\mu S_\nu+\tilde S_\mu k_{23\nu}\Big)e^{ik_{23}\cdot x}  \;,  
\end{equation}
yielding the \acr{OPE} result
\begin{equation}\label{OPE intermediate}
\cV_2\cV_3+\{Q,\cV_{23}\}=c\,\tilde S_\mu\,m_2^\mu(\epsilon_2,\epsilon_3)\,e^{ik_{23}\cdot x} \;,   
\end{equation}
where all operators above are evaluated at equal times. 

At this stage, the missing ingredient to establish the Ward identity is to prove that the right-hand side of \eqref{OPE intermediate} is $Q$-exact and, in particular, given by
\begin{equation}\label{final Ward}
c\,\tilde S_\mu\,m_2^\mu(\epsilon_2,\epsilon_3)\,e^{ik_{23}\cdot x}=-\{Q,\cV_{23}^{\rm pinch}\}\;.   
\end{equation}
To this end, we consider the vertex operator for an off-shell field
\begin{equation}\label{Voffshell}
\cV_A=A^\mu(k)\big(S_\mu+i\,c\,\dot x_\mu\big)e^{ik\cdot x}+2i\,c\,k^\mu A^\nu(k)\,S_{\mu\nu}\,e^{ik\cdot x}\;,    
\end{equation}
where in general $k^2\neq0$ and $k\cdot A(k)\neq0$. Its $Q$-commutator is readily computed, yielding
\begin{equation}\label{MaxwellQ}
\{Q,\cV_A\}=c\,\tilde S^\mu\big(-k^2A_\mu+k_\mu k\cdot A\big)\,e^{ik\cdot x} \;.   
\end{equation}
For the case at hand, $\cV_{23}^{\rm pinch} := \cV_{A_{23}}$ is of the form \eqref{Voffshell} for a composite field $A_{23}^\mu$, where the momentum $k_{23}^\mu=k_2^\mu+k_3^\mu$ is off-shell and the polarization is given by
\begin{equation}
\epsilon_{23}^\mu=\frac{1}{k_{23}^2}\,m_2^\mu(\epsilon_2,\epsilon_3) := \frac{i}{k_{23}^2}\,\big(\epsilon_2\cdot k_3\,\epsilon_3^\mu-\epsilon_3\cdot k_2\,\epsilon_2^\mu+(k_2-k_3)^\mu\,\epsilon_2\cdot\epsilon_3\big)\;.
\end{equation}
One can easily check that $k_{23}\cdot m_2(\epsilon_2,\epsilon_3)=0$ (this is essentially the Ward identity for the three-point case), so that \eqref{MaxwellQ} yields
\begin{equation}
\begin{split}
\{Q,\cV_{23}^{\rm pinch}\}&=c\,\tilde S_\mu\big(-k_{23}^2\epsilon_{23}^\mu+k_{23}^\mu k_{23}\cdot\epsilon_{23}\big)\,e^{ik_{23}\cdot x}=c\,\tilde S_\mu\big(-k_{23}^2\epsilon_{23}^\mu\big)\,e^{ik_{23}\cdot x}\\
&=-c\,\tilde S_\mu\,m_2^\mu(\epsilon_2,\epsilon_3)\,e^{ik_{23}\cdot x}\;,    
\end{split}
\end{equation}
thus proving \eqref{final Ward} and hence the Ward identity. 

Let us conclude by emphasizing that the above discussion demonstrates how the form of the composite field \eqref{A23} can also be inferred (albeit in a somewhat roundabout way) by demanding the Ward identities. Moreover, as expected, the Ward identity remains unchanged whether one uses $A_{23}^\mu$ or its gauge-transformed version $\tilde A_{23}^\mu$ \eqref{BGtilde}, since the two fields differ by a gauge transformation:
\begin{equation}
\{Q,\cV_{\tilde A_{23}}\}=\{Q,\cV_{A_{23}}\}+\{Q,[Q,\lambda_{23}]\}=\{Q,\cV_{A_{23}}\}\;,    
\end{equation}
where $\lambda_{23}(x)$ is given by \eqref{composite lambda} and we have used $\{Q,Q\}=0$ together with the graded Jacobi identity of (anti)commutators.

\section{Final remarks} \label{chap:third:sec5}
In this chapter, we have quantized the bosonic spinning particle on the open worldline, with the aim of providing a first-quantized description of gluon scattering amplitudes. To this end, we considered an open worldline of infinite proper length \cite{Daikouji:1995dz,
Laenen:2008gt, Bonocore:2020xuj, Mogull:2020sak}, with boundary conditions corresponding to zero-momentum asymptotic vacua. In this way, all external gluons (including the two at the asymptotic endpoints of the line) are described by insertions of suitable vertex operators, thus mimicking the worldsheet techniques of string perturbation theory. In particular, the infinite length of the line allowed us to bypass the LSZ procedure \cite{Mogull:2020sak,Bonezzi:2025iza}, in that the correlation functions already provide the on-shell and amputated amplitudes. Contrary to string theory, however, linear vertex operators are not sufficient to compute arbitrary amplitudes. In general, one needs nonlinear vertex operators, describing the insertion of entire subtrees on the main worldline. While this is typically achieved by using the so-called Bern-Kosower pinching rules \cite{Bern:1990cu,Bern:1991an,Bern:1992ad,Schubert:2001he}, here we have argued that these nonlinear vertex operators can be obtained as well from worldline path integrals on a semi-infinite open geometry \cite{Bonezzi:2025iza}. Finally, to test our results, we have shown how the Ward identities of the scattering amplitudes descend from the BRST invariance of the worldline correlators (see Ref.~\cite{Ahmadiniaz:2020htz} for a related discussion).

The framework presented here should be extended to arbitrary multiplicities and, possibly, to loop amplitudes. In particular, one would like to be able to prove gauge invariance of arbitrary amplitudes by extending the BRST analysis introduced here to general correlation functions. Besides the point of principle, having a first-quantized description of Yang--Mills amplitudes could help in shedding light on the so-called duality between color and kinematics in gauge theories \cite{Bern:2008qj,Bern:2019prr}, which underpins the double copy construction of gravity amplitudes \cite{Bern:2010ue,Bern:2022wqg,Adamo:2022dcm}. More specifically, it would prove beneficial to connect the off-shell algebraic approach of Refs.~\cite{Bonezzi:2022bse,Bonezzi:2024emt,Bonezzi:2024fhd} to the worldline techniques employed in Refs.~\cite{Ahmadiniaz:2021fey,Ahmadiniaz:2021ayd}, where an algorithm combining worldline integration by parts and Bern-Kosower pinching rules was shown to produce gauge theory numerators obeying color-kinematics duality.
Since the bosonic spinning particle can also accommodate gravitons in its spectrum, it would be interesting to extend the present framework to the spin-two sector of the theory. There, one expects to need multi-graviton vertices of arbitrary order, but the linear ones should essentially be the product of two gluon vertex operators, thus showing a worldline incarnation of the double copy, along the lines of Refs.~\cite{Shi:2021qsb,Bastianelli:2021rbt}.

\addcontentsline{toc}{chapter}{Appendix}
\section*{Appendix}
\begin{subappendices}

\section{Path integral in phase space} \label{chap:third:Appendix:A}
To study the propagator directly in momentum space, we wish to consider the path integral representation of the matrix element 
 \begin{equation}
\la p' | e^{-i HT} |p\ra = \int Dp Dx \  e^{iS}\;,
\label{app-pi} 
\end{equation}
where one could take $H= p^2$ as needed in the main text. We study the simpler  case 
$H=0$, which already captures all the essential aspects of our discussion. 
For simplicity, we consider just one target space dimension. 
Then, the left-hand side of \eqref{app-pi} reduces to 
\be
\la p' |p\ra = 2\pi \delta (p-p')\;,
\label{app-2}
\ee
which should be reproduced by the path integral on the right-hand side. To take into account the 
boundary conditions, that should be imposed on $p(\tau)$ only, the action takes the form
\be
S =  \int_0^T d\tau\,(- \dot p x ) \;.
\ee
Of course, one could perform the path integral by first integrating over $p$ and then over $x$, which works well 
for nontrivial quadratic Hamiltonians, like the free one, $H= p^2$. This is well-known.
Here we wish to path integrate in $x$ first, leaving the path integral on $p$ as the last one. 
To do this, we notice that the paths $x(\tau)$ should not contain any boundary conditions, so that we can 
expand them by extracting the constant path $x_0$. That is, we parametrize
\be
x(\tau)= x_0 +\tilde x(\tau) 
\ee 
with $\int_0^T d\tau\, \tilde x(\tau) =0$. 
Then, we can write
\be
\int Dp\, D\tilde x\, dx_0 \  e^{-i\int_0^T \! d\tau\,  \dot p x} = \int Dp\,  D\tilde x\, 
\underbrace{dx_0\,
\  e^{-i x_0 \int_0^T \! d\tau\,  \dot p}}_{\sim \delta(p-p')}
 e^{ -i \int_0^T \! d\tau\, \dot p  \tilde x  }\;,  
 \ee
where $p$ and $p'$ denote the boundary conditions that must be imposed on $p(\tau)$.
Of course, $\int_0^T \! d\tau\,  \dot p = p'-p$.
Then, path integrating over $\tilde x(\tau)$ produces a delta functional $\delta(\dot p) $
 leading to
\be
\int Dp\, D x \  e^{-i\int_0^T \! d\tau\, \dot p x} 
\ \sim \
\delta(p-p')
\int Dp\, \delta(\dot p) 
\ \sim \ 
\text{Det}(\partial_\tau) 
\delta(p-p') 
\ \sim \ 
2\pi \delta(p-p')\;,
 \ee
where the combined unfixed constants have been chosen to match the expected result in \eqref{app-2}.
This way of computing the path integral allows us to bypass configuration space and obtain 
results directly in momentum space. Presumably, perturbation theory can be introduced by building further 
on the above expression. 

Let us also briefly outline the derivation of the other “phase-space” path integral involved in our derivation, the one for the bosonic oscillators, 
\begin{align} \label{A.7}
Z(\alpha,\bar\alpha,\theta)=\int_{\bar\alpha(0)=\bar\alpha}^{\alpha(1)=\alpha} \!\!\!D\bar\alpha D\alpha\ e^{\int_0^1d\tau\, \bar\alpha^\mu(\partial_\tau-i\theta)\alpha_\mu(\tau)+\bar\alpha^\mu\alpha_\mu(0)}= e^{e^{-i\theta}\alpha\cdot\bar\alpha}\ ,
\end{align}
which was, for example, discussed and used in Ref. \cite{Ahmadiniaz:2015xoa}. It corresponds to the usual (Euclidean) path integral for the harmonic oscillator, though with an imaginary frequency $\omega= - i\theta$.

For $\theta=0$, the latter reduces to the path integral representation of the scalar product of the coherent states,
\begin{align}
    \langle \alpha|\bar\alpha\rangle = e^{\alpha\cdot \bar\alpha} = \int_{\bar\alpha(0)=\bar\alpha}^{\alpha(1)=\alpha} \!\!\!D\bar\alpha D\alpha\ e^{\int_0^1d\tau\, \bar\alpha^\mu\dot\alpha_\mu(\tau)+\bar\alpha^\mu\alpha_\mu(0)}\;,
\end{align}
which can be used to fix the path integral normalization as follows. Let us split the fields into  background parts -- which satisfy the equations of motion $\dot\alpha^\mu =\dot{\bar\alpha}^\mu=0$ and the boundary conditions -- and quantum fluctuations as
\begin{align}
    &\alpha^\mu(\tau)=\alpha^\mu +\kappa^\mu(\tau)\,,\quad \kappa^\mu(1)=0\;,\\
&\bar\alpha^\mu(\tau)=\bar\alpha^\mu +\bar\kappa^\mu(\tau)\,,\quad \bar\kappa^\mu(0)=0\ .
\end{align}
Thus, we get
\begin{align}
    e^{\alpha\cdot \bar\alpha} =e^{\alpha\cdot \bar\alpha} \int_{\bar\kappa(0)=0}^{\kappa(1)=0} \!\!\!D\bar\kappa D\kappa\ e^{\int_0^1d\tau\, \bar\kappa^\mu\dot\kappa_\mu(\tau)}\ ,
\end{align}
i.e.,
\begin{align}
\int_{\bar\kappa(0)=0}^{\kappa(1)=0} \!\!\!D\bar\kappa D\kappa\ e^{\int_0^1d\tau\, \bar\kappa^\mu\dot\kappa_\mu(\tau)}=1\ .
\label{eq:PI-normalization}
\end{align}
We can now apply the background splitting to the $\theta$-dependent path integral. We have,
\begin{align}
&\alpha^\mu(\tau)=e^{i\theta\tau} \big(e^{-i\theta} \alpha^\mu +\kappa^\mu(\tau)\big)\,,\quad \kappa^\mu(1)=0\;,\\
&\bar\alpha^\mu(\tau)=e^{-i\theta\tau} \big(\bar\alpha^\mu +\bar\kappa^\mu(\tau)\big)\,,\quad \bar\kappa^\mu(0)=0\ ,
\end{align}
and, in turn,
\begin{align}
Z(\alpha,\bar\alpha,\theta)=e^{e^{-i\theta}\alpha\cdot\bar\alpha}\int_{\bar\kappa(0)=0}^{\kappa(1)=0} \!\!\!D\bar\kappa D\kappa\ e^{\int_0^1d\tau\, \bar\kappa^\mu\dot\kappa_\mu(\tau)} = e^{e^{-i\theta}\alpha\cdot\bar\alpha}\ .
\label{A.15}
\end{align}
One may consult Ref. \cite{Ahmadiniaz:2015xoa} for further details.
\end{subappendices}

\newpage
\thispagestyle{empty}
\mbox{}
\newpage

\chapter*{Spin 2}
\addcontentsline{toc}{chapter}{\color{turquoise}\raisebox{0.5ex}{\rule{7.75cm}{0.4pt}} Spin 2 \raisebox{0.5ex}{\rule{7.75cm}{0.4pt}}}
\chapter{BRST quantization of the $\N=4$ spinning particle}\label{chap:fourth}
\textit{In this chapter, we turn to the description of both massless and massive spin-2 particles within the Worldline Formalism, with the aim of illustrating one of the contexts where the method arguably proves most powerful: the perturbative evaluation of the Seeley–DeWitt coefficients. As a preliminary step towards these computations, we devote the present chapter to the BRST quantization of the worldline model known as the $\N=4$ supersymmetric spinning particle, with and without a mass. The massless case is reviewed following Ref.~\cite{Bonezzi:2018box}, both to introduce the model and to highlight the subtleties relevant for the calculations of Chapter~\ref{chap:sixth}. We then move on to the massive case, presenting in detail the model employed in the computations of Chapter~\ref{chap:fifth}. Our treatment serves a twofold purpose: first, to explore different mechanisms for endowing spinning particle models with mass; and second, to formulate a first-quantized description of linearized massive gravity, both in flat and in curved backgrounds. In the latter case, we find that the nilpotency of the BRST charge is preserved only on Einstein spacetimes with vanishing cosmological constant, which thus emerges as the unique consistent background. This will be the starting point for the computations developed in the following chapters.}

\paragraph{Conventions} The target space of our worldline model is a $d$-dimensional Minkowski spacetime, equipped with the metric $\eta_{\mu\nu}=\mathrm{diag}(-,+,\cdots,+)$, which is used to raise and lower spacetime indices. Greek letters $\mu,\nu,\dots$ denote spacetime indices $(\mu,\nu=0,1,\dots,d-1)$, while capital Latin letters $I,J,\dots$ are reserved for internal $SO(\N)$ indices $(I,J=1,\dots, \N)$.\\
We further employ the following notation for the graded commutators:
\begin{equation*}
[\cdot,\cdot\}=
\begin{cases*}
\{\cdot,\cdot\} \quad \text{anticommutator if both variables are fermionic,} \\
[\cdot,\cdot] \quad \;\; \text{commutator otherwise.}
\end{cases*}
\end{equation*}

\section{$O(\N)$ massless spinning particles} \label{chap:fourth:sec1}
Before focusing on the $\mathcal{N}=4$ spinning particle, we first recall some general features of spinning particle models, which have already been discussed in the \hyperref[chap:WF]{introductory section on the Worldline Formalism}. In this section, we summarize the relevant equations for two main reasons: first, to facilitate comparison with the forthcoming massive case; and second, to re-establish our notation, since, unlike in the previous chapters, the spin degrees of freedom are now described by Grassmann-odd (fermionic) variables.

The phase space action of such models depends on the particle spacetime coordinates $x^\mu$ together with ${\cal N}$ real fermionic superpartners $\Xi^\mu_I$, and the phase-space action is given by
\begin{equation} \label{chap:fourth:action}
S=\int d\tau \left[p_\mu\dot x^{\mu}+\frac{i}{2}\Xi_{\mu}\cdot\dot\Xi^{\mu}-e\,H-i\chi^I\,q_I-a^{IJ}\,J_{IJ}\right]\ , 
\end{equation}
The canonical coordinates $(x^\mu, p_\mu, \Xi^\mu_I)\,$ upon quantization, are subject to the canonical commutation relations (CCR)
\begin{equation} \label{CR}
[x^\mu, p_\nu]=i\,\delta^\mu_\nu\ ,\quad \{\Xi^\mu_I, \Xi^\nu_J\}=\delta_{IJ}\,\eta^{\mu\nu}\ . 
\end{equation}
The worldline supergravity multiplet in one dimension $(e,\chi^I, a_{IJ})$ contains the einbein $e$, the $\N$ gravitinos $\chi$, and the gauge field $a$. On the other hand, the first-class constraints are represented by the Hamiltonian $H$, the supercharges $q_I$, and the $R$-symmetry algebra generators $J_{IJ}$. They can be expressed as
\begin{equation} \label{const}
H:=\frac{1}{2} \, p^\mu p_\mu\ , \quad q_I:=\Xi_I^\mu \, p_\mu\ , \quad J_{IJ}:=i\,\Xi_{[I}^\mu \, \Xi_{J]_\mu}\ .
\end{equation} 
Hamiltonian and supercharges together form the \acr{SUSY} algebra
\begin{equation} \label{alg1}
\{q_I, q_J\}=2\,\delta_{IJ}\,H\ ,\quad [q_I, H]=0\ .
\end{equation}
While the $R$-symmetry algebra reads
\begin{align}
\begin{split} \label{alg2}
[J_{IJ}, q_K] &= i \left( \delta_{JK}q_I -\delta_{IK}q_J \right)\ , \\
[J_{IJ}, J_{KL}] &= i \left( \delta_{JK}J_{IL} -\delta_{IK}J_{JL} -\delta_{JL}J_{IK} +\delta_{IL}J_{JK} \right)\ .
\end{split}
\end{align}
In the following, we will refer only to \eqref{alg1} as “$\N$ \acr{SUSY} algebra”, since it will be appropriate to distinguish the two algebras in light of the future BSRT procedure. It is rather convenient to work with a complex redefinition of the original fermionic variables $\Xi_I\to(\xi_i, \bar\xi^i)$, namely
\begin{equation}
\xi^\mu_i:=\tfrac{1}{\sqrt2}(\Xi^\mu_i+i\,\Xi^\mu_{i+\nicefrac{\N}{2}})\ , \quad \bar\xi^{\mu i}:=\tfrac{1}{\sqrt2}(\Xi^\mu_i-i\,\Xi^\mu_{i+\nicefrac{\N}{2}})\ , \quad \text{with} \; i=1,\dots,\tfrac{\N}{2}
\end{equation}
for the case of even $\N$, i.e. for particles with integer spin. The respective CCR \eqref{CR} become
\begin{equation}
\{\bar\xi^i_\mu, \xi_j^\nu\}=\delta^i_j\,\delta^\nu_\mu\ , \quad \{\bar\xi^i_\mu, \bar\xi^j_\nu\}=0=\{\xi_i^\mu, \xi_j^\nu\}\ . 
\end{equation}
Accordingly, both the supercharges and the $R$-symmetry generators split under the complex redefinition: we report the explicit expressions for the particle models of interest, namely the spin one and the spin two. By choosing a specific Fock vacuum, an arbitrary state $\ket{\Psi}$ in the Hilbert space is isomorphic to the wavefunction $\Psi(x,\xi)$, on which the conjugated momenta act as derivatives 
\begin{equation}
p_\mu=-i\partial_{\mu}\ , \quad \bar\xi_\mu=\frac{\partial}{\partial\xi^\mu}\ .
\end{equation}
Thus, for the case $\N=2$, the spin one model, one has the splitting $q_I \to (q, \bar q)$ and $J_{IJ} \to \mathrm{J}$ as follows
\begin{equation} \label{J}
\begin{array}{l}
q:=-i\,\xi^\mu\partial_\mu \\[2mm]
\bar{q}:=-i\,\partial^\mu\frac{\partial}{\partial\xi^\mu}
\end{array}
\ , \qquad \mathrm{J}:=\xi^\mu \frac{\partial}{\partial \xi^\mu} -\frac{d}{2}\ ,
\end{equation}
where the shift $-\tfrac{d}{2}$ in the definition of the $R$-symmetry generator is a quantum effect stemming from an antisymmetric ordering for the Grassmann variables \cite{Bastianelli:2005vk}. For the $\N=4$ case, the spin two, one has $q_I \to (q_i, \bar q^i)$ and the $SO(4)$-symmetry generators are realized maintaining manifest covariance only under a $u(2)$-subalgebra $J_{IJ} \to (\mathrm{J}_i{}^j, \mathrm{Tr}^{ij}, \mathrm{G}_{ij})$, explicitly realized as\footnote{In the following, whether a dot $\cdot$ indicates contractions on internal or spacetime indices should be clear within the context.}
\begin{equation} \label{JJ}
\begin{array}{l}
q_i:=-i\,\xi_i^\mu\partial_\mu \\[2mm]
\bar q^i:=-i\,\partial^\mu\frac{\partial}{\partial\xi_i^\mu} 
\end{array}
\ , \qquad 
\begin{array}{l}
\mathrm{J}_i{}^j:=\xi_i \cdot \frac{\partial}{\partial \xi_j} -\tfrac{d}{2}\, \delta_i^j \\[2mm]
\mathrm{Tr}^{ij}:= \frac{\partial^2}{\partial \xi_i \cdot \partial \xi_j} \\[2mm]
\mathrm{G}_{ij}:= \xi_i\cdot \xi_j
\end{array}\ .
\end{equation}
Note that the only surviving components of the $\mathrm{Tr}$ and $\mathrm{G}$ operators are those with $i \neq j$.

\section{BRST quantization} \label{chap:fourth:sec2}
Although BRST quantization was already discussed in Chapter~\ref{chap:second}, we shall review it here, since the presence of fermionic rather than bosonic variables modifies the analysis in several non-trivial respects.\footnote{We refer to \href{https://www.youtube.com/watch?v=e0QiR829ORo&list=PLx5f8IelFRgGfQMuGGOuqHtRaelyJFj0u&index=17}{this} talk by Ivo~Sachs for a brilliant exposition of the methodology \cite{SachsSeminar}.}

Let us start with the massless $\N=4$ spinning particle with a four-dimensional target space. The internal indices take values $i=1,2$ and the $\N=4$ worldline \acr{SUSY} algebra is explicitly realized as
\begin{align} \label{alg}
\{q_i, \bar q^j\}=2\,\delta_{i}^{j}\,H\ ,\quad [q_i, H]=[\bar q^i, H]=\{q_i, q_j\}=\{\bar q^i, \bar q^j\}=0\ .
\end{align}
The initial step of the BRST procedure entails an enlargement of the Hilbert space to realize ghost-antighost pairs of operators $(g^\alpha, P_\alpha)$ associated with each constraint $C_\alpha$, with opposite Grassmann parity of the latter -- that is, anticommuting ghosts for bosonic constraints and commuting ghosts for fermionic constraints -- and canonical graded commutation relation
\begin{equation} \label{CIA}
[ P_\alpha,g^\beta \}=\delta^\beta_\alpha\ .
\end{equation}
Hence, we assign the fermionic pair $(c,b)$ to the Hamiltonian, and the bosonic superghost pairs $(\bar\gamma^i, \beta_i)$ and $(\gamma_i,\bar\beta^i)$ to the supercharges $q_i$ and $\bar q^i\,$ respectively, obeying
\begin{equation} \label{3.4}
\{b,c\}=1\ , \quad [\beta_i,\bar\gamma^j]=\delta_i^j\ , \quad [\bar\beta^j,\gamma_i]=\delta_i^j\ , 
\end{equation}
with ghost number assignments
\begin{align}
&{\gh(c,\gamma_i,\bar\gamma^i)=+1}\ , \\ 
&{\gh(b,\beta_i,\bar\beta^i)=-1}\ . 
\end{align}
The second step consists of constructing the BRST operator $\Q$. It is realized as follows \cite{VanHolten:2001nj}
\begin{equation}\label{def}
\Q:=g^\alpha C_\alpha-\frac12(-1)^{\epsilon(g_\beta)} g^\beta g^\alpha f_{\alpha\beta}{}^{\gamma}P_\gamma\ ,
\end{equation}
where $\epsilon(g_\beta)$ is the Grassmann parity\footnote{We employ the following convention: the parity of $g^\alpha$ is $0$ if $g^\alpha$ is Grassmann even and $1$ if $g^\alpha$ is Grassmann odd.} of the ghost $g^\beta$. Equation \eqref{def} holds exactly when the structure functions $f_{\alpha\beta}{}^{\gamma}$ are constant,\footnote{
Recall that the constraints satisfy upon quantization a graded algebra of the form
\begin{equation*}
[ C_\alpha, C_\beta \}=f_{\alpha\beta}{}^{\gamma}C_{\gamma}
\end{equation*}
which defines the structure functions $f_{\alpha\beta}{}^{\gamma}$.} but its validity extends to more general settings as well \cite{Fecit:2025eet}. It can be derived by demanding the following properties for the BRST operator:
\begin{itemize}
\item It has to be anticommuting and of ghost number $+1$.
\item It has to act on the operators, corresponding to the original phase space variables prior to quantization, as the gauge transformations with the ghost variables replacing the gauge parameters. This, together with the latter requirement, is enough to constrain the structure to be $\Q=g^\alpha C_\alpha + \dots$
\item Finally, the BRST charge has to be nilpotent. This determines the second structure of \eqref{def} as can be checked by direct computation of $\{\Q,\Q\}$. 
\end{itemize}
Note that the BRST charge is nilpotent \emph{by construction} as long as the associated algebra is first-class. In more general cases, higher-order terms may appear and need to be determined by the nilpotency condition. Specializing to the $\N=4$ case, the BRST operator associated with the first-class algebra \eqref{alg} is
\begin{equation} \label{QQ}
\Q= c\,H+\gamma_i\,\bar q^i+\bar\gamma^i\, q_i-2\bar\gamma^i\gamma_i\, b\ .
\end{equation}
To verify its nilpotency, it is first convenient to define $\bm{\nabla}:=\gamma \cdot \bar q+\bar\gamma \cdot q$, such that for any operator of the form \eqref{QQ} one always finds the following potential obstruction-terms to its nilpotency
\begin{equation} \label{sì}
\Q^2= -2 \, \bar\gamma \cdot\gamma\,H+\bm{\nabla}^2-2c\,\cancel{\left[H,\bm{\nabla}\right]}\ , 
\end{equation}
which are vanishing in the present case since 
\begin{equation} \label{no}
\bm{\nabla}^2=\gamma_i\bar \gamma^j \{\bar q^i, q_j \} \implies -2 \, \bar\gamma \cdot\gamma\,H+\bm{\nabla}^2=0\ .
\end{equation}
Notice the critical role played by the presence of the first-class algebra \eqref{alg} in both the cancellation of \eqref{sì} and of \eqref{no}. The same does not hold anymore in more general cases. In the next section, we will illustrate how to address this issue. \\

The careful reader might wonder about the absence of a set of ghosts associated with the $so(4)$ constraints. The key idea of the BRST procedure, as first developed in Ref.~\cite{Dai:2008bh} providing a first-quantized description of the gluon from a “fermionic” worldline model, is to treat the $R$-symmetry constraints and the \acr{SUSY} ones on different footings. The formers are imposed as constraints on the BRST Hilbert space, thus defining precisely the general dependence of the string field $\Psi$ on the spacetime fields content.\footnote{Regarding terminology, we deliberately confuse “string field” and “BRST wavefunction” since $\Psi$ plays the same role in the worldline theory as it would in string field theory: its expansion as a linear combination of first-quantized states display coefficients which correspond to ordinary particle fields \cite{Zwiebach:1992ie, Gomis:1994he, Thorn:1988hm}.} It is within this restricted Hilbert space that the cohomology of the BRST charge has to be studied. The procedure remains consistent as long as the $R$-symmetry constraints \eqref{JJ} commute with the BRST charge: in order to achieve that, it is necessary to extend $(\mathrm{J}, \mathrm{Tr}, \mathrm{G})\to (\J, \gTr, \gG)$ as follows 
\begin{equation}
\begin{split}
\J_i{}^j &:= \xi_i\cdot\bar\xi^j+\gamma_i\bar\beta^j-\bar\gamma^j\beta_i-2\,\delta_i^j \ ,\\[2mm]
\gTr^{ij} &:= \bar\xi^i\cdot\bar\xi^j+\bar\gamma^i\bar\beta^j - \bar\gamma^j\bar\beta^i\ ,\\[2mm]
\gG_{ij} &:= \xi_i\cdot\xi_j+\gamma_i\beta_j - \gamma_j\beta_i\ .
\end{split} 
\end{equation}
The relevant $so(4)$ generators to be imposed as constraints on the BRST Hilbert space are the \emph{two number operators} $\J_i:=\J_i{}^i$ ($i$ not summed), namely the diagonal entries of $\J_i{}^j$, the \emph{Young antisymmetrizer} $\Y:=\J_1^2$, and finally the \emph{trace} $\gTr$, which implement the maximal reduction of the model. We collectively denote these constraints as $\mathcal{T}_\alpha:=(\J,\Y,\gTr)$. The BRST system is defined as follows
\begin{equation}\label{BRSTsystem}
\begin{split}
& \Q \, \Psi=0\;,\quad \delta\Psi=\Q \, \Lambda\;,\\[2mm]
&{\cal T}_\alpha \, \Psi=0\;,\quad {\cal T}_\alpha \, \Lambda=0\;,
\end{split} 
\end{equation}
and its consistency is ensured by $[\Q,\mathcal{T}_\alpha]=0$. This is indeed equivalent to saying that we are studying the cohomology of $\Q$ on the restricted Hilbert space of fixed $R$-charge defined by $\mathcal{H}_{\mathrm{red}}:=\ker\mathcal{T}_\alpha$. The system above will serve as the starting point for BRST quantization in both the massless and massive cases.

\subsection{Massless graviton on a flat spacetime}
The ghost vacuum $\ket{0}$ is chosen such that it is annihilated by $(b,\bar\gamma^i,\bar\beta^i)$, so that a general state $\ket{\Psi}$ in the BRST extended Hilbert space is isomorphic to the wavefunction $\Psi(x,\xi\,|\,c,\gamma,\beta)$, on which the antighosts act as derivatives, i.e. 
\begin{equation}
b=\frac{\partial}{\partial c}\ , \quad \bar\gamma^i=-\frac{\partial}{\partial \beta_i}\ , \quad \bar \beta^i=\frac{\partial}{\partial \gamma_i}\ .
\end{equation}
In the following, we collectively denote the ghost oscillators as $\g:=(c,\gamma,\beta)$. The BRST operator, making explicit the d'Alembertian $\Box=\eta^{\mu\nu}\partial_\mu\partial_\nu$ while adjusting the coefficients, acts as 
\begin{equation} \label{chap:fourth:Q}
\Q=c\,\Box+\gamma_i\bar q^i-q_i\,\frac{\partial}{\partial\beta_i}-\gamma_i\,\frac{\partial^2}{\partial\beta_i\partial c}\ ,
\end{equation}
while, regarding the $so(4)$ generators
\begin{equation}
\begin{split} \label{rel}
\J_i &= N_{\xi_i}+N_{\gamma_i}+N_{\beta_i}-1\ , \\[2mm]
\Y &= \xi_1\cdot\frac{\partial}{\partial\xi_2}+\gamma_1\frac{\partial}{\partial\gamma_2}+\beta_1\frac{\partial}{\partial\beta_2}\ ,\\[2mm]
\gTr &= \frac{\partial^2}{\partial\xi_1\cdot\partial\xi_2}+\frac{\partial^2}{\partial\gamma_1\partial\beta_2}-\frac{\partial^2}{\partial\gamma_2\partial\beta_1}\ ,
\end{split} 
\end{equation}
where we have explicitly expressed the number operators, counting the number of oscillators with a fixed flavor index, and where we solved ambiguities in the definition of $\J_i$ by choosing a symmetric ordering for the ghosts. \\

It is useful for future reference to highlight the intermediate steps of the calculation. The first condition imposed by $\Psi \in \mathcal{H}_{\mathrm{red}}$, namely $\J_i \ket{\Psi}=0$, significantly reduces the components of $\Psi(x,\xi\,|\,\g)$ to
\begin{align} \label{massless}
\Psi(x,\xi\,|\,\g)&=
a_{\mu\nu}\,\xi^\mu_1\xi^\nu_2+b_{\mu}\,\xi_1^\mu\gamma_2+C_{\mu}\,\xi_1^\mu\beta_2+d_{\mu}\,\gamma_1\xi_2^\mu+e\,\gamma_1\beta_2+f_{\mu}\,\beta_1\xi_2^\mu \nonumber \\[1mm] 
&\phantom{=}+g\,\beta_1\gamma_2+k\,\gamma_1\gamma_2+l\,\beta_1\beta_2 \nonumber \\[1mm]
&\phantom{=}+a^{\ast}_{\mu\nu}\,\xi^\mu_1\xi^\nu_2c+b^{\ast}_{\mu}\,\xi_1^\mu\gamma_2c+C^{\ast}_{\mu}\,\xi_1^\mu\beta_2c+d^{\ast}_{\mu}\,\gamma_1\xi_2^\mu c+e^{\ast}\,\gamma_1\beta_2 c+f^{\ast}_{\mu}\,\beta_1\xi_2^\mu c \nonumber \\[1mm] 
&\phantom{=}+g^{\ast}\,\beta_1\gamma_2c+k^{\ast}\,\gamma_1\gamma_2c+l^{\ast}\,\beta_1\beta_2 c\ .
\end{align}
The latter expression is essentially a Taylor expansion in powers of $c$ taking the form $\Psi=\chi+\chi^{\ast}c$, where both terms $\chi$ and $\chi^{\ast}$ of the wavefunction contain oscillators and spacetime-dependent field components. Requiring the entire wavefunction to have fixed Grassmann parity and fixed ghost number forces the field components in $\chi^{\ast}$ to have opposite parities and ghost numbers decreased by one compared to those in $\chi$. It is important to note that we are not yet in the position to correctly identify (anti)fields: we first need to reduce $\Psi$ to include only the \acr{BV} spectrum of the theory. Imposing the two remaining conditions we obtain
\begin{align}
\begin{split} \label{cond1}
&b_\mu =-d_\mu\ , \quad C_\mu=-f_\mu\ , \quad e=-g\ ,\quad k=0\ , \quad l=0\ , \\
&b^{\ast}_\mu =-d^{\ast}_\mu\ , \quad C^{\ast}_\mu=-f^{\ast}_\mu\ , \quad e^{\ast}=-g^{\ast}\ ,\quad k^{\ast}=0\ , \quad l^{\ast}=0\ ,
\end{split}
\end{align}
together with $a_{\mu\nu}=a_{\nu\mu}$ using $\Y \ket{\Psi}=0$, while from $\gTr \ket{\Psi}=0$ we get
\begin{equation} \label{cond2}
e=\tfrac{1}{2}a^\mu_\mu\ , \quad e^{\ast}=\tfrac{1}{2}a^{\ast \mu}_\mu\ .
\end{equation}
Now we can indeed interpret $\Psi$ as a spacetime Batalin--Vilkovisky string field, that contains the whole minimal \acr{BV} spectrum of pure gravity along with auxiliary fields: we assign Grassmann parity and ghost number to the component fields such that the entire wavefunction $\Psi$ has total even parity and ghost number zero. It becomes explicit with the following identifications for the fields
\begin{align} \label{renam}
a_{\mu\nu} \longrightarrow h_{\mu\nu} \quad a^{\mu}_{\mu} \longrightarrow h^{\mu}_{\mu}=:h \quad C^{\ast}_{\mu} \longrightarrow v_{\mu} \quad C_{\mu} \longrightarrow \zeta_{\mu}\ ,
\end{align}
and for the corresponding antifields
\begin{align}
a^{\ast}_{\mu\nu} \longrightarrow h^{\ast}_{\mu\nu} \quad b_{\mu} \longrightarrow v^{\ast}_{\mu} \quad b^{\ast}_{\mu} \longrightarrow \zeta^{\ast}_{\mu}\ .
\end{align}
Grassmann parities and ghost numbers can be read from Table~\ref{table}. The most general state in $\mathcal{H}_{\mathrm{red}}$ is then
\begin{align} \label{BVstringfield}
\Psi(x,\xi\,|\,\g) &= h_{\mu\nu}(x)\,\xi^\mu_1\xi^\nu_2+\tfrac12\,h(x)\,(\gamma_1\beta_2-\gamma_2\beta_1)-\tfrac{i}{2}\,v_\mu(x)\,(\xi^\mu_1\beta_2-\xi^\mu_2\beta_1)c \nonumber\\[1mm]
&\phantom{=}-\tfrac{i}{2}\,\zeta_\mu(x)\,(\xi^\mu_1\beta_2-\xi^\mu_2\beta_1) \nonumber\\[2mm]
&\phantom{=}+h^{\ast}_{\mu\nu}(x)\,\xi^\mu_1\xi^\nu_2c+\tfrac12\,h^{\ast}(x)\,(\gamma_1\beta_2-\gamma_2\beta_1)c-\tfrac{i}{2}\,v^{\ast}_\mu(x)\,(\xi^\mu_1\gamma_2-\xi^\mu_2\gamma_1) \nonumber\\[1mm]
&\phantom{=}-\tfrac{i}{2}\,\zeta^{\ast}_\mu(x)\,(\xi^\mu_1\gamma_2-\xi^\mu_2\gamma_1)c\ . 
\end{align}
It contains the graviton $h_{\mu\nu}$ and its trace $h$, an auxiliary vector field $v_\mu$, and the diffeomorphism ghost $\zeta_\mu$, while the remaining components are the corresponding antifields.

\begin{table}[!ht]
\centering
\begin{tabular}{ |c|c|c|c| } 
 \hline
 \acr{BV} role & Field & Grassmann parity & Ghost number \\
 \hline
 massless graviton & $h_{\mu\nu}$ & $0$ & $0$ \\
 trace & $h$ & $0$ & $0$ \\
 auxiliary vector & $v_\mu$ & $0$ & $0$ \\ 
 diffeomorphism ghost & $\zeta_\mu$ & $1$ & $1$ \\
 \hline
\end{tabular}
\caption{List of fields in the physical sector of the \emph{massless} $\N=4$ model with the corresponding ghost number and Grassmann parity.}
\label{table}
\end{table}
\vspace{1ex}
The final step involves evaluating the field equations of the theory, to verify that the model correctly reproduces a first-quantized representation of linearized gravity. This can be accomplished through the BRST closure equation
\begin{equation}
\Q\,\Psi(x,\xi\,|\,\g)=0
\end{equation} at ghost number zero, which produces the massless free spin two field equation
\begin{equation} \label{chap:fourth:free}
\Box h_{\mu\nu}-2\,\partial_{(\mu}\partial\cdot h_{\nu)}+\partial_\mu\partial_\nu h =0\ , 
\end{equation}
and
\begin{equation} \label{free2}
\Box h -\partial^\mu \partial^\nu h_{\mu\nu}=0\ ,
\end{equation}
which implies a vanishing linearized Ricci scalar. To conclude this section, one is left to verify the accurate reproduction of the gauge symmetry: this is achieved from the ghost number zero part of
\begin{equation}
\delta\Psi=\Q\,\Lambda\ ,
\end{equation}
where $\Lambda \in \ker\mathcal{T}_\alpha$ contains the gauge parameters of the associated symmetry while having the same functional form as $\Psi$, with overall odd parity and ghost number $-1$, i.e. 
\begin{equation}\label{Lambdaparameter}
\Lambda=i\varepsilon_\mu(x)\,(\xi^\mu_1\beta_2-\xi^\mu_2\beta_1)+\cdots\ ,
\end{equation}
where the gauge parameter $\varepsilon_\mu$ has even parity and ghost number zero. As expected, the result is
\begin{equation}
\delta h_{\mu\nu}=2\partial_{(\mu} \varepsilon_{\nu)}\ .
\end{equation}
For future reference, the closure equation at ghost number one is reported below:
\begin{equation} \label{dimenticavo}
\Box \, \zeta_\mu=0\;, \quad \partial_{(\mu} \, \zeta_{\nu)}=0\ .
\end{equation}
Equations \eqref{dimenticavo} represent a set of equations of motion for the diffeomorphism ghost $\zeta_\mu(x)$.

\subsection{Massless graviton on a curved background} \label{chap:fourth:sec3.2}
The consistent coupling of the spinning particles to more general backgrounds beyond flat spacetime is a rather delicate matter \cite{Getzler:2016fek}. In previous works, a quantization \emph{à la} Dirac has been extensively employed. However, when the \acr{SUSY} algebra fails to be first-class this method loses its validity, potentially leading to the misleading conclusion that quantization is not feasible in more general cases. As already mentioned, this limitation was evident in the case of the $\N=4$ spinning particle, for which only certain restricted backgrounds were found to be viable until it was realized that the technique of BRST quantization offers a way to explore more general backgrounds, as recently discussed in Refs.~\cite{Boffo:2022egz, Grigoriev:2021bes}. In this section, we review the exploration of such possibility.

The coupling of the massless $\N=4$ spinning particle to a curved background with metric $g_{\mu\nu}(x)$ is realized by the covariantization of the derivatives, i.e.
\begin{equation}
\hat \nabla_\mu:=\partial_\mu+\omega_{\mu\, ab}\,\xi^a\cdot\bar\xi^b\ ,\quad \text{with} \quad [\hat \nabla_\mu, \hat \nabla_\nu]=R_{\mu\nu\lambda\sigma}\,\xi^\lambda\cdot\bar\xi^\sigma\ , 
\end{equation}
where fermions carry flat Lorentz indices so that $\xi^\mu_i:=e^\mu_a(x)\,\xi^a_i$, introducing a background vielbein $e_\mu^a(x)$ and the torsion-free spin connection $\omega_{\mu\, ab}$. The covariant derivative operator $\hat \nabla_\mu$ reproduces the effect of the usual covariant (partial) derivative $\nabla_\mu$ ($\partial_\mu$) on the tensorial (scalar) components contained in the wavefunction \eqref{BVstringfield}, e.g.
\begin{equation}
\hat \nabla_\mu \Psi(x,\xi\,|\,\g) = \nabla_\mu h_{\alpha\beta}(x) \, \xi_1^\alpha\xi_2^\beta +\tfrac{1}{2} \partial_\mu h(x) \, (\gamma_1\beta_2-\gamma_2\beta_1)+\dots \ .
\end{equation}
The presence of a general background manifests itself as deformations of the original BRST system: indeed the supercharges become
\begin{equation}
\q_i:=-i\,\xi_i^a\,e^\mu_a\,\hat \nabla_\mu\ ,\quad \bar \q^i:=-i\,\bar\xi^{i\,a}\,e^\mu_a\,\hat \nabla_\mu\ .
\end{equation}
The main consequence is that the \acr{SUSY} algebra does not close anymore:\footnote{As to not further burden the notation we denote $ \bm{R}_{\mu\nu}:=R_{\mu\nu\lambda\sigma}\,\xi^\lambda\cdot\bar\xi^\sigma$ and $\bm{R}:=R_{\mu\nu\lambda\sigma}\,\xi^\mu\cdot\bar\xi^\nu\,\xi^\lambda\cdot\bar\xi^\sigma$.}
\begin{align} \label{bo}
&\{\q_i, \q_j\}=-\xi^\mu_i\xi^\nu_j\, \bm{R}_{\mu\nu} \;,\quad \{\bar \q^i, \bar \q^j\}=-\bar\xi^{\mu\,i}\bar\xi^{\nu\,j}\, \bm{R}_{\mu\nu}\;,\quad \{\q_i, \bar \q^j\}=-\delta_i^j\nabla^2-\xi^\mu_i\bar\xi^{\nu\,j}\, \bm{R}_{\mu\nu}\ , \nonumber\\[3mm]
& [\nabla^2, \q_i]=i\,\xi^\mu_i\big(2\, \bm{R}_{\mu\nu}\,\hat \nabla^\nu-\nabla^\lambda \bm{R}_{\lambda\mu}-R_{\mu\nu}\hat \nabla^\nu\big)\ ,\nonumber\\[3mm]
& [\nabla^2, \bar \q^i]=i\,\bar\xi^{\mu\,i}\big(2\, \bm{R}_{\mu\nu}\,\hat \nabla^\nu-\nabla^\lambda \bm{R}_{\lambda\mu}-R_{\mu\nu}\hat \nabla^\nu\big)\ , 
\end{align}
where the Laplacian is defined as
\begin{equation} \label{laplace}
\nabla^2:=g^{\mu\nu}\hat \nabla_\mu \hat \nabla_\nu-g^{\mu\nu}\,\Gamma^\lambda_{\mu\nu}\,\hat \nabla_\lambda\ .
\end{equation}
The BRST operator needs to be deformed as well, requiring the construction of an ansatz due to the fact that the associated algebra is no longer first-class. The general approach involves considering the same general form \eqref{QQ}, but with the inclusion of more general terms accounting for the obstruction to the first-class character of the associated algebra, namely the curvature in the present case. Therefore, possible non-minimal couplings to the curvature, collectively denoted as $\Re$, must be incorporated inside the Hamiltonian. These couplings may act as obstructions to the nilpotency of the BRST operator, and the BRST analysis aims to determine which of these terms persist to ensure a nilpotent $\Q$. Avoiding higher powers of ghost momenta and assuming that derivatives are deformed only through minimal coupling, the ansatz for the BRST charge is
\begin{equation} \label{curvedQ}
\Q=c\,\D+\bm{\nabla}+\bar\gamma \cdot\gamma\,b\ ,
\end{equation}
where $\D$ is the deformed Hamiltonian in its operatorial form
\begin{equation} 
\D:=\nabla^2+\Re\ , \quad \text{with} \quad \Re:=R_{\mu\nu\lambda\sigma}\,\xi^\mu\cdot\bar\xi^\nu\,\xi^\lambda\cdot\bar\xi^\sigma+ \kappa R\ ,
\end{equation} 
with $\kappa$ a coefficient to be determined, and the $\bm{\nabla}$ operator has been conveniently redefined as 
\begin{equation}
\bm{\nabla}:=-i\,S^\mu\hat \nabla_\mu\ , \quad \text{with} \quad S^\mu:=\bar\gamma \cdot \xi^\mu+\gamma \cdot\bar\xi^{\mu}\ .
\end{equation}
Note that only the $so(4)$ constraints ${\cal T}_\alpha$ \eqref{rel} remain unchanged, therefore the wavefunction $\Psi(x,\xi\,|\,\g)$ is still expressed as in \eqref{BVstringfield}.

The BRST analysis starts with the general expression for $\Q^2$, namely the extension of its flat spacetime counterpart \eqref{sì} with adjusted coefficients:
\begin{equation} \label{Q22}
\Q^2=\bm{\nabla}^2+\bar\gamma \cdot\gamma\,\D+c\,\left[\D, \bm{\nabla}\right]\ , 
\end{equation}
which remains valid regardless of the specific spinning particle model coupled to a general curved background. In general, \eqref{Q22} includes two independent obstructions, which in this case are given by 
\begin{align} 
\bm{\nabla}^2+\bar\gamma \cdot\gamma\,\D &= -\tfrac12\,S^\mu S^\nu\, \bm{R}_{\mu\nu}+\bar\gamma \cdot\gamma\,\Re\ , \label{first obstruction} \\ 
\left[\D, \bm{\nabla}\right] &= -iS^\mu\nabla^\lambda \bm{R}_{\lambda \mu} +i S^\mu\nabla_\mu \bm{R}+\kappa [\,R,\bm{\nabla}]\ . \label{second obstruction}
\end{align}
Recall that the BRST cohomology is defined on the reduced Hilbert space $\mathcal{H}_{\mathrm{red}}$, that is, one needs to evaluate the obstructions on $\ker{\cal T}_\alpha$. Therefore, for the sake of the cohomology, it suffices to ensure nilpotency of the BRST charge \emph{when acting} on the physical sector of the theory:
\begin{equation}
 \Q^2\stackrel{{\ker}{\cal T}_\alpha}{=}0 \quad \text{i.e.} \quad \Q^2\Psi(x,\xi\,|\,\g)\must0\ .
\end{equation}
The effect is for instance that any contribution with at least three barred oscillators is set to zero. Equations \eqref{first obstruction}--\eqref{second obstruction} get then reduced to
\begin{align}
\bm{\nabla}^2+\bar\gamma \cdot\gamma\,\D &\stackrel{{\ker}{\cal T}_\alpha}{=}-\,\bar\gamma\cdot\gamma\left(2\,R_{\mu\nu}\,\xi^\mu \cdot\bar\xi^\nu-\kappa \, R\right)\ , \label{first-first} \\
\begin{split} \label{second-second}
\left[\D, \bm{\nabla}\right] &\stackrel{\ker{\cal T}_\alpha}{=}-iS^\mu\nabla^\lambda \bm{R}_{\lambda\mu}+i \big(2 \nabla^\lambda \bm{R}_{\lambda
\mu}\gamma\cdot\bar\xi^\mu-S^\mu\nabla_\mu R_{\nu\lambda}\xi^\nu \cdot\bar\xi^\lambda\big)\\
&\phantom{\stackrel{\ker{\cal T}_\alpha}{=}}\;+i\kappa \,S^\mu\nabla_\mu R\ .
\end{split}
\end{align}
The conclusion is that the nilpotency of the BRST charge, in the massless case, is achieved only on Einstein manifolds, i.e.
\begin{equation}
\Q^2 \Psi(x,\xi\,|\,\g)=0 \; \iff \; R_{\mu\nu}=\lambda\,g_{\mu\nu}\ ,
\end{equation}
upon setting $\kappa=\tfrac{1}{2}$. Let us clarify this crucial point for the upcoming discussion: while the obstruction given in \eqref{second-second} vanishes by itself on Einstein spaces, the one in \eqref{first-first} is zero \emph{only} when acting on the physical sector:
\begin{equation}
\Q^2 \Psi(x,\xi\,|\,\g) \ni -2\lambda\,\bar\gamma\cdot\gamma\left(\xi^\mu \cdot\bar\xi_\mu-1\right) \, \Psi(x,\xi\,|\,\g)=0\ .
\end{equation}
The closure equation evaluated using the deformed BRST charge yields the correct field equations for a massless\footnote{It is a rather fascinating topic the concept of mass on a general spacetime, which is not well defined. In particular, one should be careful to call a particle \emph{massless}: see \cite{Deser:1983mm} for a discussion on the connection between gauge invariance, masslessness, and null cone propagation.} graviton on Einstein spaces, namely
\begin{align}
\nabla^2h_{\mu\nu}-2\nabla_{(\mu}\nabla \cdot h_{\nu)}+\nabla_\mu\nabla_\nu h+2 R_{\mu\alpha\nu\beta}h^{\alpha\beta} =0\ ,
\end{align}
together with
\begin{align}
( \nabla^2 +\lambda )h-\nabla^\mu\nabla^\nu h_{\mu\nu}=0\ .
\end{align}
The correct gauge symmetry $\delta h_{\mu\nu}=2\nabla_{(\mu}\varepsilon_{\nu)}$ is obtained as well.

\section{Giving mass to the graviton} \label{chap:fourth:sec3}
In this section, we present two methods to confer a mass to a spinning particle model. The first one, the Scherk--Schwarz mechanism, is analogous to the Kaluza--Klein compactification. This has already been encountered in Chapter~\ref{chap:second}, but with bosonic oscillators. In the second method, the auxiliary oscillators approach, the model is treated as a truncation of the RNS open superstring \cite{Green:2012pqa}. We employ both methods to give mass to the graviton on a flat background and subsequently discuss the results. However, before delving into the graviton case, we begin by examining the $\N=2$ scenario, in order to introduce the techniques and investigate the role of the mass as a potential obstruction to nilpotency of the BRST charge. 

\subsection{$\N=2$ massive spinning particle and the Proca Theory}
The massless $\N=2$ spinning particle has been thoroughly examined in previous works, leading to a first-quantized description of massless antisymmetric tensor fields of arbitrary rank \cite{Berezin:1976eg, Gershun:1979fb, Howe:1988ft, Howe:1989vn, Brink:1976uf}. It has been shown how to provide a first-quantized description of the photon both on a flat and on a curved spacetime \cite{Bastianelli:2005vk, Grigoriev:2021bes}. Additionally, the coupling to a non-abelian background has been explored, reproducing the non-linear Yang--Mills equations \cite{Dai:2008bh} upon BRST quantization. 

The massive scenario has undergone investigation as well: the procedure \emph{à la} Kaluza--Klein has been implemented in Refs.~\cite{Brink:1976uf, Bastianelli:2005uy} where it has been shown how to embed the model in an \emph{arbitrary curved} spacetime background. This has allowed for the utilization of this model in the context of worldline quantum field theory formalism, particularly in modeling classical scatterings of compact objects in general relativity (see \cite{Mogull:2020sak, Jakobsen:2021zvh} and related literature). More recently, the coupling to both an abelian and a non-abelian vector background field has been investigated in Ref.~\cite{Carosi:2021wbi}, where the authors first introduced the auxiliary oscillators approach.

\subsubsection{Mass improvement on flat spacetime}
The main features of the $\N=2$ model can be immediately derived from section \ref{chap:fourth:sec1} by setting $I=1,2$ and are collected in detail in Ref.~\cite{Bastianelli:2005uy}. In this chapter, we will provide a concise overview of the essential elements of the BRST quantization of the massive case on flat and, most notably, on curved background as a toy model for the subsequent analysis, which is not covered in the literature mentioned above.

The procedure consists of starting with the massless theory formulated in one dimension higher, and subsequently reducing the dimensionality of the target space through the introduction of suitable constraints \cite{Bastianelli:2014lia}. We consider the model to live in a flat spacetime of the form $\mathcal{M}_d\times S^1$ of $D=d+1$ dimensions with coordinates and worldline superpartners given by
\begin{align}
x^M = (x^\mu , x^D)\ , \quad \xi^M=(\psi^\mu,\theta)\ , \quad \bar \xi^{M}=(\bar \psi^{\mu},\bar \theta)\ ,
\end{align}
with the index splitting $M=(\mu, D)$. The idea is to gauge the compact direction $x^D$, corresponding to $S^1$, by imposing the first-class constraint 
\begin{equation}
p_D-m=0\ .
\end{equation}
This results in the phase space action \eqref{chap:fourth:action} having a leftover term $\sim \dot x^D$, which can be regarded as a total derivative and thus dropped. The CCR \eqref{CR} are realized explicitly as
\begin{equation} 
[x^\mu, p_\nu]=i\,\delta^\mu_\nu\ ,\quad \{\bar \psi^\mu, \psi^\nu\}=\eta^{\mu\nu}\ , \quad \{\bar \theta, \theta\}=1\ , 
\end{equation}
with the other (anti)commutators being zero. The $\N=2$ \acr{SUSY} constraints get modified by the presence of the mass, taking the form
\begin{align}
H=\frac12 \left(p^\mu p_\mu+ m^2\right)\ , \quad
q=p_\mu \psi^\mu + m \theta\ , \quad 
\bar q=p_\mu \bar \psi^\mu + m \bar \theta\ ,
\end{align}
while the $R$-symmetry generator \eqref{J} becomes
\begin{equation}
\mathrm{J} =\psi^\mu \bar \psi_\mu +\theta \bar \theta-\frac{d+1}{2}\ .
\end{equation}
Together they still satisfy the same $\N=2$ supersymmetry algebra despite the mass improvement
\begin{align} \label{boo}
\{q, \bar q\}=2H\ ,\quad [q, H]=[\bar q, H]=\{q, q\}=\{\bar q, \bar q\}=0\ ,
\end{align}
which successfully remains first-class. Note the key role played by the surviving fermionic coordinates coming from the extra dimension $\theta:=\xi^D$ and $\bar \theta:=\bar \xi^{D}$, which are responsible for the introduction of the mass term into the theory. Indeed, in the limit where $(\theta, \bar \theta) \to (0,0)$ the massless theory must be recovered as a consistency check. 

At this point, the BRST quantization on a four-dimensional flat spacetime proceeds smoothly, just as described for the $\N=4$ case. Upon enlarging the Hilbert space to realize the set of ghost operators $(c, \bar \gamma, \gamma)$ with relative momenta $(b,\beta, \bar \beta)$ as in \eqref{3.4}, the BRST charge is constructed as usual
\begin{equation}
\Q= c\,H+\gamma\,\bar q+\bar\gamma\, q-2\bar\gamma\gamma\, b\ ,
\end{equation}
and the $so(2)$ constraint has to be extended to include ghost contributions as follows
\begin{equation}
\J:= \psi^\mu \bar \psi_\mu +\theta \bar \theta +\gamma\bar\beta-\bar\gamma\beta-\tfrac{3}{2}\ ,
\end{equation}
with the usual prescription employed to resolve ambiguities. Nilpotency of $\Q$ is still guaranteed, despite the mass improvement, since the associated algebra is first-class
\begin{equation}
\Q^2=0\ ,
\end{equation}
and the BRST system is consistent since $[\Q,\J]=0$. The physical sector is defined as the eigenspace of $\J$ with a fixed $R$-charge $-\tfrac{1}{2}$,\footnote{This condition is equivalent to demanding that $\Psi \in{\ker}{\tilde \J}$, with $\tilde \J$ being the shifted $SO(2)$ constraint $\tilde \J = \J+\tfrac{1}{2}$. The procedure remains consistent as long as $[\Q,\tilde \J]=0$. Eventually, the shift can be interpreted as the introduction of a Chern--Simons term selecting the desired degrees of freedom, as in Ref.~\cite{Bastianelli:2005uy}.} i.e. physical states are eigenstates $\J\ket{\Psi}=-\tfrac{1}{2}\ket{\Psi}$ and are isomorphic to wavefunctions
\begin{align}
\begin{split}
\Psi(x,\psi,\theta\,|\,\g) &= A_\mu(x)\psi^\mu-i\varphi(x)\theta +\phi(x) \beta c +\Phi(x) \beta\\[1mm]
&\phantom{=}+A_\mu
^{\ast}(x) \psi^\mu c-i\varphi^{\ast}(x) \theta c+\phi^{\ast}(x)\gamma+\Phi^{\ast}(x)\gamma c\ .
\end{split}
\end{align}
Requiring $\Psi$ to be Grassmann-odd and have ghost number zero, it can be interpreted as a spacetime \acr{BV} string field displaying the complete minimal \acr{BV} spectrum of the Proca theory along with an auxiliary field. The Grassmann parities and ghost numbers of the components can be found in Table~\ref{table3}. In contrast to the massless case, the physical fields include not only the massive spin one $A_\mu$ but also the St\"uckelberg scalar $\varphi$. The inclusion of the St\"uckelberg scalar becomes necessary to restore the $U(1)$ gauge symmetry \cite{Stueckelberg:1957zz}, which is broken due to the introduction of the mass. Notably, the spectrum also includes the associated scalar ghost $\Phi$ in the spectrum.

\begin{table}[!ht]
\centering
\begin{tabular}{ |c|c|c|c| } 
 \hline
 \acr{BV} role & Field & Grassmann parity & Ghost number \\
 \hline
 massive spin one & $A_{\mu}$ & $0$ & $0$ \\
 St\"uckelberg scalar & $\varphi$ & $0$ & $0$ \\
 auxiliary scalar & $\phi$ & $0$ & $0$ \\ 
 scalar ghost & $\Phi$ & $1$ & $1$ \\
 \hline
\end{tabular}
\caption{List of fields in the physical sector of the \emph{massive} $\N=2$ model with the corresponding ghost number and Grassmann parity.}
\label{table3}
\end{table}
\vspace{1ex}
The field equations, upon solving for the auxiliary field, are
\begin{align}
\left( \Box-m^2 \right)A_\mu-\partial_\mu\partial \cdot A-m\partial_\mu\varphi&=0\ , \\
\Box \varphi+m\partial_\mu A^\mu&=0\ , 
\end{align}
which are indeed the Proca field equations with the scalar $\varphi(x)$ playing the role of St\"uckelberg field, while the gauge symmetry, from $\delta \Psi=\Q \Lambda$ with
\begin{equation} \label{gg}
\Lambda=i\Sigma(x) \beta +\cdots\ ,
\end{equation}
is
\begin{equation}
\delta A_\mu=\partial_\mu\Sigma\ , \quad \delta\varphi=-m \, \Sigma\ ,
\end{equation}
where $\Sigma(x)$ is a local gauge parameter, of even parity and with ghost number zero. 

Let us end by noting that the same results can be obtained either by employing the auxiliary oscillators approach, although it is not explicitly shown here, or with the “bosonic” worldline model of Chapter~\ref{chap:second}. 

\subsubsection{Mass obstruction on curved spacetime}
We are finally in the position to investigate whether the presence of the mass plays a role when a curved background is considered, generalizing the massless BRST analysis of Ref.~\cite{Grigoriev:2021bes}. The coupling of the massive $\N=2$ spinning particle to gravity is realized by the covariantization of the reduced model just discussed, with the mass already present in the theory. The supercharges are then deformed as follows
\begin{equation}
\q:=-i\,\psi^a\,e^\mu_a\,\hat \nabla_\mu\ +m\theta\ ,\quad \bar \q:=-i\,\bar\psi^{a}\,e^\mu_a\,\hat \nabla_\mu+m\bar\theta\ .
\end{equation}
Unlike the situation with the $\N=4$ spinning particle \eqref{bo}, the $\N=2$ algebra \eqref{boo} remains first-class regardless of a particular background, upon a suitable redefinition of the Hamiltonian 
\begin{equation} \label{H}
\D:=\nabla^2-m^2+\bm{R}\ .
\end{equation}
Indeed, the \acr{SUSY} algebra takes the form
\begin{align} 
\begin{split} \label{ALG}
\{\q, \bar \q\}=-\D\ , \quad \{\q, \q\}=\{\bar \q, \bar \q\}&=0\ , \\
[\q,\D]=[\bar \q,\D]&=0\ ,
\end{split}
\end{align}
where the vanishing of the top-right commutators is guaranteed by the cyclic identity for the Riemann tensor. The corresponding BRST operator is defined as usual $\Q=c\,\D+\bm{\nabla}+\bar\gamma \gamma\,b$, with potential obstructions incorporated into the definition of $\bm{\nabla}$ with respect to the massless case:
\begin{equation}
\bm{\nabla}:=-i\,S^\mu\hat \nabla_\mu+ m\rho\ , \quad \rho:=\bar \gamma \, \theta+\gamma \, \bar \theta\ .
\end{equation}
The existence of an associated first-class algebra \eqref{ALG} ensures the nilpotency of the BRST operator without any further conditions on the background metric,\footnote{Actually, the only requirement is that of a torsionless connection, i.e. $T^\mu_{\nu\rho}:=\Gamma^\mu_{[\nu\rho]}=0$.} which is expected from \acr{QFT} considerations. While this has been already established for the massless case, the same holds even in the massive one, as the mass does not affect the algebra, as we now shall work out explicitly. Starting from the general expression
\begin{equation}
\Q^2=\bm{\nabla}^2+\bar\gamma \gamma\,\D+c\,\left[\D, \bm{\nabla}\right]\ ,
\end{equation}
one finds the following independent obstructions 
\begin{align} 
\bm{\nabla}^2+\bar\gamma \gamma\,\D &=-\frac{1}{2}S^\mu S^\nu \bm{R}_{\mu\nu}+\cancel{\bar\gamma \gamma \, m^2}+ \bar\gamma \gamma \left(\cancel{-m^2}+ \bm{R}\right)\ , \label{uno} \\
\left[\D, \bm{\nabla}\right] &=-iS^\mu
\nabla^\lambda\bm{R}_{\lambda\mu}+iS^\mu
\nabla_\mu\bm{R}\ .\label{due}
\end{align}
In the present case, evaluating the BRST cohomology on the reduced Hilbert space, $\ker\mathcal{\tilde \J}$, has the effect of setting to zero any obstruction of the form $\mathcal{O}^{AB}\bar Z_A\bar Z_B$ with $\mathcal{O}^{AB}$ arbitrary operators since an arbitrary state in the physical sector has the form
\begin{equation}
\Upsilon_{A}(x)\,Z^A + \Omega_{B}(x)\,Z^B\,c \quad \text{for} \quad Z^A=(\psi^\mu, \theta, \gamma,\beta)\ .
\end{equation}
This is enough to conclude that indeed
\begin{equation}
\Q^2\stackrel{\ker\mathcal{\tilde \J}}{=}0 
\end{equation}
without further conditions on the background; in other words, the massive $\N=2$ spinning particle can be coupled to off-shell gravity. Remarkably, the mass plays no role in the BRST algebra, as it is trivially canceled in \eqref{uno} and commutes with any operator inside \eqref{due} being constant. It is worth noting that introducing a term proportional to the Ricci curvature inside the Hamiltonian \eqref{H} $\D \to \D +\kappa R$, would force $\kappa$ to be zero (either that or a vanishing Ricci scalar $R=0$). This is dictated by the BRST algebra \eqref{uno}--\eqref{due}, as any term of that form is incompatible with the closure of the $\N=2$ \acr{SUSY} algebra \eqref{boo}. Interestingly, the $\N=2$ spinning particle appears to select only the minimal coupling to the background, as we shall check from the equations of motion. 
From the closure equation $\Q\Psi=0$, one finds
\begin{align}
( \nabla^2-m^2 )A_\mu-R_{\mu\nu}A^\nu
+\nabla_\mu \phi&=0\ , \\
( \nabla^2-m^2 )\varphi-i m\phi&=0\ , \\
\phi+i m\varphi-i \nabla_\mu A^\mu&=0\ ,
\end{align}
which, upon eliminating the auxiliary field, yield
\begin{align}
( \nabla^2-m^2)A_\mu-\nabla_\mu\nabla_\nu A^\nu+R_{\mu\nu}A^\nu-m\nabla_\mu\varphi&=0\ , \label{proca1} \\
\nabla^2\varphi+m\nabla_\mu A^\mu&=0\ . \label{proca2}
\end{align}
Equations \eqref{proca1}--\eqref{proca2} represent the field equations of the Proca theory on a general curved spacetime in its St\"uckelberg formulation. In particular, the $\N=2$ spinning particle provides a first-quantized version of the minimal extension for a theory of a free massive vector field in curved spacetime, without the inclusion of possible non-minimal couplings to the background in the corresponding spacetime quantum field theory \cite{Belokogne:2015etf, Buchbinder:2017zaa}. Their absence is attributed to the reasons previously discussed. As a final note, the gauge symmetry, using \eqref{gg}, reads
\begin{align}
\delta A_\mu=\nabla_\mu\Sigma\ , \quad \delta\varphi=-m \, \Sigma\ .
\end{align}

\subsection{$\N=4$ massive spinning particle and linearized massive gravity}
In the remainder of this section, the methods previously discussed are exploited to give a mass to the graviton, considering a flat target spacetime for the time being.

\subsubsection{Dimensional reduction approach}
The first method proceeds along the lines of what has been shown for the $\N=2$ case, namely through the reduction of the higher-dimensional massless model. The fermionic coordinates carry a flavor index 
\begin{align}
\xi^M_i=(\psi^\mu_i,\theta_i)\ , \quad \bar \xi^{M i}=(\bar \psi^{\mu i},\bar \theta^i)\ ,
\end{align}
with CCR
\begin{equation} 
\{\bar \psi^{\mu i}, \psi^{\nu}_j\}=\delta^i_j\,\eta^{\mu\nu}\ , \quad \{\bar \theta^i, \theta_j\}=\delta^i_j\ .
\end{equation}
The same applies to the \acr{SUSY} constraints \eqref{const}
\begin{align}
H=\frac12 \left(p^\mu p_\mu+ m^2\right)\ ,\quad q_i=p \cdot \psi_i + m \theta_i\ , \quad \bar q^i=p \cdot \bar \psi^i + m \bar \theta^i\ ,
\end{align}
that satisfy the same first-class algebra 
\begin{align}
\{q_i, \bar q^j\}=2\delta_i^j H\ ,\quad [q_i, H]=[\bar q^i, H]=\{q_i, q_j\}=\{\bar q^i, \bar q^j\}=0
\end{align}
despite the mass improvement. Regarding the $so(4)$ symmetry constraints, equations \eqref{JJ} become
\begin{align}
\begin{split}
\mathrm{J}_i{}^j &=\psi_i \cdot \frac{\partial}{\partial \psi_j} +\theta_i \; \frac{\partial}{\partial \theta_j} -\frac{d+1}{2}\, \delta_i^j\ , \\
\mathrm{Tr}^{ij} &= \frac{\partial^2}{\partial \psi_i \cdot \partial \psi_j}+\frac{\partial^2}{\partial \theta_i \; \partial \theta_j}\ , \\
\mathrm{G}_{ij} &= \psi_i\cdot \psi_j+\theta_i \; \theta_j\ .
\end{split}
\end{align}
At this point, the BRST quantization proceeds as outlined in section \ref{chap:fourth:sec2}, with the BRST system defined by \eqref{BRSTsystem}. The BRST operator \eqref{chap:fourth:Q} takes the usual form when acting on wavefunctions $\Psi(x,\psi,\theta\,|\,\g)$
\begin{equation} \label{chap:fourth:Q2}
\Q=c\,(\Box-m^2)+\gamma_i\,\bar q^i-q_i\,\frac{\partial}{\partial\beta_i}-\gamma_i\,\frac{\partial^2}{\partial\beta_i\partial c}
\end{equation}
and is still nilpotent, for the same reasons outlined in the previous section. The relevant $so(4)$ generators \eqref{rel} to be imposed on the extended BRST Hilbert space are extended to commute with $\Q$, including the extra fermionic oscillators
\begin{equation} \label{Ta}
\begin{split}
\J_i &= N_{\psi_i}+N_{\theta_i}+N_{\gamma_i}+N_{\beta_i}-\frac{d-1}{2}\ , \\[2mm]
\Y &= \psi_1\cdot\frac{\partial}{\partial\psi_2}+\theta_1 \; \frac{\partial}{\partial\theta_2}+\gamma_1\frac{\partial}{\partial\gamma_2}+\beta_1\frac{\partial}{\partial\beta_2}\ ,\\[2mm]
\gTr &= \frac{\partial^2}{\partial\psi_1\cdot\partial\psi_2}+\frac{\partial^2}{\partial\theta_1 \; \partial\theta_2}+\frac{\partial^2}{\partial\gamma_1\partial\beta_2}-\frac{\partial^2}{\partial\gamma_2\partial\beta_1}\ ,
\end{split} 
\end{equation}
The physical subspace has a fixed $U(1)\times U(1)$ charge corresponding to $\tfrac{3-d}{2}$. Let us comment on the fact that only in three spacetime dimensions the wavefunction $\Psi \in \ker\J_i$, which becomes relevant when dealing with the construction of the worldline path integral.\footnote{This is the content of Chapter~\ref{chap:fifth}.} In four spacetime dimensions, physical states are charged $-\tfrac{1}{2}$, and we shall consider the shifted operator $\tilde \J = \J+\tfrac{1}{2}$ for the sake of notational simplicity. To emphasize the impact of the introduction of a mass term, we follow the same steps as in the massless case to impose the condition $\Psi \in \mathcal{H}_{\mathrm{red}}$, with the reduced Hilbert space defined as the kernel of $\tilde \J_i$ and $\Y, \gTr$ \eqref{Ta}, still collectively denoted as $\mathcal{T}_\alpha$\color{black}. From $\tilde \J_i \ket{\Psi}=0$ descends the general form of the wavefunction
\begin{align} \label{comp}
\Psi(x,\psi,\theta\,|\,\g)&=\left. \Psi\right|_{m=0} \nonumber \\[1mm]
&\phantom{=}+n\,\theta_1\gamma_2+p\,\theta_1\beta_2+q\,\gamma_1\theta_2+r\,\beta_1\theta_2+s_\mu\,\theta_1\psi_2^\mu+t_\mu\,\psi_1^\mu\theta_2 +u\,\theta_1\theta_2 \\[1mm]
&\phantom{=}+n^{\ast}\,\theta_1\gamma_2c+p^{\ast}\,\theta_1\beta_2c+q^{\ast}\,\gamma_1\theta_2c+r^{\ast}\,\beta_1\theta_2c+s^{\ast}_\mu\,\theta_1\psi_2^\mu c+t^{\ast}_\mu\,\psi_1^\mu\theta_2c+u^{\ast}\,\theta_1\theta_2c\ , \nonumber
\end{align}
where in the first line $\left. \Psi\right|_{m=0}$ denotes the massless wavefunction \eqref{massless}. The condition $\Y \ket{\Psi}=0$ produces, in addition to the “massless” contribution \eqref{cond1},
\begin{align}
\begin{split}
&n=-q\ , \quad p=-r\ , \quad s_\mu=t_\mu\ , \\
&n^{\ast}=-q^{\ast}\ , \quad p^{\ast}=-r^{\ast}\ , \quad s^{\ast}_\mu=t^{\ast}_\mu\ ,
\end{split}
\end{align}
while the remaining $\gTr \ket{\Psi}=0$ produces
\begin{align}
e=\tfrac{1}{2}a^\mu_\mu+\tfrac{1}{2}u\ , \quad e^{\ast}=\tfrac{1}{2}a^{\ast \mu}_\mu+\tfrac{1}{2}u^{\ast}\ .
\end{align}
It is now possible to identify the \acr{BV} spectrum of the theory, which becomes evident after renaming the field components as in \eqref{renam} along with 
\begin{align} \label{renam2}
s_{\mu} \longrightarrow A_{\mu} \quad u \longrightarrow \varphi \quad p \longrightarrow \phi \quad p^{\ast} \longrightarrow \Phi
\end{align}
and, for the corresponding antifields,
\begin{align}
s^{\ast}_{\mu} \longrightarrow A^{\ast}_{\mu} \quad u^{\ast} \longrightarrow \varphi^{\ast} \quad n \longrightarrow \phi^{\ast} \quad n^{\ast} \longrightarrow \Phi^{\ast}\ .
\end{align}
Requiring $\Psi$ to be Grassmann-even and have ghost number zero the field content gets assigned with the corresponding Grassmann parities and ghost numbers as reported in Table~\ref{table2}. The most general string field $\Psi$ in $\ker\mathcal{T}_\alpha$ reads
\begin{align} \label{BVstringfield2}
\Psi(x,\psi,\theta\,|\,\g) &= h_{\mu\nu}(x)\,\psi^\mu_1\psi^\nu_2+\tfrac12\,h(x)\,(\gamma_1\beta_2-\gamma_2\beta_1)-\tfrac{i}{2}\,v_\mu(x)\,(\psi^\mu_1\beta_2-\psi^\mu_2\beta_1)c \nonumber\\[1mm]
&\phantom{=}-\tfrac{i}{2}\,\zeta_\mu(x)\,(\psi^\mu_1\beta_2-\psi^\mu_2\beta_1)\nonumber\\[2mm]
&\phantom{=}-iA_\mu(x)\,(\theta_1\psi_2^\mu+\psi_1^\mu\theta_2)-\varphi(x)\,(2\theta_1\theta_2+\gamma_1\beta_2-\gamma_2\beta_1)+\phi(x)\,(\theta_1\beta_2-\theta_2\beta_1)c\nonumber\\[1mm]
&\phantom{=}+\Phi(x)\,(\theta_1\beta_2-\theta_2\beta_1)\ ,
\end{align}
where, for the sake of simplicity, the antifields content is left implicit.

\begin{table}[!ht]
\centering
\begin{tabular}{ |c|c|c|c| } 
 \hline
 \acr{BV} role & Field & Grassmann parity & Ghost number \\
 \hline
 massless graviton & $h_{\mu\nu}$ & $0$ & $0$ \\
 trace & $h$ & $0$ & $0$ \\
 auxiliary vector & $v_\mu$ & $0$ & $0$ \\ 
 diffeomorphism ghost & $\zeta_\mu$ & $1$ & $1$ \\
 St\"uckelberg vector & $A_\mu$ & $0$ & $0$ \\
 St\"uckelberg scalar & $\varphi$ & $0$ & $0$ \\
 auxiliary scalar & $\phi$ & $0$ & $0$ \\
 scalar ghost & $\Phi$ & $1$ & $1$ \\
 \hline
\end{tabular}
\caption{List of fields in the physical sector of the \emph{massive} $\N=4$ model with the corresponding ghost number and Grassmann parity in the dimensional reduction approach.}
\label{table2}
\end{table}
\vspace{1ex}
A few comments are in order. Firstly, it can be verified that the first two lines of \eqref{BVstringfield2} correspond to the field content in the massless case \eqref{BVstringfield}. This is expected since, as previously suggested,
\begin{equation}
\Psi \xrightarrow[]{(\theta, \bar \theta) \to (0,0)}\left. \Psi \right|_{m=0}
\end{equation}
up to an obvious redefinition of the scalar field $\varphi(x)$ which can be absorbed into the trace $h(x)$. Taking into account also the other two lines, it is clear that $\Psi$ contains the whole minimal \acr{BV} spectrum of linearized massive gravity, with the inclusion, with respect to the massless spectrum, of the two St\"uckelberg fields $A_\mu$ and $\varphi$, a scalar ghost field $\Phi$ and an auxiliary scalar field $\phi$ \cite{Boulanger:2018dau}.

At this point, we proceed to analyze the field equations, which are derived from the closure equation at ghost number zero using the BRST operator \eqref{chap:fourth:Q2}. The result is
\begin{subequations} \label{eq}
\begin{align}
\left(\Box-m^2\right)h_{\mu\nu}-\partial_{(\mu}v_{\nu)}&=0 \label{eq1}\ , \\
\left(\Box-m^2\right)h-2\left(\Box-m^2\right)\varphi-\partial \cdot v+2m\phi&=0\ , \label{eq5}\\
\left(\Box-m^2\right)A_\mu +\partial_\mu \phi +\tfrac{m}{2} v_\mu
&=0\ , \\
\left(\Box-m^2\right)\varphi-m\phi&=0\ , \label{eq3}\\
v_\mu -2\partial \cdot h_\mu +\partial_\mu h-2m A_\mu -2\partial_\mu \varphi&=0 \label{eq4}\ , \\
\partial \cdot A +m\varphi +\tfrac{1}{2}mh+\phi&=0\label{eq2}\ .
\end{align}
\end{subequations}
It is possible to solve \eqref{eq4} and \eqref{eq2} for the auxiliary fields $v_\mu$ and $\phi$ respectively, leading to the following equations
\begin{subequations} \label{main}
\begin{align}
&\left(\Box-m^2\right)h_{\mu\nu}-2\partial_{(\mu} \partial \cdot h_{\nu)}+\partial_\mu\partial_\nu h=2m\partial_{(\mu}A_{\nu)}+2\partial_\mu \partial_\nu \varphi\ ,\label{main2.1} \\
&\left(\Box-m^2\right)h-\partial^\mu\partial^\nu h_{\mu\nu}=2m \partial \cdot A+2\Box\varphi\ , \label{main2.2}\\
&\phantom{(}\Box A_\mu -\partial_\mu \partial \cdot A=m\left( \partial_\mu h-\partial \cdot h_\mu \right)\ ,\label{main2.3} \\
&\phantom{(}\Box\varphi +\tfrac{m^2}{2} h+m\partial \cdot A=0\label{main2.4}\ , 
\end{align}
\end{subequations}
which can be further simplified combining \eqref{main2.2} and \eqref{main2.4} to reach the following set of equations 
\begin{align}
&\left(\Box-m^2\right)h_{\mu\nu}-2\partial_{(\mu} \partial \cdot h_{\nu)}+\partial_\mu\partial_\nu h=2m\partial_{(\mu}A_{\nu)}+2\partial_\mu \partial_\nu \varphi\ , \label{final1}\\
&\phantom{(}\Box A_\mu -\partial_\mu \partial \cdot A=m\left( \partial_\mu h-\partial \cdot h_\mu \right)\ , \\
&\phantom{(}\Box h-\partial^\mu\partial^\nu h_{\mu\nu}=0\ .\label{final2}
\end{align}
Equations \eqref{final1}--\eqref{final2} correspond to the field equations of linearized massive gravity in the St\"uckelberg formalism, where, as anticipated, $\varphi(x)$ is the St\"uckelberg field and $A_\mu(x)$ is its vector counterpart.

A few comments are in order. 
\begin{itemize}
\item
Setting both St\"uckelberg fields to zero, which in literature is known as the \emph{unitary} or \emph{physical gauge}, equations \eqref{main} can be reduced to the Fierz--Pauli system
\begin{align}
\begin{split}
&\left(\Box-m^2\right)h_{\mu\nu}=0\ , \\
&\phantom{(}\partial^\nu h_{\mu\nu} =0\ , \\
&\phantom{(}h=0\ ,
\end{split} 
\end{align}
i.e. the field equations for the theory of a massive spin two field in which the mass term explicitly breaks the gauge invariance.
\item At first sight, the massless $m \to 0$ limit is rather peculiar. Instead of resulting in the free and massless spin two field equations \eqref{chap:fourth:free}, it produces the following outcome
\begin{align}
&\Box h_{\mu\nu}-2\partial_{(\mu} \partial \cdot h_{\nu)}+\partial_\mu\partial_\nu h=2\partial_\mu \partial_\nu \varphi \label{E}\ , \\
&\Box A_\mu -\partial_\mu \partial \cdot A=0\label{C}\ , \\
&\Box\varphi =0\label{chap:fourth:D}\ .
\end{align}
Equation \eqref{C} represents the field equation for a free-propagating vector field $A_\mu(x)$, describing a spin one massless particle that becomes decoupled in this limit. On the other hand, \eqref{chap:fourth:D} describes the propagation of a massless scalar field $\varphi(x)$, which, surprisingly enough, is still coupled to the wanna-be free massless graviton field $h_{\mu\nu}(x)$ in \eqref{E}. This hints at the \acr{vDVZ} discontinuity, a peculiarity associated with the Fierz--Pauli formulation of massive gravity. Notably, there exist other \acr{LMG} formulations in which this discontinuity is absent, as discussed in Refs.~\cite{Chamseddine:2018gqh, Gambuti:2021meo, deFreitas:2023ujo} and related literature. The origin of the discontinuity is traced back to the fact that the massless limit of a Fierz--Pauli graviton is not a massless graviton, but rather a massless graviton plus a coupled scalar.\footnote{The resolution lies in the so-called \emph{Vainshtein mechanism}: in a nutshell, general relativity can be recovered around massive bodies by hiding extra degrees of freedom by strong kinetic self-coupling so that they almost do not propagate. We refer to the review \cite{Babichev:2013usa} for further details.} Remarkably, our model captures this characteristic feature of massive gravity.
\end{itemize}

Let us finalize the analysis by investigating the gauge symmetries of the theory. \\

Any theory of massive gravity -- particularly the Fierz--Pauli theory -- is \emph{not} gauge invariant due to the presence of the mass term, unless redundant degrees of freedom are introduced, as is the case for the St\"uckelberg trick. To explore this fact from a first-quantized perspective, it is sufficient to consider the closure equation at ghost number one: while in the massless case, the result is a set of dynamical equations for the diffeomorphism ghost \eqref{dimenticavo}, in the massive scenario things become more intricate due to the presence of another ghost field $\Phi(x)$. The outcome manifests the following additional equations:
\begin{align}
m \, \zeta_\mu = i \partial_\mu \Phi\ , \quad m \, \Phi =0\ .
\end{align}
This indicates that both the scalar and the diffeomorphism ghosts are set to zero, which accounts for the breaking of the gauge invariance due to the presence of a mass term. To display said symmetries, one has to compute the ghost number zero component of $\delta\Psi=\Q \, \Lambda$, where $\Lambda$ is the massive extension of \eqref{Lambdaparameter}, namely
\begin{equation} \label{Lambda}
\Lambda=i\varepsilon_\mu(x)\,(\psi^\mu_1\beta_2-\psi^\mu_2\beta_1)+\Sigma(x)\,(\theta_1\beta_2-\theta_2\beta_1)+\cdots\ ,
\end{equation}
where $\varepsilon(x)$ and $\Sigma(x)$ are the two gauge parameters associated with the two gauge symmetries. The outcome is
\begin{align}
& \delta h_{\mu\nu}=2\partial_{(\mu} \varepsilon_{\nu)} \quad \delta A_\mu =-m \, \varepsilon_\mu\ , \\
& \delta A_\mu =\partial_\mu \Sigma \qquad \;\; \delta \varphi =-m \, \Sigma\ ,
\end{align}
as expected from the St\"uckelberg formulation. Therefore, the massive $\N=4$ spinning particle produces a gauge invariant formulation of \acr{LMG}.

\subsubsection{Auxiliary oscillators approach}
The main idea of the auxiliary oscillators approach consists of enlarging the BRST algebra with additional variables, whose physical interpretation can be seen as to describe the internal degrees of freedom of the $\N=4$ spinning particle. Thus, the first step involves an extension of the phase space by introducing two canonically conjugated complex bosonic variables, denoted as $(\alpha^a, \bar{\alpha}_a)$, with commutation relation 
\begin{equation}
[\bar{\alpha}_a,\alpha^b] = \delta^b_a\ .
\end{equation}
Note that the index $a$ may run over an arbitrary set, including a single value. To preserve the $\mathcal{N}=4$ worldline supersymmetry, the introduction of four complex fermionic variables $(\eta_i^a,\bar{\eta}^i_a)$ is necessary. They represent the superpartners of the $\alpha$'s, with anticommutator
\begin{equation}
\{\bar{\eta}^i_a , \eta_j^b \} =\delta^i_j \delta^b_a\ .
\end{equation}
The next step involves finding a way to deform the \acr{SUSY} constraints while ensuring the closure of the algebra. A straightforward attempt would be to extend the $\N=2$ constraints of Ref.~\cite{Carosi:2021wbi}, resulting in
\begin{align}
\begin{split}
H&=\tfrac12 \left(p^2+ m^2 \, \alpha^a\bar{\alpha}_a + m^2 \, \eta^a \cdot \bar{\eta}_a\right)\ , \\
q_i&=p \cdot \psi_i + m \, \eta_i^a\bar{\alpha}_a\ , \\
\bar q^i&=p \cdot \bar \psi^i + m \,\bar{\eta}^i_a \alpha^a\ .
\end{split}
\end{align}
However, this leads to $\left\{q_i,\bar q^j \right\} \neq 2\delta_i^j \, H$, specifically
\begin{equation} \label{prob}
\left\{q_i,\bar q^j \right\}=\delta_i^j p^2+m^2\left( \eta_i^a \bar \eta^j_a +\delta_i^j\alpha_a\bar\alpha^a\right)
\end{equation}
and there is not an obvious redefinition of the Hamiltonian that allows for a closure. Consequently, the resulting algebra is not first-class, and potential issues are anticipated in the quantization process. Indeed, the associated BRST charge $\Q= c\,H+\gamma \cdot \bar q+\bar\gamma \cdot q-2\bar\gamma \cdot \gamma\, b$ fails to be nilpotent:
\begin{equation} \label{ivo}
\Q^2=m^2 \left( \bar \gamma \cdot \eta^a \gamma \cdot \bar \eta_a -\gamma \cdot \bar \gamma \, \eta^a \cdot \bar \eta_a \right)\neq 0\ .
\end{equation}
The auxiliary oscillators approach may be a viable option, although it is necessary to follow the principles outlined in section \ref{chap:fourth:sec3.2}. One should engineer a deformation of the BRST algebra that leads to preventing the obstruction given by \eqref{ivo}, e.g. through a suitable redefinition of the Hamiltonian and of the $so(4)$ constraints. Further work is needed to explore this avenue, but such investigations are beyond the scope of the current discussion, as a promising method for dealing with a massive graviton is already at hand.

\section{Massive graviton on curved spacetimes} \label{chap:fourth:sec4}
In this section, the investigation of a first-quantized massive graviton on a curved spacetime is carried out. Previous works have addressed the challenge of coupling massive higher spin fields -- which, in the present formalism means for $\N>2$ -- but the analysis failed in going beyond (A)dS spaces \cite{Bastianelli:2014lia}. In the present section, we overcome said result with the $\N=4$ spinning particle. Let us rephrase the objective more clearly: the aim is to provide a worldline formulation of the \emph{linear} theory of massive gravity, namely of the Fierz--Pauli theory, on a curved spacetime; such a theory describes the propagation of massive spin 2 particle on a non-flat background. \\
The coupling to a generic curved background is achieved through the covariantization of the reduced model previously discussed, which results only in a mild modification of the massless model of section \eqref{chap:fourth:sec3.2}. The supercharges are deformed as
\begin{equation}
\q_i:=-i\,\psi_i^a\,e^\mu_a\,\hat \nabla_\mu\ +m\theta_i\ ,\quad \bar \q^i:=-i\,\bar\psi^{i\,a}\,e^\mu_a\,\hat \nabla_\mu+m\bar\theta^i\ ,
\end{equation}
and together with the Laplacian $\nabla^2$ form the same obstructed algebra \eqref{bo}, with the sole modification arising from the following anticommutator
\begin{equation}
\{\q_i, \bar \q^j\}=-\delta_i^j\left(\nabla^2-m^2\right)-\psi^\mu_i\bar\psi^{\nu\,j}\, \bm{R}_{\mu\nu}\ .
\end{equation}
The deformed BRST charge displays still the general form $\Q=c\,\D+\bm{\nabla}+\bar\gamma \cdot\gamma\,b$, where now
\begin{align} 
\D:=\nabla^2-m^2+\Re\ , \quad &\text{with} \quad \Re:=R_{\mu\nu\lambda\sigma}\,\psi^\mu \cdot\bar\psi^\nu\,\psi^\lambda \cdot\bar\psi^\sigma+ \tfrac{1}{2}R\ , \label{DD} \\
\bm{\nabla}:=-i\,S^\mu\hat \nabla_\mu+ m\rho\ , \quad &\text{with} \quad \rho:=\bar \gamma \cdot \theta+\gamma \cdot \bar \theta\ .
\end{align} 
The model is considered on Einstein spaces: indeed, it is expected for the massive theory to reproduce the correct results in the $m \to 0$ limit -- similarly to the flat spacetime scenario -- and the massless BRST charge is nilpotent only when $R_{\mu\nu}=\tfrac14 g_{\mu\nu}R$. It is not possible to obtain a weaker condition for the background in the massive case. Moreover, from a \acr{QFT} perspective, it is known that an Einstein spacetime is the only space on which a free (massive) graviton can consistently propagate \cite{Aragone:1971kh, Bengtsson:1994vn, Buchbinder:1999ar, Deser:2006sq}. Hence, it is natural to implement this condition right at the beginning of the analysis. \\

The $so(4)$ constraints ${\cal T}_\alpha$ are given in \eqref{Ta}, thus the wavefunction $\Psi$ can be read from \eqref{BVstringfield2}. This section aims to investigate if and how the presence of the mass affects the nilpotency of the BRST operator. In contrast to the case of the spin one particle on a general background, where nilpotency is not a concern even in the simpler massless case, the situation is different here. The massless $\N=4$ spinning particle already selects only a specific subset of available backgrounds. The starting point remains
\begin{equation}
\Q^2=\bm{\nabla}^2+\bar\gamma \cdot\gamma\,\D+c\,\left[\D, \bm{\nabla}\right]\ , 
\end{equation}
and thus, it is necessary to address the independent obstructions
\begin{align} 
\bm{\nabla}^2+\bar\gamma \cdot\gamma\,\D &= -\tfrac12\,S^\mu S^\nu\, \bm{R}_{\mu\nu}+m^2\left(\cancel{ \gamma \cdot \bar \theta \, \gamma \cdot \bar \theta + \bar \gamma \cdot \theta \, \bar \gamma \cdot \theta }\right)+\bar\gamma \cdot\gamma\,\Re\ , \label{first obstruction22} \\ 
\left[\D, \bm{\nabla}\right] &= -iS^\mu\nabla^\lambda \bm{R}_{\lambda \mu} +i S^\mu\nabla_\mu \bm{R}+\kappa [\,R,\bm{\nabla}]\ . \label{second obstruction23}
\end{align}
Just as in the case of the Proca theory, it is evident that the mass does not enter into the BRST algebra. Specifically, note that the term in \eqref{first obstruction22} that is striked out cancels due to a symmetry-based argument and also because it annihilates any state $\Psi$ in the physical sector, thereby being set to zero. Additionally, no mass term survives inside the commutator in \eqref{second obstruction23}. Consequently, the mass does not \emph{explicitly} obstruct the nilpotency. The expressions remain the same as in the massless case: once evaluated on the reduced Hilbert space $\mathcal{H}_{\mathrm{red}}$, taking Einstein manifold simplifications into account, they read
\begin{align}
\bm{\nabla}^2+\bar\gamma \cdot\gamma\,\D &\stackrel{{\ker}{\cal T}_\alpha}{=}-\frac12\,\bar\gamma\cdot\gamma\left(\psi^\mu \cdot\bar\psi_\mu-1 \right)R\ , \label{first-first2} \\
\left[\D, \bm{\nabla}\right] &\stackrel{\ker{\cal T}_\alpha}{=}0\ . \label{second-second2}
\end{align}
The leftover operator \eqref{first-first2} is vanishing when acting on the massless physical wavefunction $\Q^2\left. \Psi\right|_{m=0}=0$. However, this is \emph{not} the case for the massive graviton, as the wavefunction has changed. When acting on the physical sector, the outcome is
\begin{align}
\Q^2\Psi(x,\psi,\theta\,|\,\g)=-\tfrac{1}{2} \, \phi \left( \theta_1\gamma_2-\theta_2\gamma_1 \right)cR-\tfrac{1}{2} \, \Phi \left( \theta_1\gamma_2-\theta_2\gamma_1 \right)R\neq 0\ .
\end{align}
Let us highlight that attempts to implement possible couplings involving traces of the Riemann tensor and the newly introduced fermionic variables inside the Hamiltonian \eqref{DD}, such as
\begin{equation}
\Delta R=c_1 \, R \, \theta \cdot \bar \theta + c_2 \, R \, \theta \cdot \bar \theta \, \theta \cdot \bar \theta+ c_3 \, R_{\mu\nu} \, \psi^\mu \cdot \bar \psi^\nu \theta \cdot \bar \theta\ , 
\end{equation}
do not solve the issue. Although an appropriate tuning of the coefficients -- in particular, one finds $c_1+c_2=-2$ -- potentially addresses the first obstruction \eqref{first-first2}, these terms would disrupt the second one \eqref{second-second2}, compelling $c_1=c_2=c_3=0$. This explains why those terms have been omitted right from the beginning: as already discussed in the massless case, one cannot have fermions inside $\Delta R$. The only possibility to achieve nilpotency of the BRST operator is to consider Ricci-flat backgrounds, i.e.
\begin{equation}
R_{\mu\nu}(x)=0\ ,
\end{equation}
which immediately addresses the remaining obstruction \eqref{first-first2}, yielding a nilpotent BRST charge. This outcome resembles the scenario encountered in string theory, where consistency of the string propagation, worldsheet conformal invariance, implies Ricci flatness to leading order \cite{Lu:2003gp}. Such an unfortunate result amounts to the difficulties that arise when trying to give a mass to a graviton and trying to generalize it to a curved spacetime.

In conclusion, the field equations for a massive graviton on an Einstein background with zero cosmological constant (i.e. a Ricci-flat background) are evaluated using the deformed BRST operator incorporating the appropriate curvature terms. Upon eliminating auxiliary fields, the resulting equations of motion are
\begin{subequations} \label{aaa}
\begin{align}
&\left( \nabla^2-m^2 \right)h_{\mu\nu}-2\nabla_{(\mu}\nabla \cdot h_{\nu)}+\nabla_\mu\nabla_\nu h+2R_{\mu\alpha\nu\beta}h^{\alpha\beta}=2m \nabla_{(\mu}A_{\nu)}+2\nabla_\mu\nabla_\nu\varphi\ , \\
&\left( \nabla^2-m^2 \right)h-\nabla^\mu\nabla^\nu h_{\mu\nu}=2m \nabla \cdot A+2\nabla^2\varphi\ , \\
&\phantom{(}\nabla^2 A_\mu-\nabla_\mu\nabla\cdot A=m(\nabla_\mu h -\nabla \cdot h_\mu)\ , \\
&\phantom{(}\nabla^2 \varphi +\tfrac{m^2}{2}h+m \nabla \cdot A=0\ ,
\end{align}
\end{subequations}
which can be simplified as
\begin{align}
&\left( \nabla^2-m^2 \right)h_{\mu\nu}-2\nabla_{(\mu}\nabla \cdot h_{\nu)}+\nabla_\mu\nabla_\nu h+2R_{\mu\alpha\nu\beta}h^{\alpha\beta}=2m \nabla_{(\mu}A_{\nu)}+2\nabla_\mu\nabla_\nu\varphi\ , \label{final2.1} \\
&\phantom{(}\nabla^2 A_\mu-\nabla_\mu\nabla\cdot A=m(\nabla_\mu h -\nabla \cdot h_\mu)\ , \\
&\phantom{(}\nabla^2 h-\nabla^\mu\nabla^\nu h_{\mu\nu}=0\ .\label{final2.2}
\end{align}
Equations \eqref{final2.1}--\eqref{final2.2} correspond to the field equations governing linearized massive gravity on a Ricci-flat background within its St\"uckelberg formulation. The same equations of motion can be derived from a field theory approach: one has to take into account the most general action functional while keeping only a restricted class of non-minimal couplings to the background, namely, the ones that do not excite possible unphysical degrees of freedom \cite{Buchbinder:1999ar}. The final step involves performing the necessary St\"uckelberg tricks to restore gauge invariance. As a final remark, it is worth noting that the correct gauge symmetries can be derived from $\Lambda$ \eqref{Lambda}, yielding
\begin{align}
& \delta h_{\mu\nu}=2\nabla_{(\mu} \varepsilon_{\nu)} \quad \delta A_\mu =-m \, \varepsilon_\mu\ , \\
& \delta A_\mu =\nabla_\mu \Sigma \qquad \; \; \delta \varphi =-m \, \Sigma
\ .
\end{align}

\section{Final remarks} \label{conc}
In this chapter, we have outlined the BRST quantization of the $\mathcal{N}=4$ spinning particle, in both the massless and massive cases. This has served to introduce the models that will be employed in the following two chapters to perform perturbative calculations relevant to quantum gravity (massless, Chapter~\ref{chap:sixth}) and massive gravity (massive, Chapter~\ref{chap:fifth}).\\
The genuinely novel contribution of this chapter has been to investigate the consistent coupling of massive higher-spin fields to general curved backgrounds within the Worldline Formalism, with particular emphasis on developing a first-quantized formulation of linearized massive gravity. 
During this investigation, we have explicitly shown the BRST quantization of the massive $\N=2$ spinning particle coupled to off-shell gravity, which, to the best of our knowledge, has never been investigated before in the literature with these tools. Our results show that the model reproduces the Proca theory on curved spacetime with the specific selection of the minimal coupling to the background, improving knowledge built on previous results and offering potential utility in the context of worldline computations employing this model. Subsequently, we have addressed directly the issue of a first-quantized massive graviton, discussing the correct reproduction of the Fierz--Pauli theory on a flat spacetime through the dimensional reduction of the higher-dimensional massless $\N=4$ spinning particle. Additionally, we addressed the challenges associated with the auxiliary oscillators approach. Finally, we tried to couple the massive particle to an Einstein background. Our findings suggest that a Ricci-flat spacetime emerges as the only available background for consistency at the quantum level. Let us stress that the theory constructed here is the worldline counterpart of the \emph{linear} theory of massive gravity, namely the Fierz--Pauli theory from a \acr{QFT} perspective: section \ref{chap:fourth:sec4} has been devoted to the correct reproduction, starting from the BRST system of the spinning particle model \eqref{BRSTsystem}, of the field equations and the gauge symmetries of \acr{LMG} in the St\"uckelberg formulation. In particular, from the quantum field theory side, the same results can be derived from the Fierz--Pauli action put on a curved spacetime, which is the theory for a massive spin 2 propagating on non-flat geometry and which can be obtained as the first approximation of a general massive gravity theory around a given fixed background. Even from the quantum field theory side, the construction of interacting theories of massive gravity is much harder, and a well-defined non-linear completion of the theory has been only recently constructed: achieving such a result in the first quantization formalism -- if feasible -- would require a considerable effort. 

With these results at hand, this work may be regarded as a first step towards the realization of a first-quantized massive graviton on an Einstein spacetime with \emph{non-zero} cosmological constant, either by identifying consistent improvements in the dimensional-reduced BRST system, or fully pursuing the auxiliary oscillators procedure, contingent on resolving the seemingly simpler yet already intricate flat spacetime scenario first. As a preliminary step in both instances, an intriguing prospect involves investigating whether modifying the model could result in a first-quantized \emph{partially massless} graviton \cite{Higuchi:1986py, Deser:2001us}. There are possible extensions worth exploring, such as relaxing the constraint on the BRST Hilbert space to subalgebras, which should in principle lead to the inclusion inside the \acr{BV} spectrum of the $\N=0$ supergravity, i.e. the particle theory that has in its spectrum the graviton, the dilaton, and the antisymmetric Kalb--Ramond tensor field. Such extensions might be pursued along the lines of Ref.~\cite{Bonezzi:2020jjq}, investigating possible couplings to the background fields through deformations of the BRST charge.

\newpage
\thispagestyle{empty}
\mbox{}
\newpage

\chapter{One-loop divergences in quantum gravity}\label{chap:sixth}
\textit{In this chapter,  we employ the massless model developed in the previous chapter to carry out a perturbative computation in quantum gravity. Specifically, we evaluate the counterterms required for the renormalization of the one-loop effective action of linearized Einstein gravity. The central result of this chapter is the determination of the Seeley-DeWitt coefficient $a_3(D)$ for perturbative quantum gravity with a cosmological constant, which we evaluate on Einstein manifolds of arbitrary $D$ dimensions. This coefficient provides a gauge-invariant characterization of quantum gravity, owing to the on-shell condition satisfied by the background on which the graviton propagates. To the best of our knowledge, the explicit form of this coefficient had not been fully established in the literature. The correctness of our calculations is further cross-checked using the heat kernel method, ensuring consistency of the results. We conclude by specializing to six dimensions, where $a_3(D)$ governs the logarithmic divergences of the effective action, and compare our findings with existing results.}

\paragraph{Conventions} Whether we work in Minkowski or Euclidean signature, and in which spacetime dimension, will be specified as needed throughout the text for convenience.

\section{Worldline path integral on the circle} \label{chap:sixth:sec2}
In this section, we briefly review the representation of the one-loop effective action for pure Einstein--Hilbert gravity in the Worldline Formalism by means of the massless $\N=4$ spinning particle. The first step would be to verify that the spectrum of this worldline model coincides with that of the graviton in $D$ dimensions, particularly once the coupling to a background metric is introduced. One then evaluates the path integral of the model on the circle, which yields the one-loop effective action of the graviton as a functional of the background metric. Rather than reproducing the full derivation of the worldline representation of the effective action for pure quantum gravity, we focus here on the main conceptual steps that lead to the result, referring the reader to Ref.~\cite{Bastianelli:2019xhi} for detailed treatments.
As emphasized earlier, quantum consistency of the worldline theory requires that the background metric satisfy Einstein’s equations of motion, i.e. that the Ricci tensor is proportional to the metric 
\begin{equation} \label{1EM}
R_{\mu\nu} = \lambda g_{\mu\nu}
\end{equation}
with constant $\lambda$, thus admitting a cosmological constant of indefinite sign. The BRST construction was crucial to develop a path integral quantization delivering a worldline representation of the one-loop effective action for quantum gravity as done in Ref.~\cite{Bastianelli:2019xhi}. Path integrals for one-dimensional nonlinear sigma models, such as the one associated with the $\mathcal{N} = 4$ spinning particle in curved spaces, require counterterms that are related to the regularization scheme used in defining the path integral itself \cite{Bastianelli:2006rx, new-book}. The counterterm used in Ref.~\cite{Bastianelli:2019xhi} was the one associated with wordline dimensional regularization and was effectively valid only for $D=4$. Its extension to arbitrary $D$ dimensions was constructed in Ref.~\cite{Bastianelli:2022pqq} and it is the one that we use here below. 

Alternative worldline representations of the effective action for quantum gravity are possible, see for instance \cite{Bastianelli:2013tsa, Bastianelli:2023oyz}. They are close in spirit to the heat kernel approach
employed in Section \ref{chap:sixth:sec4}. For their formulation they need direct inputs from the associated \acr{QFT}, just as it happens in the heat kernel method. In this sense, they are not independent of the second quantized theory 
and will not be used in this section. 

After this short review, we can move on to the construction of the worldline representation of the gravitational effective action.
 
\subsection{The worldloop path integral}
The one-loop effective action $\Gamma [g_{\mu\nu}]$ for pure gravity corresponds to the path integral of the massless $\mathcal{N}=4$ spinning particle action $S[X,G;g_{\mu\nu}]$ on worldlines with the topology of the circle $S^1$ and takes the schematic form 
\begin{equation} \label{1}
\Gamma[g_{\mu\nu}] = \int_{S^1}
\frac{\mathcal{D}G\,\mathcal{D}X}{\mathrm{Vol(Gauge)}}\, {\rm e}^{-S[X,G;g_{\mu\nu}]}\ .
\end{equation}
The particle Euclidean action depends on the worldline gauge fields $G=\left( e, \chi, \bar{\chi},a\right)$ and coordinates with supersymmetric partners $X=\left(x,\psi,\bar{\psi}\right)$, while the overcounting from summing over gauge equivalent configurations is formally taken into account by dividing by the volume of the gauge group. As implied by the BRST analysis, the gauge symmetries are consistent only when the background 
metric $g_{\mu\nu}$ is on-shell, i.e. satisfies eq. \eqref{1EM}. Explicitly, the effective action \eqref{1} is related to a partially gauge-fixed version of the $\mathcal{N}=4$ spinning particle path integral $Z(T)$ through a Schwinger representation, and given by 
\begin{equation} \label{euc}
\Gamma[g_{\mu\nu}] = -\frac{1}{2}\int_{0}^{\infty}\frac{dT}{T}Z(T)
\end{equation}
where $T$ is the usual Schwinger proper time, arising from the gauge-fixing of the einbein $e$ on the circle. In the present work, we only mention some of the important technicalities, namely the gauging of a parabolic subgroup of the $SO(\mathcal{N})$ $R$-symmetry group, the choice of the aforementioned gauge fixing of the worldline action, and the regularization of the nonlinear supersymmetric sigma model, skipping the details while referring to previous work for the reader interested in them \cite{Bastianelli:2019xhi, Bastianelli:2022pqq}. We choose to focus our discussion on the perturbative computation of the path integral providing only a few remarks when needed. The partition function $Z(T)$, upon gauge fixing, takes the following explicit form 
\begin{align} \label{master}
Z(T)&=\int_{0}^{2\pi}\frac{d\theta}{2\pi}\int_{0}^{2\pi}\frac{d\phi}{2\pi}\;P(\theta,\phi) \;\int_{_{\rm PBC}}\hskip-.4cm{ D}xDaDbDc\int_{_{\rm ABC}}\hskip-.4cm D\bar{\psi}D\psi \, e^{-S[X;g_{\mu\nu}]}\ ,
\end{align}
where $P(\theta,\phi)$ is the measure on the moduli space $(\theta, \phi)$ generated by the gauge fixing and which implements the correct projection on the physical graviton Hilbert space in $D$ dimensions. The worldline variables $X=\left(x,\psi,\bar{\psi},a,b,c\right)$ now include bosonic $a$ and fermionic $(b,c)$ “metric ghosts”, introduced in order to keep translational invariance of the path integral measure and which renormalize potentially divergent worldline diagrams \cite{Bastianelli:2006rx}. Once again, the path integral over bosonic variables and metric ghosts is evaluated by fixing periodic boundary conditions, while the fermionic path integral is performed by choosing antiperiodic boundary conditions on each flavor of fermionic fields $\psi^a_{i}$, with the internal index $i$ taking values $i=1,2$. The gauge-fixed nonlinear sigma model action reads\footnote{The bosonic coordinates are understood to be shifted as $\dot{x}^{\mu}\dot{x}^{\nu}\rightarrow\dot{x}^{\mu}\dot{x}^{\nu}+ a^{\mu}a^{\nu}+b^{\mu}c^{\nu}$. This shift implements into \eqref{action} the ghost action $S_{\rm{gh}}[x,a,b,c]=\int d\tau \tfrac{1}{4T} g_{\mu\nu}(x)(a^{\mu}a^{\nu}+b^{\mu}c^{\nu})$ which allows for the exponentiation of the determinant factor hidden inside the path integral measure on a curved spacetime, i.e. 
\begin{equation*}
\mathcal{D}x=\prod_\tau d^Dx(\tau)\sqrt{g(x(\tau))}={D}x\int Da Db Dc
\; e^{-S_{\rm{gh}}}\ ,
\end{equation*}
where $Dx$, $Da$, $Db$, and $Dc$ are the standard translational invariant measures. In particular, these metric ghosts create worldline divergences that compensate for the divergences generated by correlators of the $\dot x^\mu$'s. Divergences formally cancel out and one is left with a finite theory, whose remaining ambiguities are taken care of by choosing a regularization scheme with a corresponding counterterm that remains finite.}
\begin{align} \label{action}
S[X;g_{\mu\nu}]=\int d\tau \Big[ &\frac{1}{4T}g_{\mu\nu}(x)\,\dot{x}^{\mu}\dot{x}^{\nu}+\bar{\psi}^{a i}\left(
	\delta_i^j D_\tau - \hat a_{i}^{j}\right){\psi}_{aj} -T R_{abcd}(x) \, \bar{\psi}^{a}\cdot \psi^{b} \bar{\psi}^{c}\cdot \psi^{d} -T \, \mathcal{V}(x)\Big]\ ,
\end{align}
where we use flat indices on the worldline complex fermions $\psi^{a}_{i}$ and denoted the covariant derivative with spin connection $\omega_{\mu ab}$ acting on the fermions by $D_\tau \psi^a_i = \partial_\tau \psi^a_i + \dot x^\mu \omega_{\mu}{}^a{}_b (x) \psi^b_i$. We also used a dot to indicate contraction on the internal indices and denoted
\begin{equation}
\hat a_{i}^{j}=\left(\begin{array}{cc} \theta & 0 \\ 0 & \phi \\
\end{array}\right)
\end{equation}
the gauge-fixed values of the worldline gauge fields acting on the fermions and related to the gauging of the parabolic subgroup of the $R$-symmetry group. The angles $\theta$ and $\phi$ are precisely the two leftover moduli remaining after the gauge-fixing procedure.\footnote{As discussed in Ref.~\cite{Bastianelli:2019xhi}, the gauging of the parabolic subgroup allows the one-loop measure for the path integral to be modified so that it projects \emph{exactly} onto the graviton state. Alternatively, the gauging of the entire SO($4$) group would result in the graviton plus unwanted contributions of topological nature.}

A few comments are in order. The theory described by \eqref{action} should be seen as a one-dimensional field theory living on the worldline with the bosonic fields $x^\mu(\tau)$, the embedding of the worldline into spacetime, taking values in a $D$-dimensional target space $\cal{M}$. On the worldline, one usually finds it convenient to rescale the parameter $\tau$ to take values on the finite interval $I  :=  [0, 1]$. A crucial role is played by the scalar potential term $\mathcal{V}$ of quantum origin, see Ref.~\cite{Bastianelli:2022pqq}. It is necessary since it contains the counterterm required by the regularization scheme one decides to use to define the path integral and an additional potential needed to achieve nilpotency of the BRST charge at the quantum level. The latter condition requires a value of $V_{\rm BRST}= \frac{2}{D} R$ in the Hamiltonian constraint, see Chapter~\ref{chap:fourth}. Regarding the former, in the present chapter we adopt dimensional regularization (\acr{DR}) on the worldline, as used for instance in Refs.~\cite{Bastianelli:2000dw, Bastianelli:2002fv, Bastianelli:2002qw}
 in similar contexts, while reading from \cite{Bastianelli:2011cc} the counterterm $V_{\rm CT}=-\frac{1}{4} R$ needed for the case of four supersymmetries, thus producing an effective potential
\begin{equation}
\mathcal{V}=V_{\rm BRST}+V_{\rm CT}=\left(\frac{2}{D}-\frac{1}{4}\right)R  :=  \Omega R
\end{equation}
which indeed is the one used in Ref.~\cite{Bastianelli:2022pqq}.
Finally, for computational purposes, it is convenient to rewrite the angular integrations over the moduli $\theta$ and $\phi$ in the complex plane. This can be achieved by introducing the Wilson variables $z :=  e^{i\theta}$ and $\omega :=  e^{i\phi}$, so to recast the partition function as
\begin{align} \label{chap:sixth:MS} 
Z(T)&=\oint \frac{dz}{2\pi i}\frac{d\omega}{2\pi i}\;P(z,\omega) \;\int_{_{\rm PBC}}\hskip-.4cm{ D}xDaDbDc\int_{_{\rm ABC}}\hskip-.4cm D\bar{\psi}D\psi \, e^{-S[X;g_{\mu\nu}]}\ ,
\end{align}
where the modular integration is performed over the circle $|z|=1$, with the singular point $z=-1$ pushed out of the contour,\footnote{Poles at $z=-1$ arise when computing perturbative corrections and are excluded by this prescription. The same goes for $\omega$. Discussion on the regulated contour of integration for the modular parameters can be found in Ref.~\cite{Bastianelli:2005vk}.} and with the measure on the moduli space being now
\begin{equation} \label{measure}
P(z,\omega)=\frac{1}{2}\frac{(z+1)^{D-2}}{z^{3}}\frac{(\omega+1)^{D-2}}{\omega^{3}}(z-\omega)^{2}(z\omega-1)\ .
\end{equation}
In the next section, we set up the perturbative expansion for small values of $T$ of the path integral \eqref{chap:sixth:MS}, such that the full $T^{n}$ correction is an $(n+1)$-loop expansion\footnote{Here, $(n+1)$ identifies the loop order of the worldline sigma model expansion. Crucially, the worldline path integral remains a one-loop representation of the target-space gravitational theory.} in the worldline theory, which will allow us to identify the divergences in the effective action of pure gravity.

\subsection{Setting up the perturbative expansion}
Having at hand a path integral representation for the effective action, it is possible to set up the perturbative expansion around the free theory. However, there are two issues to take care of in order to be able to perform calculations. The first involves factorizing out the zero modes in the kinetic operator in \eqref{action}. Zero modes appear perturbatively once expanding around a constant metric and when considering periodic boundary conditions. To this task, we parametrize the bosonic coordinates of the circle (interpreted as a parametrization of the particle paths in target space) as
\begin{equation}
x^{\mu}(\tau) = x_{0}^{\mu}+q^{\mu}(\tau)\ ,
\end{equation}
thus describing all loops in spacetime with a fixed base point $x^{\mu}_{0}$ (the zero mode that is integrated over only at the end) plus quantum fluctuations with vanishing Dirichlet boundary conditions, 
indicated by $q^{\mu}(\tau)$ and thus satisfying $ q^\mu(0) =q^\mu(1) =0$. Note that the fermionic coordinates have no zero modes, due to their antiperiodic boundary conditions. The second issue consists in expanding in Riemann normal coordinates centered around $x^{\mu}_{0}$, so to write the metric tensor and the spin connection as follows \cite{Bastianelli:2000dw, Muller:1997zk}
\begin{align} 
\begin{split}
g_{\mu\nu}(x(\tau)) &= g_{\mu\nu} + \frac13 R_{\alpha\mu\nu\beta} q^\alpha q^\beta +\frac16 \nabla_\gamma R_{\alpha\mu\nu\beta} q^\alpha q^\beta q^\gamma+ R_{\alpha\beta\mu\nu\gamma\delta} q^\alpha q^\beta q^\gamma q^\delta \\
&\phantom{=}+\frac{1}{315}R_{\mu\alpha\beta}{}^{\sigma}R_{\sigma\gamma\delta}{}^{\lambda}R_{\lambda\tau\epsilon\nu}q^\alpha q^\beta q^\gamma q^\delta q^\tau q^\epsilon +\mathcal{O}(q^6) \label{RNCg} 
\end{split}\\[.5em]
\begin{split}
\omega_{\mu ab}(x(\tau)) &= \frac12 R_{\alpha\mu ab} q^\alpha +\frac13 \nabla_\alpha R_{\beta\mu ab} q^\alpha q^\beta +\frac18 \nabla_\alpha \nabla_\beta R_{\gamma\mu ab} +\frac{1}{24}R^{\tau}{}_{\alpha\beta\mu} R_{\gamma\tau ab}q^\alpha q^\beta q^\gamma + \mathcal{O}(q^4) \label{RNCw}
\end{split}
\end{align}
where \begin{equation}
R_{\alpha\beta\mu\nu\gamma\delta} := \frac{1}{20} \nabla_\delta \nabla_\gamma R_{\alpha\mu\nu\beta} +\frac{2}{45}R_{\alpha\mu}{}^\sigma{}_\beta R_{\gamma\sigma\nu\delta}
\end{equation}
and where we only kept the terms needed to obtain a perturbative expansion to order $T^3$. In \eqref{RNCg}-\eqref{RNCw} and henceforth, unless specified otherwise, we intend all tensor structures to be evaluated at the initial point $x^{\mu}_{0}$, thus factorizing out their dependence upon the worldline bosonic variables $q^\mu(\tau)$. Finally, the Riemann tensor appearing in the four-fermions interaction in \eqref{action} has to be Taylor expanded as well, around the same reference point $x^{\mu}_{0}$. The perturbative expansion of the path integral \eqref{chap:sixth:MS} reads 
\begin{equation} \label{4}
Z(T)=\oint \frac{dz}{2\pi i}\frac{d\omega}{2\pi i}\;P(z,\omega) \;\int d^{D}x_{0}\frac{\sqrt{g(x_{0})}}{\left(4\pi T\right)^{\frac{D}{2}}}\Big\langle e^{-S_{\rm int}}\Big\rangle \ ,
\end{equation}
factorizing out for convenience the $\sqrt{g(x_{0})}$ arising from the free ghost part of the action after functional integrating, together with $\left(4\pi T\right)^{-\frac{D}{2}}$ arising from the free particle 
path integral. The remaining expectation value is to be evaluated using the Wick theorem on the free path integral, with the free action being given by the quadratic part of \eqref{action}, namely
\begin{equation} \label{freeaction}
S_{0}[X] = \int d\tau \left[\frac{1}{4T}g_{\mu\nu}\left( \dot{x}^{\mu}\dot{x}^{\nu}+ a^{\mu}a^{\nu}+b^{\mu}c^{\nu}	\right)+\bar{\psi}^{ai}\left(\delta_i^j \, \partial_{\tau}-\hat a_{i}\,^{j}\right)\psi_{aj}\right]\ ,
\end{equation} 
from which one obtains the worldline propagators of the theory, reported in Appendix \ref{appendixB1}. Higher order terms form the interacting action $S_{\rm int}$, to be analyzed later on. It is possible to recast \eqref{4} in a more compact form introducing the double expectation value of the interacting action $\langle \hskip -.05 cm \langle\cdots \rangle \hskip -.05cm \rangle$, namely the average over the path integral and over the moduli space parametrized by $z$ and $w$
\begin{equation}
\Big\langle \hskip -.1cm \Big\langle e^{-S_{\rm int}}\Big\rangle \hskip -.1cm \Big\rangle
 = \oint \frac{dz}{2\pi i}\frac{d\omega}{2\pi i}\;P(z,\omega) \Big\langle e^{-S_{\rm int}}\Big\rangle \label{double}\ .
\end{equation}
Identifying the expectation values defined above as
\begin{align} 
\Big\langle e^{-S_{\rm int}}\Big\rangle = \sum_{n=0}^{\infty}a_{n}(D,z,\omega)\, T^{n} \quad \longrightarrow \quad \Big\langle \hskip -.1cm \Big\langle e^{-S_{\rm int}}\Big\rangle \hskip -.1cm \Big\rangle 
= \sum_{n=0}^{\infty}a_{n}(D)\, T^{n}\ , \label{chap:sixth:2.16}
\end{align}
allows us to rearrange the path integral \eqref{4} so as to make explicit the Seeley-DeWitt coefficients arising from the perturbative expansion
\begin{align}
Z(T)=\int \frac{d^{D}x_{0}}{\left(4\pi T\right)^{\frac{D}{2}}}\sqrt{g(x_{0})}\left[a_0(D)
+ a_1(D) \,T + a_2 (D)\, T^2 + a_3 (D)\, T^3+ \mathcal{O}(T^{4})\right]\ .\label{series}
\end{align}
One can recognize that, while in the above sum $a_{0}(D,z,\omega)=1$, its projected partner 
\begin{equation}
a_{0}(D) = \langle \hskip -.05cm \langle 1 \rangle \hskip -.05cm \rangle =
\frac{1}{2}\oint \frac{dz}{2\pi i}\frac{d\omega}{2\pi i}\frac{(z+1)^{D-2}}{z^{3}}\frac{(\omega+1)^{D-2}}{\omega^{3}}(z-\omega)^{2}(z\omega-1) = \frac{D(D-3)}{2} 
\end{equation}
gives the massless graviton physical polarizations in $D$ spacetime dimensions when using the correct measure $P(z,\omega)$ in eq. \eqref{measure}. We have now made explicit our main task, namely to determine the Seeley-DeWitt coefficient $a_3(D)$ in the perturbative expansion on Einstein spaces.

\subsection{Outline of the computation} \label{sec2.3}
We will now give a brief outline of the procedure, delegating the details of both the computation and regularization of potentially divergent diagrams to Appendix~\ref{appendixB}. To systematically work out all perturbative contributions to the desired order, one has first to identify and compute the connected worldline diagrams arising from the path integral expansion. In order to do so, we report the interacting action in \eqref{4} expanded to the desired order
\begin{align} \label{S-int}
S_{\rm int}&=\int d\tau\Bigg[
 \frac{1}{4T}\bigg(\frac{1}{3}R_{\alpha\mu\nu\beta}q^\alpha q^\beta +\frac{1}{6}\nabla_\gamma R_{\alpha\mu\nu\beta}q^\alpha q^\beta q^\gamma+\frac{1}{20} \nabla_\delta\nabla_\gamma R_{\alpha\mu\nu\beta}q^\alpha q^\beta q^\gamma q^\delta \nonumber\\
&+\frac{2}{45}R_{\alpha\mu}{}^{\sigma}\,_{\beta}R_{\gamma\sigma\nu\delta}q^\alpha q^\beta q^\gamma q^\delta+\frac{1}{315}R_{\mu\alpha\beta}{}^{\sigma}R_{\sigma\gamma\delta}{}^{\lambda}R_{\lambda\tau\epsilon\nu}q^\alpha q^\beta q^\gamma q^\delta q^\tau q^\epsilon \bigg)\left(\dot{q}^\mu\dot{q}^\nu +a^\mu a^\nu +b^\mu b^\nu \right)\nonumber\\
&+ 
\left( \frac{1}{2}R_{\alpha\mu ab}q^\alpha+\frac{1}{3}\nabla_\alpha R_{\beta\mu a b}q^\alpha q^\beta +\frac{1}{8}\nabla_\alpha \nabla_\beta R_{\gamma\mu a b}q^\alpha q^\beta q^\gamma+\frac{1}{24}R^{\tau}{}_{\alpha\beta\mu}R_{\gamma\tau a b }q^\alpha q^\beta q^\gamma\right)
\bar{\psi}^a \cdot \psi^b\, \dot{q}^\mu \nonumber
\\
&-T\left(R_{abcd}+q^\alpha \nabla_\alpha R_{abcd}+\frac{1}{2}q^\alpha q^\beta \nabla_\alpha \nabla_\beta R_{abcd} \right)\bar{\psi}^{a}\cdot \psi^{b}\bar{\psi}^{c}\cdot \psi^{d}-T\,\Omega R\Bigg]\ ,
\end{align}
which we write in a more compact form as 
\begin{align}
\begin{split} 
S_{\rm int}=&\frac{1}{4T}\left(S_{\rm K1} +DS_{\rm K1} + D^2S_{\rm K1} +S_{\rm K2} +S_{\rm K3}\right)\\ &
+ S_{\rm C1}+ DS_{\rm C1} + D^2S_{\rm C1}+ S_{\rm C2}\\
& -T\left( S_{\rm F}+ DS_{\rm F} +D^2 S_{\rm F}\right)+TS_{\rm V}\ , \label{10}
\end{split}
\end{align}
where the explicit expression of each term is obtained by comparing with the previous expression, see also Appendix \ref{appendixB}. Expanding the exponential in the path integral to order $T^3$, we obtain the contributions that need to be path-averaged and computed using Wick contractions. We list them here, leaving their systematic analysis to Appendix \ref{appendixB2}, where we also show more details on the intermediate steps of computation:
\begin{align}
\begin{split} \label{terms}
e^{-S_{\rm int}} \Big|_{T^{3}}&=S_{\rm KIN}-\frac{1}{8T}S_{\rm K1}S_{\rm C1}^{2}+S_{\rm C1}S_{\rm C2}+\frac{1}{2}DS_{\rm C1}^{2}+D^{2}S_{\rm C1}S_{\rm C1}\\
&\phantom{=}+\frac{1}{2}T^{2}S_{\rm C1}^{2}S_{\rm F}+\frac{1}{6}T^{3}S_{\rm F}^{3}+\frac{1}{2}T^2DS_{\rm F}^2 +T^2D^{2}S_{\rm F}S_{\rm F}\ ,
\end{split}
\end{align}
where we have collectively denoted $S_{\rm KIN}$ the three-loop contributions arising from the pure kinetic term, see \eqref{B27}. As an illustrative example, we shall show how to compute a contribution containing all the main features of this type of calculation, namely the correction arising from the spin-connection vertex only\footnote{The subscript $01$ in \eqref{C1C2} (as well as in Appendix \ref{appendixB}) serves as a shorthand notation to indicate henceforth $\int_{01} :=  \int_{0}^{1} d\tau\int_{0}^{1} d\sigma$ in worldline integrals.}
\begin{align} \label{C1C2}
\langle S_{\rm C1} S_{\rm C2}\rangle =\frac{1}{48}R_{\alpha\mu ab}R^{\tau}{}_{\beta\lambda\nu}
R_{\rho \tau cd}\int_{01}\langle \dot{q}^{\mu}_{0}q^{\alpha}_{0}\dot{q}^{\nu}_{1}q^{\beta}_{1}q^{\lambda}_{1}q^{\rho}_{1}\rangle \langle \bar{\psi}^{a}_{0}\cdot \psi_{0}^{b} \bar{\psi}^{c}_{1}\cdot \psi_{1}^{d}\rangle\ .
\end{align}
The fermionic Wick contractions produce
\begin{equation} 
R_{\alpha\mu ab}R^{\tau}{}_{\beta\lambda\nu}
R_{\rho \tau cd} \, \langle \bar{\psi}^{a}_{0}\cdot \psi_{0}^{b} \bar{\psi}^{c}_{1}\cdot \psi_{1}^{d}\rangle
=-\del_{\rm{AF}}{}_{ji}(\tau,\sigma)\del_{\rm{AF}}{}_{ij}(\sigma,\tau)R_{\alpha\mu ab}R^{\tau}{}_{\beta\lambda\nu}R_{\rho\tau}{}^{ba}\ ,
\end{equation}
where we introduced the $\mathcal{N}=4$ fermionic propagator $\del_{\rm{AF}}$, cf.~Eq.~\eqref{AF}. Evaluating then the bosonic contractions one gets different Riemann tensor strings, which can be reduced using
\begin{align}
R^{\mu\nu\rho\sigma}R_{\mu\nu}{}^{\alpha\beta}R_{\rho\alpha\sigma\beta}=\frac{1}{2}R^{\mu\nu\rho\sigma}R_{\mu\nu}{}^{\alpha\beta}R_{\rho\sigma\alpha\beta}\;.
\end{align}
Recall that all the calculations have to be further carried out with two precautions, namely to use Einstein spaces simplifications (see Appendix \ref{appendixA}) and to perform the regularization using \acr{DR}. Then, collecting all terms one gets 
\begin{equation}
\left( \frac{1}{4}R_{\mu\nu\rho\sigma}R^{\rho\sigma\alpha\beta}R_{\alpha\beta}{}^{\mu\nu}+\frac{R}{6 D} R_{\mu\nu\rho\sigma}^{2}\right) \left(
\ione -\itwo +z\rightarrow \omega 
\right)
\end{equation}
where in the graphical representation of the worldline Feynman diagrams full dots denote vertices, an empty dot represents a derivative, a line denotes a bosonic propagator, and oriented lines represent $z$-fermionic propagators. The first diagram has to be regularized due to the singularity carried by the bosonic propagator $\ddeld$. In \acr{DR} it has to be $d$-dimensional extended as outlined in the following:
\begin{align}
& \hspace{-1cm}
\ione = \int d\tau d\sigma \ \ddeld (\tau,\sigma) \del(\tau,\sigma) \del(\tau,\tau) 
F(z,\tau,\sigma)F(z,\sigma,\tau) 
\nonumber \\
& \hspace{-.9cm}
\rightarrow \int d^{d+1}t\, d^{d+1}s \, 
{}_{\alpha}\del_{\beta}(t,s)\, \del(s,t)\del(t,t)\Tr\left( \gamma^{\alpha}F(z,t,s)\gamma^{\beta}F(z,s,t)\right)
\nonumber \\
=-&\int d^{d+1}t\, d^{d+1}s \ 
{}_{\alpha}\del(t,s)\, \del(t,s)_{\beta}\del(t,t)\Tr\left(\gamma^{\alpha} F(z,t,s)\gamma^{\beta}F(z,s,t)\right) 
\nonumber \\
-&\int d^{d+1}t\, d^{d+1}s \
{}_{\alpha}\del(t,s)\del(s,t)\del(t,t)\Tr \left(F(z,t,s)\lpartial F(z,s,t)+F(z,t,s)\rpartial F(z,s,t) \right)
\nonumber \\
\rightarrow -&\int d\tau d\sigma \, \ddel(\tau,\sigma)\deld(\tau,\sigma)\del(\tau,\tau) F(z,\tau,\sigma)F(z,\sigma,\tau)
\nonumber \\
 \hskip.3cm = -\hskip .1cm &\itwo =\frac{z}{60(z+1)^{2}} \label{2.32}
\end{align}
where in reaching the last-but-one line we used the regulated Green equation for the component $F$ of the fermionic propagator $\del_{\rm{AF} \,}$, see eqs. \eqref{AF}, \eqref{B16} and \eqref{greenEQ2} of Appendix \ref{appendixB}, and then removed the regularization by sending the extra dimensions to zero ($d\to 0$). Finally, taking into account the $\omega$ partner of the integrals in \eqref{2.32}, one gets the following result:
\begin{align}
\langle S_{\rm C1}S_{\rm C2}\rangle = \left( \frac{1}{120}R_{\mu\nu\rho\sigma}R^{\rho\sigma\alpha\beta}R_{\alpha\beta}{}^{\mu\nu}+\frac{R}{180D} R_{\mu\nu\rho\sigma}^{2}\right)\left(\frac{z}{(z+1)^{2}}+\frac{\omega}{(\omega+1)^{2}}\right)\;.
\end{align}
Once having worked out systematically all the corrections up to and including order $T^3$ related to the connected graphs, it only remains to exponentiate them and, finally, Taylor expand in $T$ so as to reach the desired order and get the full result.

\subsection{Seeley-DeWitt coefficients} \label{sec2.4}
The calculation of the various terms in the perturbative expansion delivers the following coefficients in the perturbative series \eqref{series}, including the newly found $a_3(D)$
\begin{align}
\begin{split}
a_{0}(D)&= \frac{D(D -3)}{2} \label{a0}
\end{split}\\[.5em]
\begin{split}
\label{a1} a_{1}(D)&= \frac{D^{2}-3D-36}{12} \; R
\end{split}\\[.5em]
\begin{split}
\label{a2} a_{2}(D)&=
\frac{5 D^3 -17 D^2 -354 D -720}{720D }\; R^2 +\frac{D^2-33 D+540 }{360} \; R_{\mu\nu\rho\sigma}^{2}
\end{split}\\[.5em]
\begin{split}
a_{3}(D)&= \frac{35 D^4-147 D^3-3670 D^2-13560D-30240}{90720 D^2}\;R^3+\frac{7 D^3-230 D^2+3357 D+12600}{15120 D}\;R\,R_{\mu\nu\rho\sigma}^{2}\\[.3em]
&\phantom{=}+ 
\frac{17 D^2-555 D-15120}{90720} \;R_{\mu\nu\rho\sigma}R^{\rho\sigma\alpha\beta}R_{\alpha\beta}{}^{\mu\nu}+\frac{D^2-39 D-1080}{3240}\; R_{\alpha\mu\nu\beta}R^{\mu\rho\sigma\nu}R_{\rho}{}^{\alpha\beta}{}_{\sigma}\ . \label{a3}
\end{split}
\end{align}
The above expressions are understood to be gauge-invariant, as they have been calculated specifically on Einstein spaces for the reasons previously discussed. As we shall discuss later, the newly computed coefficient $a_3(D)$ plays a key role in quantum gravity in six dimensions, as it is related to the logarithmic divergences, hence it will be useful to read its value explicitly in this case as well. In $D=6$ it reduces to
\begin{align}
\left. a_{3}\right|_{D=6}&=-\frac{799}{11340}\; R^3 +\frac{481}{1680}\; R\, R_{\mu\nu\rho\sigma}^{2}-\frac{991}{5040} \; R_{\mu\nu\rho\sigma}R^{\rho\sigma\alpha\beta}R_{\alpha\beta}{}^{\mu\nu}-\frac{71}{180}\, R_{\alpha\mu\nu\beta}R^{\mu\rho\sigma\nu}R_{\rho}{}^{\alpha\beta}{}_{\sigma}\ . \label{a3D6}
\end{align}
Also, its expression on maximally symmetric spaces (MSS) may be useful for future reference. As we are not aware of any such calculation carried out in the literature we list it here. Using the relations in Appendix \ref{appendixA}, we find that \eqref{a3} on MSS collapses into 
\begin{align} \label{94}
a_{3}^{\rm MSS}(D) = \frac{35 D^6-217 D^5-3257 D^4-9239D^3+37470D^2+183672 D-302400}{90720 (D-1)^2 D^2} \; R^3 \ ,
\end{align}
yielding in $D=6$
\begin{align}
\left.a_{3}^{\rm MSS}\right|_{D=6} =-\frac{3181}{63000}\; R^3\ .
\end{align}

\subsubsection{Ghost and Graviton separately}
For future comparison with the results of the heat kernel method calculation, it is useful to project on the degrees of freedom of the ghost and graviton respectively. To this task, we must use the right measures in the path integral \eqref{4}. These can be found generalizing to our case the worldline partition functions of Ref.~\cite{Bastianelli:2013tsa} leading us to 
\begin{align}
&P_{\rm gh}(z,\omega)=\frac{(z+1)^D}{z^2}\frac{1}{\omega }\ , \\[.5em]
&P_{\rm gr}(z,\omega)=\frac{2 (z+1)^D}{\omega z^2}+\frac{(\omega +1)^{D-2} (z+1)^{D-2} (z-\omega )^2 (\omega z-1)}{2 \omega ^3 z^3}\ .
\end{align}
Indeed, note that
\begin{equation}\label{SeparateMeas}
P_{\rm gr}(z,\omega) -2 \, P_{\rm gh}(z,\omega)=P(z,\omega)\ .
\end{equation}
It is therefore possible to obtain the contributions to the Seeley-DeWitt coefficients coming only from the ghost (graviton) projecting onto the desired Hilbert space via $P_{\rm gh}\,(P_{\rm gr})$. Let us emphasize that these results are not new \emph{per se} since the heat kernel procedure requires calculating them individually and then putting them together, as we shall see. What is new here is the possibility of obtaining them also from the worldline viewpoint of the ${\cal N}=4$ particle. Therefore, regarding the ghost we have
\begin{align}
\begin{split}
\label{a0gh} a_{0}^{\rm gh}(D)&= D
\end{split}\\[.5em]
\begin{split}
\label{a1gh} a_{1}^{\rm gh}(D)&= \frac{D+6}{6}\; R
\end{split}\\[.5em]
\begin{split}
\label{a2gh} a_{2}^{\rm gh}(D)&=\frac{5 D^2+58 D+180}{360 D}\; R^2 + \frac{D-15}{180}\; R_{\mu\nu\rho\sigma}^{2}
\end{split}\\[.5em]
\begin{split}
\label{a3gh} a_{3}^{\rm gh}(D)&=\frac{35 D^3+588 D^2+3512 D+7560}{45360 D^2}\; R^3 +\frac{7 D^2-62 D-714}{7560 D}\;R\, R_{\mu\nu\alpha\beta}^{2} \\[.3em]
&\phantom{=}+\frac{17 D-252}{45360}\;R_{\mu\nu\rho\sigma}R^{\rho\sigma\alpha\beta}R_{\alpha\beta}{}^{\mu\nu} +\frac{D-18}{1620}\;R_{\alpha\mu\nu\beta}R^{\mu\rho\sigma\nu}R_{\rho}{}^{\alpha\beta}{}_{\sigma}\ ,
\end{split}
\end{align}
while concerning the graviton 
\begin{align}
\begin{split}
\label{a0gr} a_{0}^{\rm gr}(D)&=\frac{1}{2} D (D+1) 
\end{split}\\[.5em]
\begin{split}
\label{a1gr} a_{1}^{\rm gr}(D)&= \frac{1}{12} \left(D^2+D-12\right)\; R
\end{split}\\[.5em]
\begin{split}
\label{a2gr} a_{2}^{\rm gr}(D)&=\frac{\left(5 D^2+3 D-122\right)}{720} \; R^2 +\frac{\left(D^2-29 D+480\right)}{360} \; R_{\mu\nu\rho\sigma}^{2}
\end{split}\\[.5em]
\begin{split}
\label{a3gr} a_{3}^{\rm gr}(D)&= \frac{35 D^3-7 D^2-1318 D+488}{90720 D}\;R^3+ \frac{7 D^3-202 D^2+3109 D+9744}{15120 D}\;R\, R_{\mu\nu\alpha\beta}^{2} \\[.3em]
&\phantom{=}+\frac{17 D^2-487 D-16128}{90720}\;R_{\mu\nu\rho\sigma}R^{\rho\sigma\alpha\beta}R_{\alpha\beta}{}^{\mu\nu}+\frac{D^2-35 D-1152}{3240}\;R_{\alpha\mu\nu\beta}R^{\mu\rho\sigma\nu}R_{\rho}{}^{\alpha\beta}{}_{\sigma}\ .
\end{split}
\end{align}
One can easily see that the coefficient $a_0(D)$ correctly reproduces the expected degrees of freedom in $D$ dimensions. Moreover, it is immediate to check that by summing up the contributions as prescribed by \eqref{SeparateMeas} we obtain the correct total coefficients \eqref{a0}--\eqref{a3}.

\section{Heat kernel method} \label{chap:sixth:sec3'}
The gauge-invariant coefficients computed from the $\mathcal{N}=4$ spinning particle can be obtained in an equivalent but completely independent manner by exploiting the heat kernel method. In this section, we apply it to the case of Euclidean perturbative quantum gravity, to reproduce the coefficients \eqref{a1}--\eqref{a3} and provide a strong consistency check for our computations.

\subsection{Relevant formulae}
We recall here, for ease of reference, the main equations from the \hyperref[chap:introII]{prelude to Part~II}, which will be needed to carry out the heat kernel computation. In particular, recall that an elliptic second-order differential operator can be written in the form
\begin{equation}\label{DiffOp}
\mathcal{Q}= -\nabla_{\scriptscriptstyle \! \! ({\cal A})}^2 -V 
\end{equation}
and that the associated heat kernel $K(x,x';T)$ admits, for small Euclidean time $T\to 0^+$, the coincident-point expansion
\begin{equation}
K(x,x;T)\sim (4\pi T)^{-\frac{D}{2}}\sum_{j=0}^{\infty} T^j a_j(x)\ ,
\end{equation}
in terms of the local heat kernel coefficients $a_j(x)$. These coefficients determine the divergent structure of the one-loop effective action, which can be expressed as
\begin{equation}\label{Gamma1ter}
\Gamma=-\frac{1}{2}\int_0^\infty \frac{\diff{T}}{T}\ \exp{(-T m^2)}\int \frac{\diff{^Dx}\,\sqrt{g}}{(4\pi T)^{\frac{D}{2}}}\ \operatorname{sTr}{\sum_{j=0}^{\infty} T^j a_j(x)}\ .
\end{equation}
The problem of finding the \acr{UV} divergences of the effective action is then reduced to the computation of Seeley-DeWitt coefficients for a generic theory, which has already been carried out up to $a_4(x)$ \cite{Avramidi2015}. For application to perturbative quantum gravity, we will consider here the first four coefficients, i.e. from $a_0(x)$ to $a_3(x)$. The fourth coefficient $a_3(x)$, in particular, has been computed for the first time by Gilkey \cite{Gilkey:1975iq}, and later confirmed by Avramidi through a fully covariant method \cite{Avramidi:1990je}.\footnote{It is important to note that the notation we employ here for the Riemann tensor, Ricci tensor, and scalar is the same of Ref.~\cite{Bastianelli:2022pqq, Bastianelli:2000hi}, reported in Eq.~\eqref{Notation}. Some references, notably Refs.~\cite{Gilkey:1975iq, Vassilevich:2003xt}, adopt different conventions.} 

We now list the general results for the coefficients corresponding to the operator \eqref{DiffOp} up to $a_3(x)$, as taken from \cite{Gilkey:1975iq, Bastianelli:2000hi}. Consider an exponentiated form of the heat kernel series:
\begin{equation}\label{HKconnected}
\operatorname{sTr}{\left[\sum_{j=0}^{\infty} T^j a_j(x) \right]} :=  \operatorname{sTr}{\left[\exp{\left(\sum_{j=1}^{\infty} T^j \alpha_j(x)\right)}\right]}\ .
\end{equation}
It is useful to define, for the sake of brevity, the remainder
\begin{equation}\label{HKalphabeta}
 \beta_j(x):=a_j(x)-\alpha_j(x)\ ,
\end{equation}
which is evaluated, up to the third order, as
\begin{equation}\label{BetaValues}
\beta_0=\beta_1 = 0\ ,\quad
\beta_2 = \frac{1}{2}\alpha_1^2\ ,\quad
\beta_3 = \frac{1}{6}\alpha_1^3+\alpha_1\alpha_2\ .
\end{equation}
The coefficients $\alpha_j(x)$ are given by 
\begin{align}
\begin{split}
\label{HK0} \alpha_0(x) &= \mathbbm{1}\;,
\end{split}\\[.5em]
\begin{split}
\label{HK1} \alpha_1(x) &= \frac{1}{6}R\mathbbm{1} + V\;,
\end{split}\\[.5em]
\begin{split}
\label{chap:sixth:HK2} \alpha_2(x) &= \frac{1}{6}\nabla^2\left(\frac{1}{5}R\mathbbm{1}+V\right)+ \frac{1}{180}\left(R_{\mu\nu\rho\sigma}^2-R_{\mu\nu}^2\right)\mathbbm{1} + \frac{1}{12}\Omega_{\mu\nu}^2\;,
\end{split}\\[.5em]
\begin{split} \label{HK3}
\alpha_3(x) &= \frac{1}{7!}\left[18\nabla^4 R + 17(\nabla_\mu R)^2-2(\nabla_\mu R_{\nu\sigma})^2 - 4\nabla_\mu R_{\nu\sigma}\nabla^\nu R^{\mu\sigma}+9(\nabla_\alpha R_{\mu\nu\rho\sigma})^2 -8R_{\mu\nu}\nabla^2 R^{\mu\nu}\right. \\
&\phantom{=}
+12R^{\mu\nu}\nabla_\mu \nabla_\nu R
+12R_{\mu\nu\rho\sigma}\nabla^2R^{\mu\nu\rho\sigma} 
+ \frac{8}{9}R_{\mu}{}^{\nu}R_{\nu}{}^{\sigma}R_{\sigma}{}^{\mu} - \frac{8}{3}R_{\mu\nu}R_{\rho\sigma}R^{\mu\rho\nu\sigma}
\\
&\phantom{=}\left.-\frac{16}{3}R_{\mu\nu}R^\mu{}_{\rho\sigma\tau}R^{\nu\rho\sigma\tau}+\frac{44}{9}R_{\mu\nu}{}^{\rho\sigma}R_{\rho\sigma}{}^{\alpha\beta}R_{\alpha\beta}{}^{\mu\nu}+\frac{80}{9}R_{\mu\nu\rho\sigma}R^{\mu\alpha\rho\beta}R^{\nu}{}_{\alpha}{}^{\sigma}{}_{\beta}\right]\mathbbm{1}\\
&\phantom{=}+\frac{2}{6!}\left[8(\nabla_\mu \Omega_{\nu\sigma})^2+2(\nabla^\mu \Omega_{\mu\nu})^2 + 12\Omega_{\mu\nu}\nabla^2\Omega^{\mu\nu}-12\Omega_{\mu}{}^{\nu}\Omega_{\nu}{}^{\sigma}\Omega_{\sigma}{}^{\mu}+6R_{\mu\nu\rho\sigma}\Omega^{\mu\nu}\Omega^{\rho\sigma}-4R_{\mu\nu}\Omega^{\mu\sigma}\Omega^\nu{}_{\sigma}\right.\\
&\phantom{=}\left. +6\nabla^4 V+30(\nabla_\mu V)^2+4R_{\mu\nu}\nabla^\mu\nabla^\nu V+12\nabla_\mu R\nabla^\mu V\right]\ .
\end{split}
\end{align}
They will be used in the next section.

\subsection{Euclidean quantum gravity}
Consider a $D$-dimensional Riemannian manifold $(\mathcal{M}, \bm{G})$ equipped with a metric tensor $\bm{G}$ having Euclidean signature. The starting point for our treatment of gravity is the Einstein--Hilbert action,
\begin{equation}\label{EH}
S_{\rm EH}[\bm{G}]=-\frac{1}{k^2}\int\diff{^D x}\ \sqrt{G}\left[R(\bm{G})-2\Lambda\right]\ ,
\end{equation}
where $k^2  :=  16\pi G_{\rm N}$, being $G_{\rm N}$ the Newton constant, $R(\bm{G})$ is the Ricci scalar computed from $\bm{G}$, and $G :=  \left|\det{G_{\mu\nu}}\right|$. A cosmological constant $\Lambda$ has also been included. By employing the background field method, we split the metric tensor $\bm{G}$ into a fixed classical background $\bm{g}$ and “small” quantum perturbations $\bm{h}$, namely:
\begin{equation}\label{Splitting}
G_{\mu\nu}(x) = g_{\mu\nu}(x) + h_{\mu\nu}(x)\ .
\end{equation}
As a consequence of this splitting, the action \eqref{EH} can be expanded in power series in the fluctuations $\bm{h}$. Since we are interested in the one-loop level of accuracy, we will be concerned with the second-order term in $\bm{h}$, which reads 
\begin{align}
\label{S2} S_2 &= \int \diff{^Dx} \sqrt{g} \left[-\frac{1}{4}h^{\mu\nu}\left(\nabla^2+2\Lambda-R\right)h_{\mu\nu} +\frac{1}{8}h\left(\nabla^2+2\Lambda-R\right)h -\frac{1}{2}\left(\nabla^\nu h_{\mu\nu}-\frac{1}{2}\nabla_\mu h\right)^2\right. \nonumber \\
&\phantom{= \int \diff{^Dx} \sqrt{g}+}\left. - \frac{1}{2}\left(h^{\mu\lambda}h_\lambda{}^{\nu} - hh^{\mu\nu}\right)R_{\mu\nu} -\frac{1}{2}h^{\mu\lambda}h^{\nu\rho}R_{\mu\nu\lambda\rho}\right]\ .
\end{align}
It is important to note that in \eqref{S2} the Ricci tensor $R_{\mu\nu}=R_{\mu\nu}(\bm{g})$ and scalar $R=R(\bm{g})$, as well as covariant derivatives $\nabla_\mu = \nabla_\mu(\bm{g})$, are computed with respect to the background metric $\bm{g}$. The gauge symmetries acting on $\bm{h}$ and leaving the background metric $\bm{g}$ invariant are then BRST quantized by introducing the ghost $c$ and antighost $\overline{c}$ fields, and adding to the action the Slavnov variation of the de Donder gauge-fixing function $f_\mu =\nabla^\nu h_{\mu\nu}-\frac{1}{2}\nabla_\mu h$, see for example \cite{Bastianelli:2013tsa} for further details. The final result is
\begin{equation}
S_2[\bm{h},c,\overline{c}] = S_{\rm gr}[\bm{h}]+S_{\rm gh}[c,\overline{c}]\ ,
\end{equation}
where
\begin{align}
\begin{split}
S_{\rm gr}[\bm{h}] &= \int \diff{^Dx} \sqrt{g} \left[-\frac{1}{4}h^{\mu\nu}\left(\nabla^2+2\Lambda-R\right)h_{\mu\nu} +\frac{1}{8}h\left(\nabla^2+2\Lambda-R\right)h\right. \\
&\phantom{= \int \diff{^Dx} \sqrt{g}+}\left. - \frac{1}{2}\left(h^{\mu\lambda}h_\lambda{}^{\nu} - hh^{\mu\nu}\right)R_{\mu\nu} -\frac{1}{2}h^{\mu\lambda}h^{\nu\rho}R_{\mu\nu\lambda\rho}\right]\ ,
\end{split}\\ 
\begin{split}
S_{\rm gh}[c,\overline{c}] &= \int \diff{^Dx} \sqrt{g}\ \bar{c}^\mu \left(\nabla^2 c_\mu + R_{\mu\nu}c^\nu\right)\ .
\end{split}
\end{align}
We are now able to identify from the actions $S_{\rm gr}$ and $S_{\rm gh}$ the invertible kinetic operators for the graviton and ghost fields, denoted by $F_{\mu\nu\alpha\beta}$ and $\mathcal{F}_{\mu\nu}$, respectively. By setting
\begin{equation}
S_{\rm gr}[\bm{h}] =\int \diff{^Dx}\ \sqrt{g}\ \frac{1}{2}h_{\mu\nu}F^{\mu\nu\alpha\beta}h_{\alpha\beta} \qquad\mbox{and}\qquad S_{\rm gh}[c,\overline{c}]=\int \diff{^Dx}\ \sqrt{g}\ \bar{c}_\mu \mathcal{F}^{\mu}{}_{\nu} c^\nu\ ,
\end{equation}
and exploiting the properties of Einstein spaces, we find
\begin{align}
\label{GravitonKin} F_{\mu\nu}{}^{\alpha\beta} &= -\frac{1}{2}\left(\delta_\mu^\alpha\delta_\nu^\beta+\delta_\mu^\beta\delta_\nu^\alpha\right)\nabla^2 - R_{\mu}{}^{\alpha}{}_{\nu}{}^{\beta} - R_{\mu}{}^{\beta}{}_{\nu}{}^{\alpha}\\[.3em]
\label{GhostKin} \mathcal{F}^{\mu}_{\ \nu} &= \delta^\mu_\nu \left(\nabla^2+\frac{1}{D}R\right)\ ,
\end{align}
where the graviton operator indices are raised and lowered with the DeWitt supermetric
\begin{equation}\label{DeWittSM}
\gamma^{\mu\nu\alpha\beta}  :=  \frac{1}{4}\left(g^{\mu\alpha}g^{\nu\beta}+g^{\mu\beta}g^{\nu\alpha}-g^{\mu\nu}g^{\alpha\beta}\right)\ ,\quad
\gamma_{\mu\nu\alpha\beta} = g_{\mu\alpha}g_{\nu\beta}+ g_{\mu\beta}g_{\nu\alpha} - \frac{2}{D-2}g_{\mu\nu}g_{\alpha\beta}\ .
\end{equation}

The reasons for immediately reducing to Einstein spaces --- that is, to compute the coefficients directly on-shell --- are twofold. Firstly, this allows us a direct comparison with the worldline results, even separately for ghost and graviton, since the on-shell condition is forced by the quantum consistency of the $\mathcal{N}=4$ spinning particle. Secondly, proceeding otherwise the results would not be gauge-invariant, but rather would depend on the gauge chosen, as expected for gauge theories, see for instance the recent analysis carried out in Ref.~\cite{Brandt:2022und}.

Since \eqref{GravitonKin}--\eqref{GhostKin} are elliptic second-order differential operators, they can be treated within the heat kernel expansion, and the coefficients \eqref{HK0}--\eqref{HK3} can be computed by identifying the explicit formulae for $\mathbbm{1}$, $V$ and $\Omega_{\mu\nu}$. The ghost field configuration space is $D$-dimensional, and by comparing \eqref{GhostKin} with \eqref{DiffOp}, as well as recalling that
\begin{equation}\label{GhostComm}
[\nabla_\mu, \nabla_\nu]c^\rho=R_{\mu\nu}{}^{\rho}{}_{\sigma}c^\sigma\ ,
\end{equation}
we conclude that the substitutions to be performed in the heat kernel coefficients \eqref{HK0}--\eqref{HK3} are
\begin{equation}\label{GhostSost}
\begin{cases}
\mathbbm{1}\ \leftrightarrow\ \delta^\mu_\nu\\[.3em]
V\ \leftrightarrow\ \dfrac{1}{D}R\delta^\mu_\nu\\[.5em]
(\Omega_{\mu\nu})^\rho_{\ \sigma} \ \leftrightarrow\ R_{\mu\nu}{}^{\rho}{}_{\sigma}\ .
\end{cases}
\end{equation}
Note that in this expression the indices $\mu$, $\nu$ label the different elements of the gauge field strength $\Omega_{\mu\nu}$, which are $D\times D$ matrices whose components are given by the (spacetime) indices $\rho$, $\sigma$. On the other hand, for the graviton field the configuration space is $\frac{1}{2}D(D+1)$ dimensional (space of symmetric tensors) and the substitutions to be performed are
\begin{equation}\label{GravitonSost}
\begin{cases}
\mathbbm{1}\ \leftrightarrow\ \delta_{\mu\nu}^{\ \ \ \alpha\beta}\\
V\ \leftrightarrow\ \mathcal{V}_{\mu\nu}{}^{\alpha\beta}\\
(\Omega_{\mu\nu})_{\rho\sigma}{}^{\alpha\beta} \ \leftrightarrow\ R_{\rho\sigma}{}^{\alpha\beta}{}_{\mu\nu}
\end{cases}
\end{equation}
where
\begin{align}
\label{SuperDelta} \delta_{\mu\nu}{}^{\alpha\beta} & :=  \frac{1}{2}\left(\delta_\mu^\alpha\delta_\nu^\beta+\delta_\mu^\beta\delta_\nu^\alpha\right)\\[.5em]
\label{VGraviton} \mathcal{V}_{\mu\nu}{}^{\alpha\beta} & :=  R_{\mu}{}^{\alpha}{}_{\nu}{}^{\beta} + R_{\mu}{}^{\beta}{}_{\nu}{}^{\alpha}\ ,
\end{align}
and the commutator is given by a symmetrized version of the Riemann tensor
\begin{equation}\label{GravitonComm}
[\nabla_\mu, \nabla_\nu]h_{\rho\sigma}=R_{\rho\sigma}{}^{\alpha\beta}{}_{\mu\nu}\,h_{\alpha\beta}\ ,
\end{equation}
where
\begin{equation}\label{SuperRiemann}
R_{\rho\sigma}{}^{\alpha\beta}{}_{\mu\nu} :=  \frac{1}{2}\left(\delta_\rho^\alpha R_{\sigma}{}^{\beta}{}_{\mu\nu} + \delta_\rho^\beta R_{\sigma}{}^{\alpha}{}_{\mu\nu} +\delta_\sigma^\alpha R_{\rho}{}^{\beta}{}_{\mu\nu} + \delta_\sigma^\beta R_{\rho}{}^{\alpha}{}_{\mu\nu} \right)\ .
\end{equation}
At this point, the computations are tedious but straightforward, see Appendix \ref{appendixC} for details. The final results for the ghost and graviton fields separately are precisely the coefficients \eqref{a0gh}--\eqref{a3gh} and \eqref{a0gr}--\eqref{a3gr} obtained from the Worldline Formalism. The total coefficients for the physical graviton, according to the supertrace appearing in \eqref{Gamma1ter}, as recognized also 
from \eqref{SeparateMeas}, are given by
\begin{equation}\label{TotHKgen}
\Tr{\left[a_j\right]}=\Tr{\left[a_j^{\rm gr}\right]}-2\Tr{\left[a_j^{\rm gh}\right]}\ .
\end{equation}
Again, the results obtained reproduce the ones coming from worldline computations \eqref{a1}--\eqref{a3}, providing a strong cross-check for the correctness of both.

\section{On one-loop divergences of quantum gravity} \label{chap:sixth:sec4}
The coefficients \eqref{a1}--\eqref{a3}, which include the newly computed coefficient $a_3(D)$, allow for further investigations of the issue of divergences in the quantum theory of gravity. Thus, let us review some crucial results from the literature and discuss how our newly-calculated coefficient $a_3(D)$ fits into the picture, thus providing us with additional confirmation of the validity of our result. We focus our discussion on the spacetime of dimensions $D=4$ and $D=6$, as in these cases there are no additional divergences on top of the one we have already computed (new divergences start to appear from $D=8$ onwards). The type of divergences arising in quantum gravity emerge naturally from the representation of the one-loop effective action with a short proper time expansion, which we can read both from the worldline viewpoint \eqref{series} and from the heat kernel one \eqref{Gamma1ter}:
\begin{equation}
\Gamma[g_{\mu\nu}]= -\frac{1}{2}\int_{0}^{\infty}\frac{dT}{T^{1+\frac{D}{2}}}\int \frac{d^{D}x_{0}}{\left(4\pi \right)^{\frac{D}{2}}}\sqrt{g(x_{0})}\left[a_0 + a_1 T + a_2 T^2 + a_3 T^3+ \mathcal{O}(T^{4})\right]\ . \label{divergences}
\end{equation} 
We are interested in studying the \acr{UV} divergences that arise from the $T\rightarrow0$ limit of the proper time integration. Setting $D=4$ we recognize that possible divergences arise from the coefficients $a_0, a_1,a_2$,
with $a_2$ being associated with the logarithmic divergence. In $D=6$, also $a_3$ gives rise to an additional divergence, the logarithmic one in that dimension.

One may wonder how to relate the $\frac{1}{\epsilon}$ pole of dimensional regularization in \acr{QFT}, widely present in the literature, with our situation. To address this point, it is useful to evaluate the proper time integral term by term in \eqref{divergences}, to display the gamma function dependence. We find 
\begin{equation}
\int_{0}^{\infty}\frac{dT}{T^{1+\frac{D}{2}}}T^{p} \, e^{-m^{2}T}= (m^{2})^{\frac{D}{2}-p}\, \Gamma\left(p-\frac{D}{2}\right)\ ,
\end{equation}
where $p=2,3$ correspond to our cases of interest $D=4,6$, respectively. Now, using dimensional regularization, namely taking $D=2p-2\epsilon$ and expanding the gamma function, we see the appearance of the usual $\frac{1}{\epsilon}$ pole as the leading divergent term: it corresponds precisely to the logarithmic divergences seen in dimensional regularization \cite{Schwartz:2014sze}. 

In general, one has to deal also with infrared divergences: for the sake of our discussion, they can be avoided either by introducing an upper cutoff in the proper time or by keeping a “small” mass regulator $m$, as we have done above.
 
\subsection{Pure gravity in four dimensions}
It has long been known since the pioneering work of 't Hooft and Veltman \cite{tHooft:1974toh}, that pure gravity with vanishing cosmological constant is a renormalizable theory at one-loop in $D=4$. It is free of logarithmic divergences, while other divergences are not seen in dimensional regularization, and in any case they can be eliminated by renormalization. The same does not hold in the case of a non-vanishing cosmological constant, as found by Christensen and Duff \cite{Christensen:1979iy}. Let us briefly review these statements in the light of our calculations. Setting $D=4$ in \eqref{divergences} we see that the different powers of $T$ give rise to the quartic, quadratic, and logarithmic divergences parametrized by $a_0$, $a_1$ and $a_2$, respectively. From \eqref{a2}, we read
\begin{equation}
\left. a_{2}\right|_{D=4}=-\frac{29}{40}\, R^2+\frac{53}{45} \, R_{\mu\nu\rho\sigma}^{2}\ .
\end{equation}
These numerical values for the one-loop four-dimensional logarithmic divergences of quantum gravity with non-vanishing cosmological constant coincide precisely with those calculated long ago by Christensen and Duff.\footnote{For comparison, one has to make evident the cosmological constant term with the on-shell condition \eqref{1EM}, which reads $R=4\Lambda$.} The term proportional to $R_{\mu\nu\rho\sigma}^{2}$ could be neglected, as thanks to the Chern--Gauss--Bonnet theorem for four-dimensional Einstein manifolds it is proportional to a total derivative, and thus eliminable from the effective action, but the remaining term proportional
to $R^2$ cannot be renormalized away by redefining the parameters of the Einstein--Hilbert action. The theory is not renormalizable.
 
On the other hand, setting the cosmological constant to vanish, one finds that the on-shell background satisfies $R_{\mu\nu}=0$, and thus $R=0$. The logarithmic divergence reduces to 
\begin{equation}
a_{2}\Big|_{\substack{D=4\\[0.1em] \Lambda=0}} = \frac{53}{45} \, R_{\mu\nu\rho\sigma}^{2} 
\end{equation}
which in four dimensions is a total derivative, as discussed earlier, and can be eliminated from the effective action. Thus, one recovers the result that the one-loop logarithmic divergences of pure quantum gravity without cosmological constant vanish in four dimensions. This property does not hold true anymore at two-loops, as found by Goroff and Sagnotti \cite{Goroff:1985th} and verified by van de Ven \cite{vandeVen:1991gw}.
Returning to the one-loop divergences for vanishing cosmological constant, one finds that also $a_1$ vanishes. This leaves only the quartic divergence proportional to $a_0$, which gives the number of degrees of freedom of the graviton. It requires a renormalization of the cosmological constant back to zero, which makes the theory rather unnatural in the technical sense of 't Hooft \cite{tHooft:1979rat}, but in any case renormalizable at one-loop.

For arbitrary nonzero values of the cosmological constant, the quadratic divergence related to $a_1$ is not vanishing anymore and its value at $D=4$ 
\begin{equation}
\left. a_{1}\right|_{D=4}= -\frac83\, R
\end{equation}
reproduces the gauge-invariant result already computed in Refs.~\cite{Bastianelli:2013tsa, Martini:2021slj}. It can be renormalized away by redefining the Newton constant. Finally, the coefficient $a_3$ gives rise to a finite term in the four-dimensional effective action, but its physical meaning is unclear. It is gauge invariant, but infrared divergences invalidate a local expansion of the effective action as delivered by the small proper time approximation of the heat kernel, which is useful to locate the \acr{UV} divergences.

\subsection{Pure gravity in six dimensions}
The newly computed coefficient $a_3$ \eqref{a3}, allows us to see what happens in six spacetime dimensions. Setting $D=6$ in the coefficients \eqref{a0}--\eqref{a3} we find
\begin{align}
\begin{split} 
\left. a_{0} \right|_{D=6}
&= 9 
\;, \qquad \qquad \quad
\left. a_{1}\right|_{D=6} = -\frac32 \; R
\;, \qquad \qquad \quad
\left. a_{2} \right|_{D=6}= -\frac{11}{20}\; R^2 +\frac{21}{20} \; R_{\mu\nu\rho\sigma}^{2}
\end{split}\\[.5em]
\begin{split}
\left. a_{3}\right|_{D=6} &= -\frac{799}{11340} \;R^3
+\frac{481}{1680} \;R\,R_{\mu\nu\rho\sigma}^{2}
-\frac{991}{5040} \;R_{\mu\nu\rho\sigma}R^{\rho\sigma\alpha\beta}R_{\alpha\beta}{}^{\mu\nu}
-\frac{71}{180}\; R_{\alpha\mu\nu\beta}R^{\mu\rho\sigma\nu}R_{\rho}{}^{\alpha\beta}{}_{\sigma}\ ,
\end{split}
\end{align}
that furnish the full list of one-loop divergences of quantum gravity with cosmological constant in six dimensions. We stress that these coefficients are gauge invariant, and thus any other method of calculation should reproduce these values. 

A comparison with the literature can be made by setting the cosmological constant to zero and considering the logarithmic divergence parametrized by $a_3$, which reduces to
\begin{equation} \label{a3D6v2}
a_{3}\Big|_{\substack{D=6\\[0.1em] \Lambda=0}}=-\frac{991}{5040} \; R_{\mu\nu\rho\sigma}R^{\rho\sigma\alpha\beta}R_{\alpha\beta}{}^{\mu\nu} -\frac{71}{180}\, R_{\alpha\mu\nu\beta}R^{\mu\rho\sigma\nu}R_{\rho}{}^{\alpha\beta}{}_{\sigma}\ .
\end{equation}
These two remaining terms are proportional to two invariants that are generally independent of each other. However, it turns out that in six dimensions there exists an integral relation that connects them. It involves the use of the Chern--Gauss--Bonnet theorem and the introduction of the Euler character $\chi_{\rm E}(\mathcal{M})$, as explained in Ref.~\cite{vanNieuwenhuizen:1976vb} and discussed more extensively in Appendix \ref{appendixD}. Bottom line, we can further simplify the coefficient $a_3$, which becomes 
\begin{equation} \label{DivD6}
a_{3}\Big|_{\substack{D=6\\[0.1em] \Lambda=0}}=\frac{9}{15120} \; R_{\mu\nu\rho\sigma}R^{\rho\sigma\alpha\beta}R_{\alpha\beta}{}^{\mu\nu}\; .
\end{equation}
It encodes the one-loop logarithmic divergences of pure gravity in six dimensions. We are now in the position of carrying out a comparison with the literature: our value in \eqref{DivD6} is in complete agreement with van Nieuwenhuizen's pioneering calculation \cite{VanNieuwenhuizen:1977ca}, besides a computational error in the numerical factor in his equation (81), already noted a year later by Critchley \cite{Critchley:1978kb}. 
Furthermore, confirmation of this value is also found in more recent works, see for instance \cite{Gibbons:1999qz, Dunbar:2002gu}. 

As for the general case of arbitrary cosmological constant, we are not aware of similar calculations, although they would certainly be interesting to pursue to further verify our findings.

\section{Final remarks} \label{chap:sixth:sec5}
In this work, we have investigated the computation of counterterms necessary for the renormalization of the one-loop effective action of quantum gravity with cosmological constant in arbitrary dimensions. Our results are complete for dimensions $D<8$. The counterterms have been computed on-shell, so they furnish gauge invariant quantities characteristic of the quantum theory of gravity. 

Our main contribution was the determination of the Seeley-DeWitt coefficient $a_3(D)$ of perturbative quantum gravity, which to our knowledge has never been reported in its full generality in the literature. When restricted to six dimensions, it parameterizes the logarithmic divergence which was previously known only for the case of vanishing cosmological constant. To cross-check our calculations, we have used two distinct methods: a first-quantized description of the graviton in terms of the ${\cal N}=4$ spinning particle and the time-honored heat kernel method, finding complete agreement.
 
While the utility of heat kernel methods is well-known and they keep being employed in many contexts, see for example \cite{Bastianelli:2019zrq, Casarin:2023ifl} for some recent applications to trace anomalies,
first-quantized methods for treating the graviton with the ${\cal N}=4$ spinning particle are more recent and we have championed them here to show their usefulness. In this respect, it would be interesting to extend the present analysis to the first quantized model that describes the ${\cal N}=0$ supergravity \cite{Bonezzi:2020jjq}, i.e. the particle theory that has in its spectrum the graviton, the dilaton, and the antisymmetric tensor $B_{\mu\nu}$, as well as extend the present methods to the $U(\N)$ spinning particles \cite{Marcus:1994em, Bastianelli:2009vj, Bastianelli:2011pe, Bastianelli:2012nh} to find a useful first-quantized way of describing gravitational theories on complex (K\"ahler) manifolds, and finally also consider double copy features on the worldline \cite{Bastianelli:2021rbt} to address gravitational aspects from a different perspective. \nocite{Ellis_2017}

\addcontentsline{toc}{chapter}{Appendices}
\section*{Appendices}
\begin{subappendices}

\section{Basis of invariants on Einstein manifolds}\label{appendixA}
We use the following conventions for the curvature tensors:
\begin{equation}\label{Notation}
[\nabla_\mu, \nabla_\nu] V^\lambda = 
R_{\mu\nu}{}^\lambda{}_\rho V^\rho \ , \ \ \ 
R_{\mu\nu}= R_{\lambda\mu}{}^\lambda{}_\nu 
\ , \ \ \ R= R^\mu{}_\mu > 0 \ {\rm on\ spheres.} 
\end{equation}
A $D$-dimensional Riemannian manifold without boundaries can be described through an (infinite) basis of curvature monomials $\mathcal{K}_i^n$. These are geometric invariants of order $n$ in the Riemann tensor, Ricci tensor, and scalar curvature, with two covariant derivatives counting as a Riemann tensor. They have been introduced by \cite{Fulling1992} and recently reviewed in Refs.~\cite{Decanini2007, Decanini2008}. 
 At order $n=3$ we use the basis considered in Ref.~\cite{Bastianelli:2000rs} which is made of $17$ independent invariants:
\begin{align}
\begin{array}{lll}
\mathcal{K}_1 = R^3 & 
\mathcal{K}_2 = R R_{\mu\nu}^2 & 
\mathcal{K}_3 = R R_{\mu\nu\rho\sigma}^2 \\ [3mm] 
\mathcal{K}_4 = R_\mu{}^\rho R_\rho{}^\nu R_\nu{}^\mu & 
\mathcal{K}_5 = R_{\mu\nu} R_{\rho\sigma} R^{\rho\mu\nu\sigma} & 
\mathcal{K}_6= R_{\mu\nu} R^{\mu\rho\sigma\lambda} R^\nu{}_{\rho\sigma\lambda} \\ [3mm] 
\mathcal{K}_7 = R_{\mu\nu}{}^{\rho\sigma} R_{\rho\sigma}{}^{\alpha\beta}R_{\alpha\beta}{}^{\mu\nu} \ \ \ \ &
\mathcal{K}_8 = R_{\mu\rho\sigma\nu} R^{\rho\alpha\beta\sigma} R_{\alpha}{}^{\mu\nu}{}_\beta \ \ \ \ &
\mathcal{K}_9 = R\nabla^2 R \\ [3mm]
\mathcal{K}_{10} = R_{\mu\nu}\nabla^2 R^{\mu\nu} &
\mathcal{K}_{11} = R_{\mu\nu\rho\sigma}\nabla^2 R^{\mu\nu\rho\sigma} &
\mathcal{K}_{12} = R^{\mu\nu} \nabla_\mu \nabla_\nu R \\ [3mm]
\mathcal{K}_{13} = (\nabla_\mu R_{\rho\sigma})^2 &
\mathcal{K}_{14} = \nabla_\mu R_{\nu\rho} \nabla^\nu R^{\mu\rho} &
\mathcal{K}_{15} = (\nabla_\alpha R_{\mu\nu\rho\sigma})^2 \\ [3mm]
\mathcal{K}_{16} = \nabla^2 R^2 &
\mathcal{K}_{17} =\nabla^4 R\ . &
\label{R3}
\end{array}
\end{align}
All other terms cubic in the curvature are linear combinations of the above invariants after taking into account the symmetry properties and the Bianchi identities of the Riemann tensor. 

On Einstein manifolds, the basis \eqref{R3} can be reduced further. Einstein metrics are defined by the equation
\begin{equation}
R_{\mu\nu}=\lambda g_{\mu\nu} 
\label{def-Em}
 \end{equation}
that upon contraction leads to $ R=\lambda D$. From the second Bianchi identity one finds that $R$ and $\lambda$ are constant for $D> 2$
\begin{equation}
\nabla^{\mu}R_{\mu\nu}=\frac{1}{2}\nabla_{\nu}R \quad\longrightarrow\quad (D-2) \nabla_{\nu}\lambda=0 \quad\longrightarrow\quad \nabla_{\nu}R=0 
\end{equation}
and then from \eqref{def-Em}, one finds that the Ricci tensor is also covariantly constant for $D> 2$
\begin{equation} \label{2EM}
\nabla_{\alpha}R_{\mu\nu}=0\, .
\end{equation}
Let us consider now the second Bianchi identity, namely
\begin{equation}
\nabla_{\rho}R_{\mu\nu\alpha\beta}+\nabla_{\beta}R_{\mu\nu\rho\alpha}+\nabla_{\alpha}R_{\mu\nu\beta\rho} = 0 \, ,
\end{equation}
by contracting the above identity with $g^{\rho\mu}$ and using \eqref{2EM} one gets 
\begin{equation} \label{div}
\nabla^{\mu}R_{\mu\nu\alpha\beta} = 0\ .
\end{equation}
We are now in the position to reduce the six-dimensional basis of invariants \eqref{R3} on Einstein manifolds to a minimal set of independent ones, namely
\begin{equation} \label{E3}
\mathcal{E}_1=R^{3} \;,
\hskip1cm \mathcal{E}_2= R\,R_{\mu\nu\rho\sigma}^{2}\;,
\hskip1cm \mathcal{E}_3=R_{\mu\nu\rho\sigma}R^{\rho\sigma\alpha\beta}R_{\alpha\beta}{}^{\mu\nu}\;,
\hskip1cm \mathcal{E}_4=R_{\alpha\mu\nu\beta}R^{\mu\rho\sigma\nu}R_{\rho}{}^{\alpha\beta}{}_{\sigma}\;.
\end{equation}
Indeed we have
\begin{equation}\label{EinsteinCond}
\mathcal{K}_9 = \mathcal{K}_{10} = 0\ , \qquad \mathcal{K}_2 = \frac{1}{D}\mathcal{E}_1\ ,\qquad \mathcal{K}_4 = -\mathcal{K}_5 = \frac{1}{D^2}\mathcal{E}_1\ ,\qquad \mathcal{K}_6 = \frac{1}{D}\mathcal{E}_2\ .
\end{equation}
Moreover, the only term of \eqref{R3} containing covariant derivatives and non-vanishing on Einstein manifolds, i.e. $\mathcal{K}_{11}$,\footnote{Since, up to a total derivative term, we have $\mathcal{K}_{15}=-\mathcal{K}_{11}$.} can be written as
\begin{align}
R_{\mu\nu\alpha\beta}\nabla^{2}R^{\mu\nu\alpha\beta}&=-R_{\mu\nu\alpha\beta}\nabla_{\rho}\left(
\nabla^{\beta}R^{\mu\nu\rho\alpha}+\nabla^{\alpha}R^{\mu\nu\beta\rho}
\right)=-2R_{\mu\nu\alpha\beta}\nabla_{\rho}\nabla^{\beta}R^{\mu\nu\rho\alpha}\nonumber\\
&=-2R_{\mu\nu\alpha\beta}\left(
\nabla^{\beta}\nabla_{\rho}R^{\rho\alpha\mu\nu}+R^{\beta}{}_{\lambda}R^{\lambda\alpha\mu\nu}+R_{\rho}{}^{\beta\alpha}{}_{\lambda}R^{\rho\lambda\mu\nu}+R_{\rho}{}^{\beta\mu}{}_{\lambda}R^{\rho\alpha\lambda\nu}+R_{\rho}{}^{\beta\nu}{}_{\lambda}R^{\rho\alpha\mu\lambda}
\right)\nonumber\\
&=-2R_{\mu\nu\alpha\beta}R_{\rho}{}^{\beta\alpha}{}_{\lambda}R^{\rho\lambda\mu\nu}+
\frac{2}{D}R\, R_{\mu\nu\alpha\lambda}^{2}
-4R_{\mu\nu\alpha\beta}R_{\rho}{}^{\beta\mu}{}_{\lambda}R^{\rho\alpha\lambda\nu}\nonumber\\
&=-R_{\mu\nu\alpha\beta}R^{\alpha\beta}{}_{\rho\lambda}R^{\rho\lambda\mu\nu}+\frac{2}{D}RR_{\mu\nu\alpha\beta}^{2}+4R_{\mu\nu\alpha\beta}R^{\nu\lambda\rho\alpha}R_{\lambda}{}^{\mu\beta}{}_{\rho}\nonumber\\
&=\frac{2}{D}\mathcal{E}_2 -\mathcal{E}_3 + 4\mathcal{E}_4\ ,
\end{align} 
where we made use of the second Bianchi identity, antisymmetry of the Riemann tensor and 
\begin{equation}
[\nabla_{\alpha},\nabla_{\beta}]R_{\mu\nu\rho\sigma}=R_{\alpha\beta \mu}{}^{\lambda}R_{\lambda\nu\rho\sigma}+R_{\alpha\beta\nu}{}^{\lambda}R_{\mu\lambda\rho\sigma}+R_{\alpha\beta\rho}{}^{\lambda}R_{\mu\nu\lambda\sigma}+R_{\alpha\beta \sigma}{}^{\lambda}R_{\mu\nu\rho\lambda}\ .
\end{equation}

Finally, let us discuss how the basis \eqref{E3} further simplifies on maximally symmetric spaces (MSS), which form a subset of the Einstein ones, where the Riemann tensor is given by
\begin{equation}
R_{\mu\nu\alpha\beta} = \frac{R}{D(D-1)}\left(g_{\mu\alpha}g_{\nu\beta}-g_{\mu\beta}g_{\nu\alpha}\right)\ .
\end{equation}
The invariants previously defined in \eqref{E3} all collapse to $\mathcal{E}_{1}$, and one finds
\begin{equation}
\mathcal{E}_{2} = \frac{2}{D(D-1)}\mathcal{E}_{1}\ ,\qquad \mathcal{E}_{3}=\frac{4}{D^{2}(D-1)^{2}}\mathcal{E}_{1}\ , \qquad \mathcal{E}_{4}=-\frac{D-2}{D^{2}(D-1)^{2}}\mathcal{E}_{1}\ .
\end{equation}
These relations allow us to evaluate the newly-computed coefficient $a_3(D)$ on MSS, giving \eqref{94} as a result.

\section{Worldline computations}\label{appendixB}

\subsection{Worldline propagators}\label{appendixB1}
The worldline propagators for the $\mathcal{N}=4$ spinning particle descend from the free action \eqref{freeaction}. Regarding the bosonic quantum fluctuations $q^{\mu}(\tau)$, we considered \acr{DBC} during the evaluation of the derivative expansion of the effective action, namely $q^\mu(0)=q^\mu(1)=0$. This leads to the \acr{DBC} worldline propagator defined by the two-point function 
\begin{equation} \label{A1}
\langle q^{\mu}(\tau)q^{\nu}(\sigma)\rangle = -2 Tg^{\mu\nu}(x_{0})\del_{{\rm D}}(\tau,\sigma)\ ,
\end{equation}
where
\begin{equation} \label{chap:sixth:DBC}
\del_{{\rm D}}(\tau,\sigma)= \left(\tau-1\right)\sigma\, \theta\left(\tau-\sigma \right)+\left(\sigma-1\right)\tau\,\theta\left(\sigma-\tau \right)\;.
\end{equation}
We also list the derivatives of the \acr{DBC} propagator
\begin{align}
\ddel_{\rm D}(\tau,\sigma)&=\sigma-\theta(\sigma-\tau)\label{derivata}\\
\deld_{\rm D}(\tau,\sigma)&= \tau-\theta(\tau-\sigma)\\
\ddeld_{\rm D}(\tau,\sigma)&= 1-\delta(\tau-\sigma)\\
\dddel_{\rm D}(\tau,\sigma)&=\delta(\tau-\sigma)\;.
\end{align}
Concerning the fermionic fields $\psi^{a}_{i}(\tau)$, we list here their propagator
\begin{equation} \label{AF}
\langle \psi^{a}_{i}(\sigma)\bar{\psi}^{b}_{j}(\tau)\rangle =\delta^{ab}\del_{ \rm{AF}}\,_{ij}(\tau,\sigma) =\delta^{ab}\left(
\begin{array}{cc}
 F(z,\tau,\sigma) & 0 \\
 0 & F(\omega,\tau,\sigma) \\
\end{array}
\right)
\end{equation}
where each entry in the matrix is a ${\cal N}=2$ fermionic propagator \cite{Bastianelli:2005vk} defined as
\begin{align}
F(z,\tau,\sigma) &=z^{-(\tau-\sigma)}\left(\frac{1}{z+1}\right)\left( z\, \theta(\tau-\sigma)-\theta(\sigma-\tau)\right)\\
F(\omega,\tau,\sigma) &=\omega^{-(\tau-\sigma)}\left(\frac{1}{\omega+1}\right)\left( \omega\, \theta(\tau-\sigma)-\theta(\sigma-\tau)\right)\ .
\end{align}
Next, we have the ghost variable propagators, defined as
\begin{align}
\langle a^{\mu}(\tau) a^{\nu}(\sigma)\rangle &= 2T g^{\mu\nu}(x_{0})\del_{_{\rm gh}}(\tau,\sigma)=2T g^{\mu\nu}(x_{0})\delta(\tau-\sigma)\\[1mm]
\langle b^{\mu}(\tau) c^{\nu}(\sigma)\rangle &=-4T g^{\mu\nu}(x_{0})\del_{_{\rm gh}}(\tau,\sigma) =-4T g^{\mu\nu}(x_{0})\delta(\tau-\sigma)\;.
\end{align}
As previously stated, the calculation involving these propagators may result in products/derivatives of delta distributions, which are ill-defined, but also divergent quantities such as $\delta(\tau,\tau)$, therefore one needs to regularize the path integral. In the present chapter, we chose worldline dimensional regularization, which consists in continuing the compact time direction with the addition of $d$ non-compact extra dimensions, i.e. extending the space $\tau\in[0,1]\rightarrow t^\alpha=\left(\tau,\mathbf{t}\right)\in[0,1]\times\mathbb{R}^n$. More details can be found for example in Ref.~\cite{Bastianelli:2006rx}; here we list the dimensional regularized expression of propagators, exploited during intermediate steps when performing computations. The $d+1$ extended propagators read
\begin{align}
\del_{\rm D}(t,s) &= \int \frac{d^{d}k}{(2\pi)^{d}}\sum_{m=1}^{\infty}\frac{-2}{(\pi m)^{2}+\mathbf{k}^{2}}\sin(\pi m \tau)\sin(\pi m \sigma)\,e^{i \mathbf{k}\cdot(\mathbf{t}-\mathbf{s})} \\
\del_{\rm gh}(t,s)&=\int \frac{d^{d}k}{(2\pi)^{d}}\sum_{m=1}^{\infty} 2\sin(\pi m\tau)\sin(\pi m \sigma) e^{i \mathbf{k}\cdot(\mathbf{t}-\mathbf{s})}=\delta(\tau-\sigma)\delta^{d}(t-s) \\
F(\theta,t,s)&=-i\int \frac{d^{d}k}{(2\pi)^{d}}\sum_{r\in Z+\frac{1}{2}}\frac{2\pi r\gamma^{0}+k\cdot \gamma-\theta}{(2\pi r)^{2}+\mathbf{k}^{2}-\theta^{2}}\, e^{2\pi i r(\tau-\sigma)}e^{i\mathbf{k}\cdot (\mathbf{t}-\mathbf{s})}
\end{align}
where in the dimensional regularized expressions one has $\mathbf{t}$ as $d$-dimensional vector and $\gamma^\alpha$ are the Dirac matrices in the $(d+1)$-dimensional extended space. Each extended worldline propagator satisfies a generalization of its own one-dimensional Green equation
\begin{align} \label{greenEQ}
\partial_{\mu}\partial^{\mu}\del_{\rm D}(t,s) &= \delta^{d+1}(t-s)\\
\left(\rpartial + i\theta\right)F(\theta,t,s)& = \delta_{\rm AF}(\tau-\sigma)\delta^{d}(t-s) \label{B16} \\
F(\theta,t,s) \left( -\lpartial +i\theta\right) &=\delta_{\rm AF}(\tau-\sigma)\delta^{d}(t-s)\ , \label{greenEQ2}
\end{align}
where $\delta_{\rm AF}$ is the delta distribution acting on antiperiodic functions on $[0,1]$ and where a slashed derivative is the usual contraction between derivative and gamma matrices. For computational purposes, the index contractions in $d+1$ dimensions serve mostly as a bookkeeping device to keep track of which derivative can be contracted to which vertex to produce the $(d+1)$-dimensional delta function. The delta functions in (\ref{greenEQ})--\eqref{greenEQ2} are only to be used in $d+1$ dimensions; then, by using partial integration one casts the various loop integrals in a form that can be computed by sending $d\to 0$ first. At this stage, one can use the one-dimensional propagators \eqref{chap:sixth:DBC}--\eqref{AF}, and $\gamma^0=1$ (with no extra factors arising from the Dirac algebra in $d+1$ dimensions). 

\subsection{Analysis of perturbative contributions}\label{appendixB2}
In this appendix, we will give the essential details on the evaluation of the path integral average and its subsequent modular integration \eqref{chap:sixth:2.16}, which produces the Seeley-DeWitt coefficients to be inserted in \eqref{series}. As anticipated in Section \ref{sec2.3}, in order to find all the possible contributions to order $T^3$ we have to expand the exponential with the interacting action \eqref{S-int}, written more compactly as in \eqref{10} which in particular carries the following vertices 
\begin{align}
&S_{\rm K1}=\int d\tau \; \frac{1}{3}R_{\alpha\mu\nu\beta}\, q^{\alpha}q^{\beta}\left( \dot{q}^{\mu}\dot{q}^{\nu} + \text{\rm gh}\right) \hskip.2cm , \hskip.2cm S_{\rm C1} = \int d\tau \; \frac{1}{2}R_{\alpha\mu ab}\, \dot{q}^{\mu}q^{\alpha}\, \bar{\psi}^{a}\cdot \psi^{b}\hskip.2cm ,\\
&DS_{\rm C1}=\int d\tau \; \frac{1}{3}\nabla_{\alpha}R_{\beta\mu ab}\,\dot{q}^{\mu}q^{\alpha}q^{\beta} \, \bar{\psi}^{a}\cdot \psi^{b} \hskip.2cm , \hskip.2cm D^{2}S_{\rm C1}= \int d\tau \; \frac{1}{8}\nabla_{\alpha}\nabla_{\beta}R_{\gamma\mu ab}\, \dot{q}^{\mu}q^{\alpha} q^{\beta}q^{\gamma}\, \bar{\psi}^{a}\cdot \psi^{b}\hskip.2cm , \\
&S_{\rm C2} =\int d\tau \; \frac{1}{24}R^{\tau}{}_{\alpha\beta\mu}R_{\gamma \tau ab}\,
 \dot{q}^{\mu}q^{\alpha}q^{\beta}q^{\gamma}\, \bar{\psi}^{a}\cdot \psi^{b} \hskip.2cm , \hskip.2cm S_{\rm F}= \int d\tau \; R_{abcd}\, \bar{\psi}^{a}\cdot \psi^{b}\bar{\psi}^{c}\cdot \psi^{d}\hskip.2cm , \\
&DS_{\rm F}=\int d\tau \; q^\alpha \nabla_{\alpha}R_{abcd} \, \bar{\psi}^{a}\cdot \psi^{b}\bar{\psi}^{c}\cdot \psi^{d} \hskip.2cm , \hskip.2cm D^{2}S_{\rm F}= \int d\tau \; \frac{1}{2} q^\alpha q^\beta \nabla_{\alpha}\nabla_{\beta}R_{abcd}\, \bar{\psi}^{a}\cdot \psi^{b}\bar{\psi}^{c}\cdot \psi^{d}\hskip.2cm , \\
&S_{\rm V}=-\int d\tau \; \Omega R\ . \label{B.22}
\end{align}
Through Wick contractions they result in a plethora of different terms, most of which either give rise to disconnected diagrams (so that they are easily taken care of afterward, as connected diagrams exponentiate and there is no need to compute them anew) or can easily be shown to vanish. To start with, we see that the vertex $S_{\rm V}$ in \eqref{B.22} does not depend on $q$, $\psi$ nor on $\bar \psi$, thus it can be taken out of the path integral and it remains exponentiated (as will be all connected worldline diagrams). Let's see some other illustrative examples. For instance, a disconnected diagram arises from the term
\begin{equation}
-\frac{1}{8}T \langle S_{\rm K1}S_{\rm F}^2\rangle \sim \disc \sim \text{disconnected}\;.
\end{equation}
Other diagrams can be shown to be zero exploiting (anti)symmetry of the tensor structures and/or of the resulting propagators, like
\begin{align}
-\frac{1}{32 T^2} \langle S_{\rm K1}^2 S_{\rm C1}\rangle &\sim R_{\mu\nu ab}\langle\bar{\psi}^{a}\cdot \psi^{b} \rangle=0 \\
\frac{1}{4}\langle S_{\rm K1} S_{\rm C1}S_{\rm F}\rangle &\sim R_{ab}R_{\mu\nu}{}^{ab}R_{\alpha\rho\sigma\beta}=0\ .
\end{align}
Lastly, others are simply zero after explicitly calculating the integrals, as happens for 
$\langle S_{\rm C2}S_{\rm F}\rangle $.

We report now the surviving contributions \eqref{terms} in full glory:
\begin{align*}
\langle e^{-S_{\rm int}} \big \rangle \Big|_{T^3} 
&= \langle S_{\rm KIN} \rangle \tag{A}\\
&\phantom{=}-\frac{1}{96T}R_{\alpha\mu\nu\beta}R_{\gamma\lambda ab}R_{\delta\epsilon cd}\int_{012}
\langle \dot{q}^{\mu}_{0}\dot{q}^{\nu}_{0}q^{\alpha}_{0}q^{\beta}_{0}\dot{q}^{\lambda}_{1}q^{\gamma}_{1}\dot{q}^{\epsilon}_{2}q^{\delta}_{2}
\rangle \langle \bar{\psi}^{a}_{1}\cdot \psi_{1}^{b} \bar{\psi}^{c}_{2}\cdot \psi_{2}^{d} \rangle \tag{B} \\
&\phantom{=}+\frac{1}{48}R_{\alpha\mu ab}R^{\tau}{}_{\beta\lambda\nu}
R_{\rho \tau cd}\int_{01}\langle \dot{q}^{\mu}_{0}q^{\alpha}_{0}\dot{q}^{\nu}_{1}q^{\beta}_{1}q^{\lambda}_{1}q^{\rho}_{1}\rangle \langle \bar{\psi}^{a}_{0}\cdot \psi_{0}^{b} \bar{\psi}^{c}_{1}\cdot \psi_{1}^{d}\rangle \tag{C} \\
\tag{D} \label{D} &\begin{aligned} 
\begin{split}
\phantom{=}+\frac{1}{18}\nabla_{\alpha}R_{\beta\mu ab}\nabla_{\lambda}R_{\rho\nu cd}\int_{01} \langle \dot{q}^{\mu}_{0}q^{\alpha}_{0}q^{\beta}_{0}\dot{q}^{\nu}_{1}q^{\lambda}_{1}q^{\rho}_{1}\rangle \, \langle \bar{\psi}^{a}_{0}\cdot \psi^{b}_{0}\bar{\psi}^{c}_{1}\cdot \psi^{d}_{1}\rangle \\
\phantom{=} +\frac{1}{16}\nabla_{\alpha}\nabla_{\beta}R_{\lambda\mu ab} R_{\rho \nu cd}\int_{01} \langle \dot{q}^{\mu}_{0}q^{\alpha}_{0}q^{\beta}_{0}q^{\lambda}_{0}\dot{q}^{\nu}_{1}q^{\rho}_{1}\rangle \, \langle \bar{\psi}^{a}_{0}\cdot \psi^{b}_{0}\bar{\psi}^{c}_{1}\cdot \psi^{d}_{1}\rangle 
\end{split}
\end{aligned}\\
&\phantom{=} +\frac{T}{8}R_{\alpha\mu ab}R_{\beta\nu cd}R_{efgh}\int_{012}\langle \dot{q}^{\mu}_{0}\dot{q}^{\nu}_{1}q^{\alpha}_{0}q^{\beta}_{1}\rangle \langle \bar{\psi}^{a}_{0}\cdot \psi^{b}_{0}\bar{\psi}^{c}_{1}\cdot\psi^{d}_{1}\psi^{e}_{2}\cdot \bar{\psi}^{f}_{2}\psi^{g}_{2}\cdot \bar{\psi}^{h}_{2} \rangle \tag{E}\\
&\phantom{=}+\frac{1}{6}R_{abcd}R_{efgh}R_{lmno}\int_{012}\langle \bar{\psi}^{a}_{0}\cdot \psi^{b}_{0}\, \bar{\psi}^{c}_{0}\cdot \psi^{d}_{0}
\bar{\psi}^{e}_{1}\cdot \psi^{f}_{1}\, \bar{\psi}^{g}_{1}\cdot \psi^{h}_{1} \bar{\psi}^{l}_{2}\cdot \psi^{m}_{2}\, \bar{\psi}^{n}_{2}\cdot \psi^{o}_{2}\rangle \tag{F} \\
\tag{G} \label{G} &\begin{aligned} 
\begin{split}
&\phantom{=}+ \frac{T^2}{2} \nabla_{\alpha}R_{abcd}\nabla_{\beta}R_{efgh} \int_{01} \langle \dot{q}^{\alpha}_{0}\dot{q}^{\beta}_{1}\rangle \langle \bar{\psi}^{a}_{0}\cdot \psi^{b}_{0}\, \bar{\psi}^{c}_{0}\cdot \psi^{d}_{0}
\bar{\psi}^{e}_{1}\cdot \psi^{f}_{1}\, \bar{\psi}^{g}_{1}\cdot \psi^{h}_{1} \rangle \\
&\phantom{=} +\frac{T^2}{2}\nabla_{\alpha}\nabla_{\beta}R_{abcd}R_{efgh}\int_{01} \langle \dot{q}^{\alpha}_{0}\dot{q}^{\beta}_{0}\rangle \langle \bar{\psi}^{a}_{0}\cdot \psi^{b}_{0}\, \bar{\psi}^{c}_{0}\cdot \psi^{d}_{0}
\bar{\psi}^{e}_{1}\cdot \psi^{f}_{1}\, \bar{\psi}^{g}_{1}\cdot \psi^{h}_{1} \rangle\ .
\end{split}
\end{aligned}
\end{align*}
We decided to keep both the ghost fields and the term $S_{\rm KIN}$ implicit so as not to further burden the notation, and we combined the two terms of pure spin connection with covariant derivatives $\frac{1}{2}DS_{\rm C1}^{2}+D^{2}S_{\rm C1}S_{\rm C1}$ in eq. \eqref{D} and the two terms of the Taylor expansion of the four-fermions action $\frac{1}{2}T^2DS_{\rm F}^2 +T^2D^{2}S_{\rm F}S_{\rm F}$ in eq. \eqref{G}, for reasons that will become clear shortly thereafter. Below we illustrate the main steps of the calculation for each contribution. 
\begin{enumerate}[label=(\Alph*)]
\item The first contribution we called $S_{\rm KIN}$ to indicate economically the sum of all terms arising \emph{only} from the pure kinetic part of the interacting action. In the notation previously presented it would read
\begin{equation}
S_{\rm KIN}=-\frac{1}{384 T^3}S_{\rm K1}^3+\frac{1}{32 T^2}DS_{\rm K1}^2+\frac{S_{\rm K1}}{16T^2}\left( S_{\rm K2}+D^2S_{\rm K1} \right) -\frac{1}{4T}S_{\rm K3}\ .
\label{B27}
\end{equation}
It has already been computed in Ref.~\cite{Bastianelli:2000dw} (eq.\ 20) and can be read out, translated in our basis \eqref{E3}, from the exponential of connected diagrams as
\begin{equation}
\langle S_{\rm KIN} \rangle=\frac{T^{3}}{7!}	\left(	-\frac{16}{9}\frac{\mathcal{E}_{1}}{D^{2}}+\frac{2}{3}\frac{\mathcal{E}_{2}}{D}+\frac{17}{9}\mathcal{E}_{3}+\frac{28}{9}\mathcal{E}_{4}	\right)\ .
\end{equation}
\item Regarding the contribution obtained by coupling the bosonic kinetic term and the spin connection, once the contractions have been evaluated, one can use integration by parts and \acr{DR} to reduce integrals to a set of independent ones in $z$, modulo their $\omega$ partner.\footnote{This should also be understood for forthcoming worldline integrals when the fermionic integrands depend \emph{only} on the Wilson variable $z$.} We list here such integrals
\begin{align}
&\int_{012} \ddel(\tau,\sigma)\deld(\tau,\sigma)\ddel(\tau,\rho)\deld(\tau,\rho)F(z,\sigma,\rho)F(z,\rho,\sigma)=-\frac{z}{120\left(1+z	\right)^{2}}\ , \\
&\int_{012} \ddeld(\tau,\rho)\ddel(\tau,\tau)\deld(\tau,\sigma)\del(\sigma,\rho)F(z,\sigma,\rho)F(z,\rho,\sigma)=-\frac{1}{720}\frac{z}{(z+1)^{2}}\ ,
\end{align}
so that the final result for $-\frac{1}{8T}\langle S_{\rm K1}S^{2}_{\rm SC1} \rangle$ reads
\begin{equation}
T^{3}\frac{\mathcal{E}_{2}}{D}\left[-\frac{1}{90}\frac{z}{(z+1)^{2}}-\frac{1}{90}\frac{\omega}{(\omega+1)^{2}}		\right]+T^{3}\mathcal{E}_{3}\left[-\frac{z}{60\left(1+z	\right)^{2}}-\frac{\omega}{60\left(1+\omega	\right)^{2}}\right]\ .
\end{equation}
\item The contribution of pure spin connection has already been evaluated in Section \ref{sec2.3}; we report here the result for completeness:
\begin{equation}
\langle S_{\rm C1}S_{\rm C2} \rangle=\left(\frac{1}{180}\frac{\mathcal{E}_{2}}{D}+\frac{1}{120}\mathcal{E}_{3}\right)\left[\frac{z}{(z+1)^{2}}+\frac{\omega}{(\omega+1)^{2}}\right]\ .
\end{equation}
\item The contribution of pure spin connection with covariant derivatives is made up of the sum of two pieces: using Einstein manifolds simplifications one can collect the whole result, with only the following diagrams to be evaluated
\begin{align}
&\int_{01} \ddeld(\sigma,\tau)\del(\sigma,\tau)\del(\sigma,\tau)F(z,\sigma,\tau)F(z,\tau,\sigma) = \frac{z}{45(z+1)^{2}}\ ,\\
&\int_{01} \ddeld(\sigma,\tau)\del(\sigma,\tau)\del(\sigma,\sigma)F(z,\sigma,\tau)F(z,\tau,\sigma)=\frac{z}{60(z+1)^{2}}\ ,
\end{align}
getting the final result
\begin{equation}
\left\langle \frac{1}{2}DS_{\rm C1}^{2}+D^{2}S_{\rm C1}S_{\rm C1} \right\rangle=-\frac{T^{3}}{360}\left[ \frac{z}{(z+1)^{2}}+\frac{\omega}{(\omega+1)^{2}}\right]\left(2\frac{\mathcal{E}_{2}}{D}-\mathcal{E}_{3}+4\mathcal{E}_{4}\right)\ .
\end{equation}
\item This is produced from the coupling of spin connection to the four-fermions vertex. The list of independent diagrams one needs to evaluate is
\begin{align}
&\int_{012} \ddeld(\sigma,\tau)\del(\sigma,\tau)F(z,\rho,\tau)F(z,\tau,\rho)F(z,\rho,\sigma)F(z,\sigma,\rho)=-\frac{z^{2}}{12(z+1)^{4}}\ ,\\
&\int_{012} \ddeld(\sigma,\tau)\del(\sigma,\tau)F(z,\rho,\tau)F(z,\tau,\sigma)F(z,\sigma,u)F(z,\rho,\rho) =-\frac{z(z-1)^{2}}{48(z+1)^{4}}\ ,\\
&\int_{012} \ddel(\sigma,\tau)\deld(\sigma,\tau)F(z,\sigma,\rho)F(z,\rho,\sigma)F(\omega,\tau,\rho)F(\omega,u,\tau)=-\frac{z\omega}{12(z+1)^{2}(\omega+1)^{2}}\ ,
\end{align}
such that the final result reads
\begin{align}
\begin{split}
\left\langle \frac{1}{2}T^{2}S_{\rm C1}^{2}S_{\rm F} \right\rangle &= T^{3}\frac{\mathcal{E}_{2}}{D}\left[\frac{z\left(z-1	\right)^{2}}{12 \left(z+1\right)^{4}} +	\frac{\omega\left(\omega-1	\right)^{2}}{12 \left(\omega+1\right)^{4}}	\right]\\
&\phantom{=}+T^{3}\mathcal{E}_{3}\left[	\frac{z^{2}}{12\left(z+1\right)^{4}}+\frac{\omega^{2}}{12\left(\omega+1\right)^{4}}+\frac{z\omega}{3\left(	z+1\right)^{2}\left(\omega+1	\right)^{2}} \right] \ .
\end{split}
\end{align}
\item The one arising from the pure fermionic vertex is quite tricky to evaluate at once. To simplify the calculation, it is possible to exploit the subtle double copy structure underlying the $\mathcal{N}=4$ spinning particle, rewriting this term as a sum of contributions coming from the two copies of the $\mathcal{N}=2$ particles, i.e. for the two values of the internal index $i=1,2$. This allows us to rewrite the
above term as
\begin{align}
\begin{split}
\frac{1}{6} S_{\rm F}^{3} &= \frac{1}{6} S_{1}^{3}(z) +\frac{1}{6} S_{1}^{3}(\omega) +\frac{1}{6} S_{\rm mix}^{3} + \frac{1}{2}S_{1}(z)S_{2}^{2}(\omega)+\frac{1}{2}S_{1}(z)^{2}S_{2}(\omega) \\ 
&\phantom{=}+\frac{1}{2}S_{1}^{2}(z)S_{\rm mix}+\frac{1}{2}S_{1}(z)S_{\rm mix}^{2}+\frac{1}{2}S_{2}^{2}(\omega)S_{\rm mix}+\frac{1}{2}S_{2}(\omega)S_{\rm mix}^{2}+S_{1}(z)S_{2}(\omega)S_{\rm mix}\ .
\end{split}
\end{align}
In the previous expression, we defined for simplicity the actions
\begin{align}
S_{1}(z) &= \int d\tau\, TR_{abcd}\,\bar{\psi}^{1a}\psi_{1}^{b}\bar{\psi}^{1c}\psi_{1}\,^{d}\ , \\
S_{\rm mix} &= \int d\tau \left( TR_{abcd}\,\bar{\psi}^{a1}\psi^{b}_{1}\bar{\psi}^{2c}\psi^{d}_{2} + 1\leftrightarrow 2\right)\ ,
\end{align}
where we explicated the flavor index. Once performed the contractions with Mathematica, based on the \emph{xTensor} package \cite{garcia_2002}, we further simplify the result by reducing tensor structures using the Bianchi identity as follows
\begin{align}
\label{RiemID3} &R_{\alpha\beta}{}^{\rho\sigma}R^{\alpha\mu\beta\nu}R_{\rho\mu\sigma\nu} =\frac{1}{4}\mathcal{E}_3\ ,\\[.5em]
\label{RiemID4} &R_{\mu\alpha\nu\beta}R^{\mu\rho\nu\sigma}R^{\alpha}{}_{\sigma}{}^{\beta}{}_{\rho} =-\frac{1}{4}\mathcal{E}_3+\mathcal{E}_4\ .
\end{align}
Once evaluated the fermionic diagrams our final answer for $\langle\frac{1}{6}T^{3}S_{\rm F}^{3} \rangle$ is
\begin{align}
\begin{split}
&\phantom{=+}\frac{\mathcal{E}_{1}}{D^{2}} \left[-\frac{(-1 + z)^2 z (1 + (-8 + z) z)}{6 (1 + z)^6}-\frac{(-1 + \omega)^2 \omega (1 + (-8 + \omega) \omega)}{6 (1 + \omega)^6}\right] \\[.3em]
&\phantom{=}+\frac{\mathcal{E}_{2}}{D}\left[	-\frac{(z-1)^2 z^2}{(z+1)^6}-\frac{(\omega-1)^2 \omega^2}{(\omega+1)^6}-\frac{2 (-1 + z)^2 z \omega }{(1 + z)^4 (1 + \omega )^2} -\frac{2 z (-1 + \omega )^2 \omega }{(1 + z)^2 (1 + \omega )^4}\right] \\[.3em]
&\phantom{=}+\mathcal{E}_{3}\left[\frac{z^4-2 z^3+z^2}{6 (z+1)^6}+ \frac{\omega ^4-2 \omega ^3+\omega ^2}{6 (\omega +1)^6} -\frac{2 z^2 \omega \ }{(1 + z)^4 (1 + \omega )^2}- \frac{2 z \omega ^2 }{(1 + \ z)^2 (1 + \omega )^4} \right] \\[.3em]
&\phantom{=}+\mathcal{E}_{4}\left[	 -\frac{4 \omega ^3}{3 (\omega +1)^6}-\frac{4 z^3}{3 (z+1)^6}-\frac{4 \omega z}{3(\omega +1)^2 (z+1)^2(\omega+1)^{2}} \right]\ .
\end{split}
\end{align}
\item Finally, one has the term coming from the Taylor expansion of the four-fermions interaction, made of two pieces. Just as in the \eqref{D} case, one can collect them into a single one working on the tensorial structure and using Einstein manifolds simplifications. The resulting diagrams to be evaluated are
\begin{align}
&\int_{01} \del(\tau ,\tau ) F(z,\sigma ,\tau )F(z,\sigma ,\tau ) F(z,\tau ,\sigma ) F(z,\tau ,\sigma )=-\frac{z^2}{6 (z+1)^4}\ ,\\
&\int_{01} \del (\tau ,\tau ) F(z,\sigma ,\tau ) F(z,\tau ,\sigma )F(\omega ,\sigma ,\tau ) F(\omega ,\tau ,\sigma )=-\frac{\omega z}{6 (\omega +1)^2 (z+1)^2}\ ,
\end{align}
since the partner diagrams $\del (\tau,\tau )\rightarrow \del (\tau,\sigma )$ can easily be seen to be proportional to the above. The final answer for $\langle \frac{1}{2}T^2DS_{\rm F}^2 +T^2D^{2}S_{\rm F}S_{\rm F} \rangle$ is then 
\begin{equation}
T^{3}\left[	\frac{z^{2}}{12\left(z+1\right)^{4}}+\frac{\omega^{2}}{12\left(\omega+1\right)^{4}}+\frac{z\omega}{3\left(	z+1\right)^{2}\left(\omega+1	\right)^{2}} \right] \left(2\frac{\mathcal{E}_{2}}{D}-\mathcal{E}_{3}+4\mathcal{E}_{4}\right)\ .
\end{equation}
\end{enumerate}

\subsection{Summing up all the pieces}
We can collect the result of all the computations above as follows
\begin{equation}
\alpha_{3}(z,\omega,D) = c_{0}(z,\omega)\,\frac{\mathcal{E}_{1}}{D^{2}}+c_{1}(z,\omega)\,\frac{\mathcal{E}_{2}}{D} +c_{2}(z,\omega)\, \mathcal{E}_{3}
+c_{3}(z,\omega)\,\mathcal{E}_{4}\ ,
\end{equation}
with the coefficients
\begin{align}
\begin{split}
c_{0}(z,\omega)&= -\frac{1}{2835}
-\frac{z \left(z^2-8 z+1\right) (z-1)^2}{6 (z+1)^6}-\frac{(\omega -1)^2 \omega \left(\omega ^2-8 \omega +1\right)}{6 (\omega +1)^6}\ ,
\end{split}\\[.5em]
\begin{split}
c_{1}(z,\omega)&=\frac{1}{7560}-\frac{z}{90 (z+1)^2}-\frac{\omega }{90 (\omega +1)^2}+\frac{z (z-1)^2}{12 (z+1)^4}+\frac{(\omega -1)^2 \omega }{12 (\omega +1)^4}+\frac{z^2}{6 (z+1)^4}+\frac{\omega ^2}{6 (\omega +1)^4} \\[.3em]
&\phantom{=}-\frac{z^2(z-1)^2}{(z+1)^6}-\frac{(\omega -1)^2 \omega ^2}{(\omega
 +1)^6}-\frac{2 \omega z (z-1)^2}{(\omega +1)^2 (z+1)^4}-\frac{2 (\omega -1)^2 \omega z}{(\omega
 +1)^4 (z+1)^2}+ \frac{2 z\omega}{3\left( z+1 \right)^2\left( \omega +1\right)^2}\ ,
\end{split}\\[.5em]
\begin{split}
c_{2}(z,\omega)&=\frac{17}{45360} -\frac{z}{180 (z+1)^2}-\frac{\omega }{180 (\omega +1)^2} +\frac{(z-1)^2 z^2}{6(z+1)^6} +\frac{(\omega -1)^2 \omega ^2}{6 (\omega +1)^6} \\
&\phantom{=}-\frac{2 \omega z^2}{(\omega +1)^2 (z+1)^4}-\frac{2 \omega ^2 z}{(\omega +1)^4(z+1)^2}\ ,
\end{split}\\[.5em]
\begin{split}
c_{3}(z,\omega)&=\frac{1}{1620}-\frac{z}{90(z+1)^2}-\frac{\omega }{90 (\omega +1)^2}+ \frac{z^2}{3 \left( z+1 \right)^4} + \frac{\omega^2}{3 \left( \omega+1 \right)^4} + -\frac{4 \omega ^3}{3 (\omega +1)^6}-\frac{4 z^3}{3 (z+1)^6}\ .
\end{split}
\end{align}
At this point, in order to get the full $T^3$ correction, we first need to recall the connected diagrams coming from the first- and second-order expansion of the path integral, namely
\begin{align}
\alpha_{1}(z,\omega,D)&=\left(\frac{5}{12} +\Omega-\frac{z}{(z+1)^{2}}-\frac{\omega}{(\omega+1)^{2}}\right) R\ , \\[.5em]
\alpha_{2}(z,\omega,D)&=\left(
-\frac{1}{180}+\frac{1}{2}\left[\frac{z(z-1)^{2}}{(z+1)^{4}}+\frac{\omega(\omega-1)^{2}}{(\omega+1)^{4}}
\right]\right)\frac{R^{2}}{D} \nonumber\\[.3em]
&\phantom{=}+\left(\frac{1}{180}+\frac{1}{2} \left[\frac{\omega ^2}{(\omega +1)^4}+\frac{z^2}{(z+1)^4}+\frac{4 \omega z}{(\omega +1)^2 (z+1)^2}\right]-\frac{1}{12} \left[\frac{\omega }{(\omega
 +1)^2}+\frac{z}{(z+1)^2}\right] \right)R_{\mu\nu\rho\sigma}^{2}\nonumber \\[.3em]
& :=  \beta_{1}(z,\omega)\, \frac{R^{2}}{D}+\beta_{2}(z,\omega)\,R_{\mu\nu\rho\sigma}^{2}\ ,
\end{align}
which are in complete accordance with the results of Refs.~\cite{Bastianelli:2019xhi, Bastianelli:2022pqq} and indeed reproduce the $a_1$ and $a_2$ coefficients \eqref{a1}--\eqref{a2}. Now we can move to the exponentiation of all the connected diagrams, including the new results. The expectation value $\big\langle e^{-S_{\rm int}}\big\rangle$ in \eqref{chap:sixth:2.16} can be compactly written as
\begin{align}
\begin{split}
\Big\langle e^{-S_{\rm int}}\Big\rangle &=\exp\Bigg[
T\,\alpha_{1}\, R+T^{2}\,\left(\beta_{1}\, \frac{R^{2}}{D}+\beta_{2}\,R_{\mu\nu\rho\sigma}^{2}	\right) \\
&\phantom{=}+ T^{3}\,\left( c_{0}\,\frac{R^{3}}{D^{2}}+c_{1}\,\frac{R\, R_{\mu\nu\rho\sigma}^{2}}{D}+ c_{2}\,R_{\mu\nu\rho\sigma}R^{\rho\sigma\alpha\beta}R_{\alpha\beta}{}^{\mu\nu}
+ c_{3}\,R_{\alpha\mu\nu\beta}R^{\mu\rho\sigma\nu}R_{\rho}{}^{\alpha\beta}{}_{\sigma} \right) + {\cal O}(T^4) \Bigg]\ .
\end{split}
\end{align}
The final step consists of Taylor expanding the exponential to the desired order, namely
\begin{equation}
\Big\langle e^{-S_{\rm int}}\Big\rangle\Big|_{T^{3}}= T^{3}\left[ \left( \frac{\alpha_{1}^{3}}{6}+\frac{\alpha_{1}\beta_{1}}{D}+\frac{c_{0}}{D^{2}}\right)\,\mathcal{E}_1+\left( \alpha_{1}\beta_{2}+\frac{c_{1} }{D}\right)\,\mathcal{E}_2 +c_{2}(z,\omega)\, \mathcal{E}_{3} +c_{3}(z,\omega)\,\mathcal{E}_{4} \right]+ \mathcal{O}(T^{4})\ .
\end{equation}
We obtained our final answer for the path integral average \eqref{chap:sixth:2.16}. In the end, one is left only with performing the modular integration, i.e. the double expectation value $\langle \hskip -.05 cm \langle e^{-S_{\rm int}} \rangle \hskip -.05cm \rangle$ defined in \eqref{double}. In doing so, we can choose the appropriate measure $P(z,\omega)$ to project only on the ghost, graviton, or total coefficients. This gives our final results of Section \ref{sec2.4}, in particular the newly-computed coefficient
\begin{align}
a_{3}(D)&= \frac{35 D^4-147 D^3-3670 D^2-13560D-30240}{90720 D^2}\;\mathcal{E}_1 +\frac{7 D^3-230 D^2+3357 D+12600}{15120 D}\;\mathcal{E}_2 \nonumber\\[.3em] 
&\phantom{=}+ \frac{17 D^2-555 D-15120}{90720}\;\mathcal{E}_3 +\frac{D^2-39 D-1080}{3240}\; \mathcal{E}_4\ .
\end{align}

\section{Heat kernel computations}\label{appendixC}
The fourth heat kernel coefficient for the ghost field is computed from the general formula \eqref{HK3}, by performing the substitutions \eqref{GhostSost}. It is convenient to write
\begin{equation}\label{GhostHK-split}
\alpha_3^{\rm gh}(x)  :=  \frac{1}{7!}\bm{A}_{\rm gh}[R, R_{\mu\nu}, R_{\mu\nu\rho\sigma}] + \frac{2}{6!}\bm{B}_{\rm gh}[R, R_{\mu\nu}, R_{\mu\nu\rho\sigma}, \Omega_{\mu\nu}, V]\ ,
\end{equation}
where $\bm{A}_{\rm gh}$ and $\bm{B}_{\rm gh}$ are two (involved) functions of the metric invariants and of the gauge field strength, as reported in \eqref{HK3}. Let us start from $\bm{A}_{\rm gh}$: six of the first eight terms vanish identically since they are all proportional to covariant derivatives of $R$, $R_{\mu\nu}$ or $R_{\mu\nu\rho\sigma}$, except from the two proportional to $(\nabla_\alpha R_{\mu\nu\rho\sigma})^2$ and $R_{\mu\nu\rho\sigma}\nabla^2R^{\mu\nu\rho\sigma}$. Recalling that $\Tr{[\delta^\mu_\nu]}=D$ and using \eqref{EinsteinCond}, we are therefore left with
\begin{align}
\begin{split}\label{Agh}
\Tr\left[\bm{A}_{\rm gh}\right] &= D\left(3R_{\mu\nu\rho\sigma}\nabla^2R^{\mu\nu\rho\sigma}-\frac{208}{9}R_{\mu}{}^{\nu}R_{\nu}{}^{\sigma}R_{\sigma}{}^{\mu} + \frac{64}{3}R_{\mu\nu}R_{\rho\sigma}R^{\mu\rho\nu\sigma}-\frac{16}{3}R_{\mu\nu}R^\mu{}_{\rho\sigma\tau}R^{\nu\rho\sigma\tau}\right.\\[.3em]
&\phantom{=}\left.+\frac{44}{9}R_{\mu\nu}{}^{\rho\sigma}R_{\rho\sigma}{}^{\alpha\beta}R_{\alpha\beta}{}^{\mu\nu}+\frac{80}{9}R_{\mu\nu\rho\sigma}R^{\mu\alpha\rho\beta}R^{\nu}{}_{\alpha}{}^{\sigma}{}_{\beta}\right)\\[.3em]
&= -\frac{16}{9D}\mathcal{E}_1+\frac{2}{3}\mathcal{E}_2 +\frac{17D}{9}\mathcal{E}_3 +\frac{28D}{9}\mathcal{E}_4\ .
\end{split}
\end{align}
We now compute $\bm{B}_{\rm gh}$. Since $(\Omega_{\mu\nu})^\rho{}_{\sigma} = R_{\mu\nu}{}^{\rho}{}_{\sigma}$, the second term of $\bm{B}_{\rm gh}$ in \eqref{HK3} vanishes identically; moreover, as $V\propto R$, the same occurs for the last four, leaving us with
\begin{align}
\begin{split}\label{Bgh}
\Tr\left[\bm{B}_{\rm gh}\right] &= 4\Tr{\left[\Omega_{\mu\nu}\nabla^2\Omega^{\mu\nu}\right]}-12\Tr{\left[\Omega_{\mu}{}^{\nu}\Omega_{\nu}{}^{\sigma}\Omega_{\sigma}{}^{\mu} \right]}+ 6\Tr{\left[R_{\mu\nu\rho\sigma}\Omega^{\mu\nu}\Omega^{\rho\sigma}\right]}-4 \Tr{\left[R_{\mu\nu}\Omega^{\mu\sigma}\Omega^{\nu}{}_{\sigma}\right]}\\[.3em]
&= -\frac{4}{D}\mathcal{E}_2- 2\mathcal{E}_3 -4\mathcal{E}_4\ ,
\end{split}
\end{align}
where all the traces are computed by explicit substitution, according to \eqref{GhostSost},
\begin{align}
\Tr{\left[\Omega_{\mu}{}^{\nu}\Omega_{\nu}{}^{\sigma}\Omega_{\sigma}{}^{\mu}\right]} &= \Tr{\left[R^{\alpha\ \ \nu}_{\ \beta\mu}R^{\beta\ \ \sigma}_{\ \gamma\nu}R^{\gamma\ \ \mu}_{\ \delta\sigma}\right]} = R^{\alpha\ \ \nu}_{\ \beta\mu}R^{\beta\ \ \sigma}_{\ \gamma\nu}R^{\gamma\ \ \mu}_{\ \alpha\sigma} = -\mathcal{E}_4\ ,\\[.5em]
\Tr{\left[R_{\mu\nu\rho\sigma}\Omega^{\mu\nu}\Omega^{\rho\sigma}\right]} &= R_{\mu\nu\rho\sigma} \Tr{\left[R^{\alpha\ \mu\nu}_{\ \beta} R^{\beta\ \rho\sigma}_{\ \gamma}\right]} = R_{\mu\nu\rho\sigma}R^{\alpha\ \mu\nu}_{\ \beta} R^{\beta\ \rho\sigma}_{\ \alpha}= -\mathcal{E}_3\ ,\\[.5em]
\Tr{\left[R_{\mu\nu}\Omega^{\mu\sigma}\Omega^{\nu}{}_{\sigma}\right]} &= R_{\mu\nu} \Tr{\left[R^{\alpha\ \mu\sigma}_{\ \beta}R^{\beta\ \nu}_{\ \gamma\ \sigma}\right]} = R_{\mu\nu} R^{\alpha\ \mu\sigma}_{\ \beta}R^{\beta\ \nu}_{\ \alpha\ \sigma}= -\frac{1}{D}\mathcal{E}_2\ .
\end{align}
Going back to \eqref{GhostHK-split} we conclude that
\begin{equation}\label{GhostAlpha}
\Tr{\left[\alpha_3^{\rm gh}(x)\right]} = -\frac{1}{2835 D}\mathcal{E}_1+\frac{D-84}{7560 D}\mathcal{E}_2 +\frac{17 D-252}{45360}\mathcal{E}_3 +\frac{D-18}{1620}\mathcal{E}_4\ ,
\end{equation}
which however does not provide the full coefficient for the ghost, since we still have to add the term $\beta_3$ defined in \eqref{BetaValues}, which turns out to be
\begin{align}
\begin{split}
\beta_3^{\rm gh} &= \frac{1}{6}\left(\alpha_1^{\rm gh}\right)^3+\alpha_1^{gh}\alpha_2^{\rm gh} \\
&= \frac{1}{6}\left[\delta^\tau_\alpha\left(\frac{1}{6}+\frac{1}{D}\right)R\right]^3 + \left[\delta^\tau_\alpha\left(\frac{1}{6}+\frac{1}{D}\right)R\right]\left[\frac{1}{180}\left(R_{\mu\nu\rho\sigma}^2-R_{\mu\nu}^2\right)\delta^\alpha_\gamma+\frac{1}{12}\Omega_{\mu\nu}^2\right]\ ,
\end{split}
\end{align}
and taking the trace
\begin{equation}\label{Beta3Ghost}
\Tr{\left[\beta_3^{\rm gh}(x)\right]} = \frac{(5D^2+54D+180)(D+6)}{6480D^2}\mathcal{E}_1+\frac{(D+6)(D-15)}{1080D}\mathcal{E}_2\ .
\end{equation}
The overall result for the ghost is
\begin{align}
\begin{split}
\Tr{\left[a_3^{\rm gh}(x)\right]} &= \frac{35D^3+588D^2+3512D+7560}{45360D^2}\ \mathcal{E}_1 \\[.5em]
&\phantom{=} + \frac{7D^2-62D-714}{7560D}\mathcal{E}_2 +\frac{17D-252}{45360}\mathcal{E}_3 +\frac{D-18}{1620}\mathcal{E}_4
\end{split}
\end{align}
and corresponds to \eqref{a3gh}. To compute the fourth heat kernel coefficient for the graviton, we start again from the general formula \eqref{HK3} and perform the substitutions \eqref{GravitonSost}. It is once again more time convenient to split
\begin{equation}\label{GravitonHK-split}
\alpha_3^{\rm gr}(x)  :=  \frac{1}{7!}\bm{A}_{\rm gr}[R, R_{\mu\nu}, R_{\mu\nu\rho\sigma}] + \frac{2}{6!}\bm{B}_{\rm gr}[R, R_{\mu\nu}, R_{\mu\nu\rho\sigma}, \Omega_{\mu\nu}, V]
\end{equation}
as we did in \eqref{GhostHK-split} for the ghost. By inspecting more closely the substitution rules \eqref{GhostSost} and \eqref{GravitonSost}, though, it is clear that $\bm{A}_{\rm gr}$ and $\bm{A}_{\rm gh}$ differ only by the trace of the identity operator $\mathbbm{1}$. Using \eqref{Agh}, we find in particular
\begin{equation}\label{Agr}
\Tr\left[\bm{A}_{\rm gr}\right] = \frac{D+1}{2}\, \Tr\left[\bm{A}_{\rm gh}\right]
= (D+1)\left(-\frac{8}{9D}\mathcal{E}_1+ \frac{1}{3}\mathcal{E}_2 +\frac{17D}{18}\mathcal{E}_3 +\frac{14D}{9}\mathcal{E}_4\right)\ ,
\end{equation}
while for the computation of $\bm{B}_{\rm gr}$ we can repeat the previous observations, and consider the only non-vanishing terms
\begin{align}
\begin{split}\label{Bgr}
\Tr\left[\bm{B}_{\rm gr}\right] &= 4\Tr{\left[\Omega_{\mu\nu}\nabla^2\Omega^{\mu\nu}\right]}-12\Tr{\left[\Omega_{\mu}{}^{\nu}\Omega_{\nu}{}^{\sigma}\Omega_{\sigma}{}^{\mu}\right]}\\[.3em]
& \phantom{=}+ 6\Tr{\left[R_{\mu\nu\rho\sigma}\Omega^{\mu\nu}\Omega^{\rho\sigma}\right]}-4 \Tr{\left[R_{\mu\nu}\Omega^{\mu\sigma}\Omega^{\nu}{}_{\sigma}\right]}+30\Tr{\left[(\nabla_\mu V)^2\right]}\\[.5em]
&=-\frac{4(D+47)}{D}\ \mathcal{E}_2- 2(D-43)\ \mathcal{E}_3 -4(D+92)\ \mathcal{E}_4
\end{split}
\end{align}
where we need to compute the traces
\begin{align}
\Tr{\left[\Omega_{\mu}{}^{\nu}\Omega_{\nu}{}^{\sigma}\Omega_{\sigma}{}^{\mu}\right]} &= -\left(D+2\right) \mathcal{E}_4\ ,\\[.3em]
\Tr{\left[R_{\mu\nu\rho\sigma}\Omega^{\mu\nu}\Omega^{\rho\sigma}\right]} &= -(D+2)\ \mathcal{E}_3\ ,\\[.3em]
\Tr{\left[R_{\mu\nu}\Omega^{\mu\sigma}\Omega^{\nu}{}_{\sigma}\right]} &= -\frac{D+2}{D}\ \mathcal{E}_2\ ,\\[.3em]
\Tr{\left[\Omega_{\mu\nu}\nabla^2\Omega^{\mu\nu}\right]} &= -(D+2)\left(\frac{2}{D}\mathcal{E}_2-\mathcal{E}_3+4\mathcal{E}_4\right)\ ,\\[.3em]
\Tr{\left[(\nabla_\alpha \mathcal{V}_{\mu\nu}{}^{\rho\sigma})^2\right]} &= -3\left(\frac{2}{D}\mathcal{E}_2-\mathcal{E}_3+4\mathcal{E}_4\right)\ .
\end{align}
Going back to \eqref{GravitonHK-split} we conclude that
\begin{align}
\begin{split}\label{GravitonAlpha} \Tr{\left[\alpha_3^{\rm gr}(x)\right]} &=-\frac{D+1}{5670 D}\ \mathcal{E}_1+\frac{D^2-167 D-7896}{15120 D}\ \mathcal{E}_2\\[.5em]
&\hspace{.5cm} +\frac{17 D^2-487 D+21672}{90720}\ \mathcal{E}_3 +\frac{D^2-35 D-3312}{3240}\ \mathcal{E}_4\ .
\end{split}
\end{align}
The last step is to compute the term $\beta_3$ defined in \eqref{BetaValues}, which is the sum of the two terms:
\begin{align}
\frac{1}{6}\left(\alpha_1^{\rm gr}\right)^3 &= \frac{1}{6} \left(\frac{1}{216}R^3\delta_{\mu\nu}^{\ \ \ \alpha\beta} + \frac{1}{12}R^2\mathcal{V}_{\mu\nu}^{\ \ \ \alpha\beta} + \frac{1}{2}R \, \mathcal{V}_{\mu\nu}^{\ \ \ \rho\sigma}\mathcal{V}_{\rho\sigma}^{\ \ \ \alpha\beta}+\mathcal{V}_{\mu\nu}^{\ \ \ \rho\sigma}\mathcal{V}_{\rho\sigma}^{\ \ \ \lambda\tau}\mathcal{V}_{\lambda\tau}^{\ \ \ \alpha\beta}\right)\ ,\\
\alpha_1^{\rm gr}\alpha_2^{\rm gr} &= \left(\frac{1}{6}R\delta_{\mu\nu}^{\ \ \ \alpha\beta}+\mathcal{V}_{\mu\nu}^{\ \ \ \alpha\beta}\right) \left[\frac{1}{180}\left(R_{\mu\nu\rho\sigma}^2-\frac{1}{D}R^2\right)\delta_{\alpha\beta}^{\ \ \ \lambda\tau} + \frac{1}{12}\left(\Omega_{\rho\sigma}^2\right)_{\alpha\beta}^{\ \ \ \lambda\tau}\right]+\frac{1}{6}\mathcal{V}_{\mu\nu}^{\ \ \ \alpha\beta}\nabla^2\mathcal{V}_{\alpha\beta}^{\ \ \ \lambda\tau}\ ,\nonumber
\end{align}
where the following traces are to be computed:
\begin{align}
\Tr{\left[\delta_{\mu\nu}^{\ \ \ \alpha\beta}\right]} &=\frac{1}{2}D(D+1)\ ,\\[.3em]
\Tr{\left[\mathcal{V}_{\mu\nu}^{\ \ \ \alpha\beta}\right]} &= -R\ ,\\[.3em]
\Tr{\left[\mathcal{V}_{\mu\nu}^{\ \ \ \rho\sigma}\mathcal{V}_{\rho\sigma}^{\ \ \ \alpha\beta}\right]} &=3R^2_{\mu\nu\rho\sigma}\ ,\\[.3em]
\Tr{\left[\mathcal{V}_{\mu\nu}^{\ \ \ \rho\sigma}\mathcal{V}_{\rho\sigma}^{\ \ \ \lambda\tau}\mathcal{V}_{\lambda\tau}^{\ \ \ \alpha\beta}\right]} &= -8\mathcal{E}_4 - \mathcal{E}_3\ ,\\[.3em]
\Tr{\left[\mathcal{V}_{\mu\nu}^{\ \ \ \alpha\beta}\left(\Omega_{\rho\sigma}^2\right)_{\alpha\beta}^{\ \ \ \lambda\tau}\right]} &= \frac{2}{D}\mathcal{E}_2 +3\mathcal{E}_3\ ,\\[.3em]
\Tr{\left[\mathcal{V}_{\mu\nu}^{\ \ \ \alpha\beta}\nabla^2\mathcal{V}_{\alpha\beta}^{\ \ \ \lambda\tau}\right]} &= 3\left(\frac{2}{D}\mathcal{E}_2-\mathcal{E}_3+4\mathcal{E}_4\right)\ ,
\end{align}
leading to
\begin{align}
\begin{split}\label{Beta3Graviton}
\Tr{\left[\beta_3^{\rm gr}(x)\right]} &= \frac{5 D^3-D^2-186 D+72}{12960 D}\ \mathcal{E}^3_1\\[.5em]
&\phantom{=} + \frac{D^3-29 D^2+468 D+2520}{2160 D}\ \mathcal{E}_2^3 - \frac{5}{12}\mathcal{E}_3^3+\frac{2}{3}\mathcal{E}_4^3\ .
\end{split}
\end{align}
In the end,
\begin{align}
\begin{split}
\Tr{\left[a_3^{\rm gr}(x)\right]} &= \frac{35 D^3-7 D^2-1318 D+488}{90720 D}\ \mathcal{E}_1 +\frac{7 D^3-202 D^2+3109 D+9744}{15120 D}\ \mathcal{E}_2 \\[.5em]
&\phantom{=}+ \frac{17 D^2-487 D-16128}{90720}\ \mathcal{E}_3 +\frac{D^2-35 D-1152}{3240}\ \mathcal{E}_4\ ,
\end{split}
\end{align}
which corresponds to \eqref{a3gr}. By summing the ghost and graviton coefficients according to \eqref{TotHKgen}, we end up with
\begin{align}
\Tr{\left[a_3(x)\right]} &= \frac{35 D^4-147 D^3-3670 D^2-13560 D-30240}{90720 D^2}\ \mathcal{E}_1 +\frac{7 D^3-230 D^2+3357 D+12600}{15120 D}\ \mathcal{E}_2 \nonumber\\[.5em]
&\phantom{=} +\frac{17 D^2-555 D-15120}{90720}\ \mathcal{E}_3+\frac{D^2-39 D-1080}{3240}\ \mathcal{E}_4\ ,
\end{align}
which corresponds to \eqref{a3}.

\section{Topological terms in even dimensions}\label{appendixD}
In order to display the cancellation of the one-loop divergences of pure gravity in four spacetime dimensions, and analogously their resilience in six, we exploit the Chern--Gauss--Bonnet theorem, which allows us to compute the Euler character $\chi_{\rm E}(\mathcal{M})$ of a manifold $\mathcal{M}$ as a volume integral of the 2-form $\mathcal{R}^{\mu\nu} :=  R^{\mu\nu}_{\ \ \ \alpha\beta}\,\diff{x}^\alpha\wedge\diff{x}^\beta$, namely
\begin{equation}\label{EulerChar}
\chi_{\rm E}(\mathcal{M})=\frac{1}{(4\pi)^d}\int_{\mathcal{M}} \varepsilon_{\mu_1\nu_1\dots\mu_d\nu_d} \mathcal{R}^{\mu_1\nu_1}\wedge\dots\wedge\mathcal{R}^{\mu_d\nu_d}\ ,
\end{equation}
where $D :=  2d$ is the dimension of the manifold, assumed here to be even.\footnote{If $D$ is odd, the integral \eqref{EulerChar} vanishes, so that the theorem does not provide a useful way of computing $\chi_{\rm E}(\mathcal{M})$.} In local coordinates, \eqref{EulerChar} becomes
\begin{equation}\label{EulerChar2}
\chi_{\rm E}(\mathcal{M})=\frac{1}{2(4\pi)^d}\int \diff{^Dx}\sqrt{g}\ \frac{D!}{D} \delta^{\alpha_1}_{[\mu_1} \delta^{\beta_1}_{\nu_1}\dots\delta^{\alpha_d}_{\mu_d} \delta^{\beta_d}_{\nu_d]} R^{\mu_1\nu_1}_{\ \ \ \ \ \alpha_1\beta_1}\dots R^{\mu_d\nu_d}_{\ \ \ \ \ \alpha_d\beta_d}\ .
\end{equation}
It is possible to prove that $\chi_{\rm E}(\mathcal{M})$ defined in this way does not depend on the metric settled upon $\mathcal{M}$, and is fixed only by the global topology of the manifold. For $D=2$ ($d=1$), \eqref{EulerChar2} becomes
\begin{equation}\label{EulerD2}
\left. \chi_{\rm E}(\mathcal{M})\right|_{D=2}= \frac{1}{16\pi} \int \diff{^2x} \sqrt{g}\ R\ ,
\end{equation}
which is proportional to the Einstein--Hilbert action, while at $D=4$ ($d=2$) the Euler character reads
\begin{equation}\label{EulerD4}
\left. \chi_{\rm E}(\mathcal{M})\right|_{D=4} = \frac{1}{32\pi^2} \int \diff{^4x} \sqrt{g}\ \left(R^2-4R_{\mu\nu}R^{\mu\nu}+R_{\mu\nu\rho\sigma}R^{\mu\nu\rho\sigma}\right)\ .
\end{equation}
On Einstein spaces the first two terms in the integrand of \eqref{EulerD4} cancel off, and therefore we are left with
\begin{equation}\label{EulerD4bis}
\left. \chi_E(\mathcal{M})\right|_{D=4} = \frac{1}{32\pi^2} \int \diff{^4x} \sqrt{g}\ R_{\mu\nu\rho\sigma}R^{\mu\nu\rho\sigma} :=  \int \diff{^4x} \sqrt{g}\ E_4\ .
\end{equation}
Therefore, the third heat kernel coefficient $\Tr{\left[a_2(x)\right]}\propto R_{\mu\nu\rho\sigma}^2$ is proportional to the Euler density $E_4$ and hence is a total derivative, which can be neglected in the effective action. This result is no more true when $\Lambda\neq 0$, even if we drop the total derivative term corresponding to Euler density. 

In dimension $D=6$ ($d=3$) the Euler character is \cite{Groh2012}
\begin{equation}\label{EulerD6}
\left. \chi_{\rm E}(\mathcal{M})\right|_{D=6} = \frac{1}{384\pi^3} \int \diff{^6x} \sqrt{g}\ \left(4\mathcal{K}_1-48\mathcal{K}_2+64\mathcal{K}_4+96\mathcal{K}_5+12\mathcal{K}_3-96\mathcal{K}_6+16\mathcal{K}_7-32\mathcal{K}_8\right)\ ,
\end{equation}
which on Einstein spaces reduces to
\begin{equation}\label{EulerD6bis}
\left. \chi_{\rm E}(\mathcal{M})\right|_{D=6} = \frac{1}{384\pi^3} \int \diff{^6x} \sqrt{g}\ \left(\frac{4}{9}\mathcal{E}_1-4\mathcal{E}_2+16\mathcal{E}_3+32\mathcal{E}_4\right)\ .
\end{equation}
The condition $\Lambda=0$, that is $R=0$, forces $\mathcal{E}_1=\mathcal{E}_2=0$, so that \eqref{EulerD6bis} eventually becomes
\begin{equation}\label{EulerD6tris2}
\left. \chi_{\rm E}(\mathcal{M})\right|_{D=6} = \frac{1}{384\pi^3} \int \diff{^6x} \sqrt{g}\ \left(16\mathcal{E}_3+32\mathcal{E}_4\right) :=  \int \diff{^6x} \sqrt{g}\ E_6\ .
\end{equation}
This shows that, up to a total derivative term
\begin{equation}
2 \mathcal{E}_4 =-\mathcal{E}_3\ . 
\end{equation}
Thus, at dimension $D=6$, even with $\Lambda=0$, the perturbative quantum gravity effective action is not free of logarithmic divergences. 
\end{subappendices}

\newpage
\thispagestyle{empty}
\mbox{}
\newpage

\chapter{Heat kernel coefficients in massive gravity}\label{chap:fifth}
\textit{In this chapter, we employ the massive model developed in Chapter~\ref{chap:fourth} to carry out a perturbative computation in massive gravity. Specifically, we evaluate the counterterms required for the renormalization of the one-loop effective action of linearized massive gravity. As a preliminary step, we construct the worldline path integral on the one-dimensional torus, while projecting onto the desired degrees of freedom. Once this is achieved, we obtain the correct expression for the one-loop divergences in arbitrary spacetime dimension $D$. The central result of this chapter is the determination of the Seeley–DeWitt coefficient $a_3(D)$, which, to the best of our knowledge, had not been previously computed in the literature.}

\paragraph{Conventions} Whether we work in Minkowski or Euclidean signature, and in which spacetime dimension, will be specified as needed throughout the text for convenience.

\section{Extracting degrees of freedom} \label{chap:fifth:sec2}
In this section, the quantization on the circle of the free massive $\mathcal{N}=4$ spinning particle with various gaugings of the SO(4) R-symmetry is analyzed. The aim is to determine how the path integral extracts physical degrees of freedom from the Hilbert space. Once this is established, it will be possible to isolate specifically the degrees of freedom associated with the massive graviton.

To set the notation, we summarize here the key features of the model. The graded phase space of the \emph{massless} $\mathcal{N}=4$ supersymmetric worldline model consists of bosonic $(x^M,p_M)$ and fermionic $(\Xi^M_I)$ coordinates, where $M=0,\dots, D$ is a spacetime vector index and $I=1,2,3,4$ is a SO(4) internal index. The target space is a $(D+1)-$dimensional Minkowski spacetime  $\mathcal{M}_{D+1}$ and $t$ denotes a parameter that labels positions along the worldline, which is embedded in spacetime by the functions $x^M(t)$. The phase space action is given by
\begin{equation} \label{act}
S=\int d t \left[p_M\dot x^{M}+\frac{i}{2}\,\Xi_{M}^I \dot\Xi^{M}_I-\frac{e}{2}\,\mathcal{H}-i \, \mathcal{X}^I Q_I\right]\ ,
\end{equation}
with $(e, \mathcal{X})$ being a one-dimensional supergravity multiplet enforcing the first-class constraints $(\mathcal{H}, Q)$ that generate through Poisson brackets the $\N=4$ superalgebra on the worldline
\begin{equation}
\{Q_I, Q_J\}_{\mathrm{PB}}=-2i\,\delta_{IJ}\,\mathcal{H}\ ,\quad \{Q_I, \mathcal{H}\}_{\mathrm{PB}}=0\ .
\end{equation}
The latter algebra is computed by using the graded Poisson brackets of the phase space coordinates $\{x^M, p_N\}_{\mathrm{PB}}=\delta^M_N$ and $\{\Xi_{M}^I, \Xi^{N}_J\}_{\mathrm{PB}}=-i\delta_M^N \delta^I_J$, fixed by the symplectic term of the action \cite{new-book}. The constraints $\mathcal{H}:=p^2$ and $Q_I:=\Xi_I^M p_M$ must be introduced to ensure the mass-shell condition and to remove unphysical degrees of freedom, eliminating negative norm states. From the latter higher-dimensional theory, the lower-dimensional \emph{massive} model is derived as follows. First, it is convenient to take complex combinations of the original four real Grassmann variables $\Xi^M_I(t)$ and the four gravitinos $\mathcal{X}^I(t)$ as follows ($i=1,2$)
\begin{align}
\begin{split} \label{complex}
\xi^M_i&:=\tfrac{1}{\sqrt2}(\Xi^M_i+i\,\Xi^M_{i+2})\ , \quad \bar\xi^{M i}:=\tfrac{1}{\sqrt2}(\Xi^M_i-i\,\Xi^M_{i+2})\ , \\
\chi_i&:=\tfrac{1}{\sqrt2}(\mathcal{X}_i+i\,\mathcal{X}_{i+2})\ , \quad \bar\chi^{i}:=\tfrac{1}{\sqrt2}(\mathcal{X}_i-i\,\mathcal{X}_{i+2})\ .
\end{split}
\end{align}
Then, employing the so-called Scherk--Schwarz mechanism \cite{Scherk:1979zr}, the model is dimensionally reduced on a flat spacetime of the form $\mathcal{M}_D \times S^1$. In practice, one gauges the compact direction $x^D$, corresponding to $S^1$, by imposing the first-class constraint $p_D=m$ while setting
\begin{align} \label{DEC}
x^M = (x^\mu , x^D)\ , \quad p_M = (p_\mu , p_D)\ , \quad \xi^M_i=(\psi^\mu_i,\theta_i)\ , \quad \bar \xi^{M i}=(\bar \psi^{\mu i},\bar \theta^i)\ .
\end{align}
The worldline phase space action of the massive $\mathcal{N}=4$ spinning particle is given by
\begin{equation} \label{chap:fifth:action}
S=\int d t \left[p_\mu\dot x^{\mu}+i\bar\psi_{\mu}\cdot\dot\psi^{\mu}+i\bar \theta \cdot \dot \theta-\frac{e}{2}\,H-i\chi \cdot \bar q-i\bar \chi \cdot q\right]\ , 
\end{equation}
where a dot indicates a contraction of the internal indices. The worldline action \eqref{chap:fifth:action} displays many symmetries. Specifically, the local symmetries are worldline reparametrizations generated by the Hamiltonian $(H)$ and four worldline supersymmetries generated by the supercharges $(q, \bar q)$, where
\begin{equation} \label{chap:fifth:const}
H:=p^\mu p_\mu+m^2\ , \quad q_i:=\psi_i^\mu \, p_\mu+m\theta_i\ , \quad \bar q^i:=\bar \psi^{i\mu} \, p_\mu+m\bar \theta^i\ .
\end{equation}
Let us stress that their presence is \emph{essential} to describe relativistic massive particles in target space as they ensure unitarity: for this very reason said symmetries have been made local. The aforementioned worldline supergravity multiplet $(e,\chi,\bar \chi)$ acts as a set of Lagrange multipliers for the first-class constraints \eqref{chap:fifth:const}. 

Upon quantization, the worldline coordinates obey the following (anti)commutation relations fixed by their classical Poisson brackets
\begin{equation}
[x^\mu, p_\nu]=i\,\delta^\mu_\nu\ ,\quad \{\bar \psi^{\mu i}, \psi^{\nu}_j\}=\delta^i_j\,\eta^{\mu\nu}\ , \quad \{\bar \theta^i, \theta_j\}=\delta^i_j\ .
\end{equation}
By choosing a Fock vacuum annihilated by $(\bar \psi^i_\mu, \bar \theta^i)$, a generic state $\ket{\Omega}$ in the Hilbert space can be identified with the wavefunctions
\begin{equation}\label{wavefunction}
\Omega(x,\psi_i,\theta_i)=\sum_{n_1,n_2=0}^D \; \sum_{m_1,m_2=0}^1\Omega_{\mu[n_1]\vert \nu[n_2]}(x)\,\psi_1^{\mu_1} \dots \psi_1^{\mu_{n_1}}\theta_1^{m_1}\,\psi_2^{\nu_1} \dots \psi_2^{\nu_{n_2}}\theta_2^{m_2}\ ,
\end{equation}
namely a collection of tensor fields with the symmetries of $(n_1,n_2)$ bi-forms. We used the condensed notation for antisymmetrized indices $\mu[n]:=[\mu_1 \dots \mu_n]$ and a vertical bar to separate indices with no symmetry relations.

The spectrum \eqref{wavefunction} contains way too many states, which is reflected on the corresponding BRST system, found to be consistent only upon a suitable truncation of the BRST Hilbert space. While this works for BRST cohomology, the problem reappears at the one-loop level, since in principle all the unwanted states may propagate in the loop. Our task is thus to find a way to implement the projection onto the massive gravity sector. The first step in this direction consists of restricting the spectrum to the $n_1+m_1=1$, $n_2+m_2=1$ subspace. Indeed, it has been shown \cite{Bastianelli:2019xhi} that the massless sector contains the massless NS-NS spectrum of closed string theory, and corresponds to the level $(n_1,n_2\,\vert\,m_1,m_2)=(1,1\,\vert\,0,0)$ fields $\Omega_{\mu\vert\nu}$. The latter decomposes into a graviton, an antisymmetric Kalb--Ramond two-form, and a dilaton 
\begin{equation}
\Omega_{\mu\vert\nu}(x)=h_{\mu\nu}(x)+B_{\mu\nu}(x)+\delta_{\mu\nu} \, \phi(x)\ ,
\end{equation}
with the graviton identified with the symmetric and traceless component. Due to the mass improvement, we anticipate the spectrum in the level $(m_1,m_2)\neq(0,0)$ to include also the associated St\"uckelberg fields $(\Omega_\mu, \Omega_\nu, \Omega)$, specifically two massless vector fields and a scalar \cite{Stueckelberg:1957zz, Boulanger:2018dau}, allowing for the propagation of the massive graviton and the massive Kalb--Ramond degrees of freedom alongside the dilaton. This will be confirmed through the analysis of the worldline path integral.

In order to implement the aforementioned projection, the R-symmetry of the model has to be appropriately exploited. Notice that the action \eqref{act} has a manifest global SO(4) symmetry that rotates the fermions, generated by the fermion bilinears
\begin{equation}
\mathrm{J}_{IJ}:=i\,\Xi_{I}^M \, \Xi_{J M}\ ,
\end{equation}
while in the complex basis \eqref{complex} and with the decomposition \eqref{DEC} only the subgroup U(2) $\subset$ SO(4) is kept manifest as rigid symmetry of the action \eqref{chap:fifth:action}. As a result, the SO(4) generators split as $\mathrm{J}_{IJ} \sim (J_i^j,K^{ij},G_{ij})$ and their explicit realization is\footnote{See Appendix B of Ref.~\cite{Bonezzi:2018box} for a detailed derivation.}
\begin{align} 
\begin{split} \label{chap:fifth:gen}
J_i^j &:=\psi_i \cdot {\bar{\psi}}^j+\theta_i \, \bar{\theta}^j\ , \\
K^{ij} &:= \bar{\psi}^i \cdot \bar{\psi}^j+\bar{\theta}^i \, \bar{\theta}^j\ , \\
G_{ij} &:= \psi_i\cdot \psi_j+\theta_i \, \theta_j\ ,
\end{split}
\end{align}
where we used a dot $\cdot$ to indicate contraction on spacetime indices. $K^{ij}$ is the so-called trace operator while $G_{ij}$ implements the insertion of the metric; they both vanish for $i=j$ \cite{Bastianelli:2014lia, Bonezzi:2018box}. On the other hand, $J_i^j$ generate the U(2) subgroup, and upon quantization become quantum operators\footnote{We used hats to stress that the expression refers to operators in that given order; however, throughout the text, we will often avoid the use of hats for quantum operators, so as not to burden the notation further.}
\begin{equation}
\hat J_i^j = \hat \psi_i \cdot \hat {\bar{\psi}}^j+\hat \theta_i \, \hat{\bar{\theta}}^j-\tfrac{D+1}{2}\,\delta_i^j\ ,
\end{equation}
where we employed the Weyl ordering to resolve ambiguities, matching the path integral regularization used. It is worth noticing that the shift $\tfrac{D+1}{2}$ is a quantum ordering effect.

The crucial point is that it is possible, although not strictly necessary, to make the R-symmetry local. Differently from worldline supersymmetries, the R-symmetry does not have to be gauged for unitarity; however, it can be used to perform algebraic projections on the desired spectrum by appropriately gauging only a specific subgroup $\R$ $\subset$ SO(4). At the level of the worldline action, this corresponds to introducing the appropriate worldline gauge fields\footnote{Here we use $\R$ as a mere label.} $a^{\mathcal{R}}(t)$, by the insertion of a term
\begin{equation}
S_{\R}=-\int d t \; a^{\mathcal{R}}\,\mathrm{J}_{\mathcal{R}}\ ,
\end{equation}
acting as Lagrange multipliers for the classical constraints $\mathrm{J}_{\mathcal{R}}$. In the following, we seek to unveil the massive gravity content with different choices of the $\R$ subgroup, which extracts degrees of freedom from the worldline path integral. 

\subsubsection*{Setting up the worldloop path integral}
The worldloop is defined as
\begin{equation} \label{chap:fifth:1}
\Gamma = \int_{S^1}
\frac{DG\,DX}{\mathrm{Vol(Gauge)}}\, \mathrm{e}^{-S_{\mathrm{E}}[X,G]}\ ,
\end{equation}
where we denoted the worldline gauge fields $G=\left( e, \chi, \bar{\chi},a\right)$ and the coordinates with supersymmetric partners $X=\left(x,\psi,\bar{\psi},\theta,\bar \theta\right)$. The action appearing in \eqref{chap:fifth:1} is the action in Euclidean configuration space ($S_{\mathrm{E}}=-iS$), obtained by a Wick-rotation to Euclidean time $t \to -i\tau$ accompanied by the Wick rotations of the gauge fields $a_{\R}\to -ia_{\R}$, just as done in Ref.~\cite{Bastianelli:2007pv} for general $\N$. From now on we will drop the subscript on $S_{\mathrm{E}}$ as no confusion should arise. The overcounting from summing over gauge equivalent configurations, which causes the path integral to diverge, is formally taken into account by dividing by the volume of the gauge group. To regularize the path integral one has to follow a gauge-fixing procedure. We use the Faddeev--Popov method to extract the volume of the gauge group and to gauge-fix completely the supergravity multiplet up to some moduli while evaluating the determinants stemming from the associated $\acr{\Phi\Pi}$ ghosts \cite{Faddeev:1967fc}, as we will outline in the following. 

The einbein is gauge-fixed to a constant, namely $e(T)=2T$, where $T$ is the usual Schwinger proper time, while the gravitinos $(\chi, \bar{\chi})$ are antiperiodic and gauge-fixed to zero, leaving no additional moduli. The path integral \eqref{chap:fifth:1} multiplied by $-\tfrac{1}{2}$ corresponds to the \acr{QFT} effective action and can be recast in the following form
\begin{equation} \label{chap:fifth:euc}
\Gamma = -\frac{1}{2}\int_{0}^{\infty}\frac{dT}{T}\,\mathrm{e}^{-m^2T} \,Z_{\R}(T)\ ,
\end{equation} 
where the integration over the Schwinger proper time arises from the gauge-fixing of the einbein $e$. The explicit expression of the density $Z_{\R}(T)$ and of the gauge-fixing conditions for the worldline gauge fields $a_{\R}$ depends on the subset $\R$ being gauged. We start our analysis with the simplest case of gauging the $\R=$ U(1) $\times$ U(1) subgroup to set the grounds for the general case. 

\subsection{Gauging of the U(1) $\times$ U(1) subgroup}
The U(1) $\times$ U(1) subgroup corresponds to the generators
\begin{equation} \label{U(1)}
J_i:=\psi_i \cdot \bar{\psi}^i + \theta_i \, \bar{\theta}^i \quad \text{($i$ not summed)}\ .
\end{equation}
The gauging of the U(1) $\times$ U(1) subgroup is realized by means of two abelian worldline gauge fields $a_i(t)$. At the level of the worldline action, this corresponds to
\begin{equation}
S_{\mathrm{U(1)}\times \mathrm{U(1)}}=-\int d t \; a^i (J_i -q_i )\ ,
\end{equation}
where two independent Chern--Simons couplings $q_i=\tfrac{3-D}{2}$, that convert the classical constraints $(J_i -q_i )$ into the operatorial constraints $(\hat N_i -2)$,\footnote{With $\hat N_i:=\hat N_{\psi_i} + \hat N_{\theta_i}=\hat \psi_i \cdot \hat{\bar{\psi}}^i+\hat \theta_i \, \hat{\bar{\theta}}^i$ being the number operators counting the number of fermionic oscillators with a fixed flavor index. The U(1) $\times$ U(1) generators explicitly read $\hat J_i:=\hat N_i-\tfrac{D+1}{2}$.} have been included to project on the gravity sector. The whole supergravity multiplet is gauge-fixed as 
\begin{equation}
\tilde G=\left(T, 0, 0,\tilde a_{i}\right) \quad \text{with} \quad \tilde a_{i}:= \left(\begin{array}{c} \alpha \\ \beta \\
\end{array}\right)\ ,
\end{equation}
where $\alpha, \beta \in [0,2\pi]$ are two angles representing additional moduli related to the gauge fields. Inserting the $\acr{\Phi\Pi}$ determinants to eliminate the volume of the gauge group, and setting appropriately the overall normalization, the partition function in \eqref{chap:fifth:euc} explicitly reads
\begin{equation} \label{chap:fifth:path}
Z_{\mathrm{U(1)}\times \mathrm{U(1)}}(T)=\int_0^{2\pi}\frac{d\alpha}{2\pi}\int_0^{2\pi}\frac{d\beta}{2\pi}\,\left(2\cos\tfrac{\alpha}{2}\right)^{-2}\left(2\cos\tfrac{\beta}{2}\right)^{-2} \Tr\left[\mathrm{e}^{-T\hat H}\mathrm{e}^{i\alpha(\hat N_{\psi_1}+\hat N_{\theta_1}-2)+i\beta(\hat N_{\psi_2}+\hat N_{\theta_2}-2)}\right]\ .
\end{equation}
A few comments are in order. The two gauge fields $a_i$ produce the integration over the angular moduli $(\alpha,\beta)$, while the \acr{SUSY} ghosts account for the cosine factors. The path integral over the “matter” sector
\begin{equation}
\int_{_{\rm PBC}}\hskip-.4cm{ D}x\int_{_{\rm ABC}}\hskip-.4cm D\bar{\psi}D\psi\int_{_{\rm ABC}}\hskip-.4cm D\bar{\theta}D\theta \, \mathrm{e}^{-S_{\mathrm{gf}}[X,\tilde G]}
\end{equation}
has been put into an operatorial form as a trace -- with $\hat H = \hat p^2$ for the free theory and with $\hat N$ being the operators counting the number of oscillators with a fixed flavor index -- over the Hilbert space consisting of the states contained in the Taylor coefficients of the wavefunctions $\Omega(x,\psi_i,\theta_i)$. The path integration over bosonic variables is evaluated by fixing periodic boundary conditions, while the fermionic path integral is performed by choosing antiperiodic boundary conditions on each flavor of fermionic fields. Finally, the gauge-fixed action reads
\begin{equation} \label{actiongf}
S_{\mathrm{gf}}[X,\tilde G]=\int d\tau \left[ \frac{1}{4T}\dot{x}^{\mu}\dot{x}_{\mu}+m^2T+\sum_{i=1,2}\bar{\psi}_i^{\mu}(\partial_\tau +i \tilde a_{i}){\psi}^i_{\mu}+\sum_{i=1,2}\bar{\theta}_{i}(\partial_\tau +i \tilde a_{i}){\theta}^i +i\sum_{i=1,2}\tilde a^i \, q_i\right]\ .
\end{equation}
Having set up the worldloop, we are in the position of analyzing the degrees of freedom of the wordline model. Firstly, note that the Hilbert space can be decomposed in terms of the eigenvalues $(n_1,n_2)$ and $(m_1,m_2)$ of the pairs of number operators $(\hat N_{\psi_1},\hat N_{\psi_2})$ and $(\hat N_{\theta_1},\hat N_{\theta_2})$ respectively. Consequently, the trace can be decomposed in terms of 
\begin{align}
\begin{split} \label{N}
N_1 &=n_1+m_1\ , \\
N_2 &=n_2+m_2\ ,
\end{split}
\end{align}
reproducing the double grading of the \emph{massless} $\mathcal{N}=4$ spinning particle \cite{Bastianelli:2019xhi}
\begin{equation}
t_{N_1,N_2}(T):=\Tr_{N_1,N_2}\left[\mathrm{e}^{-T\hat H}\right]\ .
\end{equation}
The path integral \eqref{chap:fifth:path}, using the Wilson line variables $z:= \mathrm{e}^{i\alpha}$ and $\omega:= \mathrm{e}^{i\beta}$, becomes\footnote{The modular integration is performed over the circle $|z|=1$, with the singular point $z=-1$ pushed out of the contour, the same goes for $\omega$. Discussion on the regulated contour of integration for the modular parameters can be found in Ref.~\cite{Bastianelli:2005vk}.}
\begin{equation}
Z_{\mathrm{U(1)}\times \mathrm{U(1)}}(T)=\oint \frac{dz}{2\pi i z}\oint \frac{d\omega}{2\pi i \omega}\,\frac{z}{(z+1)^2}\frac{\omega}{(\omega+1)^2} \sum_{N_1,N_2} t_{N_1,N_2}(T) \, z^{N_1-2}\omega^{N_2-2}\ .
\end{equation}
Then, we can trace back the contributions from the $\theta$s through the following identification 
\begin{equation} \label{dec}
\sum_{N_1,N_2} t_{N_1,N_2}  :=  \sum_{n_1,n_2=0}^D \; \sum_{m_1,m_2=0}^1 \mathbbmss{T}\,_{n_1,n_2}^{m_1,m_2}\ ,
\end{equation}
where $\mathbbmss{T}\,_{n_1,n_2}^{m_1,m_2}$ denotes the trace restricted to some specific eigenvalues $(n_1,n_2\,\vert\,m_1,m_2)$ of the fermionic number operators. Let us highlight how the “massless to massive decomposition” works. Upon modular integration, the massless partition function includes only the following contributions
\begin{equation}
Z_{\mathrm{U(1)}\times \mathrm{U(1)}}=t_{1,1}-2\,t_{1,0}-2\,t_{0,1}+4\,t_{0,0}\ ,
\end{equation}
while the other possible values of $(N_1, N_2)$ yield zero. Keeping in mind the decomposition \eqref{N}--\eqref{dec}, in the following denoted with arrows, we get the massive improvements listed below
\begin{align}
\begin{split}
t_{1,1} &\xlongrightarrow{m} \quad \mathbbmss{T}\,_{1,1}^{0,0}+\mathbbmss{T}\,_{1,0}^{0,1}+\mathbbmss{T}\,_{0,1}^{1,0}+\mathbbmss{T}\,_{0,0}^{1,1}\ , \\[2mm]
t_{1,0} &\xlongrightarrow{m} \quad \mathbbmss{T}\,_{1,0}^{0,0} +\mathbbmss{T}\,_{0,0}^{1,0}\ ,\\[2mm]
t_{0,1} &\xlongrightarrow{m} \quad \mathbbmss{T}\,_{0,1}^{0,0} +\mathbbmss{T}\,_{0,0}^{0,1}\ ,\\[2mm]
t_{0,0} &\xlongrightarrow{m} \quad \mathbbmss{T}\,_{0,0}^{0,0}\ .
\end{split}
\end{align}
The above partition function can then be decomposed into its irreducible spacetime components. The degrees of freedom for the free theory are given by
\begin{equation}
\mathbbmss{T}\,_{n_1,n_2}^{m_1,m_2}=\frac{1}{(4\pi T)^{D/2}}\binom{D}{n_1}\binom{D}{n_2}\ ,
\end{equation}
where the factor $\left(4\pi T\right)^{-D/2}$ corresponds to the free particle position and will be omitted in the following. On the other hand, the binomials count the number of degrees of freedom and correspond to the transverse polarizations of the tensor $\Omega_{\mu\vert\nu}$, yielding
\begin{equation}
Z_{\mathrm{U(1)}\times \mathrm{U(1)}}= Z_{h_{\mu\nu}}+Z_{A_{\mu}}+Z_\varphi+Z_{B_{\mu\nu}}+Z_{C_{\mu}}+Z_\phi=(D-1)^2\ , 
\end{equation}
where 
\begin{equation}
Z_{h_{\mu\nu}}+Z_{A_{\mu}}+Z_\varphi=\frac{(D+1)(D-2)}{2}
\end{equation}
corresponds to a \emph{massive} graviton, i.e. a massless graviton $Z_{h_{\mu\nu}}$ along with the two St\"uckelberg fields, a massless vector field $Z_{A_{\mu}}$ and a scalar field $Z_\varphi$\footnote{Eventually, one could identify them as a single \emph{massive} vector St\"uckelberg carrying $D-1$ degrees of freedom.}
\begin{align}
Z_{h_{\mu\nu}} & :=  \mathbbmss{T}\,_{1,1}^{0,0}-\mathbbmss{T}\,_{2,0}^{0,0}-2 \, \mathbbmss{T}\,_{1,0}^{0,0}=\tfrac{D(D-3)}{2}\ , \\
Z_{A_{\mu}} & := \mathbbmss{T}\,_{1,0}^{0,1}-2 \, \mathbbmss{T}\,_{0,0}^{1,0}=D-2\ ,\\
Z_\varphi & :=  \mathbbmss{T}\,_{0,0}^{1,1}=1\ ,
\end{align}
while 
\begin{equation}
Z_{B_{\mu\nu}}+Z_{C_{\mu}} =\frac{(D-1)(D-2)}{2}
\end{equation}
is the massive Kalb--Ramond field, corresponding to the massless contribution plus the massless St\"uckelberg vector
\begin{align}
Z_{B_{\mu\nu}} & :=  \mathbbmss{T}\,_{2,0}^{0,0}-2 \, \mathbbmss{T}\,_{0,1}^{0,0}+3 \, \mathbbmss{T}\,_{0,0}^{0,0}=\tfrac{(D-2)(D-3)}{2}\ ,\\
Z_{C_{\mu}} & :=  \mathbbmss{T}\,_{0,1}^{1,0}-\mathbbmss{T}\,_{0,0}^{0,1}=D-2\ ,
\end{align}
and with
\begin{equation}
Z_\phi  :=  \mathbbmss{T}\,_{0,0}^{0,0}=1
\end{equation}
being the dilaton. One can check that the $(m_1, m_2)=(0,0)$ sector correctly reproduces the massless spectrum of the Hilbert space contained in the $\mathcal{N}=4$ spinning particle model, which coincides with the massless NS-NS sector of closed strings, while the $(m_1,m_2) \neq (0,0)$ sector corresponds to the massive improvements, namely the associated St\"uckelberg fields.

\subsection{Full SO(4) group gauging}
In the following, we gauge the entire R-symmetry group. The analysis will show that this choice produces not only the massive graviton in the spectrum but also some unwanted contributions. One has to find a way to project the latter away to finally construct a worldline path integral specifically for the massive graviton. 

The gauging of the full set of generators is achieved through a one-dimensional Yang--Mills field $a^{IJ}(t)$ acting as a Lagrange multiplier in \eqref{chap:fifth:action}. Explicitly
\begin{equation}
S_{\mathrm{SO(4)}}=-\frac{1}{2}\int d t \; a^{IJ} \mathrm{J}_{IJ}\ ,
\end{equation}
or, taking into account the splitting \eqref{chap:fifth:gen},
\begin{equation}
S_{\mathrm{SO(4)}}= -\int d t \left( a_i^j J_j^i+\frac12 a_{ij} K^{ij}+\frac12 a^{ij} G_{ij}\right)\ .
\end{equation}
Note that, contrary to the previous case, there is no room for Chern--Simons couplings here. It is necessary then to restrict the analysis to $D=3$ spacetime dimensions at first, to correctly reproduce a graviton state. This is related to the BRST quantization of Chapter~\ref{chap:fourth}, where it was discussed how the physical wavefunction of the spinning particle lies in the kernel of the operator $\hat{\mathcal{J}}_i^i$ -- which is the quantum operator corresponding to \eqref{U(1)} including the contribution from the BRST bosonic superghosts -- only in three spacetime dimensions. Even if it has been shown how to reproduce a first-quantized massive gravity theory in four spacetime dimensions by demanding the physical states to have a fixed U(1) $\times$ U(1) charge of $-\tfrac{1}{2}$, in the present case the latter condition could be satisfied only upon adding a Chern--Simons term, which is not possible when the full gauging is considered. At the end of the analysis, it shall then be discussed how to overcome this obstruction. 

The starting point is the path integral \eqref{chap:fifth:euc}, with the gauge-fixing $\tilde G=\left(T, 0, 0,\tilde a_{i}^{j}\right)$ producing now the following action
\begin{equation} \label{azione}
S_{\mathrm{gf}}[X,\tilde G]=\int d\tau \left[ \frac{1}{4T}\dot{x}^{\mu}\dot{x}_{\mu}+m^2T+\bar{\psi}^{i \mu}\left(\delta_{ij} \partial_\tau +i \tilde a_{ij}\right){\psi}^j_{\mu}+\bar{\theta}^{i}\left(\delta_{ij} \partial_\tau +i \tilde a_{ij}\right){\theta}^j\right]\ , 
\end{equation}
with
\begin{equation}
\tilde a_{i}^{j}:=\left(\begin{array}{cc} \alpha & 0 \\ 0 & \beta \\
\end{array}\right)\ .
\end{equation}
In the present case, one has to introduce the non-abelian $\acr{\Phi\Pi}$ ghosts associated with the whole SO(4) group, whose path integration modifies the path integral \eqref{chap:fifth:path} with the inclusion of the sine factors previously calculated in Ref.~\cite{Bastianelli:2007pv}. 
In terms of Wilson variables the path integral explicitly reads\footnote{Note that we factored out the $\tfrac{1}{4}$ also stemming from the path integral over the SO(4) $\acr{\Phi\Pi}$ ghosts for simplicity.}
\begin{equation}
Z_{\mathrm{SO(4)}}(T)=\frac{1}{4}\oint \frac{dz}{2\pi i z}\oint \frac{d\omega}{2\pi i \omega}\,\frac{z}{(z+1)^2}\frac{\omega}{(\omega+1)^2} \, p(z,\omega)\sum_{N_1,N_2} t_{N_1,N_2}(T) \,z^{N_1-2}\omega^{N_2-2}\ ,
\end{equation}
where $p(z,\omega)$ is the measure on the moduli space arising from the full gauging and reads
\begin{equation} \label{p}
p(z,\omega):= 4-2z\omega-2\left(\frac{z}{\omega}+\frac{\omega}{z}\right)+z^2+\omega^2-\frac{2}{z\omega}+\frac{1}{z^2}+\frac{1}{\omega^2}\ .
\end{equation}
To unveil the projection of the various components of the full measure, it is possible to follow the “massless to massive decomposition” described in the previous subsection. Take as an illustrative example the first monomial of \eqref{p}. It yields the following contribution upon modular integration
\begin{align}
4 \; &\xlongrightarrow{\oint\oint} \quad t_{1,1}-2\,t_{1,0}-2\,t_{0,1}+4\,t_{0,0} \\[2mm]
&\xlongrightarrow{m} \quad \mathbbmss{T}\,_{1,1}^{0,0}+\mathbbmss{T}\,_{1,0}^{0,1}+\mathbbmss{T}\,_{0,1}^{1,0}+\mathbbmss{T}\,_{0,0}^{1,1}-2\mathbbmss{T}\,_{1,0}^{0,0}-2\mathbbmss{T}\,_{0,1}^{0,0} -2\mathbbmss{T}\,_{0,0}^{1,0}-2\mathbbmss{T}\,_{0,0}^{0,1}+4\mathbbmss{T}\,_{0,0}^{0,0} \nonumber \\[2mm]
& :=  \quad \;\;\; Z_{\mathrm{U(1)}\times \mathrm{U(1)}}\ . \nonumber
\end{align}
The first arrow implies a modular integration, while the second one denotes the massless to massive decomposition as previously discussed. As for the remaining monomials, the outcome is as follows:
\begin{align}
-2z\omega \; &\xlongrightarrow{\oint\oint} \quad -\tfrac{1}{2}\,t_{0,0} \; \xlongrightarrow{m} \quad -\tfrac{1}{2}\, \mathbbmss{T}\,_{0,0}^{0,0} \;  :=  \quad -\tfrac12\, Z_\phi\ , \\[3mm]
-2\left(\frac{z}{\omega}+\frac{\omega}{z}\right) \;&\xlongrightarrow{\oint\oint} \quad -(t_{2,0}-2\,t_{1,0}+3\,t_{0,0}) \\[2mm]
&\xlongrightarrow{m} \quad -\left( \mathbbmss{T}\,_{2,0}^{0,0}+\mathbbmss{T}\,_{1,0}^{1,0}-2\,\mathbbmss{T}\,_{1,0}^{0,0}-2\,\mathbbmss{T}\,_{0,0}^{1,0}+3\mathbbmss{T}\,_{0,0}^{0,0}\right) \nonumber \\[2mm]
& :=  \quad \;\;\; -\left( Z_{B_{\mu\nu}} + Z_{C_{\mu}} \right)\ , \nonumber \\[3mm]
z^2+\omega^2\;&\xlongrightarrow{\oint\oint} \quad 0\ ,
\end{align}
for the first half of $p(z,\omega)$. It is evident that, as for the massless case, the above monomials are enough to construct a measure capable of projecting only into the massive gravity sector. We list here the contributions stemming from the other monomials for completeness: we have
\begin{align}
-\frac{2}{z\omega}\;&\xlongrightarrow{\oint\oint} \quad -\tfrac12\, t_{2,2}-2\,t_{1,1}-\tfrac92\,t_{0,0}+2\,t_{2,1}-3\,t_{2,0}+6\,t_{1,0} \\[2mm]
&\xlongrightarrow{m} \quad -\tfrac12\,\mathbbmss{T}\,_{2,2}^{0,0}-\tfrac12\,\mathbbmss{T}\,_{2,1}^{0,1}-\tfrac12\,\mathbbmss{T}\,_{1,2}^{1,0}-\tfrac12\,\mathbbmss{T}\,_{1,1}^{1,1} 
-2\,\mathbbmss{T}\,_{1,1}^{0,0}-2\,\mathbbmss{T}\,_{1,0}^{0,1}-2\,\mathbbmss{T}\,_{0,1}^{1,0}-2\,\mathbbmss{T}\,_{0,0}^{1,1} \nonumber \\[1mm]
&\phantom{\xlongrightarrow{m}} \; \quad -\tfrac92\,\mathbbmss{T}\,_{0,0}^{0,0}
+2\,\mathbbmss{T}\,_{2,1}^{0,0}+2\,\mathbbmss{T}\,_{2,0}^{0,1}+2\,\mathbbmss{T}\,_{1,1}^{1,0}+2\,\mathbbmss{T}\,_{1,0}^{1,1}
-3\,\mathbbmss{T}\,_{2,0}^{0,0}-3\,\mathbbmss{T}\,_{1,0}^{1,0}
+6\,\mathbbmss{T}\,_{1,0}^{0,0}+6\,\mathbbmss{T}\,_{0,0}^{1,0} \nonumber \\[2mm]
& :=  \quad \;\;\; -\tfrac12\,Z_{\A_{2,2}}-Z_{\A_{2,1}}-\tfrac12\,Z_{\A_{1,1}}\ , \nonumber \\[3mm]
\frac{1}{z^2}+\frac{1}{\omega^2}\;&\xlongrightarrow{\oint\oint} \quad \tfrac12\, t_{3,1}-\,t_{2,1}+\tfrac32\,t_{1,1}-5\,t_{1,0}-\,t_{3,0}+2\,t_{2,0}+4\,t_{0,0} \\[2mm]
&\xlongrightarrow{m} \quad +\tfrac12\,\mathbbmss{T}\,_{3,1}^{0,0}+\tfrac12\,\mathbbmss{T}\,_{3,0}^{0,1}+\tfrac12\,\mathbbmss{T}\,_{2,1}^{1,0}+\tfrac12\,\mathbbmss{T}\,_{2,0}^{1,1}
-\,\mathbbmss{T}\,_{2,1}^{0,0}-\,\mathbbmss{T}\,_{2,0}^{0,1}-\,\mathbbmss{T}\,_{1,1}^{1,0}-\,\mathbbmss{T}\,_{1,0}^{1,1} \nonumber \\[1mm]
&\phantom{\xlongrightarrow{m}} \; \quad +\tfrac32\,\mathbbmss{T}\,_{1,1}^{0,0}+\tfrac32\,\mathbbmss{T}\,_{1,0}^{0,1}+\tfrac32\,\mathbbmss{T}\,_{0,1}^{1,0}+\tfrac32\,\mathbbmss{T}\,_{0,0}^{1,1}
-5\,\mathbbmss{T}\,_{1,0}^{0,0}-5\,\mathbbmss{T}\,_{0,0}^{1,0}
-\,\mathbbmss{T}\,_{3,0}^{0,0}-\,\mathbbmss{T}\,_{2,0}^{1,0} \nonumber \\[1mm]
&\phantom{\xlongrightarrow{m}} \; \quad +2\,\mathbbmss{T}\,_{2,0}^{0,0}+2\,\mathbbmss{T}\,_{1,0}^{1,0}
+4\,\mathbbmss{T}\,_{0,0}^{0,0} \nonumber \\[2mm]
& :=  \quad \;\;\; \tfrac12\,Z_{\A_{3,1}}+\tfrac12\,Z_{\A_{3,0}}+\tfrac12\,Z_{\A_{2,1}}+\tfrac12\,Z_{\A_{2,0}}\ . \nonumber
\end{align}
The above contributions can be interpreted in terms of bi-forms $\A_{p,q}$ with corresponding partition functions 
\begin{equation}
Z_{\A_{p,q}}=\sum_{k,l=0}^{p,q}(-)^{k+l}(k+1)(l+1)\,t_{p-k,q-l} 
\end{equation}
previously analyzed in Ref.~\cite{Bastianelli:2019xhi}. Indeed, they coincide with the same topological contributions of the massless path integral, $\A_{2,2}$ and $\A_{3,1}$, along with the respective massive improvements, which contribute to the effective action on non-trivial backgrounds and have to be projected out. It is immediate to see that the following polynomial
\begin{equation} \label{P}
P_{(3)}(z,\omega) := 4-2\left(\frac{z}{\omega}+\frac{\omega}{z}\right)-4z\omega+2\left(z^2+\omega^2 \right)= \frac{2 (z-\omega )^2 (z\omega-1)}{z\omega}
\end{equation}
does the work indeed, producing
\begin{equation}
P_{(3)}(z,\omega)\xlongrightarrow{\oint\oint} \quad Z_{\mathrm{U(1)}\times \mathrm{U(1)}} - Z_{B_{\mu\nu}} - Z_{C_{\mu}}-Z_\phi=Z_{h_{\mu\nu}}+Z_{A_{\mu}}+Z_\varphi\ .
\end{equation}
The measure \eqref{P} coincides with the one implemented to construct the path integral for the \emph{massless} graviton, although they work in different spacetime dimensions. This should not come as a surprise in retrospect: our analysis has shown that the introduction of the mass ensures that every single contribution to the partition function gets “St\"uckelberged” while evaluating the trace on the larger Hilbert space, without the arising of unexpected terms. Consequently, the same measure eliminates both the unwanted contributions and their St\"uckelberg companions at once.

As discussed in Ref.~\cite{Bastianelli:2019xhi}, the measure \eqref{P} can be related to a precise gauging of the R-symmetry group, specifically the gauging of the \emph{parabolic subgroup} of SO(4) (see \cite{Bastianelli:2009eh} for its application in worldline models for higher spin particles). It consists of the subgroup generated by $J_i^j$ and the trace $K^{ij}$ while excluding the insertion of the metric $G_{ij}$, which produces a measure
\begin{equation}
P_{\mathrm{par}}(z,\omega):= \frac{2 (z-\omega )^2 (z\omega-1)}{z^{3/2}\omega^{3/2}}\ .
\end{equation}
This choice leaves room for a Chern--Simons term in the Euclidean action, which can be chosen to correctly reproduce the whole measure, i.e.
\begin{equation}
S_{\mathrm{CS}}=i q\int d\tau \, a_i^i \quad \text{with} \quad q=-\frac{1}{2}\ ,
\end{equation}
which results in a modification of $P_{\mathrm{par}}(z,\omega)$ through the following multiplicative term 
\begin{equation}
P_{\mathrm{CS}}(z,\omega):=z^{1/2}\omega^{1/2} \quad \implies \quad P_{(3)}(z,\omega)=P_{\mathrm{par}}(z,\omega)P_{\mathrm{CS}}(z,\omega)\ .
\end{equation}
As previously discussed, the measure $P_{(3)}(z,\omega)$ fails in going beyond three spacetime dimensions and one has to find a way to improve it. Notably, the presence of the Chern--Simons term allows us to go to \emph{arbitrary dimensions} tuning the \acr{CS} coefficient appropriately, i.e. 
\begin{equation}
q \longrightarrow q +\frac{3-D}{2}\ ,
\end{equation}
with the latter improvement producing the operatorial constraints $(\hat N_i -2)$ in arbitrary dimensions. The correct measure \eqref{P} becomes
\begin{equation} \label{PD}
P(z,\omega):=\frac{2 (z-\omega )^2 (z\omega-1)}{z\omega} \, z^{\tfrac{D-3}{2}}\omega^{\tfrac{D-3}{2}}\ ,
\end{equation}
which is of course different from the massless $D$-dimensional measure. The phase space action with the parabolic gauging reads
\begin{equation} \label{action2}
S=\int d t \left[p_\mu\dot x^{\mu}+i\bar\psi_{\mu}\cdot\dot\psi^{\mu}+i\bar \theta \cdot \dot \theta-\frac{e}{2}\,H-i\chi_i\,\bar q^i-i\bar \chi^i\, q_i-\frac12 a_{ij} K^{ij}-a_i^j(J_j^i-q\delta_j^i)\right]\ .
\end{equation}

\section{One-loop massive gravity in the Worldline Formalism} \label{chap:fifth:sec3}
In this section, we perform the quantization of the corresponding non-linear sigma model which couples a massive spin 2 particle to background gravity. It will lead to the computation of the counterterms necessary for the renormalization of the one-loop effective action of massive gravity using the Worldline Formalism.

To achieve a representation of the \acr{QFT} effective action of massive gravity from \eqref{chap:fifth:1} we need to couple the massive $\mathcal{N}=4$ spinning particle to a curved target space metric $g_{\mu\nu}(x)$. As a result, the action \eqref{action2} gets covariantized through the deformation of the worldline \acr{SUSY} charges as follows
\begin{align}
\begin{split} \label{cov1}
q_i &\longrightarrow\mathcal{q}_i:=-i\,\psi_i^a\,e^\mu_a\, \pi_\mu\ +m\theta_i \\
\bar q^i &\longrightarrow\mathcal{\bar q}^i:= -i\,\bar\psi^{i\,a}\,e^\mu_a\, \pi_\mu+m\bar\theta^i
\end{split}
\end{align}
with the covariant momentum being
\begin{equation}
\pi_\mu:=p_\mu-i\omega_{\mu \, ab}\,\psi^a \cdot \bar\psi^{b}\ .
\end{equation}
Worldline fermions carry flat Lorentz indices so that $\psi^\mu_i:=e^\mu_a(x)\,\psi^a_i$, introducing a background vielbein $e_\mu^a(x)$ and the torsion-free spin connection $\omega_{\mu\, ab}$. The correct deformation of the Hamiltonian requires more consideration: the spinning particle coupled to gravity does not exhibit a first-class algebra, specifically the following anticommutator does not close:\footnote{The covariant derivatives are defined as $\hat \nabla_\mu:=\partial_\mu+\omega_{\mu\, ab}\,\psi^a\cdot\bar\psi^b$ and are related to the covariant momenta $\pi_\mu$ through $\hat \nabla_\mu=i g^{\frac{1}{4}} \pi_\mu g^{-\frac{1}{4}}$. The metric determinant factors account for a self-adjoint operator \cite{Bastianelli:2006rx}. The Laplacian is defined as
\begin{equation*}
\nabla^2:=g^{\mu\nu}\hat \nabla_\mu \hat \nabla_\nu-g^{\mu\nu}\,\Gamma^\lambda_{\mu\nu}\,\hat \nabla_\lambda\ ,
\end{equation*}
with $\Gamma^\lambda_{\mu\nu}$ being the Christoffel symbols \cite{Bastianelli:2008nm, Bonezzi:2018box}. For notational simplicity, we have used non-hermitian operators, keeping in mind that hermiticity is obtained by a similarity transformation $A \to g^{\frac{1}{4}} A g^{-\frac{1}{4}}$ on the quantum variables.}
\begin{equation}
\{\mathcal{q}_i,\mathcal{\bar q}^j \}=-\delta_i^j\left(\nabla^2-m^2\right)-R_{\mu\nu\lambda\sigma}\,\psi^\mu_i\bar\psi^{\nu\,j}\psi^\lambda\cdot\bar\psi^\sigma\ ,
\end{equation}
hence it is not immediate to identify a suitable deformed Hamiltonian. The BRST analysis of Chapter~\ref{chap:fourth} indicates to consider the following expression 
\begin{equation} \label{chap:fifth:H}
H:=\nabla^2-m^2+R_{\mu\nu\lambda\sigma}\,\psi^\mu \cdot\bar\psi^\nu\psi^\lambda \cdot\bar\psi^\sigma\ .
\end{equation}
The latter Hamiltonian is necessary to achieve nilpotency of the BRST charge on the relevant physical subspace of the full BRST Hilbert space, in the \emph{massive} worldline model. Furthermore, it turns out that the background metric has to be on-shell with cosmological constant set to zero, i.e.
\begin{equation} \label{Ricci}
R_{\mu\nu}(x)=0\ .
\end{equation}
Note that within the \emph{massless} BRST quantization of Chapter~\ref{chap:fourth} one has to introduce also a non-minimal coupling to the scalar curvature $\tfrac{2}{D}R$ inside the Hamiltonian \eqref{chap:fifth:H}, since for the pure gravitational case a non-zero cosmological constant is admitted. In the present case, such a coupling is inevitably zero.

In the following, we will evaluate perturbatively the massive path integral considering the BRST system as the starting point and keeping in mind that the results can be trusted only upon projection onto Ricci-flat manifolds \eqref{Ricci}. Indeed, as previously commented, the presence of a non-trivial gravitational background obstructs the first-class character of the constraints algebra, and a more appropriate way of thinking about the model is to consider it as a genuine BRST system from the start, regardless of being derived from a gauge-invariant classical predecessor.

The one-loop effective action $\Gamma [g_{\mu\nu}]$ of massive gravity corresponds to the worldloop path integral of the massive $\mathcal{N}=4$ spinning particle action $S[X, G;\,g_{\mu\nu}]$, with schematic form 
\begin{equation}
\Gamma [g_{\mu\nu}]= \int_{S^1}
\frac{DG\,DX}{\mathrm{Vol(Gauge)}}\, \mathrm{e}^{-S[X,G;\,g_{\mu\nu}]}\ ,
\end{equation}
where the full action $S[X,G;\,g_{\mu\nu}]$ is the one in \eqref{action2} with the suitable covariantizations \eqref{cov1}--\eqref{chap:fifth:H}. The gauging of the parabolic subgroup, the gauge-fixing $\tilde G=\left(T, 0, 0,\tilde a_{i}^{j}\right)$ and the path integration over the $\acr{\Phi\Pi}$ ghost proceeds as outlined in the previous sections. Factorizing out the exponential of the mass, the worldloop path integral becomes 
\begin{equation} \label{euc2}
\Gamma [g_{\mu\nu}]= -\frac{1}{2}\int_{0}^{\infty}\frac{dT}{T}\,\mathrm{e}^{-m^2T} \,Z(T)\;,
\end{equation}
where the partition function is
\begin{align} \label{chap:fifth:PATH}
Z(T)=\frac{1}{4}
\oint \frac{dz}{2\pi i z}\oint \frac{d\omega}{2\pi i \omega}\,\frac{z}{(z+1)^2}\frac{\omega}{(\omega+1)^2} \, P(z,\omega)
\int_{_{\rm PBC}}\hskip-.4cm{ D}x\int_{_{\rm ABC}}\hskip-.4cm D\bar{\psi}D\psi\int_{_{\rm ABC}}\hskip-.4cm D\bar{\theta}D\theta \; \mathrm{e}^{-S_{\mathrm{gf}}[X,\tilde G;\,g_{\mu\nu}]}\ .
\end{align}
The gauge-fixed nonlinear sigma model action reads\footnote{To correctly define the path integral, one has to adopt a regularization scheme. In this chapter, we exploit the calculations of Chapter~\ref{chap:sixth}, where dimensional regularization on the worldline was adopted. This choice produces a counterterm $\mathrm{V}_{\rm CT}=\frac{1}{4} R$ for the case of four worldline supersymmetries \cite{Bastianelli:2011cc}. However, such a term is vanishing upon going on-shell \eqref{Ricci}. Let us mention further that, in order to perform calculations, it is necessary to include “metric ghosts” to keep translational invariance of the path integral measures and to renormalize potentially divergent worldline diagrams \cite{Bastianelli:2006rx}.}
\begin{align}
\begin{split} \label{action3}
S_{\mathrm{gf}}[X,\tilde G;\,g_{\mu\nu}]&=\int d\tau \Big[ \frac{1}{4T}g_{\mu\nu}(x)\,\dot{x}^{\mu}\dot{x}^{\nu}
+\bar{\psi}^{a i}\left(
	\delta_i^j \mathcal{D}_\tau +i \tilde a_{i}^{j}\right){\psi}_{aj}+\bar{\theta}^{i}\left(\delta_i^j \partial_\tau +i \tilde a_{i}^{j}\right){\theta}_j \\
 &\phantom{=\int d\tau \Big[}-T R_{abcd}(x) \, \bar{\psi}^{a}\cdot \psi^{b} \bar{\psi}^{c}\cdot \psi^{d}\Big]\ ,
 \end{split}
\end{align}
where we denoted the covariant derivative with spin connection $\omega_{\mu \,ab}(x)$ acting on the fermions by
\begin{equation}
\mathcal{D}_\tau \psi^a_i := \partial_\tau \psi^a_i + \dot x^\mu \omega_{\mu}{}^a{}_b \, \psi^b_i\ .
\end{equation}
At this point, the perturbative evaluation of the path integral has to be treated with care: in particular, one has to factorize out the zero modes and expand in Riemann normal coordinates, as carefully discussed in Chapter~\ref{chap:sixth} for the massless counterpart of Eq.~\eqref{chap:fifth:PATH}. We will avoid repeating the same considerations here, as the technicalities remain unchanged. The partition function results in 
\begin{equation} \label{chap:fifth:4}
Z(T)=\oint \frac{dz}{2\pi i}\oint \frac{d\omega}{2\pi i}\;
\mu (z,\omega) \;\int d^{D}x \, \frac{\sqrt{g(x)}}{\left(4\pi T\right)^{\frac{D}{2}}}\Big\langle \mathrm{e}^{-S_{\rm int}}\Big\rangle \ ,
\end{equation}
where the expectation value -- with normalization one, i.e. $\langle 1 \rangle = 1$, and propagators given in Appendix \ref{appendixB1} -- has to be evaluated using the Wick theorem on the free path integral, with the free action given by the quadratic part of \eqref{action3}, while higher
order terms form the interacting action $S_{\rm int}$, for which the expansion in powers of proper time $T$ has to be truncated to the desired order. In \eqref{chap:fifth:4} we kept track inside of $\mu(z,\omega)$ of all the modular factors; these are 
\begin{enumerate}[label=(\roman*)]
 \item the parabolic measure along with the $D$-dimensional Chern--Simons term, together giving rise to $P(z,\omega)$ \eqref{PD},
 \item the $\frac{z}{(z+1)^2}\frac{\omega}{(\omega+1)^2}$ factors corresponding to the gauging of worldline supersymmetries, 
 \item the poles $\frac{1}{z}\frac{1}{\omega}$ arising from the integral measures over the moduli,
 \item the $\tfrac{1}{4}$ factor previously factored out,
 \item finally, the path integrations over the fermionic coordinates $\int D\bar{\psi}D\psi\int D\bar{\theta}D\theta$ are responsible for the following extra factors:
\begin{align} \label{mu}
\mu (z,\omega):=\frac{1}{4}\frac{1}{(z+1)^2}\frac{1}{(\omega+1)^2}\, P(z,\omega)\; \frac{(z+1)^{D+1}}{z^{(D+1)/2}}\frac{(\omega+1)^{D+1}}{\omega^{(D+1)/2}}\ .
\end{align}
In particular, the factors $\frac{(z+1)^D}{z^{D/2}}\frac{(\omega+1)^D}{\omega^{D/2}}$ comes from the normalization of the fermionic path integral over the $\psi$s \cite{Bastianelli:2005vk}. The extra $\frac{z+1}{z^{1/2}}\frac{\omega+1}{\omega^{1/2}}$ instead is justified by the fact that the $\theta$s are free in \eqref{action3} and can be integrated over producing an additional determinant \cite{Bastianelli:2005uy}.
\end{enumerate}
The whole measure can be recast in the following form
\begin{equation}
\mu (z,\omega) =\frac{1}{2}\frac{(z+1)^{D-1}}{z^{3}}\frac{(\omega+1)^{D-1}}{\omega^{3}}(z-\omega)^{2}(z\omega-1)\ ,
\end{equation}
and corresponds to a shift $D \to D+1$ of its massless counterpart. To make explicit the Seeley-DeWitt coefficients arising from the perturbative expansion, one can introduce the double expectation value
\begin{equation}
\Big\langle \hskip -.1cm \Big\langle \mathrm{e}^{-S_{\rm int}}\Big\rangle \hskip -.1cm \Big\rangle
 = \oint \frac{dz}{2\pi i}\oint \frac{d\omega}{2\pi i}\; \mu(z,\omega) \Big\langle \mathrm{e}^{-S_{\rm int}}\Big\rangle\ .
\end{equation}
The Seeley-DeWitt coefficients $a_{n}(D)$ parameterize the divergences as $\Big\langle \hskip -.1cm \Big\langle \mathrm{e}^{-S_{\rm int}}\Big\rangle \hskip -.1cm \Big\rangle=\sum_{n=0}^{\infty}a_{n}(D)\, T^{n}$, therefore the partition function can be rewritten as follows
\begin{equation}
Z(T)=\int d^{D}x \, \frac{\sqrt{g(x)}}{\left(4\pi T\right)^{\frac{D}{2}}} \; \sum_{n=0}^{\infty}a_{n}(D)\, T^{n}\ .
\end{equation}

\subsection{One-loop divergences}
Let us start by checking the computation of the correct degrees of freedom of a massive graviton in $D$ spacetime dimensions. This is given by the double expectation value of the Seeley-DeWitt coefficient $a_{0}(D,z,\omega)=1$. Its projected partner
\begin{equation}
a_{0}(D) = \langle \hskip -.05cm \langle 1 \rangle \hskip -.05cm \rangle = \oint \frac{dz}{2\pi i}\oint \frac{d\omega}{2\pi i}\; \mu (z,\omega) = \frac{(D+1)(D-2)}{2} \overset{D=4}{=} 5
\end{equation}
gives indeed the massive graviton's physical polarizations. This confirms that the measure $\mu (z,\omega)$ correctly projects only onto the degrees of freedom of a massless graviton $(h_{\mu\nu})$ plus the St\"uckelberg fields $(A_\mu,\varphi)$ introduced in order to restore gauge invariance.

Let us comment on the fact that if we were to insert inside \eqref{mu} the measure $P_{(3)}(z,\omega)$ without the improved $D$-dimensional Chern--Simons \eqref{P}, we would get the correct degrees of freedom only in three spacetime dimensions while failing to go beyond that. 

The computation of the higher-order Seeley-DeWitt coefficients proceeds exactly as in the massless case since the $\theta$s have been integrated away. The only care one should take is to use the improved measure \eqref{mu}. With these prescriptions and with the worldline diagrams of Chapter~\ref{chap:sixth}, it is possible to evaluate the counterterms up to the third order, i.e.
\begin{align} \label{chap:fifth:series}
Z(T)=\int \frac{d^{D}x}{\left(4\pi T\right)^{\frac{D}{2}}}\sqrt{g(x)} \; \left[a_0(D)
+ a_1(D) \,T + a_2 (D)\, T^2 + a_3 (D)\, T^3+ \mathcal{O}(T^{4})\right]\ .
\end{align}
The results on Ricci-flat spaces are reported as follows. The exponentiation of all the connected diagrams and the subsequent Taylor expansion to the desired order yields
\begin{align}
\Big\langle \mathrm{e}^{-S_{\rm int}}\Big\rangle &=1+T^2\; \alpha_{2} \, R_{\mu\nu\rho\sigma}^{2} +T^3 \left( \beta_{3}\, R_{\mu\nu\rho\sigma}R^{\rho\sigma\alpha\beta}R_{\alpha\beta}{}^{\mu\nu} +\gamma_{3}\,R_{\alpha\mu\nu\beta}R^{\mu\rho\sigma\nu}R_{\rho}{}^{\alpha\beta}{}_{\sigma} \right) + \mathcal{O}(T^{4})\ .
\end{align}
where the curvature invariants of order $k$ in the Riemann tensor are multiplied by the coefficients $\alpha_{k}(z,\omega, D)$, $\beta_{k}(z,\omega, D)$ and $\gamma_{k}(z,\omega, D)$, explicitly given by
\begin{align}
\alpha_{2}&=\frac{1}{180}+\frac{1}{2} \left(\frac{\omega ^2}{(\omega +1)^4}+\frac{z^2}{(z+1)^4}+\frac{4 \omega z}{(\omega +1)^2 (z+1)^2}\right)-\frac{1}{12} \left(\frac{\omega }{(\omega+1)^2}+\frac{z}{(z+1)^2}\right)\ , \\[3mm]
\begin{split}
\beta_{3} &=\frac{17}{45360} -\frac{z}{180 (z+1)^2}-\frac{\omega }{180 (\omega +1)^2} +\frac{(z-1)^2 z^2}{6(z+1)^6} +\frac{(\omega -1)^2 \omega ^2}{6 (\omega +1)^6} \\[1.5mm]
&\phantom{=}-\frac{2 \omega z^2}{(\omega +1)^2 (z+1)^4}-\frac{2 \omega ^2 z}{(\omega +1)^4(z+1)^2}\ ,
\end{split} \\[3mm]
\gamma_{3}&=\frac{1}{1620}-\frac{z}{90(z+1)^2}-\frac{\omega }{90 (\omega +1)^2}+ \frac{z^2}{3 \left( z+1 \right)^4} + \frac{\omega^2}{3 \left( \omega+1 \right)^4} + -\frac{4 \omega ^3}{3 (\omega +1)^6}-\frac{4 z^3}{3 (z+1)^6}\ .
\end{align}
One can immediately recognize the unprojected Seeley–DeWitt coefficients multiplying the respective powers of proper time. The final step consists of performing the modular integrals. The calculation of the various terms in the perturbative expansion delivers the following coefficients in the perturbative series \eqref{series}
\begin{align}
a_{0}(D)&=\frac{(D+1)(D-2)}{2}\ , \label{chap:fifth:a0} \\[.3em]
a_{2}(D)&=\frac{D^2-31 D+508}{360} \; R_{\mu\nu\rho\sigma}^{2}\ , \label{chap:fifth:a2} \\[.3em]
a_{3}(D)&=\frac{17 D^2-521 D-15658}{90720} \;R_{\mu\nu\rho\sigma}R^{\rho\sigma\alpha\beta}R_{\alpha\beta}{}^{\mu\nu}+\frac{D^2-37 D-1118}{3240}\; R_{\alpha\mu\nu\beta}R^{\mu\rho\sigma\nu}R_{\rho}{}^{\alpha\beta}{}_{\sigma}\ . \label{chap:fifth:a3}
\end{align}
The coefficients \eqref{chap:fifth:a0}--\eqref{chap:fifth:a3}, including the newly computed coefficient $a_3(D)$ on Ricci-flat spaces, allow for further investigations of the issue of divergences in the quantum theory of massive gravity. The type of
divergences arising emerge naturally from the representation of the one-loop effective action with a short proper time expansion; from \eqref{chap:fifth:series} we have
\begin{equation}
\Gamma[g_{\mu\nu}]= -\frac{1}{2}\int_{0}^{\infty}\frac{dT}{T^{1+\frac{D}{2}}} \, \mathrm{e}^{-m^2T}\int \frac{d^{D}x}{\left(4\pi \right)^{\frac{D}{2}}}\sqrt{g(x)}\left[a_0 + a_1 T + a_2 T^2 + a_3 T^3+ \mathcal{O}(T^{4})\right]\ . \label{chap:fifth:divergences}
\end{equation} 
While the \acr{IR} divergences are absent due to the presence of the mass, playing the role of a regulator, the \acr{UV} divergences\footnote{To relate the $\frac{1}{\epsilon}$ pole of dimensional regularization in \acr{QFT} with our result, one has to evaluate the proper time integral term by term in \eqref{chap:fifth:divergences}, to display the gamma function dependence.} arise from the $T\to 0$ limit of the proper time integration.
In four spacetime dimensions, the different powers of T give rise to the quartic, quadratic, and logarithmic divergences parametrized by $a_0, a_1,a_2$, respectively. In \acr{QFT} dimensional
regularization only the logarithmic divergences are visible. From \eqref{chap:fifth:a2} we have
\begin{equation}
\left. a_{2} \right|_{D=4}=\frac{10}{9} \; R_{\mu\nu\rho\sigma}^{2}\ . \label{a2D=4}
\end{equation}
The latter numerical value for the one-loop four-dimensional logarithmic divergence of massive gravity coincides precisely with that calculated in Ref.~\cite{Dilkes:2001av}. In four dimensions $R_{\mu\nu\rho\sigma}^{2}$ is a total derivative and can be eliminated from the effective action. Thus, one may conclude that the one-loop logarithmic divergences of massive gravity without cosmological constant vanish. More generally, the $a_2$ coefficient in $D$ dimensions has been recently evaluated in Ref.~\cite{Ferrero:2023xsf} and is correctly reproduced by our result \eqref{chap:fifth:a2}. Finally, it is worth noting that the coefficient $a_3$ gives rise to a finite term in the four-dimensional effective action.

In $D=6$, the coefficient $a_3$ provides an additional divergence, the logarithmic one in that dimension
\begin{equation}
\left. a_{3} \right|_{D=6} =
-\frac{649}{3240} \;R_{\mu\nu\rho\sigma}R^{\rho\sigma\alpha\beta}R_{\alpha\beta}{}^{\mu\nu}-\frac{163}{405}\; R_{\alpha\mu\nu\beta}R^{\mu\rho\sigma\nu}R_{\rho}{}^{\alpha\beta}{}_{\sigma}\ .
\end{equation}
This coefficient should be obtained by any other method of calculation with the same value. While the two curvature invariants in the previous expression are generally independent of each other, in six dimensions there exists an integral relation that connects them \cite{vanNieuwenhuizen:1976vb} and the result can be expressed as follows
\begin{align}
\left. a_{3} \right|_{D=6} = 
\frac{1}{1080} \;R_{\mu\nu\rho\sigma}R^{\rho\sigma\alpha\beta}R_{\alpha\beta}{}^{\mu\nu}\ ,
\end{align}
encoding the one-loop logarithmic divergences of massive gravity in six dimensions. Let us also mention that a more recent computation based on heat kernel techniques has provided the complete set of coefficients \eqref{chap:fifth:a0}--\eqref{chap:fifth:a3} in Einstein spaces, thereby overcoming the previous restriction to Ricci-flat backgrounds \cite{Farolfi:2025knq}. 
Interestingly, the results obtained there are in good agreement with ours, even beyond the expected regime of validity of the massive $\mathcal{N}=4$ spinning particle model. Specifically, we verified that terms proportional to the Ricci scalar, which are not reported here and vanish upon projection to a Ricci-flat background, completely agree with Eq.~(4.9) and Eq.~(4.10) of \cite{Farolfi:2025knq}. The same holds for the $R\,R_{\mu\nu\rho\sigma}^{2}$ coefficient in their Eq.~(4.11), while the only exception is represented by a numerical prefactor in front of the $R^3$ invariant. In conclusion, the difference between the two results amounts to
\begin{equation}
    \Delta a_3(D):=a^{\N=4}_3(D)-a^{\text{\cite{Farolfi:2025knq}}}_3(D)=\frac{(D+12)^3 }{432 D^3}R^3\ ,
\end{equation}
where $a^{\N=4}_3(D)$ denotes the coefficient \eqref{chap:fifth:a3} calculated on Einstein spaces with non-vanishing Ricci tensor from the massive $\N=4$ model. This suggests the interesting possibility that a slight modification of the model may lead to matching those results.

\section{Final remarks} \label{chap:fifth:sec4}
In this chapter, we have realized the worldline path integral on the circle of the massive $\N=4$ spinning particle, providing a model capable of describing a massive graviton propagating on Ricci-flat spacetimes of arbitrary dimensions. The key point in the derivation has been to realize that the gauging of a parabolic subgroup of the SO(4) R-symmetry group, together with a suitable Chern--Simons coupling, which worked for the massless model, is able to reproduce the correct result despite the mass improvement. The analysis indicates that the degrees of freedom extracted by the path integral undergo a “St\"uckelbergization”, leading to immediate identification of the massless graviton together with the associated St\"uckelberg fields, a vector and a scalar, within the spectrum of the particle model. We then applied the model to furnish a worldline representation of the effective action of massive gravity, reproducing the divergences of one-loop linearized massive gravity with vanishing cosmological constant in arbitrary dimensions. We computed the counterterms on-shell, and we believe that they could serve as a benchmark for verifying alternative approaches to massive gravity. We checked the correct reproduction of the heat kernel coefficients $a_n(D)$ for $n=0,1,2$ comparing our results with those present in the literature. Finally, our main contribution was the determination of the Seeley-DeWitt coefficient $a_3(D)$, which to our knowledge has never been computed in the literature. 

There are several promising directions for future exploration. One avenue involves relaxing the R-symmetry constraints to allow for the propagation in the loop of the $\N=0$ supergravity, i.e. not only the graviton but also the dilaton and the Kalb--Ramond field, provided that the associated BRST system is first correctly realized, following the methodology outlined in Ref.~\cite{Bonezzi:2020jjq}. Another avenue of interest lies in extending these constructions to more exotic spaces, such as non-commutative spaces, thereby realizing a covariant path integral for (spinning) particle models \cite{Bonezzi:2012vr, Franchino-Vinas:2018qyk}, or complex spaces, by employing the worldline models known as U$(\N)$ spinning particles \cite{Bastianelli:2009vj, Bastianelli:2012nh}, which may offer a fruitful first-quantized description of gravitational theories on K\"ahler manifolds once coupled to a curved background metric.

\newpage
\thispagestyle{empty}
\mbox{}
\newpage



\backmatter

\appendix
\include{chapters/appendixA}
\include{chapters/appendixB}

\addcontentsline{toc}{part}{\hyperref[bib]{Bibliography}}
\printbibliography\label{bib}

\newpage
\thispagestyle{empty}
\mbox{}
\newpage
\thispagestyle{empty}
\vspace*{4cm} 
\begin{center}
    {\hypersetup{hidelinks}\href{https://en.wikipedia.org/wiki/Intentionally_blank_page}{\textsc{This page is intentionally left blank.}}}
\end{center}

\end{document}